%% file: Thesis_Julien_de_Wit.tex
\documentclass[12pt,twoside]{mitthesis}
\usepackage{lgrind}
\usepackage{textcomp}

\include{package_and_setup}

\begin{document}

\include{Cover/cover}

\pagestyle{plain}

	\vspace{-2cm}\tableofcontents
	\newpage
	\listoffigures
	\newpage
	\listoftables

	\include{Chapter1/chap1}
	\include{Chapter2/chap2}
	\include{Chapter3/chap3}

	\include{Chapter4/chap4}

	\include{Chapter5/chap5}

	
\appendix
		\include{AppendixA/appa}
		\include{AppendixB/appb}
		\include{AppendixC/appc}
		\include{AppendixD/appd}


	\begin{singlespace}
	\bibliography{Biblio/Biblio}
	\bibliographystyle{alphanum}
	\end{singlespace}

\end{document}

%% file: package_and_setup.tex
\usepackage{amsmath,amssymb}
\usepackage{graphicx}
\usepackage{aecompl}
\graphicspath{{.}{images/}}
\usepackage{dblfloatfix}
\usepackage{array}
\usepackage{multirow}
\usepackage{afterpage}
\usepackage{natbib}
\usepackage{subfig}
\usepackage{movie15}
\usepackage{wrapfig}
\usepackage{minipage-marginpar}
\usepackage[stable]{footmisc}

\usepackage{times}

\usepackage{hyperref}  
\usepackage{epstopdf}
\usepackage{color}
\definecolor{linkcol}{rgb}{0,0,0.4} 
\definecolor{citecol}{rgb}{0.5,0,0} 
\hypersetup
{
pdftitle="Maps and Masses of Transiting Exoplanets",
pdfauthor="Julien de WIT", 
pdfsubject="PhD Thesis", 
pdfmenubar=true, 
pdfhighlight=/O, 
pdfpagemode=None, 
pdfpagelayout=SinglePage, 
pdffitwindow=true, 
linkcolor=black,
}

%% file: Cover/cover.tex
\title{Maps and Masses of Transiting Exoplanets:\\ \hspace{-0.7cm}{\large{Towards New Insights into Atmospheric and Interior Properties of Planets}\hspace{-0.7cm}}}

\author{Julien de Wit}
      \prevdegrees{B.Sc. Eng., Universit\'{e} de Li\`{e}ge (2008) \\
                   M.Res., Institut Sup\'{e}rieur de l\textquotesingle A\'{e}ronautique et de l\textquotesingle Espace (2010) \\ 
                   M.Eng., Institut Sup\'{e}rieur de l\textquotesingle A\'{e}ronautique et de l\textquotesingle Espace  (2011)\\ 
                   M.Sc. Eng., Universit\'{e} de Li\`{e}ge (2011)}
\department{Department of Earth, Atmospheric and Planetary Sciences}

\degree{Doctor of Philosophy}

\degreemonth{September}
\degreeyear{2014}
\thesisdate{July 31, 2014}


\supervisor{Sara Seager}{Professor of Planetary Sciences\\Professor of Physics\\Class of 1941 Professor}

\chairman{Robert D. Van der Hilst}{Schlumberger Professor of Earth Sciences \\ Head, Department of Earth, Atmospheric and Planetary Sciences}

\maketitle



\cleardoublepage
\setcounter{savepage}{\thepage}
\begin{abstractpage}
\input{Abstract/abstract}
\end{abstractpage}


\cleardoublepage

\section*{Acknowledgments}

MIT is a very special place, probably the most unique and vivid place of Cambridge, MA. My time at MIT has been a unique \textit{life} experience: challenging, puzzling, inspiring, rewarding, and particularly formative. This experience would have been void without the presence and support of so many people. Among those, Catherine Adans-Dester. Her support, kindness, patience, dedication, and love have been at the root of this venture and its successful outcome. The first expression of my gratitude---deep heartfelt---goes to her.

I want to thank my PhD advisor, Sara Seager, for introducing me into her fascinating world. Coming to MIT and embracing the field of exoplanetology changed the course of my engineer's life, as well as my expectations of it---hundreds of pale blue dots appear now closer than ever. I am also thankful for receiving the research freedom needed to suit my curiosity and for the opportunity to be surrounded with amazing colleagues.

Thank you to these amazing colleagues that I am lucky to call friends. Countless thanks to Andras Zsom, Vlada Stamenkovi{\'c}, Nikole Lewis, Brice-Olivier Demory, and Stephen Messenger for enthusiastic collaborations, insightful discussions, and wise advice. Thank you also to Mary Knapp, William Bains, Renyu Hu, Leslie Rogers, and Bjoern Benneke.

I am very grateful to the numerous scientists who shared their personal and/or scientific insights along the way. Many thanks to Daniel Cziczo, Heather Knutson, Hilke Schlichting, and Benjamin Weiss for sharing your expertise and serving on my committee. A particular acknowledgment goes to Micha\"{e}l Gillon for introducing me to observations with the \textit{Spitzer Space Telescope}, exoplanetary datasets, and to HD\,189733b (what a ``Rosetta Stone''!). Thanks to Eric Agol, Adrian Belu, Zachory Berta-Thompson, Andrew Collier Cameron, Giuseppe Cataldo, Bryce Croll, Ian Crossfield, Pierre Ferruit, Kevin Heng, Nikku Madhusudhan, and Amaury Triaud. Thank you to the anonymous reviewers of the papers related to this thesis for their constructive comments. Thank you to the teams behind \textit{MMIRS} and the \textit{Magellan Telescopes} for the fantastic experiences I had at Las Campanas Observatory (thank you Brice for the opportunity!).

I would also like to express my gratitude towards the institutions that supported me financially over the past years. Their support ensured my research freedom. I am grateful to \textit{Wallonie-Bruxelles International} (W.B.I.), the \textit{Belgian American Educational Foundation} (B.A.E.F.), and the \textit{Grayce B. Kerr Fund} for their support in the form of fellowships. I am also thankful to the \textit{Belgian Senate} for its support in the form of the Odissea Prize. I also owe a debt of gratitude to the \textit{Duesberg-Baily Thil Lorrain Foundation} for its support during my first visit at MIT in 2010, where I had the chance to be part of the Seager Exoplanet Theory and Computation Group and to conceive \textit{MassSpec}'s concept. 

Many thanks to all the people who have been making my life of researcher easier on a daily basis. Many thanks to Geri-Lyn Bowen, Derrick Duplessy, Robyn Jepson, Vicki McKenna, Carol Sprague, Jacqueline Taylor, Linda Meinke, and Scott Blomquist. Thank you to Monique Jacquemin for her continuous support and unrestricted kindness. Thank you to the artists who have supported my creativity and enthusiasm over days and nights---Coldplay, Michael Bubl{\'e}, Daft Punk, Pharrel Williams, U2, OneRepublic, Bob Marley, The Black Eyed Peas, Francis Cabrel, Jamiroquai, Moby, Ozark Henry,\dots

Finally, I want to thank my/our family and friends for their unwavering love and support. From Verviers, Li\`{e}ge, Toulouse, and Boston, face-to-face or via Skype, you have been sources of motivation, joy, inspiration, and reflection. You have given more color to my life, sharpened my mind, broadened my perspectives, sustained my curiosity, and refined my dreams. Thank you for being part of my life. 

\begin{center}
\rule{3cm}{0.4pt}
\end{center}

Part of this thesis is based on observations made with the \textit{Spitzer Space Telescope}, which is operated by the Jet Propulsion Laboratory, California Institute of Technology, under contract to NASA. Thank you to all the people who made and keep observatories like \textit{Spitzer} such wonderful scientific adventures.

Part of this thesis is based on intensive uses of \textit{MATLAB} and \textit{Mathematica}. Thanks to \textit{MathWork} and \textit{Wolfram Research} for the exquisite software they provide the (MIT) community.


%% file: Abstract/abstract.tex
%
%
%

With over 1800 planets discovered outside of the Solar System in the past two decades, the field of exoplanetology has broadened our perspective on planetary systems. Research priorities are now moving from planet detection to planet characterization. In this context, transiting exoplanets---planets that cross in front of their star from our point-of-view---are of special interest due to the wealth of data made available by their orbital configuration. In this thesis, I introduce two methods, and their Markov chain Monte Carlo implementations, to gain new insights into the atmospheric and interior properties of exoplanets.

The first method aims to map an exoplanet's atmosphere based on the eclipse scanning---which is obtained while a planet is occulted by its host star. Ultimately temperature, composition, and circulation patterns could be constrained in three-dimensions from these maps, a significant asset for informing atmospheric models. I introduce the basics of eclipse mapping, its caveats (particularly, the correlation between the planet's shape, brightness distribution, and four system parameters), and a framework to mitigate the caveats' effects via global analyses including transits, phase curves, and radial velocity measurements. I use this method to create the first two-dimensional map and the first cloud map of an exoplanet for the hot-Jupiters HD\,189733b and Kepler-7b, respectively.

The second method, \textit{MassSpec}, aims to determine transiting planet masses and atmospheric properties solely from transmission spectra, i.e. the starlight filtered by a planet's atmosphere during transits. Determination of an exoplanet's mass is key to understanding its basic properties, including its potential for supporting life. To date, mass constraints for exoplanets are predominantly based on radial velocity measurements, which are not suited for planets with low masses, large semi-major axes, or those orbiting faint or active stars. I demonstrate that a planet's mass has to be accounted for by atmospheric retrieval methods to ensure unbiased estimates of atmospheric properties. Utilizing \textit{MassSpec}, the \textit{James Webb Space Telescope} (launch date: 2018) could determine the mass and atmospheric properties of  half a dozen Earth-sized planets in their host's habitable zones \textit{over its lifetime}, which could lead to the first identification of a habitable exoplanet.


%% file: Chapter1/chap1.tex
\chapter{Introduction}

\vspace{-0.7cm}

Planets outside of the Solar System, known as exoplanets, have been an object of investigation for more than two centries \citep[e.g.,][]{Jacob1855}. Yet, the first exoplanet around a main-sequence star was detected less than two decades ago \citep{Mayor1995}. Since then, the field of exoplanetology has significantly and irreversibly broadened our perspective on planetary systems with more than 1800 planets found in over 1100 systems different from ours\footnote{For an up-to-date list, refer to the Extrasolar Planets Encyclopaedia \citep[\url{http://exoplanet.eu},][]{Schneider2011} and the Exoplanet Data Explorer \citep[\url{http://exoplanets.org/},][]{Wright2011}. 
}. We can now place observational constraints on the planet population in our galaxy\footnote{At least 100 billion planets orbit stars in the Milky Way \citep{Cassan2012}. As many as 40$\%$ of those planets are Earth-sized planets orbiting in the habitable zones of Sun-like stars and stars with later spectral type \citep{Petigura2013,Swift2013}.} and contemplate the results of our search for habitable worlds\footnote{ Discovery publications include  \cite{Vogt2010,Wordsworth2011,Borucki2012,Anglada2013,Tuomi2013a,Tuomi2013b,Barclay2013,Borucki2013}.}, which recently reached a climax with the first Earth-sized exoplanet found in a habitable zone\footnote{The habitable zone is the region around a star within which planetary-mass objects with sufficient atmospheric pressure can support liquid water at their surfaces.} \citep{Quintana2014}. A significant amount of theoretical, observational, and engineering efforts is still required to characterize distant worlds, understand their climates, assess their habitability, and, hence, identify habitable worlds. While the next generation of observatories is on its way, it is not clear yet \textit{what are the next scientific achievements} future observatories will bring within our reach and, most importantly, \textit{how to reach them}.

\section{First Steps towards Worlds beyond the Solar System}

\subsection{Detections}

\begin{figure}
   \centering

  \begin{center}
    \vspace{-0.4cm}\hspace{-0.cm}\includegraphics[trim = 20mm 00mm 20mm 00mm,clip,width=15cm,height=!]{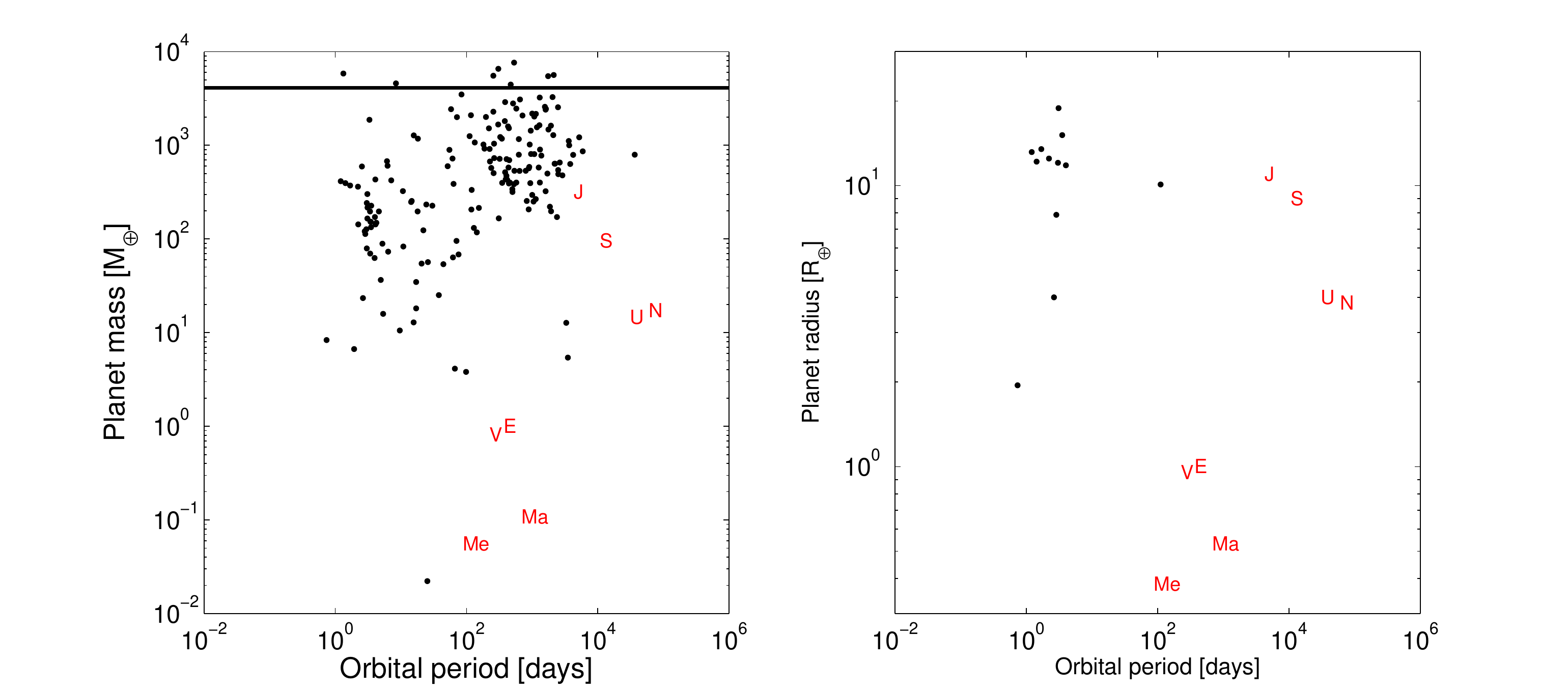}
  \end{center}
  \vspace{-0.5cm}
  \caption[Planets detected as of 2005.]{Planets detected as of 2005. The planets of the Solar System are shown in red and the exoplanets as black dots. The black line on the period-mass plot represents the brown-dwarf transition, the mass around which a planet-sized body can fuse deuterium.}
  \label{fig:planets_detected_by_2005}
    \end{figure}

Planet detections rely largely on indirect detection methods because most exoplanets cannot be directly detected with current observatories. Exoplanets are too faint and/or too close to their stars to be imaged. Over the first decade following \cite{Mayor1995}'s discovery, most of the exoplanets found were hot Jupiters\footnote{Hot Jupiters are Jupiter-sized planets heated by their close stars---within 0.1 astronomical unit (AU).} \citep[see Figure\,\ref{fig:planets_detected_by_2005} and e.g.,][and references therein]{Marcy2005}. The large mass and the proximity of hot Jupiters to their hosts favored their indirect detection by the radial velocity (RV) method\footnote{The leading initiatives for planet discovery using RV are the \textit{Geneva Extrasolar Planet Search} (\textit{ELODIE}, \textit{CORALIE}, \textit{HARPS}, and \textit{HARPS-N} spectrographs) and the \textit{Lick-Carnegie Exoplanet Survey} (\textit{HIRES} spectrometer). The first initiative led to the first planet discovery around a main sequence star, the second contributed over 70$\%$ of the discoveries performed before 2010.}, which searches for the Doppler shift of a star's spectrum as it orbits the planet-star common center of mass. The RV method contributed $\sim30\%$ of the planet discoveries to date. The most prolific planet discovery technique is the transit method\footnote{Leading initiatives for planet discovery using transits include \textit{HATNet Project}, \textit{CoRoT}, and \textit{WASP}.\vspace{-0.5cm}} ($\sim60\%$ of the planets discovered to date), which searches for the brightness drop of a star as a planet transits, i.e., passes in front of it. One mission, the \textit{Kepler} mission, has changed the planet-detection game as it detected 961 confirmed exoplanets and more than 2900 planet candidates \citep{Borucki2010,Borucki2011a,Borucki2011,Batalha2013,Burke2014} via monitoring over 145,000 main-sequence stars for over 3 years.

The transit and the RV methods are complementary in two ways: (1) in terms of detection biases and (2) for characterization purposes. (1) The sensitivity of the transit method depends mainly on the planet-to-star area ratio, while the sensitivity of the RV method depends on the planet-to-star mass ratio and distance. In other words, the detection of smaller planets is easier with the transit method. (2) Transit and RV measurements constrain the planet-to-star area and (minimum) mass ratios, respectively \citep[e.g.,][]{Winn2010}. Hence, the observation of a planet with both techniques enables the first step towards its characterization via its bulk density. A planet's density can yield its bulk composition \citep{Seager2007,Fortney2007}, whether it is a gas giant or a rocky exoplanet potentially suitable for life, as we know it.

   \begin{figure}[!h]
   \centering

  \begin{center}
    \vspace{-0.4cm}\hspace{-0.cm}\includegraphics[trim = 20mm 00mm 20mm 00mm,clip,width=15cm,height=!]{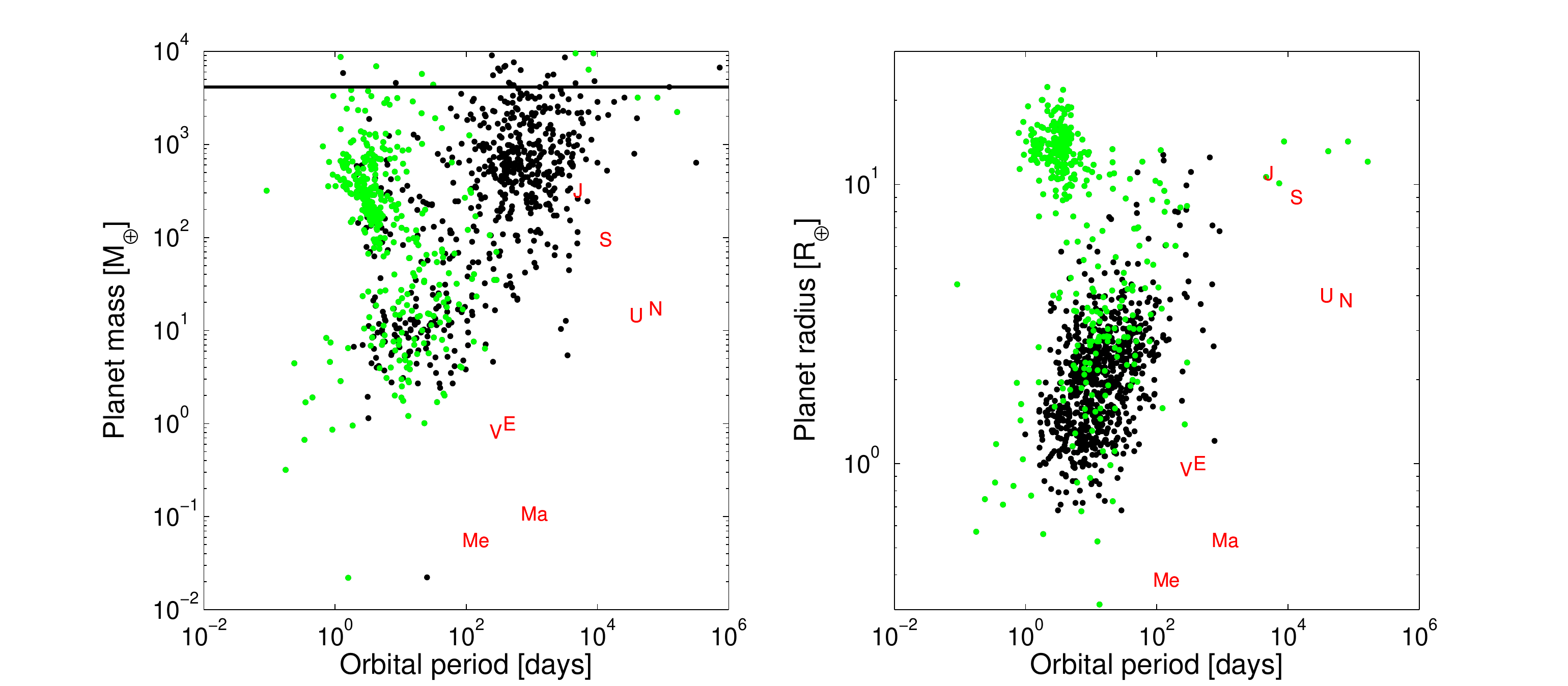}
  \end{center}
  \vspace{-0.5cm}
  \caption[Planets with constrained bulk density as of May 2014.]{Planets with constrained bulk density as of May 2014. The exoplanets with constrained bulk density are shown as green dots and the other exoplanets known as black dots. The panels show the same data type as Figure\,\ref{fig:planets_detected_by_2005}.}
  \label{fig:planets_with_density}
    \end{figure}


\subsection{Characterization}
\label{sec:charact}

The community determined the bulk composition of $\sim 20\%$ of the exoplanets known (Figure\,\ref{fig:planets_with_density}). Simultaneously, complementary techniques were developed to further the characterization of exoplanets. As an example, one of these techniques lead the detection of atmospheric components---atomic and/or molecular---for about half a percent of the detected exoplanets (i.e., 10). Now, how can we learn so much about a planet when its whole planetary system appears to us as a single bright point in the sky?

\begin{figure}[!htbp]
  \begin{center}
    \includegraphics[trim = 20mm 00mm 20mm 00mm,clip,width=15cm,height=!]{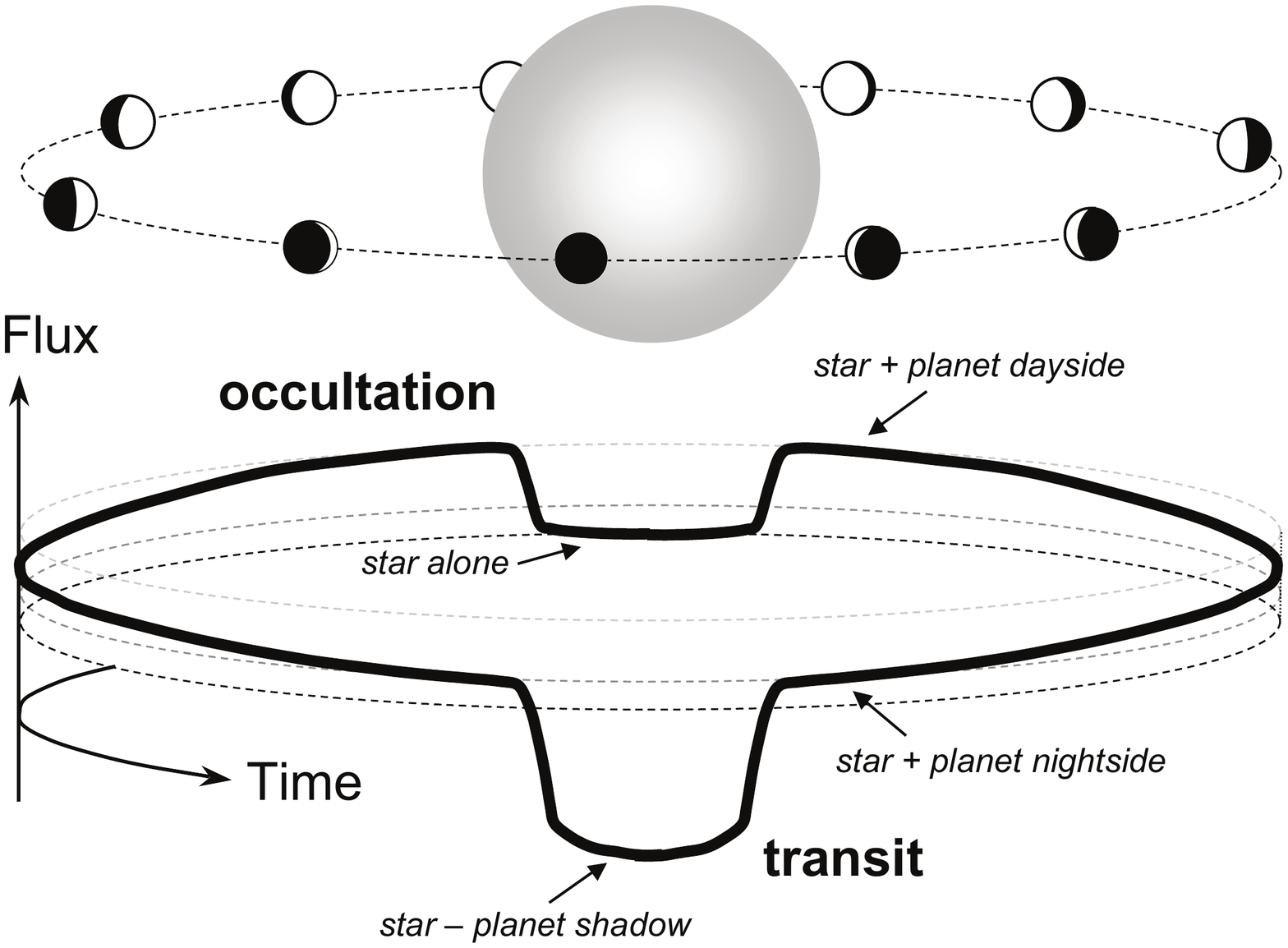}
  \end{center}
  \vspace{-0.75cm}
  \caption[Basics about transiting exoplanets.]{Basics about transiting exoplanets. During a transit, the flux observed drops because the planet blocks a fraction of the stellar disk---hence, the flux drop relates to the stellar brightness and the planet-to-star area ratio. Then the flux rises as the planet’s dayside comes into view---most exoplanets detected so far always face their host with the same hemisphere (dayside), they are tidally-locked as the Moon is to Earth. During an occultation, the flux drops relates to the emerging planetary flux. Then the flux drops as the planet’s nightside comes into view. Figure from \cite{Winn2010}.}
  \label{fig:transitoccultation}
\end{figure}

\textbf{Transiting exoplanets} are, among the exoplanets detected to date, the ones that can be most extensively characterized with current technology\footnote{\textit{Non-transiting} exoplanets can also be characterized, but to another degree. In particular, the cross-correlation technique successfully implemented by \cite{Snellen2010} allows for the detection of atmospheric components and the measurement of atmospheric winds and orbital inclination, which yield the planet mass when combined with RV---see, e.g., applications to the hot Jupiters $\tau$\,Bo{\"o}tis b \citep{Brogi2012,Rodler2012}, HD\,179949 b \citep{Brogi2014}, and the young {$\beta$} Pictoris b \citep{Snellen2014}.}. 
Beyond derivation of the exoplanet orbital inclination and density \citep[e.g.,][]{Winn2010}, transiting exoplanets are key objects 
because their atmospheres are observationally accessible through transit transmission and occultation emission spectrophotometry \citep[see e.g.,][and references therein]{Seager2010a} and phase curve measurements (Figure\,\ref{fig:transitoccultation}).

\begin{wrapfigure}{r}{0.4\textwidth}
   \begin{center}
   \begin{minipage}[c]{0.4\textwidth}
    \vspace{-1.0cm}\includegraphics[width=7.cm,height=!]{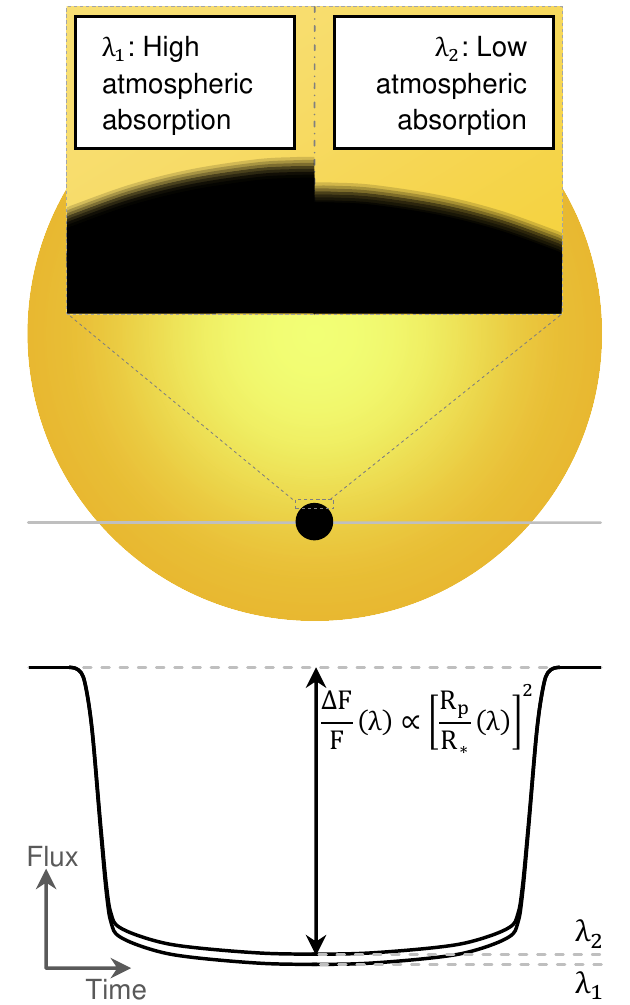}  
  \end{minipage}\hfill
  \end{center}
  \vspace{+0.5cm}
  \hspace{-0.6\textwidth}\begin{minipage}[c]{\textwidth}
  \caption[Basics of transmission spectroscopy.]{Basics of transmission spectroscopy. Transit-depth variations, $ \frac{\Delta F}{F} (\lambda)$, induced by the wavelength-dependent opacity of a transiting planet's atmosphere. The stellar disk and the planet are not resolved; the flux variation of a point source is observed.}
  \label{fig:transmission_spectrum_intro}
  \end{minipage}
\end{wrapfigure}

 \vspace{+0.0cm} \textbf{Transmission Spectroscopy} 
 studies the wavelength-dependent flux drop during a planetary transit \citep{Seager2000}. While a planet transits, part of its star's light is going through its atmosphere (Figure\,\ref{fig:transmission_spectrum_intro}). Hence, this light can show absorption features from the atmospheric components, which affect the atmosphere's opacity and its wavelength dependence. At a wavelength with high atmospheric absorption, $\lambda_1$, the transit depth is slightly deeper than at a wavelength with lower atmospheric absorption, $\lambda_2$---due to the more extended opaque atmospheric annulus. Also, the planet appears larger as a relative flux-drop, $ \frac{\Delta F}{F} (\lambda)$, is associated with an effective planetary radius, $R_p(\lambda) = R_{\star}\sqrt{\frac{\Delta F}{F} (\lambda)}$.
\newline
\newline
\newline
\newline

  \vspace{-0.0cm}

\newpage
 \textbf{Emission Spectroscopy}
  studies the wavelength-dependent flux drop during a planetary occultation. Observations of the occultation depth at different wavelengths constrain the hemisphere-integrated emission spectrum of an exoplanet. The emission spectrum of an exoplanet constrains its brightness temperature, but also its atmospheric composition. As atmospheric components affect the wavelength dependence of the atmospheric optical depth\footnote{The optical depth expresses the quantity of light removed from a radiation beam by extinction as it passes through a medium. In particular, the ratio of incoming intensity to emerging intensity is $e^{-\tau(\lambda)}$, where $\tau$ is the optical depth.}, different wavelengths probe different atmospheric layers with different brightness temperatures. In other words, the emission spectrum of an exoplanet translates primarily into constraints on the temperature and composition structure of its atmosphere. 

 \textbf{Phase curve measurements} relate to the observations of a planetary system's flux variations with the planetary phase, which provide information about the planets' brightness distribution. For most exoplanets detected, the rotation rate is synchronized with the orbital period\footnote{Close-in rocky planets can be locked in configurations different from 1:1 \citep[e.g.,][]{Makarov2012,Makarov2014}.} because close-in exoplanets are easier to detect and subject to strong tidal friction. Hence those planets always show the same face to their stars (dayside), similarly to the Moon with the Earth. As a result, along their orbits different planetary hemispheres are visible, which modulate the planetary flux contribution (Figure\,\ref{fig:transitoccultation}). Hence, monitoring the system flux modulation over an orbital period for synchronized planets can be translated into constraints on their longitudinal brightness distribution.

\subsubsection*{Overview of Exoplanet Characterization Achievements}

The \textit{Hubble Space Telescope} (\textit{HST}) and the \textit{Spitzer Space Telescope} \citep[\textit{Spitzer},][]{Werner2004}---\textit{HST}'s cousin in the infrared---have played major roles in furthering the degree to which exoplanets can be characterized. \textit{HST} yielded the first detection of an exoplanet atmosphere by revealing its sodium absorption feature in transmission \citep{Charbonneau2002}. \textit{Spitzer} yielded the first (broadband) emission spectrum of an exoplanet \citep{Deming2005,Charbonneau2005}, which indicated a temperature inversion in the target's atmosphere \citep{Knutson2008}. \textit{Spitzer} also led to the first map of an exoplanet based on its phase curve measurement \citep{Knutson2007}. \textit{HST}'s observations revealed for the first time the atmospheric escape from an exoplanet \citep{LecavelierDesEtangs2010} and the interaction between an exoplanet's atmosphere and stellar activity \citep{Lecavelier2012}. Recently, the Wide Field Camera 3 (WFC3) on \textit{HST} provided the most exquisite exoplanetary transmission spectra between 1 and 1.7$\mu$m revealing the 1.4$\mu$m absorption feature of water for a few hot Jupiters \citep{Deming2013,Huitson2013,Mandell2013} and a flat signal within the measurement uncertainties for others \citep{Berta2012,Knutson2014,Ranjan2014,Kreidberg2014}.

\section{Context}

By the time this thesis started, the quality of exoplanetary datasets was in rare cases sufficient to yield longitudinal maps of exoplanets and detections of atmospheric constituents using transmission spectroscopy. The perspectives of new observatories\footnote{Including the \textit{James Webb Space Telescope} \citep[\textit{JWST}; launch date 2018,][]{Clampin2010}, the \textit{Exoplanet Characterisation Observatory} (\textit{EChO})\footnotemark
, the \textit{European Extremely Large Telescope} (\textit{E-ELT}), the \textit{Thirty Meter Telescope} (\textit{TMT}), and the \textit{Giant Magellan Telescope} (\textit{GMT}) .}\footnotetext{\textit{EChO} was a M3 mission candidate of the European Space Agency \citep{Tinetti2012}. While \textit{EChO} was not funded, such a mission could still revolutionize our understanding of exoplanet atmospheres. For that reason, we will emphasize in this thesis the benefit of an \textit{EChO}-class mission in the context of exoplanet atmosphere characterization. For brevity, we henceforth refer to such a future \textit{EChO}-class mission as ``\textit{EChO}''.} and unprecedentedly informative datasets resonated with the field's goals to characterize exoplanets, probe their atmospheres, and assess their habitability. Yet, it was not clear \textit{what are} the new insights into the atmospheric and interior properties of exoplanets that will soon be within our reach and \textit{how to reach them}.

As an example, the assessment of a planet's habitability necessitates the determination its surface conditions---i.e., temperature, pressure, composition, and gravity. In most cases, the atmospheric layers at the surface level cannot be probed. Hence, atmospheric models are needed to extrapolate the atmospheric conditions from the layers probed by transmission and/or emission spectroscopy to deeper layers, possibly including the surface. Transmission spectroscopy constrains the atmospheric conditions at relatively low pressures (typically below $0.1$ bar) and over the planetary limbs---i.e., not a planet's 3D climate. Similarly, emission spectroscopy constrains the \textit{hemisphere-integrated} conditions of a planet's atmosphere---i.e., degenerate constraints on atmospheric models. In other words, 3D maps of exoplanetary atmospheres would be a valuable asset to extrapolate the temperature, pressure, and composition of unprobed layers. In addition 3D maps would provide feedback to atmospheric models and may reveal atmospheric processes or forcing factors yet unmodeled. The first method introduced in this thesis aims to address that need. The determination of a planet's surface gravity is also required to assess its habitability. The surface gravity is derived from the planet mass, which is  traditionally determined by the RV method. Yet, the RV method is mainly effective for massive planets around relatively bright and quiet stars. Hence the masses of small---e.g., Earth-sized---planets could remain out of reach, preventing from their in-depth characterization, including the assessment of their habitability. The second method introduced in this thesis aims to determine the planetary mass in an alternative way to the RV method.

\newpage
\section{Thesis Overview}

The continual increase in data quality of exoplanet observations drives the development of new methods to further characterize distant worlds. The main objective of this thesis is to investigate theoretically and practically the question ``\textit{What are the new insights into the atmospheric and interior properties of exoplanets within reach and how to reach them?}'' 

In Chapter\,\ref{chap:mapping}, I introduce how transiting exoplanets can be mapped while their stars cannot be resolved. I present the degeneracies intrinsic to exoplanet mapping and the framework proposed to mitigate those degeneracies while performing consistent mapping. I apply my mapping method to the hot Jupiters HD\,189733b and Kepler-7 b, respectively in the infrared and in the visible.

In Chapter\,\ref{chap:mass}, I introduce \textit{MassSpec}, a method to determine the mass of transiting exoplanets based solely on their transmission spectra. \textit{MassSpec} simultaneously and self-consistently constrains the mass and the atmospheric properties of an exoplanet, and allows mass measurements for transiting planets for which the radial velocity method fails. I demonstrate \textit{MassSpec}'s feasibility theoretically and highlight the fact that a planet's mass has to be accounted for by atmospheric retrieval methods to ensure retrieval quality. I also show the good agreement between the mass \textit{MassSpec} retrieves for HD\,189733b from transmission spectroscopy with that from RV measurements.

In Chapter\,\ref{chap:perspectives}, I investigate the prospects of the methods introduced in Chapters\,\ref{chap:mapping} and\,\ref{chap:mass}. I introduce short-term goals such as mapping for the first time an exoplanet's atmosphere in 3D with the \textit{Spitzer Space Telescope} and weighing the hottest exoplanet known with the \textit{Hubble Space Telescope}. I discuss long-term goals for the next decade such as obtaining time-dependent 3D maps of exoplanetary atmospheres and assessing the habitability of Earth-sized planets with the \textit{James Webb Space Telescope} and/or an \textit{EChO}-class mission.

In Chapter\,\ref{chap:conclusion}, I summarize the thesis' results and conclusions. Appendix\,\ref{app:TF} introduces the basics of rational functions. In Appendix\,\ref{sec:generalizedtransmissionspectra}, I present the first step towards a generalization of the transmission spectrum equations. I use these new equations in Appendices\,\ref{sec:testtransmissionspectrummodels} and\,\ref{sec:quicktransmissionspectrummodels} to introduce a simple validation test and an efficient implementation procedure for numerical transmission spectrum models, respectively. 

%% file: Chapter2/chap2.tex

\renewcommand\thefootnote{\fnsymbol{footnote}}
\chapter[Mapping Planets' Atmospheres While Their Stars Cannot Be Resolved]{Mapping Planets' Atmospheres While Their Stars Cannot Be Resolved\footnote{Work published in \textit{Astronomy and Astrophysics}, Volume 548, A 128 (2012), see \cite{deWit2012}, and in \textit{The Astrophysical Journal Letters}, Volume 776, Issue 2, L25 (2013), see \cite{Demory2013}.}}
\label{chap:mapping}
\renewcommand\thefootnote{\arabic{footnote}}
\addtocounter{footnote}{-1}

\vspace{-0.7cm}

Until recently, observations of extra-solar planetary systems primarily constrained their orbital configurations and, to a lesser extent, some planetary atmospheres---accessible through transmission and emission spectrophotometry \citep[see e.g.,][and references therein]{Seager2010a}. Although significant theoretical developments have been achieved in modeling exoplanet atmospheres by combining hydrodynamic flow with thermal forcing \cite{Showman2009,Rauscher2010,Dobbs-Dixon2010,Rauscher2012} and/or with ohmic dissipation \citep[e.g.,][]{Batygin2011,Heng2012,Menou2012}, other potential forcing factors are yet to be modeled\footnote{As an example, the magnetic star-planet interactions can also have a significant role in this matter. Nevertheless,  magnetic interactions have to date been only observed at the stellar surface, in the form of chromospheric hot spots rotating synchronously with the companions \citep[e.g.,][]{Shkolnik2005,Lanza2009}.}---or observed. The observation of specific spatial features within  an exoplanet atmosphere, such as hot spots or cold vortices, is essential for constraining its temperature, composition, and circulation patterns. Therefore, mapping exoplanet atmospheres is key for gaining further insights into their physics. 

Eclipses have proven to be powerful tools for ``spatially resolving'' distant objects, including binary stars \citep[e.g.,][]{Warner1971} and accretion disks \citep[e.g.,][]{Horne1985}. Previous theoretical studies investigated the potential of occultations in order to disentangle exoplanetary atmospheric circulation regimes \citep[e.g.,][]{William2006,Rauscher2007}.

In this Chapter, we introduce how transiting exoplanets can be mapped while their stars cannot, we present the degeneracies intrinsic to exoplanet mapping and the framework proposed to mitigate those degeneracies and perform mapping consistently. We also present two applications of our method in this Chapter: first, the 2D map of the hot-Jupiter HD\,189733b \citep[based on][]{deWit2012}; second, the 1D map of the hot-Jupiter Kepler-7 b \citep[based on the author's contribution to][]{Demory2013}.

\section{Concept}

\begin{figure}
   \centering

  \begin{center}
    \vspace{-0.4cm}\hspace{-0.cm}\includegraphics[trim = 80mm 80mm 90mm 75mm,clip,width=13.5cm,height=!]{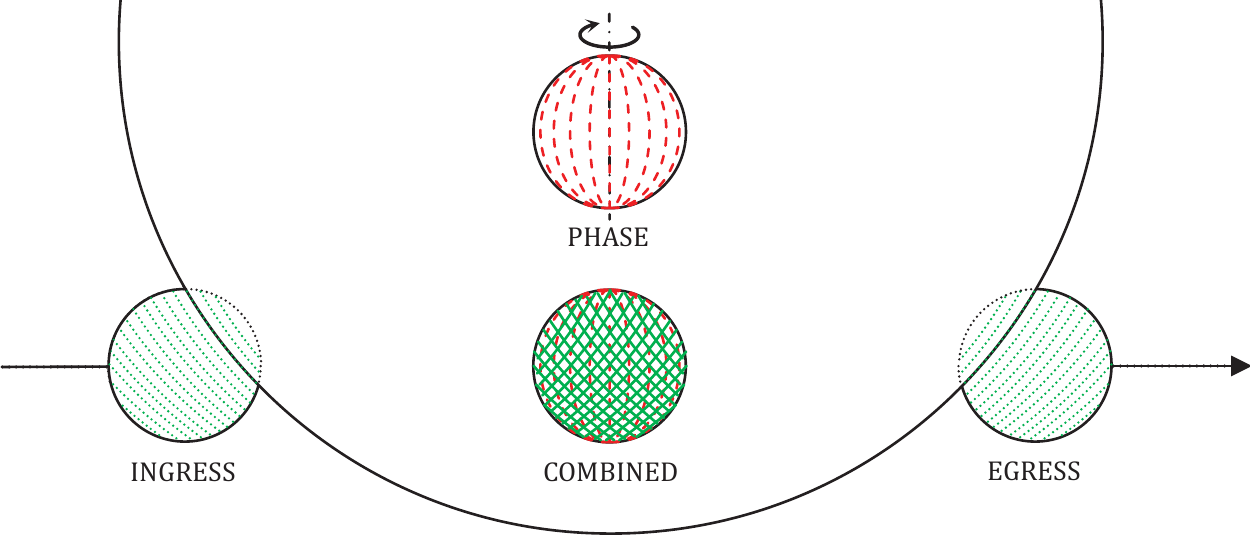}
  \end{center}
  \caption[Schematic description of the different scanning processes observable for an occulted exoplanet.]{Schematic description of the different scanning processes observable for an occulted exoplanet. The green dotted lines indicate the scanning processes during its occultation ingress/egress. The red dashed lines indicate the scanning process that results from the exoplanet rotation. The component labeled ``combined'' shows the specific grid generated by these three scanning processes.}
  \label{fig:scanning_processes}
    \end{figure}

The monochromatic light curve of a transiting and occulted exoplanet with a non-zero impact parameter can enable a 2D surface brightness map of its day side. Such an exoplanet is scanned through several processes that provide flux measurements of different planetary slices, which can be recombined into a map (see Figure \,\ref{fig:scanning_processes}).  First, the exoplanet is gradually masked/revealed by its host star during occultation ingress/egress (``eclipse scanning''). Secondly, the exoplanet rotation provides its phase-dependent hemisphere-integrated flux---i.e., its phase curve (see Section\,\ref{sec:charact}). A planet's phase curve constrains its brightness distribution (hereafter, BD) in longitudinal slices---as long as the exoplanet spin is close to the projection plane, e.g., for a transiting and synchronized exoplanet. The three scanning processes (ingress, egress, and phase curve) provide complementary pieces of information that constrain the target's BD over a specific ``grid'' (see the component labeled ``\textit{combined}'' in Figure\,\ref{fig:scanning_processes}) because they scan the planet over different directions. 

\section{Degeneracies}
\label{sec:degeneracies}

Eclipse mapping is subject to two levels of degeneracy.

\subsection{Eclipse Mapping: an Intrinsically-Underdetermined Problem\ldots}
 First, eclipse mapping is an underdetermined problem. The reason is that, conceptually, the map is composed of ``$N^2$''-grid cells constrained by the measured fluxes in $2N$ slices during the target's ingress and egress (compare the component labeled ``\textit{combined}'' to the components labeled ``\textit{ingress}'' and ``\textit{egress}'' in Figure\,\ref{fig:scanning_processes}). Therefore, different brightness maps can be solutions to the same observations, hence the intrinsic degeneracy. To account adequately for this intrinsic degeneracy, mapping methods have to be able investigate the different type of maps that can fit a given dataset. For that purpose, the analysis method we introduce in Section\,\ref{sec:analysismethod} use multiple types of brightness models to reveal different types of brightness patterns.

\subsection{\ldots with Multiple Non-Intrinsic Contributing Factors}

Secondly, eclipse mapping is based on the shape of a target's occultation ingress/egress. However, the shape of occulation ingress/egress is not solely affected by its brightness distribution. To propose a consistent methodology for mapping exoplanets, it is important to investigate the possible contributors to the shape of an occulation ingress/egress. Those possible contributors are:
\begin{itemize}
 \item the planet's projected shape at conjunctions,
 \item the planet's brightness distribution,
 \item four of the planetary system parameters\footnote{In each set of parameters sufficient to model the planetary system orbits, four of them can affect the occulation shape. The four presented here relates to the parameter set introduced in Equation\,\ref{eq:conventionalparameterset}.}:
 \begin{itemize}
  \item the planet's orbital eccentricity, $e$,
  \item the planet's periapsis argument, $\omega$,
  \item the planet's impact parameter, $b=a/R_{\star}\cos i$ (where $a$ is the exoplanet semi-major axis and $i$ the orbital inclination),
  \item the stellar density, $\rho_\star$.
  \end{itemize} 
\end{itemize} 

\begin{figure}[!p]
   \centering

  \begin{center}
    \vspace{-1cm}\hspace{-0.cm}\includegraphics[trim = 70mm 50mm 85mm 95mm,clip,width=14cm,height=!]{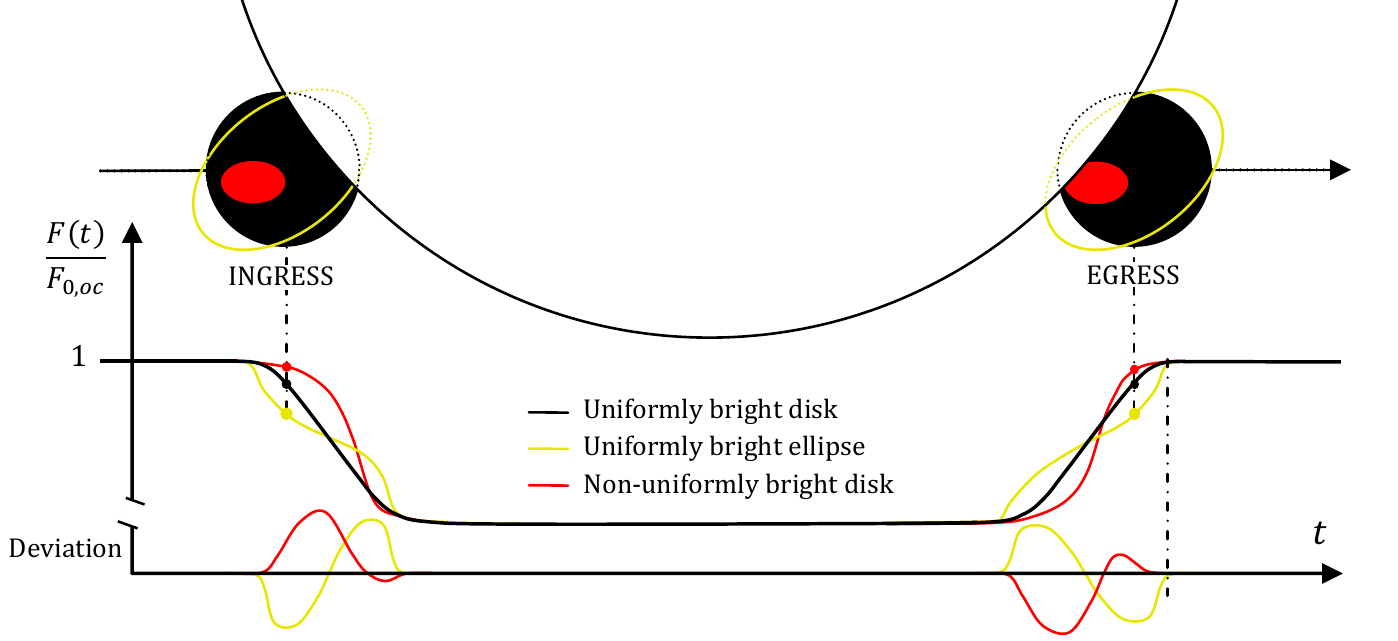}
  \end{center}
  \vspace{-0.5cm}
  \caption[Schematic description of the effect of the shape or the brightness distribution of an exoplanet on its occultation shape.]{Schematic description of the effect of the shape or the brightness distribution of an exoplanet on its occultation shape. The red curve indicates the occultation photometry for a non-uniformly bright disk (hot spot in red). The yellow curve indicates the occultation photometry for an oblate exoplanet (yellow ellipse). Both synthetic scenarios
show specific deviations from the occultation photometry of uniformly bright disk (black curve) in the occultation ingress/egress.}
  \label{fig:shape_brightness}
	\vspace{-0.5cm}
    \end{figure}

We present in Figure\,\ref{fig:shape_brightness} a schematic description of the dependency of the occultation shape on the planet's shape (yellow) and brightness distribution (red). Both synthetic scenarios show specific deviations from the occultation photometry of uniformly bright disk (black curve) in the occultation ingress/egress (see Figure\,\ref{fig:shape_brightness}, bottom panel). 

\begin{figure}[p]
   \centering

  \begin{center}
    \vspace{-0.0cm}\hspace{-0.cm}\includegraphics[trim = 00mm 05mm 00mm 05mm,clip,width=13cm,height=!]{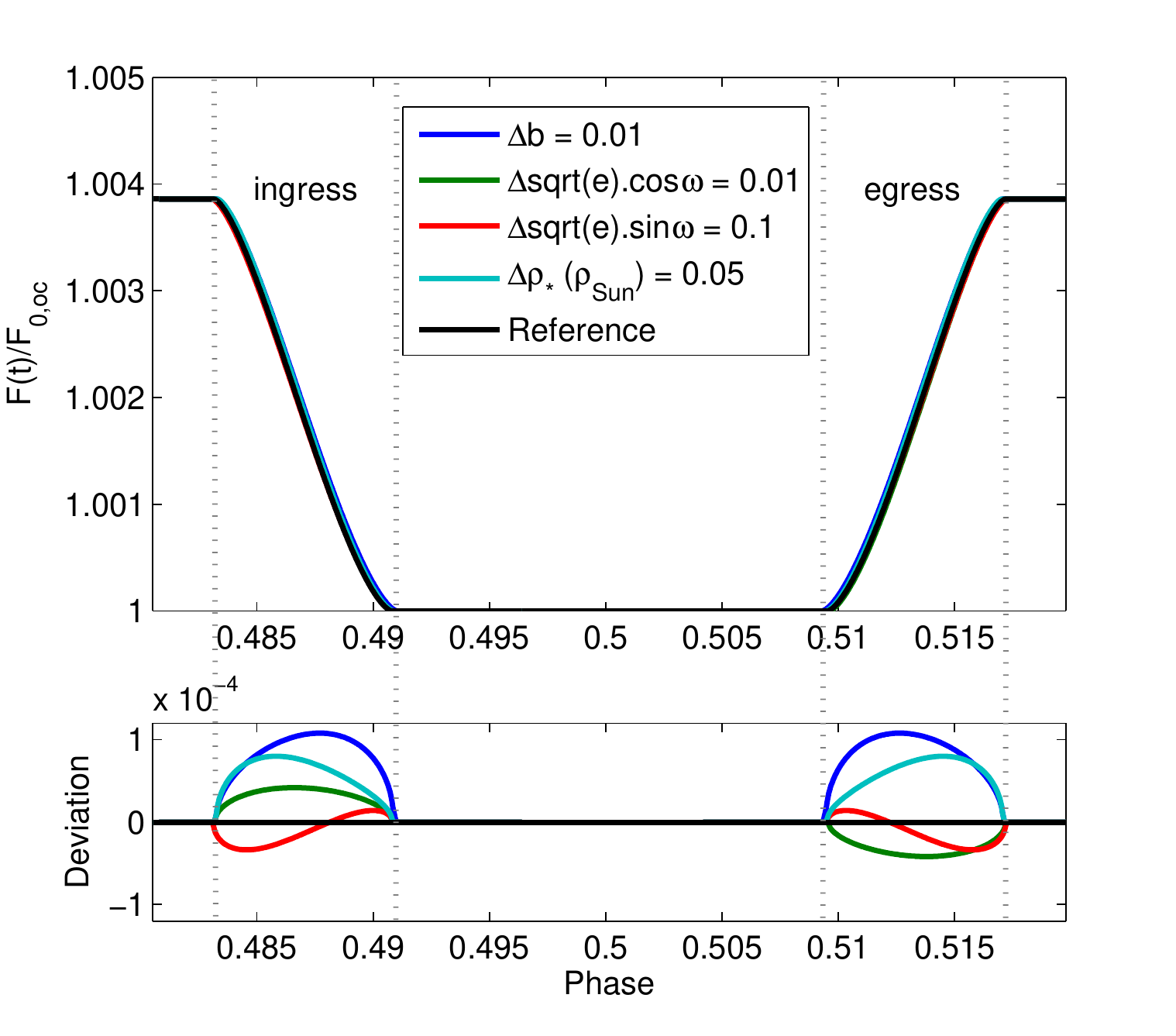}
  \end{center}
  \vspace{-0.5cm}
  \caption[Effects of system parameters on an occultation ingress/egress shape.]{Effects of system parameters on an occultation ingress/egress shape. \textit{Top panel:} Simulated occultation using the system parameters of the hot Jupiter HD\,189733b (black curve) and the simulated occultations when perturbing system parameters ($b$ in blue, $\sqrt{e}\cos\omega$ in green, $\sqrt{e}\sin\omega$ in red, and $\rho_{\star}$ in cyan). \textit{Bottom panel:} Deviations in occultation ingress/egress from the reference model.}
  \label{fig:deviation_in_ingress_egress}
\vspace{-1cm}
    \end{figure}

We present in Figure\,\ref{fig:deviation_in_ingress_egress} the dependency of the occultation shape on $e$, $\omega$, $b$, and $\rho_\star$. The top panel of Figure\,\ref{fig:deviation_in_ingress_egress} presents a simulated occultation using the system parameters of the hot Jupiter HD\,189733b (black curve) and the simulated occultations when perturbing each of the system parameters. Although the effects may be considered as of second order---the deviations from the \textit{reference} simulation are barely visible on the top panel, they actually corresponds to typical deviations expected in ingress/egress in order to perform mapping (compare with the bottom-right panel of Figure\,\ref{fig:in_eg_structures}). 

\begin{figure}[!p]
   \centering

  \begin{center}
    \vspace{-1cm}\hspace{-0.cm}\includegraphics[trim = 00mm 00mm 00mm 00mm,clip,width=14cm,height=!]{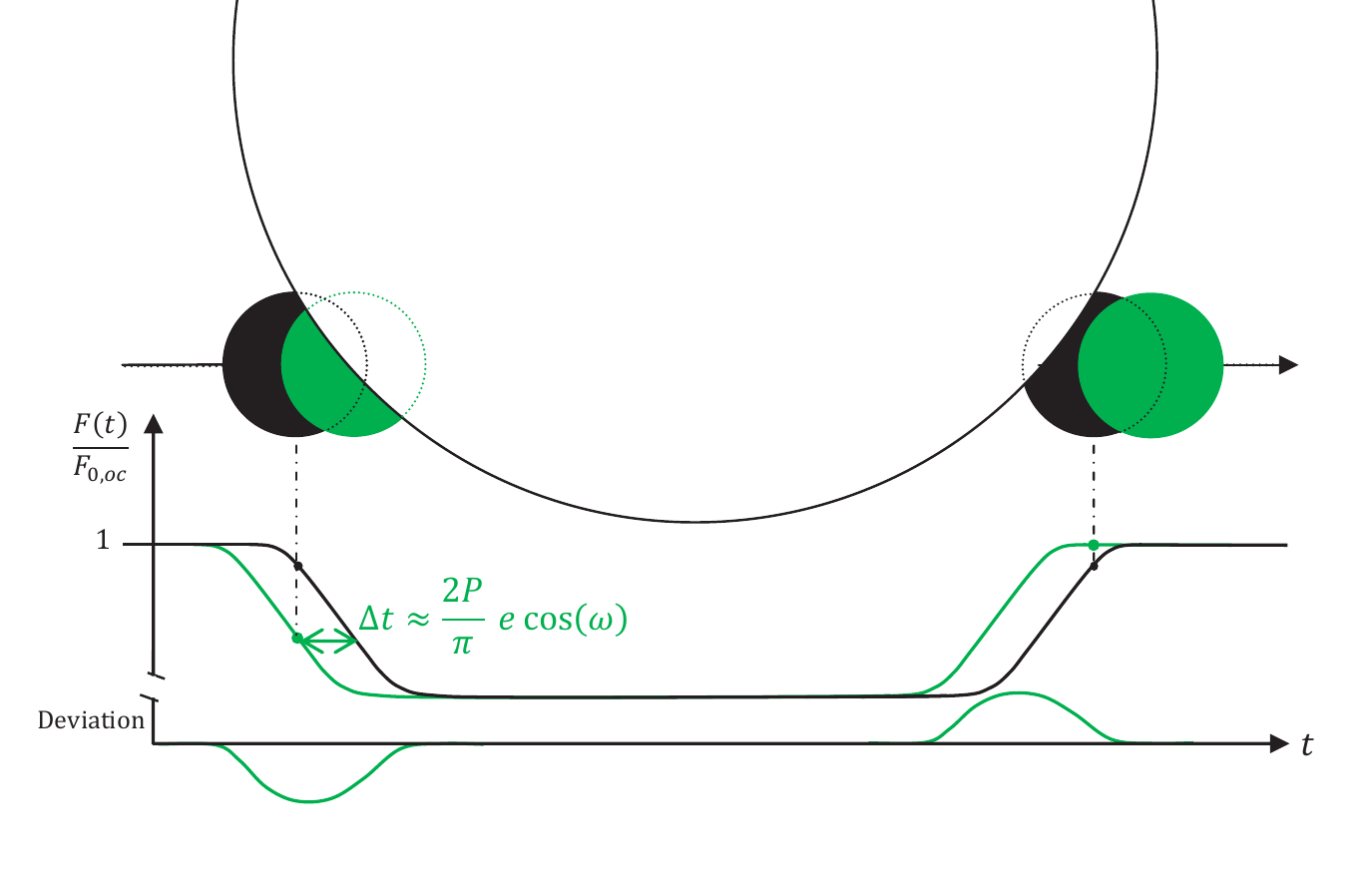}
  \end{center}
  \vspace{-1cm}
  \caption[Schematic description of the effect of $e\cos\omega$ on an occultation shape.]{Schematic description of the effect of $e\cos\omega$ on an occultation shape. The green curve indicates the occultation photometry for a decreased $e\cos\omega$, which translates into a decreased timing-offset of the occultation ($\Delta t$).}
  \label{fig:shape_brightness_ec}
	\vspace{-0.5cm}
    \end{figure}
    
    \begin{figure}[!p]
   \centering

  \begin{center}
    \vspace{-0cm}\hspace{-0.cm}\includegraphics[trim = 00mm 00mm 00mm 00mm,clip,width=14cm,height=!]{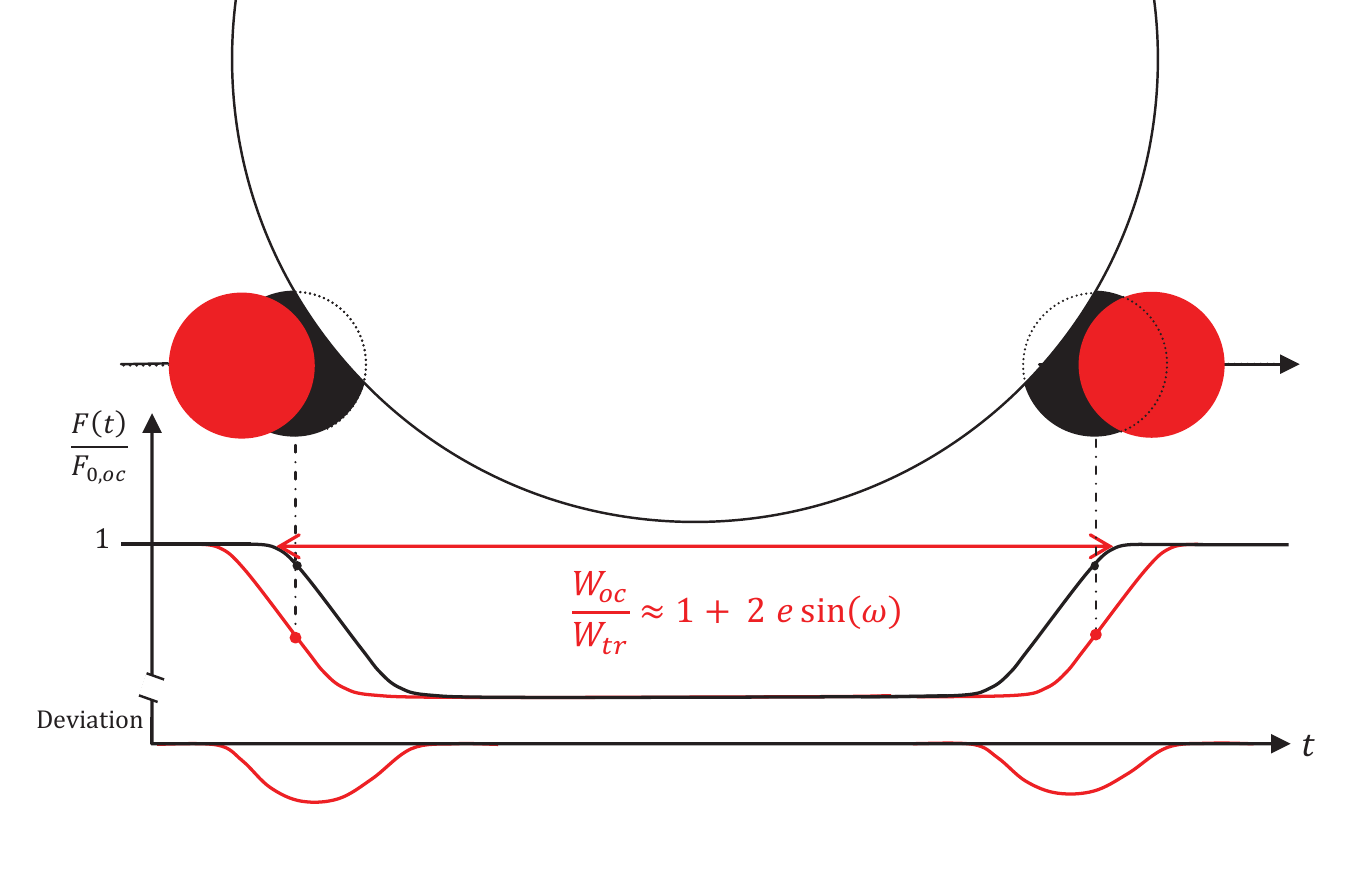}
  \end{center}
  \vspace{-1cm}
  \caption[Schematic description of the effect of $e\sin\omega$ on an occultation shape.]{Schematic description of the effect of $e\sin\omega$ on an occultation shape. The red curve indicates the occultation photometry for an increased $e\sin\omega$, which translates into an increased occultation duration $W_{oc}$---in particular an increased ratio occultation-to-transit durations $\frac{W_{oc}}{W_{tr}} $.}
  \label{fig:shape_brightness_es}
	\vspace{-0.5cm}
    \end{figure}
    
    \begin{figure}[!p]
   \centering

  \begin{center}
    \vspace{-1cm}\hspace{-0.cm}\includegraphics[trim = 00mm 00mm 00mm 00mm,clip,width=14cm,height=!]{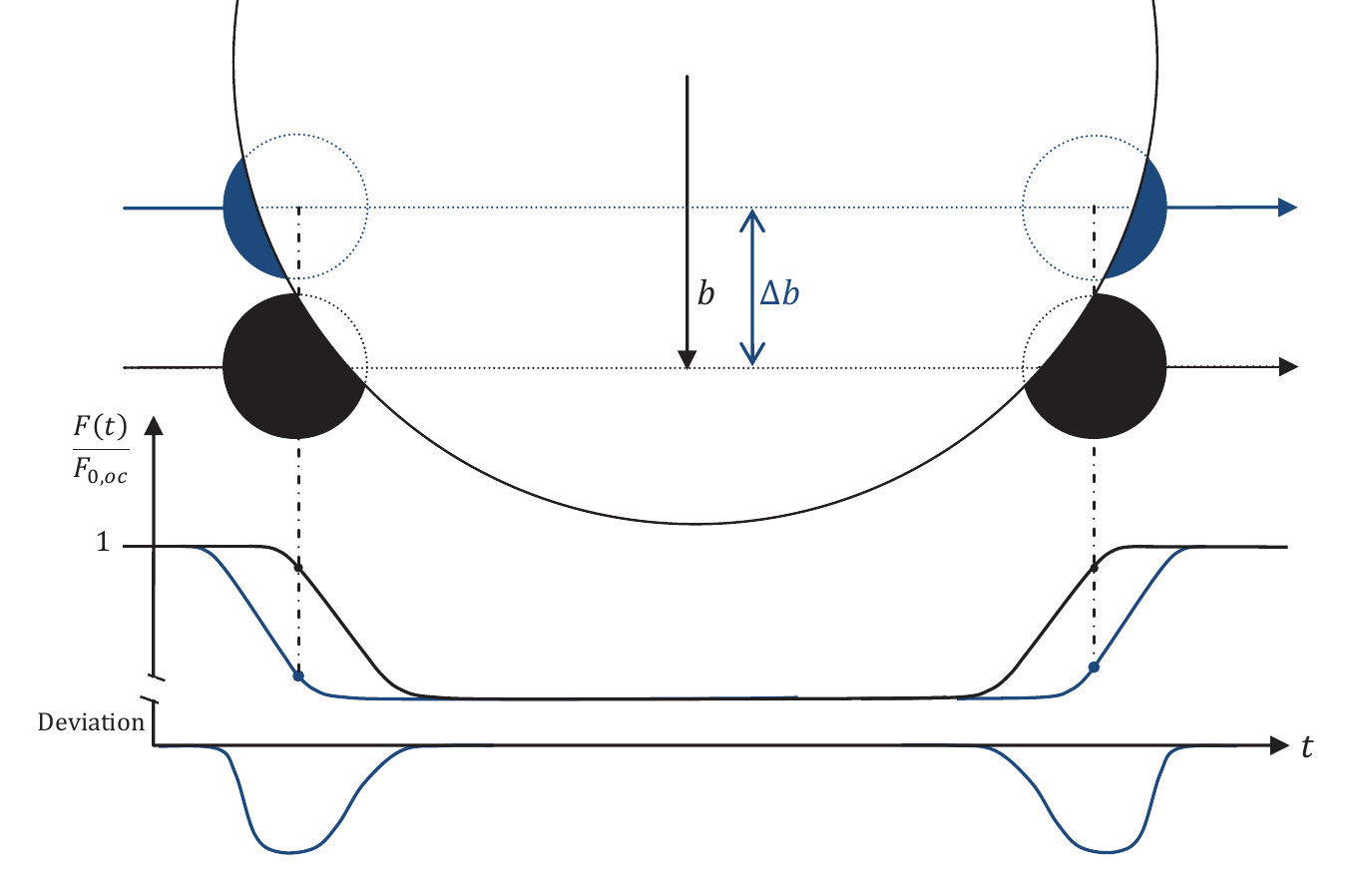}
  \end{center}
  \vspace{-1cm}
  \caption[Schematic description of the effect of the impact parameter on an occultation's shape.]{Schematic description of the effect of the impact parameter ($b$) on an occultation's shape. The red curve indicates the occultation photometry for a decreased $b$. A smaller impact parameter yields a longer eclipse duration and less gradual ingress/egress.}
  \label{fig:shape_brightness_b}
	\vspace{-0.5cm}
    \end{figure}
    
    \begin{figure}[!p]
   \centering

  \begin{center}
    \vspace{-0cm}\hspace{-0.cm}\includegraphics[trim = 00mm 00mm 00mm 00mm,clip,width=14cm,height=!]{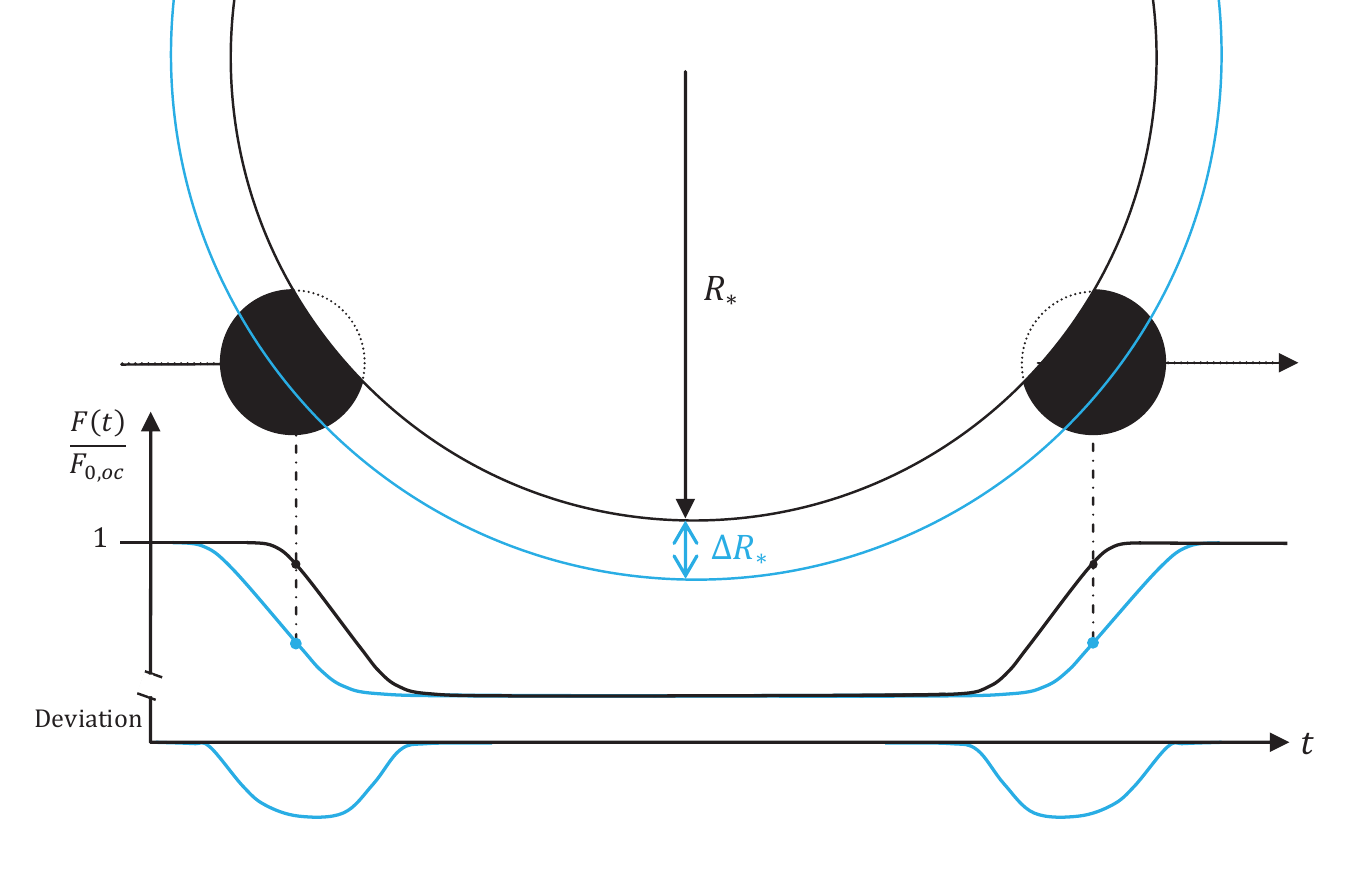}
  \end{center}
  \vspace{-1.cm}
  \caption[Schematic description of the effect of the stellar radius on an occultation shape.]{Schematic description of the effect of the stellar radius ($R_\star$) on an occultation shape---representative here of the effect of the stellar density as the stellar mass is not independently constrained by a system light curve. The red curve indicates the occultation photometry for an increased $R_\star$. A larger stellar radius yields a longer eclipse duration and more gradual ingress/egress.}
  \label{fig:shape_brightness_r}
	\vspace{-0.5cm}
    \end{figure}

Each of these deviations can be explained by one of the following effects. $e$ and $\omega$ enable both to shift the occultation by
 \begin{eqnarray}	
      \Delta t \approx \frac{2P}{\pi} e \cos\omega,
       \label{DeltaToc}
\end{eqnarray}
(hereafter ``uniform time offset''\footnote{The uniform time offset measures the time lag between the observed secondary eclipse and that predicted by a planet with spatially uniform emission \cite{William2006}.}) and to change its duration
 \begin{eqnarray}	
      \frac{W_{oc}}{W_{tr}} \approx 1 + 2 e \sin\omega,
       \label{Durations}
\end{eqnarray}
(Figures\,\ref{fig:shape_brightness_ec} and\,\ref{fig:shape_brightness_es}, respectively). Hence increasing $e\cos\omega$ primarily leads to an earlier occultation (negative deviation in ingress and positive deviation in egress) and increasing $e\sin\omega$, to an increase of the occultation duration (negative deviation in ingress and negative deviation in egress). $b$ affects the orbital inclination, hence, both the time of first and last contacts and the occultation duration (Figure\,\ref{fig:shape_brightness_b}). Decreasing $b$ primarily leads to a negative deviation in ingress and positive deviation in egress. Finally, $\rho_{\star}$ affects directly the stellar radius and hence the time of first and last contact (Figure\,\ref{fig:shape_brightness_r}). Decreasing $\rho_{\star}$ primarily leads to a positive deviation in ingress and positive deviation in egress. Note that also $b$ and $\rho_{\star}$ increase the occultation duration when decreased, their effects on the ingress/egress are different. While the ingress and egress slopes are steeper than the reference case (black) when decreasing $b$, their are more gradual when decreasing $\rho_{\star}$. Hence, each of the four system parameters has a unique effect on the occultation shape.

\subsection{Steps Towards Mitigating Degeneracies}
\label{sec:dealingwithdegeneracies}

 The various shapes of the deviations and their symmetry/asymmetry reveal that combinations of perturbations of the system parameters can adequately mimic the effect of a non-uniformly bright exoplanet (see Figure\,\ref{fig:shape_brightness}). In other words, $e$, $\omega$, $\rho_{\star}$, and $b$ are correlated with the projected shape and brightness distribution of an exoplanet in the framework of occulation dataset.

\subsubsection{Avoid Unassessed Assumptions}
 
	A first step toward mitigating the effect of the degeneracies is to avoid assumptions that relate to one of the correlated parameters mentioned above. An assumption relative to one of those parameters would affect directly the estimates of the other parameters, possible biasing these estimates (see application in Section\,\ref{sec:uniform}). In other words, assumptions such as ``the target's orbit is circularized'' (i.e., $e = 0$) or ``the target is a sphere'' will prevent analysis methods from the proper exploration of the parameter space. We highlight here that assuming a planet to be uniformly-bright while fitting high-SNR occultations is also inadequate. For that reason, it is crucial to make a parsimonious use of assumptions unless they can be assessed.

\subsubsection{Perform Global Analysis}
		Assumptions regarding the planet's shape, brightness, or system parameters can be assessed using complementary data to occultation light curves. Note that complementary occultation data obtained in different spectral bands can also help in disentangling the contributions of the system parameters to the occultation shape---as these are wavelength-independent.
		
\textbf{Transit Light Curves} provide complementary constraints on the system parameters (including $b$, $\sqrt{e}\sin\omega$, and $\rho_{\star}$) and on the projected shape of a companion at conjunction \cite{Barne2009,Carter2010}. Furthermore, the significance of shape-induced deviations in transit is about one order of magnituded larger than the shape-induced deviations in occultation\footnote{The effect ratio of shape-induced signals in occultation to in transit is proportional to the planetary hemisphere-averaged relative brightness $<I_p>/<I_\star>$ in the spectral band covered}. Note that constraining an exoplanet shape requires \textit{a priori} knowledge on the host star limb-darkening \citep[e.g.,][]{Claret2011} and on the transit timing to avoid overfitting its possible signature in the transit ingress/egress---the transit timing is independently constrained by RV measurements.

\textbf{Phase Curves} provide complementary constraints on a companion brightness and shape. Hence phase curves are beneficial to mitigate both levels of degeneracy: \textbf{(1)} it provides complementary constraints on the planet's brightness distribution (different scanning direction, Figure\,\ref{fig:shape_brightness}) and \textbf{(2)} it constrains the planet's shape. A planet's shape affect its phase curve in a way similar to a tidally distorted star induces ellipsoidal light variations \citep{Russell1952,Kopal1959}. The contributions of the shape and the BD of an exoplanet to its phase curve may be disentangle because of their different periods \citep[see e.g.,][]{Faigler2011}. As an example, for a synchronized exoplanet, the shape-induced modulation has a period of $ P/2 $---twice the same projection an orbital period, while the brightness-induced modulation as a period of $ P $.

\textbf{RV Measurements} provide complementary constraints on the orbital configuration, including on $e$ and $\omega$ (see Section\,\ref{sec:RVmodel}).

\subsubsection{Coping with Multiple Solutions}
	
	Global analysis of a complementary dataset is key to constrain consistently the possible contributing factors to the shape of an occultation. However, the complex multidimensional correlation of the problem may produce multiple solutions---their disentanglement depending on the data SNR. Additional arguments could then be used \textit{a posteriori} to disentangle further between the different solutions proposed by the retrieval method. For example, consideration from atmospheric physics or statistics \citep[e.g., the Bayesian Information Criteria,][]{Gelman2004} should be investigated.

\section{Methodology}
\label{sec:methodo}
We present in this section the retrieval method developed to map a transiting planet, details regarding the data reduction are provided in Section\,\ref{sec:map18733b}. The method can also be used to analysize traditionnal datasets such as RV measurements and/or individual eclipses.

\subsection{Analysis Method}
\label{sec:analysismethod}
 
We use an adaptive Markov Chain Monte Carlo \citep[MCMC; see e.g.][]{Gregory2005,Ford2006} algorithm. MCMC is a Bayesian inference method based on stochastic simulations that sample the posterior probability distribution (PPD) of adjusted parameters ($\textbf{x}$) for a given fitting model, $f_{sim}(\textbf{x})$. More specifically, our implementation uses Keplerian orbits\footnote{Our orbit model also includes the host-star's movement around the system center of gravity} and models photometry using the light-curve models introduced in Section\,\ref{sect:lightcurvemodel} and RV data using the model introduced in Section\,\ref{sec:RVmodel}. In addition, each photometric time-serie is corrected for systematics (see Section\,\ref{sec:systematics}).

\paragraph{Basics about MCMC Algorithms}
MCMC algorithms generate a chain in the parameter space based on a succession of stochastic jumps submitted to rules that ensure convergence of the chain towards the PPD. The basics of a chain generation can be described by the following steps \citep{Ford2005}:
\begin{enumerate}
\item 	Initialize the chain with some $\textbf{x}_n$ ($n=0$).

\item 	Evaluate $f_{sim}(\textbf{x}_n)$ and the associated chi-squared $\left(\chi^2(\textbf{x}_n)\right)$.

\item 	Generate a trial state $\textbf{x}'$ according to a transition kernel\footnote{Refer to e.g., \cite{Gregory2005} and reference therein for further details on the transition-kernel properties required to ensure the good behavior of the MCMC chain.} $q\left(\textbf{x}'|\textbf{x}_n \right)$ (e.g., a Gaussian function centered on $\textbf{x}_n$).

\item 	Evaluate $f_{sim}(\textbf{x}')$ and the associated $\chi^2(\textbf{x}')$.

\item 	Estimate the probability of transition from $\textbf{x}_n$ to $\textbf{x}' \propto \exp{\left(-\frac{1}{2}\left(\chi^2(\textbf{x}')-\chi^2(\textbf{x}_n)\right)\right) }$.

\item 	Draw a random number from a uniform distribution between 0 and 1.

\item 	If this number is below the probability of transition $\textbf{x}_{n+1} = \textbf{x}'$, else $\textbf{x}_{n+1} = \textbf{x}_n$.

\item 	$n=n+1$ and back to step 3 unless $n=n_{limit}$.

\end{enumerate}

In case of \textit{a priori} knowledge concerning the distribution of parameters, the  $\chi^2$ expression---whose evaluation conditions the acceptance of the attempted jump at step 5---is modified to include this \textit{a priori} information. As an example, if $M_{star}$ is known \textit{a priori} to be of $M_{\star,0} \pm \sigma_{M_\star}$ under normal distribution---i.e., $\textbf{N}(M_{\star,0},\sigma_{M_\star})$, then the chi-squared function includes this \textit{a priori} knowledge in the form of the following additional term:  $\left(\frac{M_{\star,n}-M_{\star_0}}{\sigma_{M_\star}}\right)^2$. 

\subsubsection{``Conventional'' Part of our Parameter Set}

In order to model ``conventional''\footnote{``Conventional'' refers here to models or methods that assume uniformly-bright planets.} light curves, the required parameter set ($\textbf{x}$) has to constrain the area and brightness ratios of the companions as well as their orbits. Several choice of parameter sets are possible for modeling system light curves, but it is necessary to aim for one with a mitigated correlation between the parameters to enhance the method efficiency \citep[see discussion in][]{Ford2006}. The conventional part of our implementation uses
\begin{eqnarray}
\left\lbrace \frac{R_p^2}{R_{\star}^2}, b, T_T, P, \sqrt{e}\cos(\omega), \sqrt{e}\sin(\omega), \rho_{\star}, M_{\star}, \frac{<I_p>}{<I_\star>}\right\rbrace,
      \label{eq:conventionalparameterset}
\end{eqnarray}
where $R_p^2/R_{\star}^2$ is the planet-to-star area ratio, $b$ is the impact parameter ($=a/R_{\star}\cos i$ where $a$ is the exoplanet semi-major axis and $i$ is the orbital inclination), $T_T$ is the time of minimum light (i.e., the transit center), $P$ is the orbital period, $e$ and $\omega$ are the eccentricity and the periapsis argument, $\rho_{\star}$ is the stellar density\footnote{We choose to use the stellar density, $\rho_{\star}$, instead of the transit duration (from the first to last contact), $W$, as jump parameter; because it relates directly to the orbital parameters \citep{Seager2003}.}, $M_{\star}$ is the stellar mass, and $<I_p>/<I_\star>$ planetary hemisphere-averaged relative brightness.

We used the following jump parameters\footnote{Jump parameters are the model parameters that are randomly perturbed at each step of the MCMC method.} $\frac{R_p^2}{R_{\star}^2}$, $b$, $T_T$, $P$, $\sqrt{e}\cos(\omega)$, $\sqrt{e}\sin(\omega)$, and  $\rho_{\star}$ and assumed a uniform prior distribution for all these jump parameters. Because the stellar mass is not constrained by the targeted dataset, the method uses a Gaussian prior from the literature.

\subsubsection{Specificity of the Linear Parameters}
\label{sec:linearparam}

$<I_p>/<I_\star>$ is not part of our set of jump parameters because it is a linear parameter of our model. Hence, it can be determined by linear least squares minimization at each step of the MCMC. For this purpose, we employed the Singular Value Decomposition (SVD) method \citep{Press92}. Distinguishing between linear and non-linear parameters enables to increase significantly the efficiency of an MCMC implementation by reducing the dimension of the space to probe---i.e., the number of jump parameters. Therefore, the linear parameters of our model are not considered as jump parameters.

We found that in order to use consistently the SVD method within such a stochastic framework, it is optimal to perturb the linear-coefficient estimates using the covariance matrix derived by the SVD. Perturbing the estimates prevents the SVD from mitigating systematically the $\chi^2$ increase due to inadequate jumps in the parameter space via the use of the best estimates of the linear parameters.

\subsection{Light-Curve Models}
\label{sect:lightcurvemodel}

\subsubsection{Stellar Contribution: Transit Model} 

We model the stellar contribution assuming a quadratic limb-darkening law for the star as in \cite{Mandel2002}. We draw the limb-darkening coefficients from the theoretical tables of  \cite{Claret2011} based on the host-star spectroscopic parameters---effective temperature, $T_{eff}$, surface gravity, $\log g$, and metallicity, $\left[\frac{Fe}{H}\right]$. We add this \textit{a priori} knowledge as a Bayesian penalty to our merit function, using as additional ``conventional'' jump parameters the combinations $c_1 = 2u_1+u_2 $ and $ c_2 = u_1-2u_2 $, as described in \cite{Gillon2010}.

\subsubsection{Planetary Contribution}
\label{sec:planet_contribution}
	We model the planetary contribution by performing a numerical integration of the observed exoplanet flux. Model approximations include: ignoring the time variability of the target atmosphere \citep[in line with atmospheric models, e.g.,][]{Cooper2005}---because it is currently necessary to combine multiple observations to obtain a sufficient signal-to-noise ratio (SNR); ignoring the planet limb darkening; and assuming the target's rotational period to be synchronized with its orbital period \citep[synchronization occurs over $\sim$\,$10^6$\,yr for hot Jupiters, e.g.,][]{Winn2010}. 
	
	From the brightness distribution and the orbits, we model the planet's flux temporal evolution---which accounts for the light-travel time across the system---by sampling its surface with a grid of $2N$ points in longitude ($\phi$) and $N+1$ points in latitude ($\theta$). We fix N to 100 to mitigate numerical effects up to $10^{-3}$ the secondary eclipse depth, i.e. at least two orders of magnitude below the current photometric precision.

\paragraph{Non-Uniform Brightness Models}

As discussed in Section\,\ref{sec:degeneracies} it is \textit{critical} to enable the mapping method to investigate different types/groups of maps in order to cope consistently with the degeneracy intrinsic to eclipse mapping. Therefore, we use three groups of brightness models: \textbf{(1)} 1D models, \textbf{(2)} spherical harmonics, and \textbf{(3)} toy models. \textbf{(1)} The 1D models are preferably used when a planet can solely be mapped from its phase-curve---i.e., when the available mapping information is mainly longitudinal (see Section\,\ref{sec:K7b}). The 1D models can be subdivided in two model families: one using $n$
longitudinal bands with fixed positions on the dayside \citep[``beach-ball'' similar to the ones introduced in][]{Cowan2009} and another
using longitudinal bands whose positions and widths are
jump parameters in the MCMC fit. \textbf{(2)} The spherical harmonics require two additional jump parameters  for the direction (longitude and latitude) and a linear parameter for the amplitude, per harmonic. The generalized formulation of this brightness distribution model is;
\begin{eqnarray}	
      \Gamma_{SH,d}(\phi,\theta) & = & \sum_{l = 0}^d I_l Y_l^0\left(\phi-\Delta\phi_l,\theta-\Delta\theta_l\right),
       \label{SH}
\end{eqnarray}
where $Y_l^0\left(\phi-\Delta\phi_l,\theta-\Delta\theta_l\right)$ is the real spherical harmonic of degree $l$ and order $0$, directed to $\Delta\phi_l$ in longitude and $\Delta\theta_l$ in latitude. $I_l$ is the $l^{th}$-harmonic amplitude that can be estimated at each step of the MCMC using the perturbed SVD method. As an example, the additional parameters for a dipolar fitting model compared to the conventional analysis method are: $\Delta\phi_1$ and $\Delta\theta_1$ as jump parameters and $I_1$ as a linear coefficient---$I_0$ is the amplitude of the uniformly bright mode, included in the conventional analysis method (as $<I_p>/<I_\star>$ in Equation\,\ref{eq:conventionalparameterset}).

\textbf{(3)} The toy models are developed specifically to model/retrieve one---or several---cold or hot spot. Their expression can be written as;
\begin{eqnarray}	
      \Gamma_1(\phi,\theta) & = &  I_1 \phi_\circ^\alpha \exp\left[-\phi_\circ^\beta\right] \cos {}^\gamma\theta_\circ + I_0 , \label{BM1}\\
      \Gamma_2(\phi,\theta) & = & 
	\left\{{
   \begin{array}{c c}
    I_1\cos {}^\alpha \phi_\circ \cos {}^\gamma \theta_\circ + I_0 & \mbox{if } \phi_\circ\geq0 
  	\\
  	I_1\cos {}^\beta \phi_\circ \cos {}^\gamma\theta_\circ + I_0 & \mbox{if } \phi_\circ<0 
  \end{array}}
  	\right.
  	,\label{BM2}\\
      \Gamma_3(\phi,\theta) & = & 
	\left\{{
   \begin{array}{c c}
    I_1\exp\left[-\left(\frac{\phi_\circ}{\alpha}\right)^2\right]\exp\left[-\left(\frac{\theta_\circ}{\gamma}\right)^2\right] + I_0 & \mbox{if } \phi_\circ\geq0 
  	\\
  	I_1\exp\left[-\left(\frac{\phi_\circ}{\beta}\right)^2\right]\exp\left[-\left(\frac{\theta_\circ}{\gamma}\right)^2\right] + I_0 & \mbox{if } \phi_\circ<0
  \end{array}}
  	\right.
  	,\label{BM3}
\end{eqnarray}
where $\phi_\circ$ and $\theta_\circ$ are respectively the longitude and latitude relative to the position of the model extremum, i.e. $\phi_\circ = f(\phi,\Delta\phi,\alpha,\beta,\gamma)$ and $\theta_\circ = f(\theta,\Delta\theta,\alpha,\beta,\gamma)$. $\Delta\phi$ and $\Delta\theta$ are respectively the longitudinal and the latitudinal shift of the model peak from the substellar point. $\alpha,\beta$ and $\gamma$ parametrize the shape of the hot/cold spot. These toy models add to the conventional analysis method five jump parameters ($\Delta\phi,\Delta\theta,\alpha,\beta,\gamma$) and a linear coefficient ($I_1$) per spot modeled.

\newpage
\subsection{Systematics and Correlated Noises}
\label{sec:systematics} 

\begin{figure}
  \begin{center}
  	\subfloat[][]
    {\label{fig:r22809088raw}\includegraphics[trim = 00mm 105mm 00mm 00mm,clip,width=13.5cm,height=!]{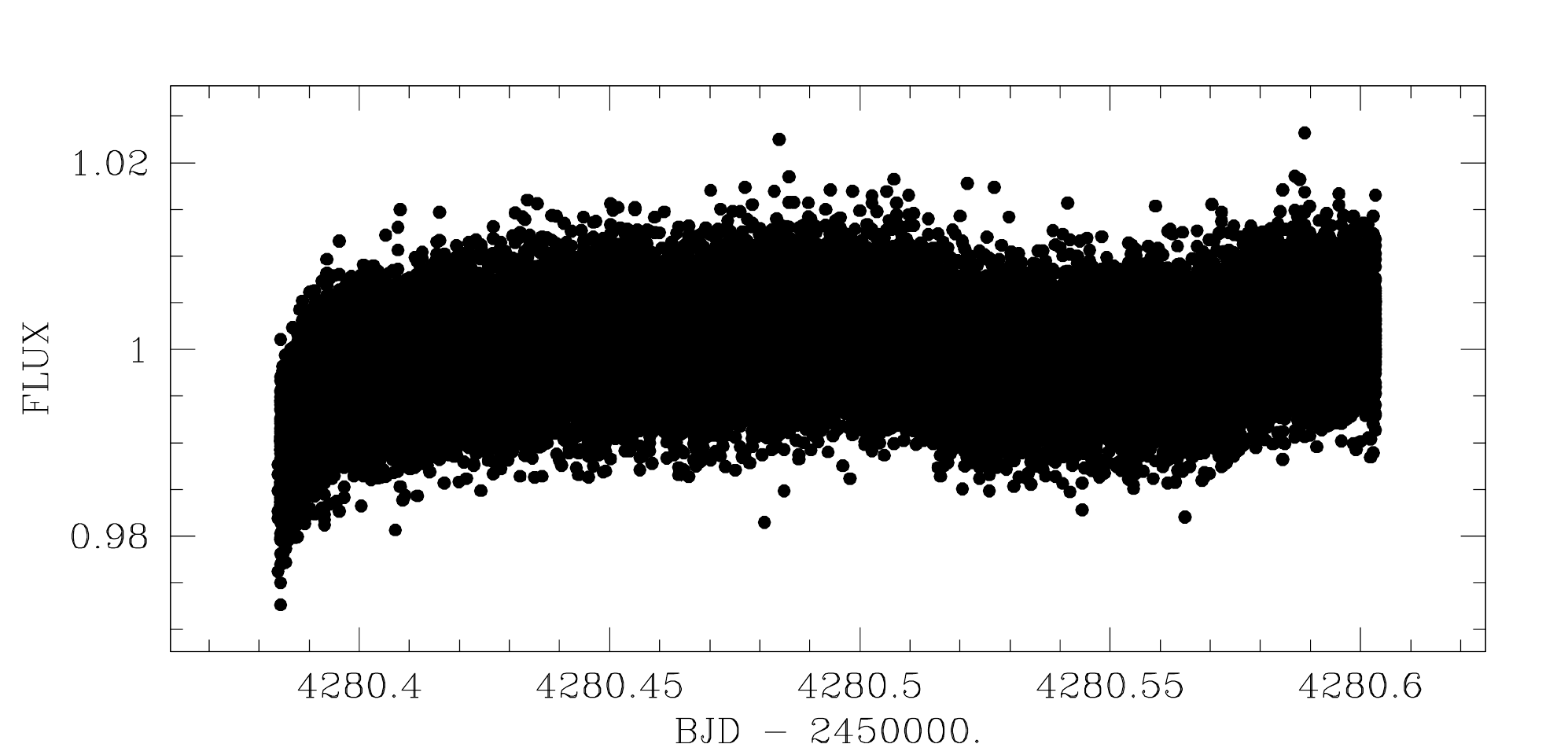}}\\
     \vspace{-0.5cm}     
    \subfloat[][]
   {\label{fig:r22809088fitted}\includegraphics[trim = 00mm 110mm 00mm 00mm,clip,width=13.5cm,height=!]{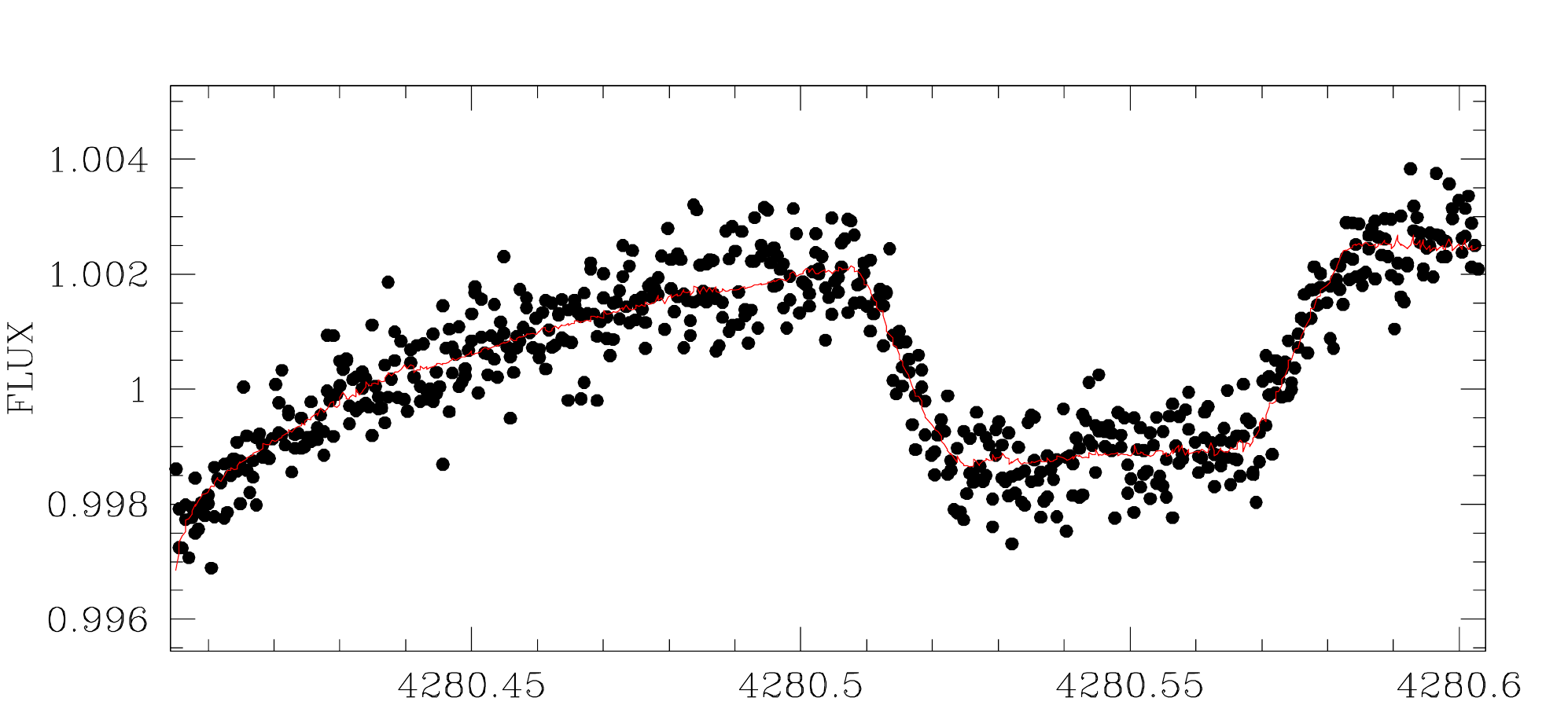}}\\
    \vspace{-0.5cm}     
    \subfloat[][]
   {\label{fig:r22809088detrended}\includegraphics[trim = 00mm 110mm 00mm 00mm,clip,width=13.5cm,height=!]{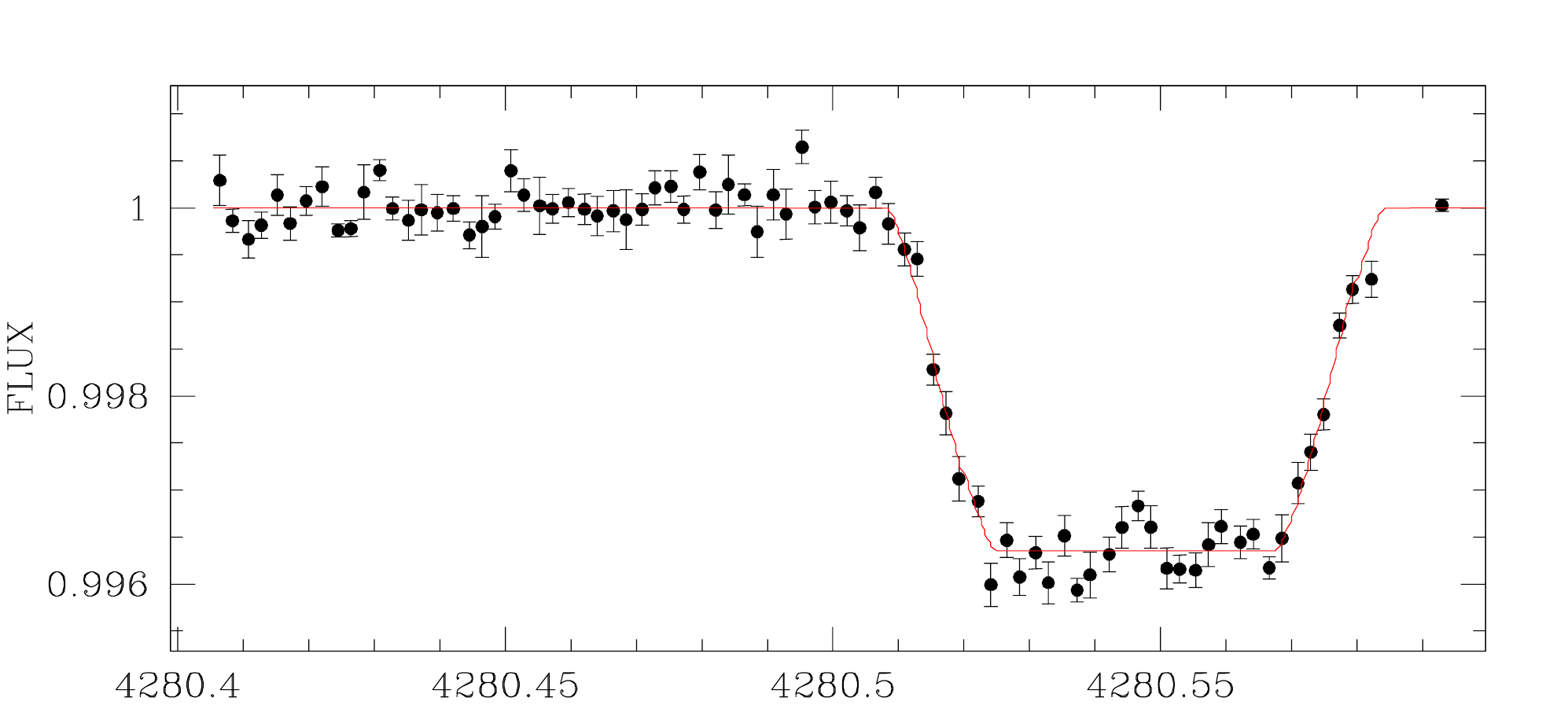}}\\
    \vspace{-0.2cm}
    \end{center}
  \caption[Light curve: noise and instrumental systematics]{Light curve: noise and instrumental systematics. Raw time series (dark symbols) of the AOR r22809088 (see Section\,\ref{sec:map18733b}) and the best fit (red curves). \subref{fig:r22809088raw} presents the raw time series obtained using aperture photometry, \subref{fig:r22809088fitted} the raw time series binned 64 times and fitted. \subref{fig:r22809088detrended} presents the ``detrended'' time series binned in 3.5min intervals and fitted.}
  \label{fig:r22809088rawtodetrended}
\end{figure}

Observations do not contain solely the signals targeted and described above. As an example, the time-dependent sensitivity of the telescope detector can significantly contribute to the observations. Figure\,\ref{fig:r22809088rawtodetrended} presents an example of raw light curves (panel a), binned down to reveal an eclipse and the asymptotic behavior of the measured electron flux that is due to the time-dependent sensitivity of the detector (panel b), and the binned and corrected light curve that contains solely the targeted signals---the stellar and planetary fluxes---(panel c).

\vspace{-0.0cm}

\subsubsection{Instrumental Systematics}

The analysis method is currently capable of analyzing reduced data from \textit{Infrared Array Camera} \citep[IRAC:][]{Fazio2004} on board of  the \textit{Spitzer Space Telescope} \citep{Werner2004}. Therefore, the only instrumental systematics currently implemented relate to \textit{Spitzer}/IRAC---or are systematics with standard functional forms. Hence, applications to datasets from other facilities  may require the use of data reduced and corrected for systematics (e.g., see Section\,\ref{sec:K7b}).

IRAC instrumental systematic variations of the observed flux, such as pixel-phase or detector ramp, are well-documented \citep[e.g.,][and references therein]{D'esert2009}. 

\begin{figure}

    \hspace{-1cm}\hspace{-0.cm}\includegraphics[trim = 00mm 00mm 00mm 00mm,clip,width=17cm,height=!]{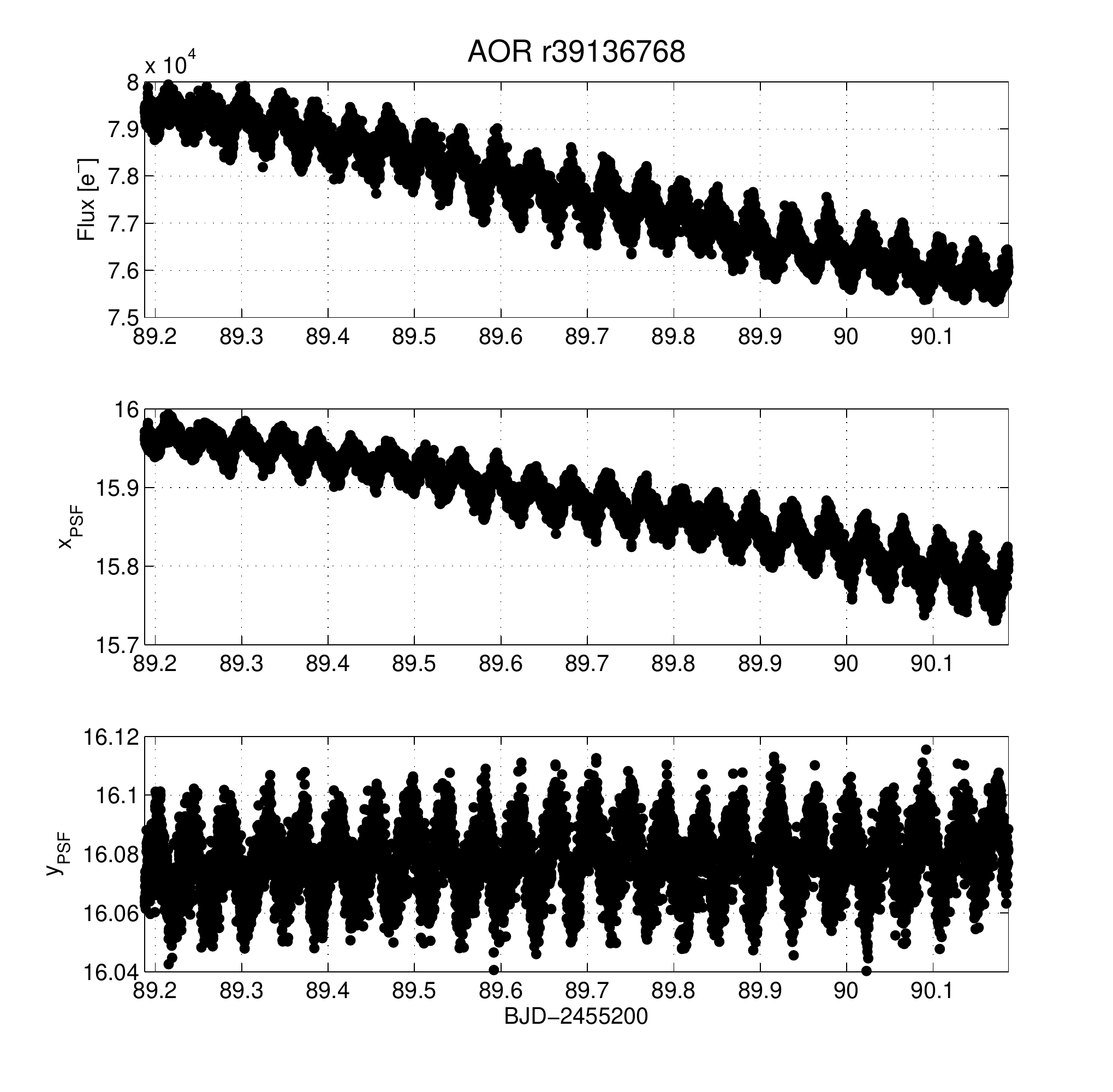}

  \vspace{-0.0cm}
  \caption[Intrapixel-sensitivity induced flux variations.]{Intrapixel-sensitivity induced flux variations (Data from \textit{Spitzer}/IRAC's AOR r39136768). \textit{Top panel:} Flux measured. \textit{Central panel:} PSF position in pixel along the x-axis of the detector. \textit{Bottom panel:} PSF position in pixel along the y-axis of the detector. The measured flux is strongly correlated with the target PSF position.}
  \label{fig:pixelphase}

\end{figure}

\textbf{Intrapixel-Sensitivity Induced Flux Variations} are due to change in position of the target's point-spread function (PSF) over a detector with non-uniform intrapixel sensitivity. Hence, these flux variations are strongly correlated with the PSF position (see Figure\,\ref{fig:pixelphase}). For IRAC, only the channels at 3.6 and 4.5 \,$\mu$m---both using $InSb$-based detectors---have shown significant effects of intrapixel-sensitivity induced flux variations. Most analysis methods perform a parametric correction based on a polynomial of the stellar centroid position \citep{Charbonneau2005,Charbonneau2008,Knutson2007,Gillon2010b}. Here, we implement the same ``pixel mapping'' method as in \cite{Lewis2013}, which was introduced in \cite{Ballard2010}. The pixel mapping method consists in deriving at each step of the MCMC the sensitivity function, $W(x_i,y_i)$---where $x_i$/$y_i$ are the stellar centroid positions in the $i^{th}$ frame---from the residuals using a Gaussian low-pass filter. In practice, the pixel map is derived from
\begin{eqnarray}
W(x_i,y_i) & = & \frac{\sum_{j\neq i}^{n_d}K_i(j) \times F_{0,j}}{\sum_{j\neq i}^{n_d}K_i(j)}
      , \label{sensitivityfunction}
\end{eqnarray}
where 
\begin{eqnarray}
K_i(j) & = & \exp{\left( -\frac{(x_j-x_i)^2}{2\sigma_x^2}-\frac{(y_j-y_i)^2}{2\sigma_y^2}\right) }
      , \label{gaussianfilter}
\end{eqnarray}
is the Gaussian low-pass filter, $F_{0,j}$ is the flux measured in the $j^{th}$ frame, and $\sigma_x$ and $\sigma_y$ are  the widths of the Gaussian filter in the x and y directions. In theory, $n_d$ is the number of frame, but we adopt here the same procedure as in \cite{Lewis2013} to prevent the pixel mapping to be computationally prohibitive---see their Appendix B. 

\textbf{Flux Ramp} is due to the time dependence of a detector response. Most IRAC's data exhibit a ramp-like behavior (see e.g., Figure\,\ref{fig:r22809088fitted}). For IRAC 8\,$\mu$m-channel---using $Si:As$-based detector, this effect is due to charge-trapping \citep{Knutson2007,Agol2010}. For IRAC 3.6 and 4.5\,$\mu$m-channels, the physical origin of the ramp is not well-documented yet, although is can be adequately correct \citep[see e.g.,][]{Lewis2013}. Our method includes several functional forms to correct for flux ramps: linear, quadratic, and logarithmic functions of time.

\subsubsection{Possible Additional Systematics}

Our method can also account for possible low-frequency noise sources (e.g., instrumental and/or stellar) with time-dependent polynomials.

\subsubsection{Correlated Noise}

We take into account the remaining correlated noise following a procedure similar to \cite{Winn2008} for obtaining reliable error bars on our parameters. For each light curve, we estimate the standard deviation of the best-fitting solution residuals for time bins ranging from 3.5 to 30 minutes in order to assess their deviation to the behavior of white noise with binning. For that purpose, the following factor $\beta_{red}$ is determined for each time bin:

\begin{eqnarray}
\beta_{red} & = & \frac{\sigma_N}{\sigma_1}\sqrt{\frac{N(M-1)}{M}}
      , \label{Bred}
\end{eqnarray}
where $N$ is the mean number of points in each bin, $M$ is the number of bins, and $\sigma_1$ and $\sigma_N$ are respectively the standard deviation of the unbinned and binned residuals. The largest value obtained with the different time bins is used to multiply the error bars of the measurements.

\subsection{Radial Velocity Measurements}
\label{sec:RVmodel}
The retrieval method also enables fitting for radial velocity (RV) measurements as they provide additional constraints that are key to mitigate the degeneracies intrinsic to eclipse mapping (see Section\,\ref{sec:dealingwithdegeneracies}). The method can fit for RV data separately, using adequate priors on the system parameters not constrained by such dataset (namely, $R_p^2/R_{\star}^2$, $b$, $\sqrt{e}\cos(\omega)$, $\sqrt{e}\sin(\omega)$, $\rho_{\star}$ and  $M_{\star}$, in our parameter set). 

The method fits for the radial velocity projection
\begin{eqnarray}
	v_r & = & V_Z + \dot{V_Z} t +K(\cos(\omega+f)+e\cos(\omega))
	,\label{eq:vit_radial_global} 
\end{eqnarray}
where $V_Z$ is the motion of the system barycenter, $\dot{V_Z}$ is the long term velocity drift---that accounts for the possible effect of an yet-undetected companion with a large orbital period, $K$ is the velocity semi-amplitude, $\omega$ is the argument of the periapsis and $f$ is the true anomaly. $K$ can be expressed in terms of the system parameters as follow
\begin{eqnarray}
K & = & \frac{M_{p}}{(M_{p}+M_{\star})^{\frac{2}{3}}}\frac{\sin(i)}{\sqrt{1-e^2}}(\frac{2\pi G}{P})^{\frac{1}{3}}
      ,\label{eq:rvsemiamplitude}
\end{eqnarray}
where $M_{p}$ and $M_{\star}$ are the planet and host-star masses respectively, $i$ is the orbital inclination, $e$ is the orbital eccentricity, $G$ is the universal gravitational constant, and $P$ is the orbital period \citep{Winn2010}. 

There are no additional ``conventional'' jump parameter required to fit RV data simultaneously to light curves. The reason is that the other additional parameters $V_Z$, $\dot{V_Z}$, and $K$ are linear parameters and can hence be derived using SVD (see Section\,\ref{sec:linearparam}). Note that $K$ leads directly to $M_p$.

The current version of our method does not fit for the Rossiter-McLaughlin effect \citep{Rossiter1924,McLaughlin1924,Queloz2000,Triaud2009}, hence measurements obtained during transit have to be discarded. 

\subsection{In Practice: Fitting Model \& Data Quality}
\label{sec:in_practice}
\paragraph*{Fitting Model}
In practice, the fitting model, $f_{sim}(\textbf{x})$, in our MCMC implementation works as follow. At each step of a chain, \textbf{(1)} the system orbits are simulated, \textbf{(2)} the different modes (stellar/planetary contributions and the systematics) that compose the simulated light curve and/or the radial velocity data are simulated and \textbf{(3)} the mode's amplitude are determined based on their simulated light curve by least squares minimization using the perturbed SVD method.

\paragraph*{Data Quality}
Eclipse mapping requires neat light-curves. The occultation shapes need to be monitored properly, both in terms of time sampling and measurement precision. We advocate that the minimum requirements to perform eclipse mapping are:
\begin{enumerate}
\item Time bins with a duration $\lesssim 5\%$ of the ingress/egress duration (i.e., $\gtrsim 20$ bins to sample the ingress/egress). 
\item A RMS per time bin $\lesssim 3\%$ of the target's occultation depth.
\end{enumerate} 
Examples of datasets fulfilling such requirements are shown in Figures\,\ref{fig:in_eg_structures} and\,\ref{fig:3d_map_spitzer_hd189733b_lc}. We show in Section\,\ref{sec:mappingpotential} that such requirements are sufficient to enable the localization of the peak in the dayside brightness distribution with an uncertainty $\lesssim3^{\circ}$ in longitude and $\lesssim$10$^{\circ}$ in latitude and a precision better than 17$^{\circ}$ on the latitudinal extent of the hot-spots and better than 8$^{\circ}$ on the longitudinal one, i.e., a precision relevant for comparisons with three-dimensional hot-Jupiter GCMs \citep{Showman2011,Heng2014}.

\newpage
\section{Application I: HD\,189733 b}

\label{sec:map18733b}

The subset of exoplanet infrared (IR) observations aiming at mapping exoplanets is growing. Currently, the \textit{Spitzer Space Telescope} \citep{Werner2004} has observed thermal phase curves for a dozen different exoplanets  as well as the IR occultations of over fourty exoplanets. Among these, HD\,189733b \citep{Bouchy2005} is arguably one of the most favorable transiting exoplanet for detailed observational atmospheric studies; in particular, because its K-dwarf host is the closest star to Earth with a transiting hot Jupiter. This means the star is bright and the eclipses are relatively deep yielding favorable signal-to-noise ratio (SNR). As such, HD\,189733b represents a ``Rosetta Stone'' for the field of exoplanetology with one of the highest SNR secondary eclipses \citep{Deming2006,Charbonneau2008}, phase-curve observations \citep{Knutson2007,Knutson2009,Knutson2012} and, consequently, numerous atmospheric observational characterizations \citep[e.g.,][]{Grillmair2007,Pont2007,Tinetti2007,Redfield2008,Swain2008,Madhusudhan2009,D'esert2009,Deroo2010,Sing2011,Gibson2011,Huitson2012}. Although HD\,189733b's atmospheric models are in qualitative agreement with observations, important discrepancies remain between simulated and observed light curves and emission spectra \citep[see e.g.,][Figures 8 and 10]{Showman2009}. In addition, discrepancies exist between several published inferences---in particular molecular detections---which emphasize the importance of data reduction and analysis procedures \citep[e.g.,][]{D'esert2009,Gibson2011}. Hence, we undertake a global analysis of all HD\,189733's public photometry obtained with the \textit{Spitzer Space Telescope} for assessing the validity of published inferences. 

In this Section, we present the first secondary eclipse scanning of an exoplanet showing a deviation from the occultation of a uniformly bright disk at the 6\,$\sigma$ level, which we obtain from the archived \textit{Spitzer}/IRAC 8-$\mu$m data of the star HD\,189733. We apply the method introduced in Section\,\ref{sec:methodo} to disentangle the possible contributing factors to the anomalous eclipse shape. As a result, we perform a new step toward mapping distant worlds by constraining consistently HD\,189733b's system parameters, shape, and brightness distribution at 8\,$\mu$m.

At the time of submission to \textit{Astronomy \& Astrophysics}, we learned about a similar study by \cite{Majeau2012}, hereafter M12, using the same data but different frameworks for data reduction and analysis. Our study differs from M12 in three main ways. First, we find a deviation from the occultation of a uniformly bright disk at the 6\,$\sigma$ level in contrast to the $\lesssim$\,3.5\,$\sigma$ level deviation in the phase-folded light curves from \cite{Agol2010} (see their Figure\,12) used in M12. Secondly, this deviation has multiple possible contributing factors and our study provides a framework for constraining consistently these contributing factors (see Section\,\ref{sec:methodo}). Thirdly, and related to the second point, we do not constrain \textit{a priori} the system parameters to the best-fit of a conventional analysis, nor the orbital eccentricity to zero; instead, we estimate the system parameters simultaneously with the BD. We compare the methodologies and the results in Section\,\ref{sec:compare_2_M12}.

This Section begins with a summary description of the {\it Spitzer} 8-$\mu$m data. In Section\,\ref{sec:data}, we present our data reduction and conventional data analysis---i.e., for which we model HD\,189733b as a uniformly bright disk. This first analysis reveals that HD\,189733b's occultation has an anomalous shape. In Section\,\ref{sec:results&discussI}, we present and assess the robustness of the anomaly in HD\,189733b's occultation. In Section\,\ref{sec:ecl}, we analyze HD\,189733b's 8-$\mu$m photometry---transits, occultations, and phase curve---to investigate the origin of HD\,189733b's occultation shape. We purposely focus on HD\,189733b's photometry to point out  the complementary insights gained from a global analysis including the RV measurements in Section\,\ref{sec:compl_analysis}. We discuss in Section\,\ref{sec:discussionII} the robustness of our results.  

The prospects of our mapping method are introduced in Chapter\,\ref{chap:perspectives}, Section\,\ref{sec:mappingpotential}.

\subsection{Data Reduction \& Conventional Analysis}
\label{sec:data}

We present here the conventional data reduction and data analysis performed for determining HD\,189733's system parameters (Equarion\,\ref{eq:conventionalparameterset}) based solely on the \textit{Infrared Array Camera} \citep[IRAC:][]{Fazio2004} 8-$\mu$m eclipse photometry. We emphasize that we simultaneously analyze the whole data set, instead of combining each separately-analyzed eclipse events. Such approach helps to mitigate the effects of noise by extracting simultaneously the information common to multiple light curves. Therefore, it also enables the detection of previously-unnoticed signals in the data. We begin with a summary description of the data sets; then, we introduce the analysis method and the physical models used for the parameter determination.

\subsubsection{Data Description and Reduction}

The eight secondary eclipses and the six transits of HD\,189733b\footnote{Data available in the form of Basic Calibrated Data (BCD) on the \textit{Infrared Science Archive} : \url{http://
sha.ipac.caltech.edu/applications/Spitzer/SHA//}} used in this study are described by
\textit{Astronomical Observation Requests} (hereafter AOR) in Table\,\ref{tab:AOR}.
The data were obtained from November 2005 to June 2008 with IRAC at 8\,$\mu$m. 
These are calibrated by the \textit{Spitzer} pipeline version S18.18.0. The S18.18.0 version enables improvements in the quality 
of data reduction over the original published data sets that used older {\it Spitzer} pipeline versions\footnote{\url{http://irsa.ipac.caltech.edu/data/SPITZER/docs/irac/iracinstrumenthandbook/73}}. Each AOR is composed of frames; each of which corresponds to 64 individual subarray images of 32x32 pixels.

\begin{table*}[!b]
	\caption{HD\,189733 b's AOR description}
	\label{tab:AOR}

	\hspace{-1.4cm}\footnotesize{\begin{tabular}{c|ccc|cc}
		\hline\hline
		\textbf{AORKEY($\mathbf{\bigoplus^{\mathrm{a}}}$)} & \textbf{PI}  & \textbf{Publication Ref.} & \textbf{Data sets$^{\mathrm{b}}$ (64x)} & \textbf{Exposure time [s]}  & \textbf{Aperture [px]}\\
		\hline
			$16343552$(O)	&	D. Charbonneau	&	\cite{Charbonneau2008}	&	$1359$ & $0.1$ & $3.6$ \\ 
			$20673792$(P)	&	D. Charbonneau	&	\cite{Knutson2007}	&	$1319$ & $0.4$ & $4.8$ \\ 
			$22808832$(O)	& & & & & \tabularnewline
			$22809088$(O)	& & & & & \tabularnewline
			$22809344$(O)	& & & & & \tabularnewline
			$22810112$(O)	& & & & & \tabularnewline
			$24537600$(O)	& & & & & \tabularnewline
			$27603456$(O)	&\multirow{-6}*{E. Agol} & \multirow{-6}*{\cite{Agol2010}} & \multirow{-6}*{$690$} & \multirow{-6}*{$0.4$}& \multirow{-6}*{$4.8$} \tabularnewline \hline
			$22807296$(T)	& & & & & \tabularnewline
			$22807552$(T)	& & & & & \tabularnewline
			$22807808$(T)	& & & & & \tabularnewline
			$24537856$(T)	& & & & & \tabularnewline
			$27603712$(T)	& & & & & \tabularnewline
			$27773440$(T)    & \multirow{-6}*{E. Agol} & \multirow{-6}*{\cite{Agol2010}} & \multirow{-6}*{$690$} & \multirow{-6}*{$0.4$}& \multirow{-6}*{$4.8$} \tabularnewline 
		\end{tabular}}	 
\begin{list}{}{}
\item[$^{\mathrm{a}}$] {AORKEY target: T, O or P respectively transit, occultation or phase curve.}
\item[$^{\mathrm{b}}$] {Present AOR are composed of data sets, each data set corresponds to 64 individual subarray images of 32x32 px.}
\end{list}

\end{table*}

The data reduction consists in converting each AOR into a light curve, hence each images into a flux measurement. For that purpose, we follow the procedure introduced in \cite{Gillon2010a}, which
is performed individually for each AOR. We first convert fluxes from \textit{Spitzer} units of specific intensity (MJy/sr) to photon counts; then, we perform aperture photometry on each image with the $\tt{IRAF}$\footnote{IRAF is distributed by the National Optical Astronomy Observatory, which is operated by the Association of Universities for Research in Astronomy, Inc., under cooperative agreement with the National Science Foundation.} $\tt{/DAOPHOT}$ software \citep{Stetson1987}. We estimate the PSF center by fitting a Gaussian profile to each image. We estimate the best aperture radius (see Table\,\ref{tab:AOR}) based on the instrument point-spread function (PSF\footnote{\url{http://ssc.spitzer.caltech.edu/irac/psf.html}}), HD189733b's, HD189733's, HD\,189733B's and the sky background flux contributions. For each image, we correct the sky background by subtracting from the measured flux its mean contribution in an annulus extending from 10 to 16 pixels from the PSF center. Then, we discard:
\begin{itemize}
\item the first 30 minutes of each AOR for allowing the detector sensitivity stabilization,
\item the few significant outliers to the bulk of the \textit{x-y} distribution of the PSF centers using a 10\,$\sigma$ median clipping,
\item and, for each subset of 64 subarray images, the few measurements with discrepant flux values, background and PSF center positions using a 10\,$\sigma$ median clipping.
\end{itemize}
Finally, the light curve corresponding to an AOR is the time series of the flux values averaged across each subset of 64 subarray images; while the photometric errors are the standard deviations on the averaged flux from each subset.

\subsubsection{Photometry Data Analysis}

See ``conventional'' model in Section\,\ref{sect:lightcurvemodel}. We use the following Gaussian prior for HD\,189733's mass: $M_{\star} = 0.84\pm0.06 M_{\odot}$ \citep{Southworth2010}. We first estimate the system parameters setting $e = 0$---similarly to \cite{Agol2010}, based on the small inferred value of $\left\lbrace e\cos \omega,e\sin \omega \right\rbrace$ and theoretical predictions advocating for HD\,189733b's orbital circularization \citep[e.g.,][]{Fabrycky2010}. We chose to perform a first analysis under the circularized-orbit assumption to point out qualitatively the influence of unassessed assumptions in the context of a correlated parameter space (see Sections\,\ref{sec:degeneracies} and\,\ref{sec:results&discussI}).

\subsection{HD\,189733b's Anomalous Occultation Shape}
\label{sec:results&discussI}

\subsubsection{Significance}

\begin{figure}

    {\hspace{-2.5cm}\includegraphics[trim = 30mm 80mm 50mm 91mm,clip,width=!,height=8cm]{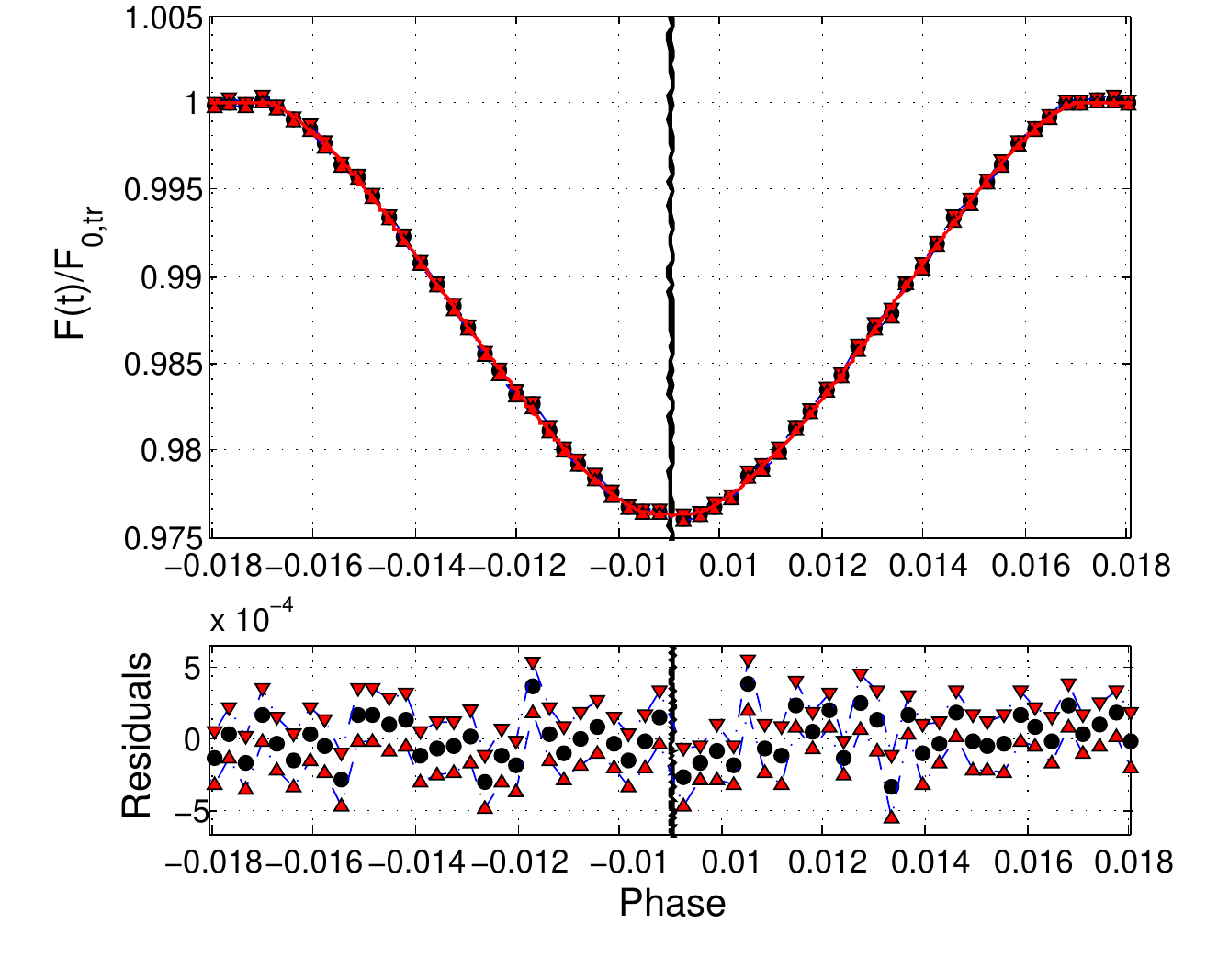}\hspace{-0.5cm}\includegraphics[trim = 30mm 80mm 50mm 91mm,clip,width=!,height=8cm]{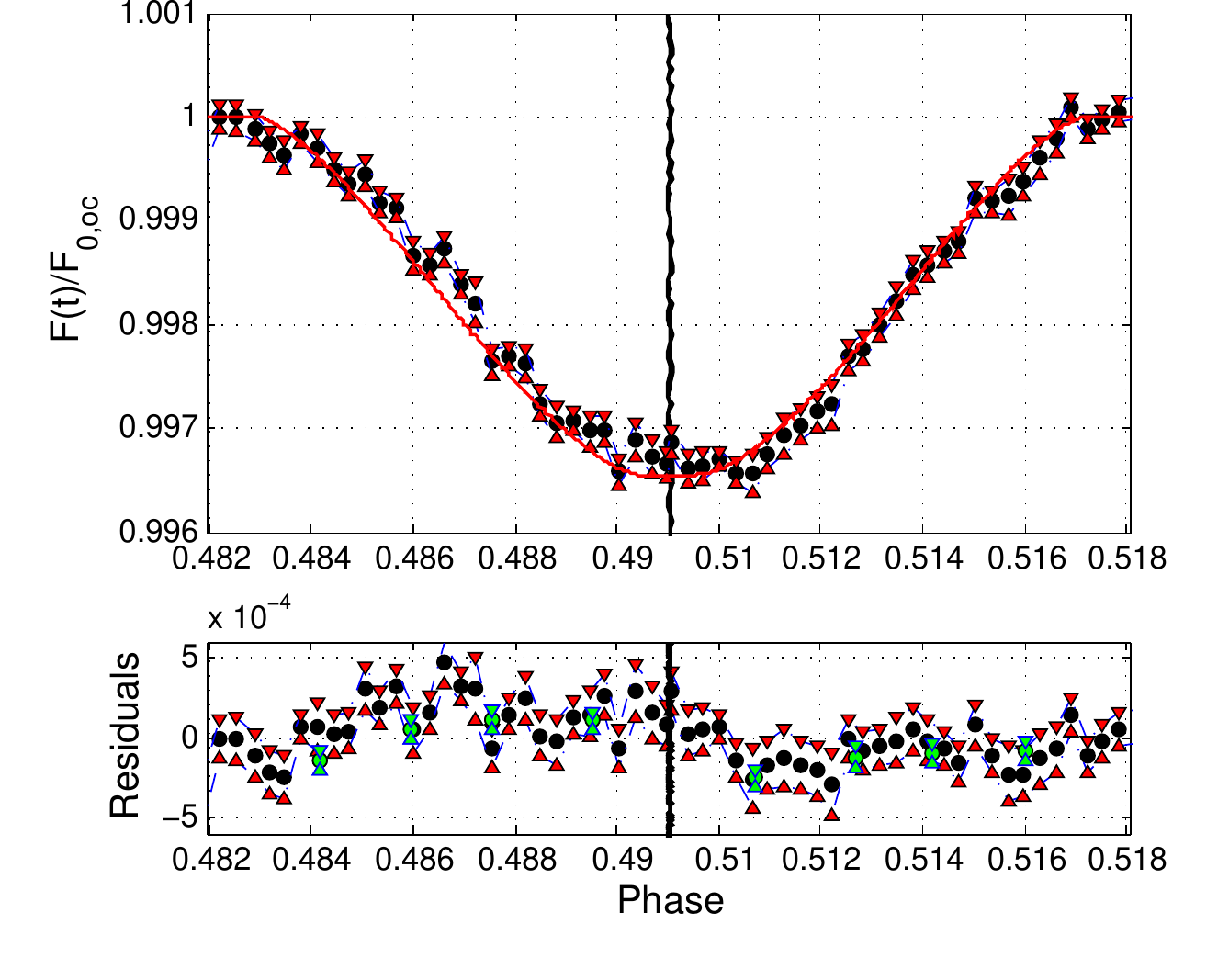}\hspace{-2.cm}}

  \vspace{-0.8cm}
  \caption[\textit{Spitzer}/IRAC 8-$\mu$m HD\,189733b's transit and occultation ingress/egress photometry]{\textit{Spitzer}/IRAC 8-$\mu$m HD\,189733b's transit and occultation ingress/egress photometry binned per 1 minute and corrected for the systematics (black dots) with their 1\,$\sigma$ error bars (red triangles) and the best-fitting eclipse model superimposed (in red). The green dots present the residuals from \cite{Agol2010} (their Figure 12), obtained using an average of the best-fit models from 7 individual eclipse analyses. \textit{Left:} Phase-folded transits show no significant deviation to the transit of a disk during ingress/egress. \textit{Right:} Phase-folded occultation ingress/egress deviate from the eclipse of a uniformly bright disk, highlighting the secondary eclipse scanning of HD\,189733b's dayside. \textit{Top:} Phase-folded and corrected eclipse photometry. \textit{Bottom:} Phase-folded residuals.}
  \label{fig:in_eg_structures}
\end{figure}

We detect an anomalous shape for HD\,189733b's occultation (see Figure\,\ref{fig:in_eg_structures}, bottom-right panel). In particular, we find that its occulation deviate from the one of a uniformly bright disk with a 6.2\,$\sigma$ significance\footnote{We determine the significance of this structure as
 $\sqrt{\sum_{i \in ingress} Y_i/\sigma_i - \sum_{i \in egress} Y_i/\sigma_i } $, where $Y_i$ and $\sigma_i$ are the flux measurement residual and its standard deviation at time $i$.}. 
 
 The IRAC 8-$\mu$m photometry of the eclipse ingress/egress corrected for the systematics and binned per 1 minute are shown in Figure\,\ref{fig:in_eg_structures}, with the best-fitting eclipse model superimposed (solid red line). The error bars (red triangles) are rescaled by $\beta_{red}$ ($\sim$1.2) to take into account the correlated noise effects on our detection. In addition, we take advantage of the MCMC framework to account for the uncertainty induced by the systematic correction on the phase-folded light curves. For that purpose, we use the posterior distribution of the accepted baselines to estimate their median instead of using the best-fit model---which has no statistical significance. We also propagate the uncertainty of the systematic correction using the standard deviation of each bin from the baseline posterior distribution---errorbars increased by up to 20\%.
 
 We include in Figure\,\ref{fig:in_eg_structures} the residuals from \cite{Agol2010} (their Figure 12). These are obtained using an average of the best-fit models from 7 individual eclipse analyses---not from a simultaneous analysis of all eclipses. For further comparison, the structure detected in HD\,189733b's occultation can be translates into a uniform time offset of 37$\pm$6 sec ($\sim$6\,$\sigma$), in agreement with \cite{Agol2010} estimate of 38$\pm$11 sec ($\sim$3.5\,$\sigma$), light travel time deduced.

\subsubsection{Robustness}

We test the robustness of our results against various effects including the baseline models, the limb darkening, HD\,189733b's circularized-orbit assumption and the AORs. Although below we mainly discuss the robustness of the structure in occultation, the robustness of system parameter estimates is assessed simultaneously, but tackled in detail in Section\,\ref{sec:uniform}.

As mentioned in Section\,\ref{sec:systematics} , \cite{Agol2010} advocate using a double-exponential for modeling the detector ramp; but we use the quadratic function of $\log(dt)$ introduced by \cite{Charbonneau2008}. The reason is that by taking advantage of our Bayesian framework we show that the quadratic function of $\log(dt)$ is the most adequate for correcting the present AORs. In particular, we use two different information criteria\footnote{We use as information criteria the BIC and the AIC \citep[see, e.g.,][]{Gelman2004}. The penalty term for additional model parameters is larger in the BIC than in the AIC.} that prevent from overfitting based on the likelihood function and on a penalty term related to the number of parameters in the fitting model.  We obtain both a higher BIC ($\Delta$\,BIC $\sim$90) and a higher AIC ($\Delta$\,AIC $\sim$1.3) with the double-exponential; this means the additional parameters do not improve the fit enough, according to both criteria. Hence, the most adequate ramp model to correct these AORs is the less complex quadratic function of $\log(dt)$. Nevertheless, we assess the robustness of our results to different baseline models including the double-exponential ramp, phase-pixel corrections, sinusoidal terms, and polynomials. These different MCMC simulations do not significantly affect the anomalous shape found in occultation ingress/egress to within  0.5\,$\sigma$. The reason is that the typical time scale of the baseline models is much larger than the timescale of the structures detected in occultation ingress/egress. Therefore, baseline models have a mitigated influence on the ingress/egress shape (and hence on the retrieved maps), and \textit{vice versa}. 

We assess the influence of the priors on the limb-darkening coefficients; for that purpose, we perform MCMC simulations with no priors on $u_1$ and $u_2$ (see Section\,\ref{sec:systematics}). Again, we observe no significant influence to within 0.5\,$\sigma$. We emphasize the necessity of precise and independent constraints on the limb-darkening coefficient, even if these might appear to be well-constrained by high SNR transit photometry.

We assess the influence of assuming HD\,189733b's orbital circularization. For that purpose, we perform MCMC simulations with a free eccentricity. We observe no significant influence to within 0.5\,$\sigma$ for the jump parameters, except for the stellar density, the impact parameter, $\sqrt{e}\cos \omega$, and $\sqrt{e}\sin \omega$ (see Section\,\ref{sec:degeneracies}). In particular, we observe a net drop of the anomalous shape significance because of the potential for $\sqrt{e}\cos \omega$ and $\sqrt{e}\sin \omega$ to compensate partially for deviations in ingress/egress.

Finally, we also validate the independence of our results to the set of AORs included by analyzing different subsets of seven out of the eight eclipses. We observe relative significance decrease of $\sim$\,$ \sqrt{7/8} $; which are consistent with a significance drop due to a reduction of the sample size. In particular, it shows that the anomalous shape in occultation ingress/egress is not due to one specific AOR. 

\subsubsection{Possible Contributing Factors: HD\,189733b's shape}

HD\,189733b's occultation shape is in agreement with the expected signature of the offset hot spot indicated by HD\,189733b's phase curve \citep{Knutson2007}. However, it could also be due to HD\,189733b's shape or biased estimates of $b$ and $\rho_\star$ resulting from the circularized-orbit assumption (see Section\,\ref{sec:degeneracies}).

 We can assume that HD\,189733b's projection at conjunctions is a disk for two reasons. \textbf{(1)} The transit residuals show no anomalous structure (Figure\,\ref{fig:in_eg_structures}, bottom-left panel). This is in agreement with current constraints on the HD\,189733b oblateness \citep[projected oblateness below 0.056, $95\%$ confidence,][]{Carter2010} and wind-driven shape \citep[expected to introduce a light curve deviation below 10\,ppm, see][]{Barne2009}. In particular, our transit residuals constrain HD\,189733b's oblateness below 0.0267 in case of a projected obliquity of 45$^\circ$ and below 0.147 in case of a projected obliquity of 0$^\circ$---$95\%$ confidence, Figure 1 of \cite{Carter2010}. No significant deviation in transit ingress/egress means that HD\,189733b's shape-induced effects in occultation ingress/egress are negligible. Indeed, shape-induced effect in occultation ingress/egress are expected to be about one order of magnitude lower than in transit (effect ratio $\varpropto <I_p>/<I_{\star}>$, see Section\,\ref{sec:dealingwithdegeneracies}). \textbf{(2)} HD\,189733b's phase curve shows mainly a $ P $-modulation---while shape-induced modulation are expected to have a period of $ P/2 $; what is in agreement with our previous constraint on HD\,189733b's shape. We therefore assume HD\,189733b to be spherical, for a further analysis of the present data.

\subsection{Non-Conventionnal Analysis of HD\,189733b's Photometry} 
\label{sec:ecl}

Our second analysis aims to investigate the origin of HD\,189733b's occultation shape using its IRAC 8-$\mu$m photometry---transits, occultations, and phase curve. We discuss in Section\,\ref{sec:dealingwithdegeneracies} the need for global analyses to consistently probe the parameter space while relaxing unverified assumptions possibly biasing the parameter estimates. However, we purposely focus the following analysis on HD\,189733b's photometry to emphasize latter the complementary insights gained while performing a global analysis including the RV measurements.

\subsubsection{HD\,189733b's Photometric Data}

In this second analysis, we use HD\,189733b's corrected and phase-folded eclipses from Section\,\ref{sec:results&discussI} (Figure\,\ref{fig:in_eg_structures}) and its phase curve corrected for stellar variability \citep[][Figure 11]{Agol2010}. As discussed in Section\,\ref{sec:results&discussI}, the transit and the phase curve enable to constrain respectively HD\,189733b's shape at conjunctions and its brightness distribution. Note that we discard the first third of the phase curve since it is strongly affected by the detector stability.

To constrain the system parameters, we choose to use HD\,189733b's phase-folded transits instead of using the system parameter estimates from Section\,\ref{sec:data} (see Table\,\ref{tab:BFP}, column 2) for two reasons. First, these estimates are under the form of 1D Gaussian distribution while light curves provide a complex posterior probability distribution (PPD) over the whole parameter space. Secondly, these estimates are affected by HD\,189733b's circularized-orbit assumption (see implications in Section\,\ref{sec:results}). For those reasons, it is relevant to simultaneously analyse  HD\,189733b's scans and estimate the system parameters, to be consistent with our methodology of a ``global'' analysis and to avoid propagation of bias through inadequate priors, i.e., inadequate assumptions (Section\,\ref{sec:dealingwithdegeneracies}).

\subsubsection{Analysis Method}

See Section\,\ref{sec:analysismethod} and note that because we use the phase-folded light curves, the main constraint on the orbital period is missing. We, therefore, use a uniform prior on $P$ based on the conventional-analysis PPD for $P$. In particular, we use a large prior (with an arbitrary symmetric extension of 10\,$\sigma_P$, where $\sigma_P$ is the estimated uncertainty on $P$, see Table\,\ref{tab:BFP}) to prevent our results to be affected by the assumption underlying this estimate. Note however that this has no influence on our further results---the reason is that the orbital period is highly-constrained by the transit epochs and, therefore, is not affected by secondary effects on the occultation shape.

\subsubsection{Results}
\label{sec:results}

For HD\,189733's system, we find that relaxing the eccentricity constraint and using more complex brightness distributions (BDs) lead to lower
stellar---and, hence, planetary---density and a more localized and latitudinally-shifted hot spot. We find that the more complex HD\,189733b's brightness model, the larger the eccentricity, the lower the densities, the larger the impact parameter and the more localized and latitudinally-shifted the hot spot estimated. As discussed in Section\,\ref{sec:degeneracies}, the correlation between the planet brightness(/shape) and the system parameters $ e $-$ \omega $-$ b $-$\rho_{\star}$ is of primary importance for data of sufficient quality. We present in this section our results for increasing model complexity to gain insight into the influence of the model underlying assumptions---e.g., the circularized-orbit assumption and the uniformly-bright exoplanet assumption.

We gather the system parameter estimates for different fitting models in Table\,\ref{tab:BFP}; it shows the median values and the $68\%$ probability interval for our jump parameters. We compute our estimates based on the posterior probability distribution (PPD) of global MCMC simulations, i.e., not as a weighted mean of individual transit or eclipse analyses. Our conventional analysis estimates are in good agreement with previous studies \citep[e.g.,][]{Winn2007,Triaud2009,Agol2010}. We discuss further the system parameter estimates obtained from our second analysis. 

\begin{table}
\caption[HD\,189733's system parameters.]{Fit properties for different fitting models of HD\,189733's photometry in the \textit{Spitzer}/IRAC 8-$\mu $m channel \label{tab:BFP}}
	\setlength{\extrarowheight}{4pt}
	\hspace{-2.2cm}\footnotesize{\begin{tabular}{c|c c|c c|c c}
	
	\hline\hline
	\multirow{2}{*}{\textbf{Parameters (units)}} & \multicolumn{2}{c}{\textbf{Uniform brightness}} \vline & \multicolumn{2}{c}{\textbf{Unipolar brightness}} \vline & \multicolumn{2}{c}{\textbf{Multipolar brightness}}\\
	  & $e = 0$ & $e$ free & $\Gamma_{SH,1}$ & $\Gamma_{2}$ & $\Gamma_{SH,2}$ & $\Gamma_{SH,3}$\\
	\hline

			$b (R_\star)$	&	$0.6576\pm^{0.0021}_{0.0021}$ & $0.6579\pm^{0.0021}_{0.0024}$ & $0.6598\pm^{0.0038}_{0.0024}$ & $0.6719\pm^{0.0063}_{0.0072}$ & $0.6609\pm^{0.0059}_{0.0031}$ & $0.6683\pm^{0.0071}_{0.0074}$
			  \\
			$\sqrt{e}\cos\omega$	& - &  $0.0043\pm^{0.0054}_{0.0027}$ & $0.0007\pm^{0.0032}_{0.0019}$ & $-0.0002\pm^{0.0008}_{0.0006}$ & $0.0001\pm^{0.0019}_{0.0018}$ & $0.0004\pm^{0.0013}_{0.0010}$
			 \\
			$\sqrt{e}\sin\omega$	& - & $-0.008\pm^{0.032}_{0.045}$ &  $0.016\pm^{0.066}_{0.034}$  &  $0.142\pm^{0.029}_{0.046}$  &  $0.046\pm^{0.065}_{0.055}$  & $0.121\pm^{0.036}_{0.064}$
			\\
			$\rho_\star (\rho_\odot)$	& $1.916\pm^{0.015}_{0.016}$ &			$1.932\pm^{0.018}_{0.015}$ &			$1.918\pm^{0.018}_{0.031}$ &			$1.816\pm^{0.058}_{0.048}$ &			$1.909\pm^{0.023}_{0.051}$ &			$1.845\pm^{0.061}_{0.058}$   
			\\
			\hline
			$\Delta$\,BIC / $\Delta$\,AIC &	$0 / 0$ &	$0.9 / -0.4$ & $-170.1 / -173.3$ & $-167.2 / -172.3$ & $-167.4 / -172.0$ & $-170.0 / -175.8$  \\
			\hline
			\hline
			\multicolumn{7}{c}{$R_p^2/R_\star^2 = 0.024068\pm^{0.000049}_{0.000049}$ ; $P (days) = 2.2185744\pm^{0.0000003}_{0.0000003}$ ;}\\
			\multicolumn{7}{c}{$T_0 (hjd-2453980) = 8.803352\pm^{0.000058}_{0.000061}$ and $F_p/F_\star|_{8\mu m} = 0.0034117\pm^{0.000037}_{0.000037}$}

	\end{tabular}}
	\vspace{-0.3cm}
\end{table}

\paragraph{Assuming HD\,189733b to Be Uniformly Bright}
\label{sec:uniform}

 We show here the similar effects of the circularized-orbit assumption and the uniformly-bright exoplanet assumption. Both assumptions prevent from exploring dimensions of the parameter space and, therefore, lead to more localized, and possibly biased, PPD.
 
\begin{figure}[!p]
   
     \centering
  
  \subfloat[]{\hspace{-08mm}\label{fig:uniform_e_ddp}\includegraphics[trim = 35mm 85mm 35mm 85mm,clip,width=0.55\textwidth,height=!]{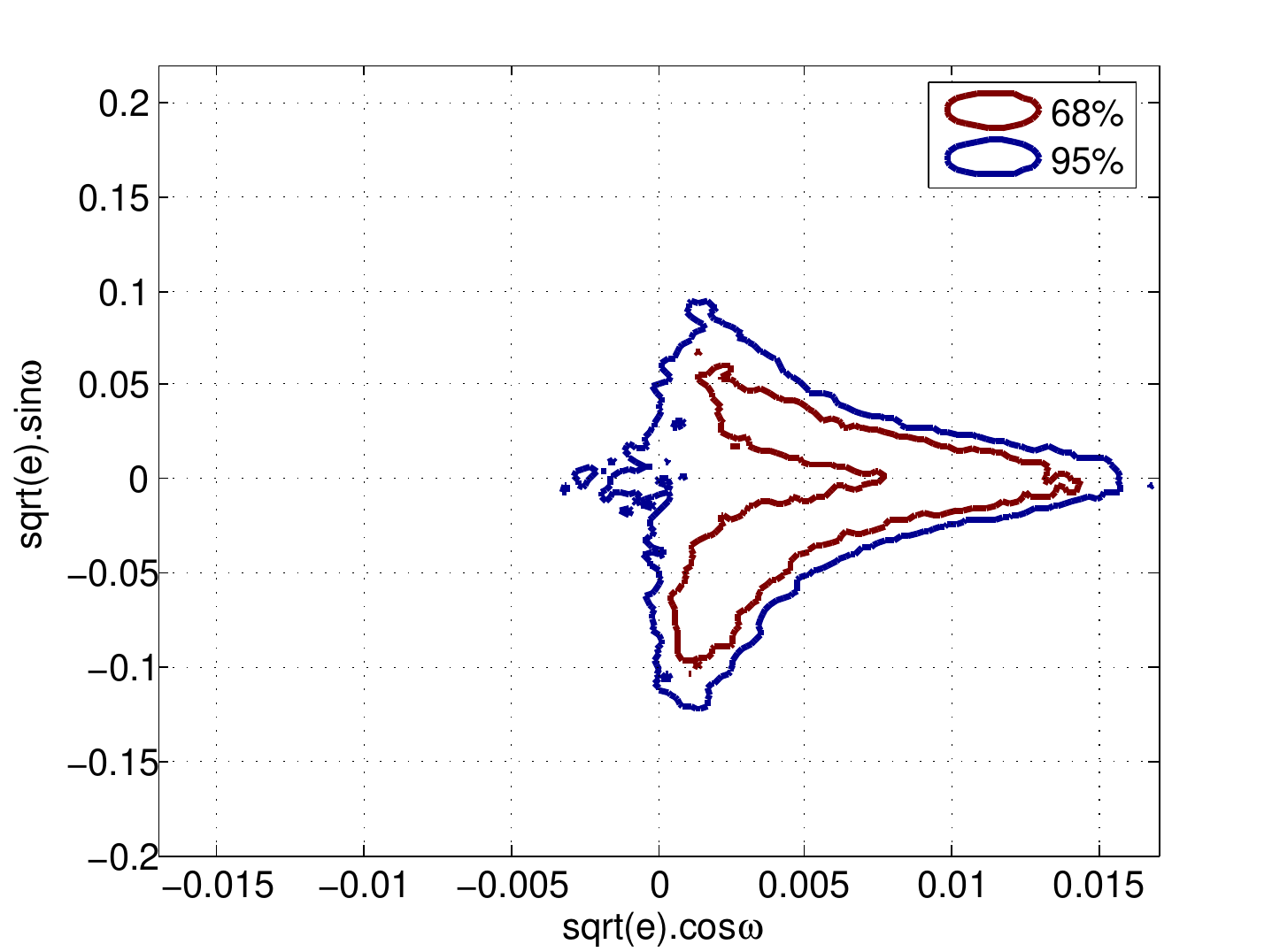}}
  ~
  \subfloat[]{\label{fig:uniform_brho_ddps}\includegraphics[trim = 35mm 85mm 35mm 85mm,clip,width=0.55\textwidth,height=!]{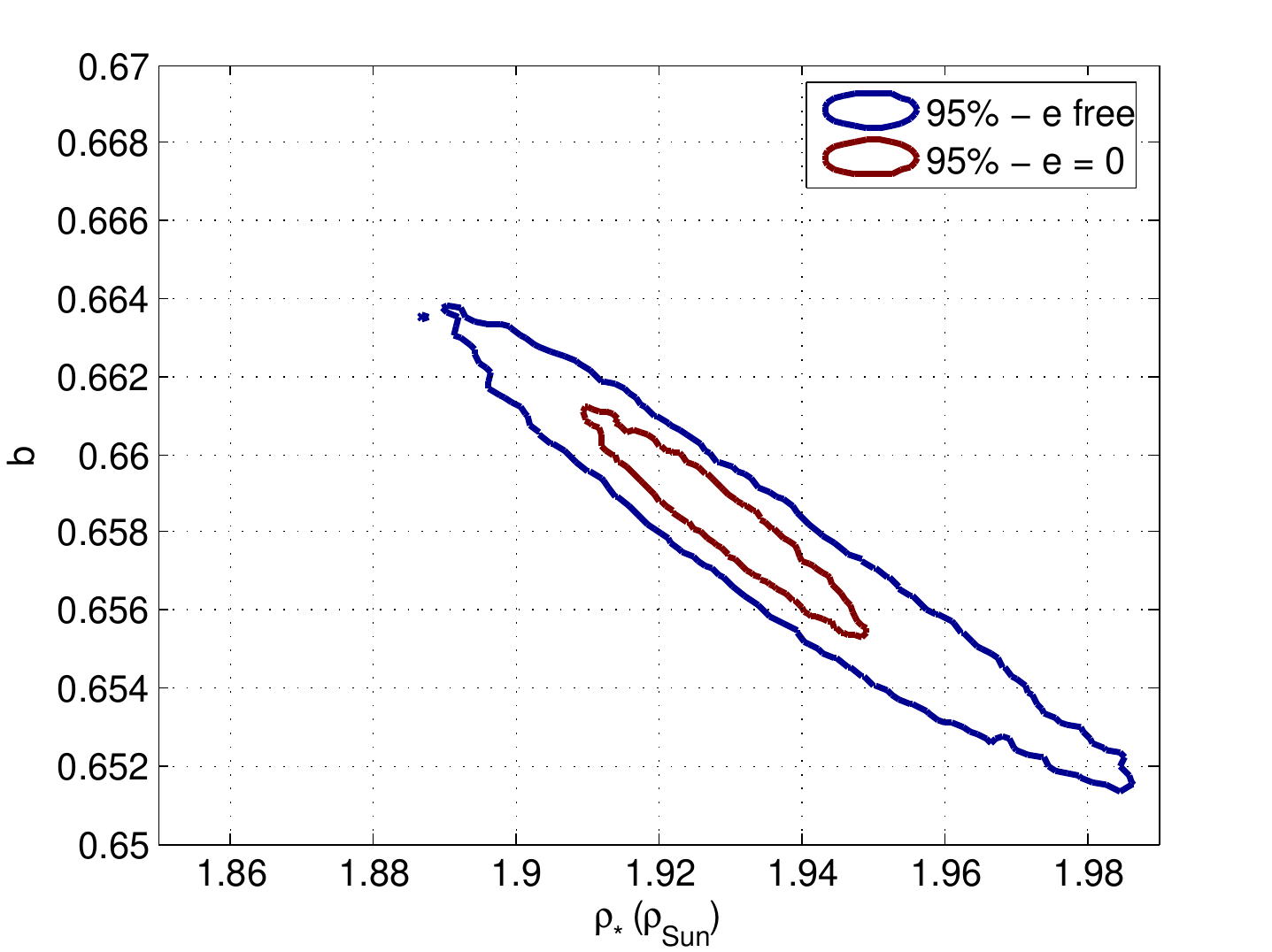}}
  
  \caption[Effect of assuming an exoplanet to be uniformly bright and its orbit to be circularized.]{Effect of assuming an exoplanet to be uniformly bright and its orbit to be circularized. \subref{fig:uniform_e_ddp} Marginal posterior probability distribution (PPD, $68\%$- and $95\%$-confidence intervals) of $\sqrt{e}\cos\omega$ and $\sqrt{e}\sin\omega$ that shows an unusual correlation. \subref{fig:uniform_brho_ddps} Marginal PPD ($95\%$-confidence intervals) of $\rho_{\star}$ and $b$ that highlights the increase of adequate solutions enabled by the additional dimensions of the parameter space probed when relaxing the circularized orbit assumption.}
  \label{fig:correl_uniform}
\end{figure}

 We first present an unusual correlation between $\sqrt{e}\cos \omega$ and $\sqrt{e}\sin \omega$ (see Figure\,\ref{fig:uniform_e_ddp}).
It emerges from the partial compensation of the anomalous occultation enabled primarily by adequate combinations of $\sqrt{e}\cos \omega$ and $\sqrt{e}\sin \omega$. As introduced in Section\,\ref{sec:degeneracies}, these parameters enable both to shift the occultation and change its duration (see Equations\,\ref{DeltaToc} and\,\ref{Durations}, respectively).

\begin{figure}[p]
   \centering
   
      \vspace{-1cm}\includegraphics[trim = 35mm 85mm 35mm 85mm,clip,width=0.65\textwidth,height=!]{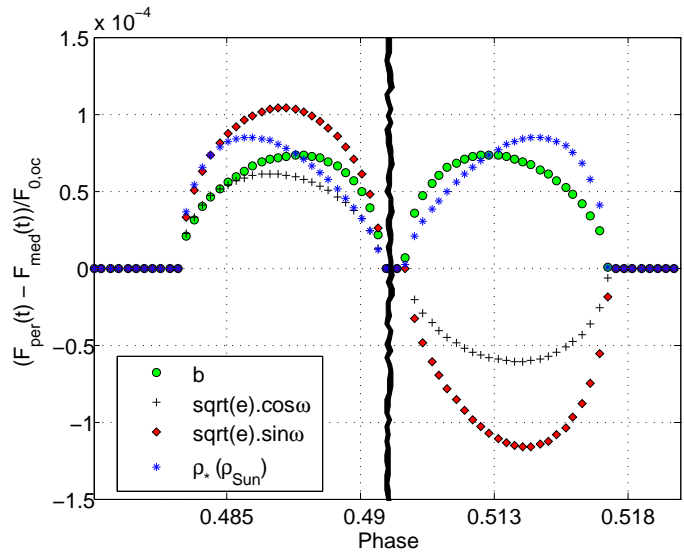}

  \caption[Deviations in occultation ingress/egress from the median-fit model for the individual perturbations of $b$, $\sqrt{e}\cos\omega$, $\sqrt{e}\sin\omega$ and $\rho_{\star}$.]{Deviations in occultation ingress/egress from the median-fit model for the individual perturbations of $b$, $\sqrt{e}\cos\omega$, $\sqrt{e}\sin\omega$ and $\rho_{\star}$ by their estimated uncertainty (see Table\,\ref{tab:BFP}, column 2). It outlines that the system parameters enable compensation of an anomalous occultation that emerges from, e.g., a non-uniformly-bright exoplanet (see Figure\,\ref{fig:shape_brightness}) possibly leading to biased estimates of the system parameters (Section\,\ref{sec:degeneracies}).}
  \label{fig:Impact_syst_param}
\end{figure}

 Similarly, $\rho_{\star}$ and $b$ are also affected by the occultation shape. We emphasize this point in Figure\,\ref{fig:Impact_syst_param} (to be compare with Figure\,\ref{fig:deviation_in_ingress_egress}). Figure\,\ref{fig:Impact_syst_param} shows the deviations in occultation ingress/egress induced by individual perturbations of $b$, $\sqrt{e}\cos\omega$, $\sqrt{e}\sin\omega$ and $\rho_{\star}$ by their estimated uncertainty (see Table.\,\ref{tab:BFP}, column 3). Because these parameters affect an occulation shape, there exist adequate combinations of these parameters that can mimic an anomalous occultation within the measurement precision (see Figure\,\ref{fig:in_eg_structures}, bottom-right panel, and Figure\,\ref{fig:Impact_syst_param}). Therefore, the circularized-orbit assumption and the uniformly-bright exoplanet assumption also affect the marginal PPD of $\left\lbrace\rho_{\star},b\right\rbrace$ by inhibiting the exploration of dimensions of the parameter space. We present in Figure\,\ref{fig:uniform_brho_ddps} the extension of the $\left\lbrace\rho_{\star},b\right\rbrace$ marginal PPD that results from relaxing the circularized-orbit assumption. Similarly, we expect that the relaxation of the uniformly-bright exoplanet assumption would significantly affect the system-parameter PPD---compare Figures\,\ref{fig:shape_brightness} and \,\ref{fig:Impact_syst_param}. That is why we advocate in Section\,\ref{sec:dealingwithdegeneracies} for performing global analyses that constrain simultaneously the possible contributing factors to the shape of an occultation and, thus, prevent the influence of unverified assumptions.

 \paragraph{Assuming HD\,189733b to Be Non-Uniformly Bright}
 
\begin{figure}
   
    \hspace{-1.5cm}\includegraphics[angle = -90, trim = 40mm 00mm 40mm 00mm,clip,width=18cm,height=!]{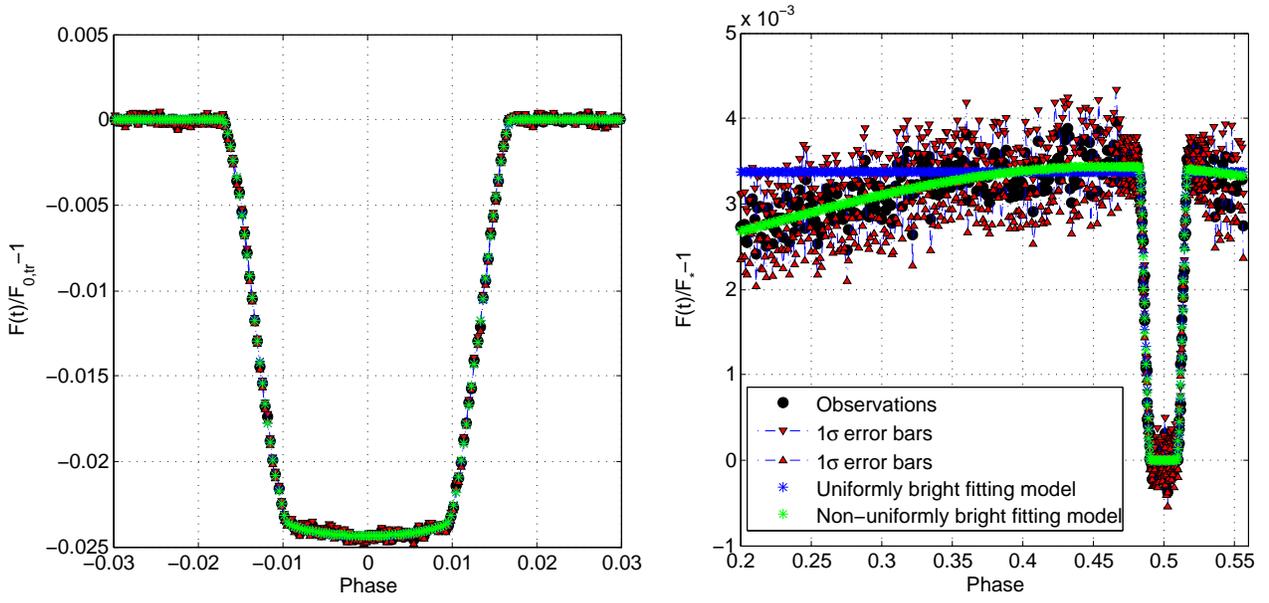}
  \vspace{-0.5cm}
  \caption[Phase-folded IRAC 8-$\mu$m HD\,189733b's photometry.]{Phase-folded IRAC 8-$\mu$m HD\,189733b's photometry binned per 1 minute and corrected for the systematics (black dots) with their 1\,$\sigma$ error bars (red triangles) and the best-fitting eclipse models superimposed. \textit{Left:} Phase-folded transits. \textit{Right:} Phase curve and phase-folded occultations that show the benefit of using non-uniform brightness model.}
  \label{fig:fits}

\end{figure}

 We emphasize here the effect of relaxing the uniformly-bright exoplanet assumption on the fit improvement and the system-parameter PPD. We present in Figure\,\ref{fig:fits} the phase-folded IRAC 8-$\mu$m photometry of HD\,189733b, corrected for the systematics with the best-fitting eclipse models for a uniformly (blue) and a non-uniformly (green) bright exoplanet superimposed. The best-fitting non-uniformly bright eclipse model is shown for the $\Gamma_{SH,1}$ model---chosen arbitrarily, as the non-uniformly-bright models provide similar fits (see Table\,\ref{tab:BFP}). In particular, these models are significantly more adequate according to both the BIC and the AIC, see Table\,\ref{tab:BFP} (odds ratio: $\sim$\,$10^{36}$). 
 
 First, we introduce the results obtained using unipolar (i.e., with one spot on the planetary dayside) BDs to gain insight into the influence of relaxing the uniformly-bright exoplanet assumption. Then, we introduce the results obtained using multipolar BDs to assess the validity of trends observed in the unipolar-model results.

 \paragraph*{Unipolar Brightness Distribution:}
 \label{sec:unipolar}
		
\begin{figure}
  \centering
  
  \subfloat[]{\hspace{-08mm}\label{fig:101_e_ddp}\includegraphics[trim = 35mm 85mm 35mm 92mm,clip,width=0.55\textwidth,height=!]{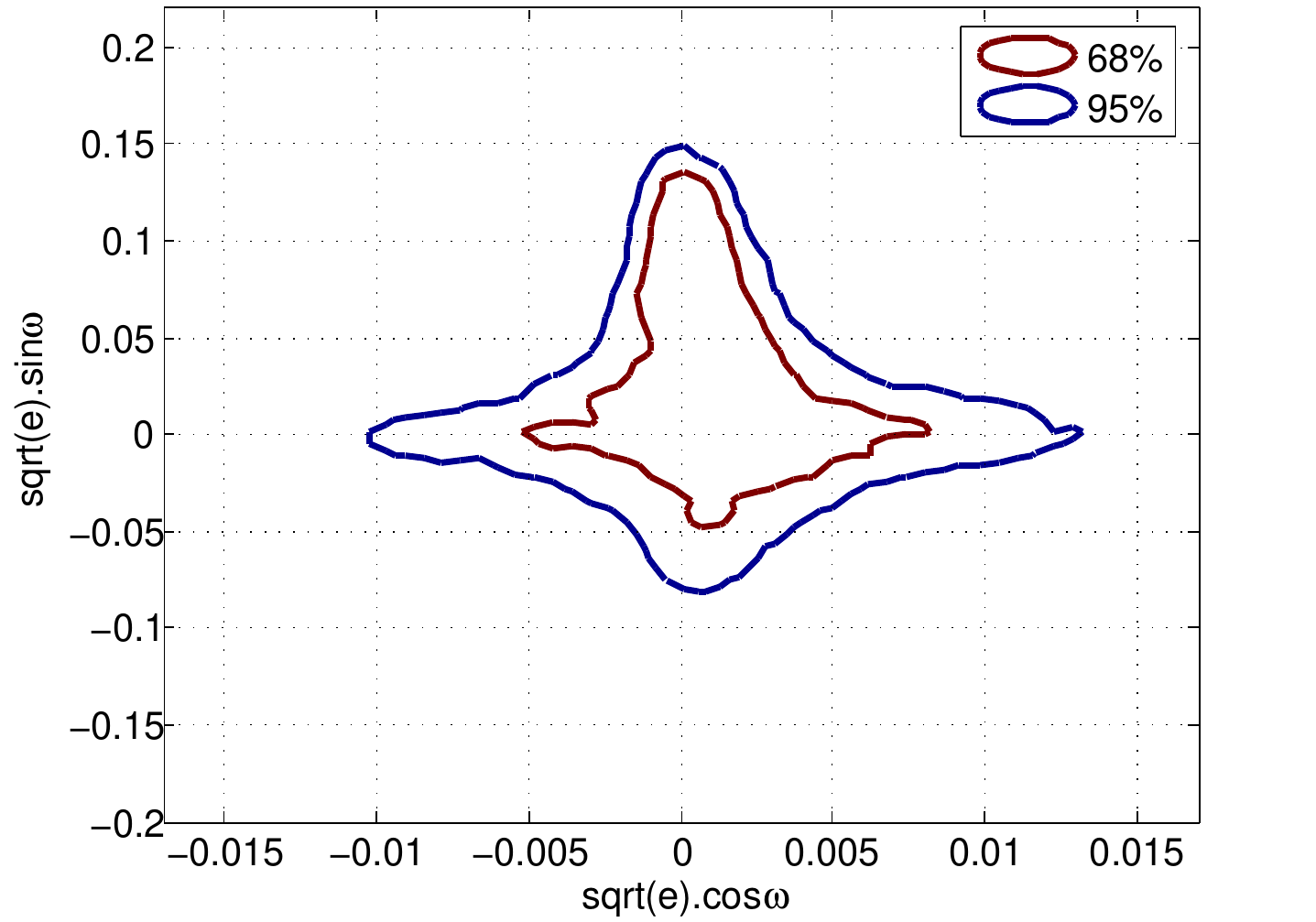}}
  ~ 
  \subfloat[]{\label{fig:3_e_ddp}\includegraphics[trim = 35mm 85mm 35mm 92mm,clip,width=0.55\textwidth,height=!]{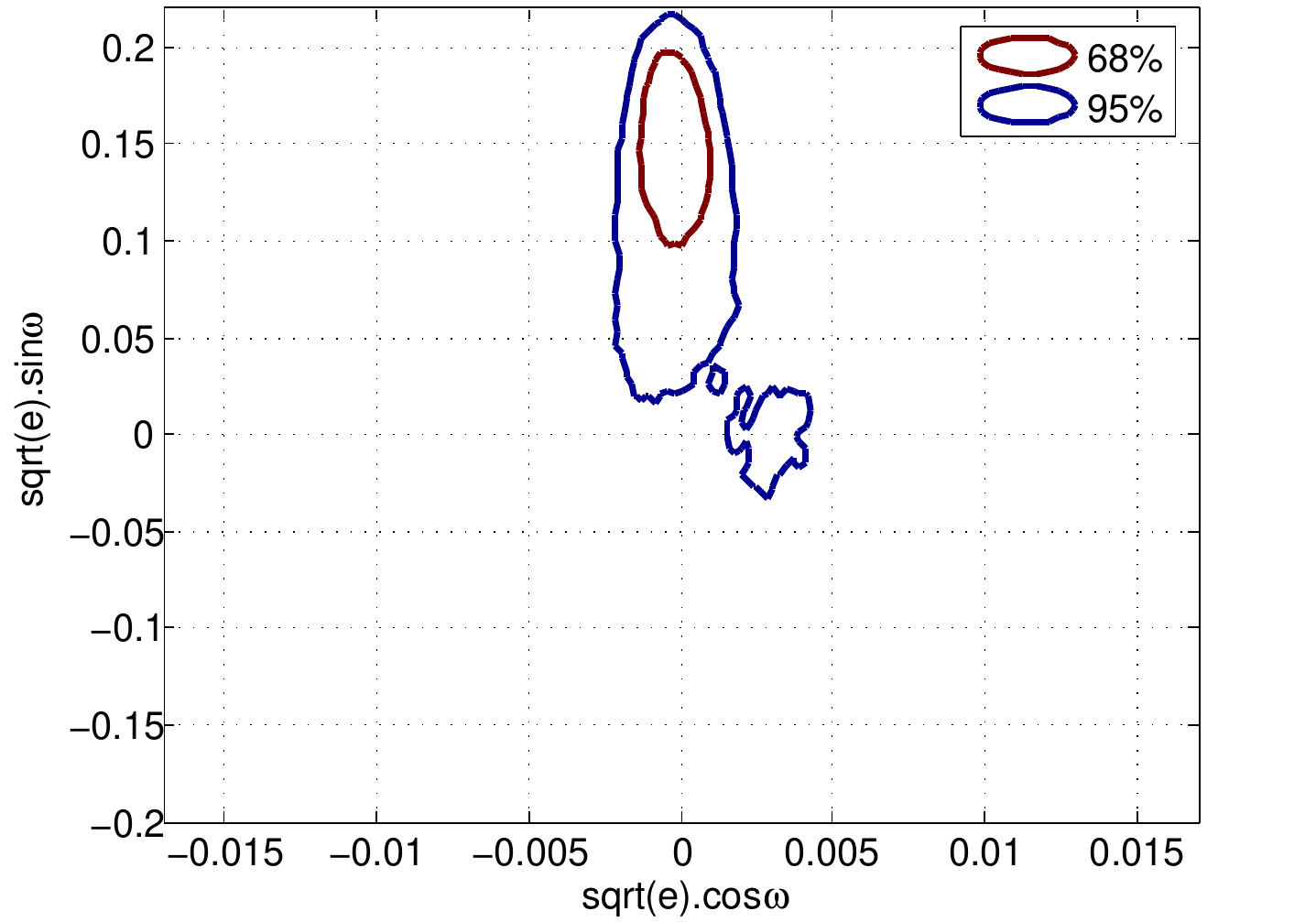}}

   \subfloat[]{\hspace{-08mm}\label{fig:101_erho_ddps}\includegraphics[trim = 35mm 85mm 35mm 92mm,clip,width=0.55\textwidth,height=!]{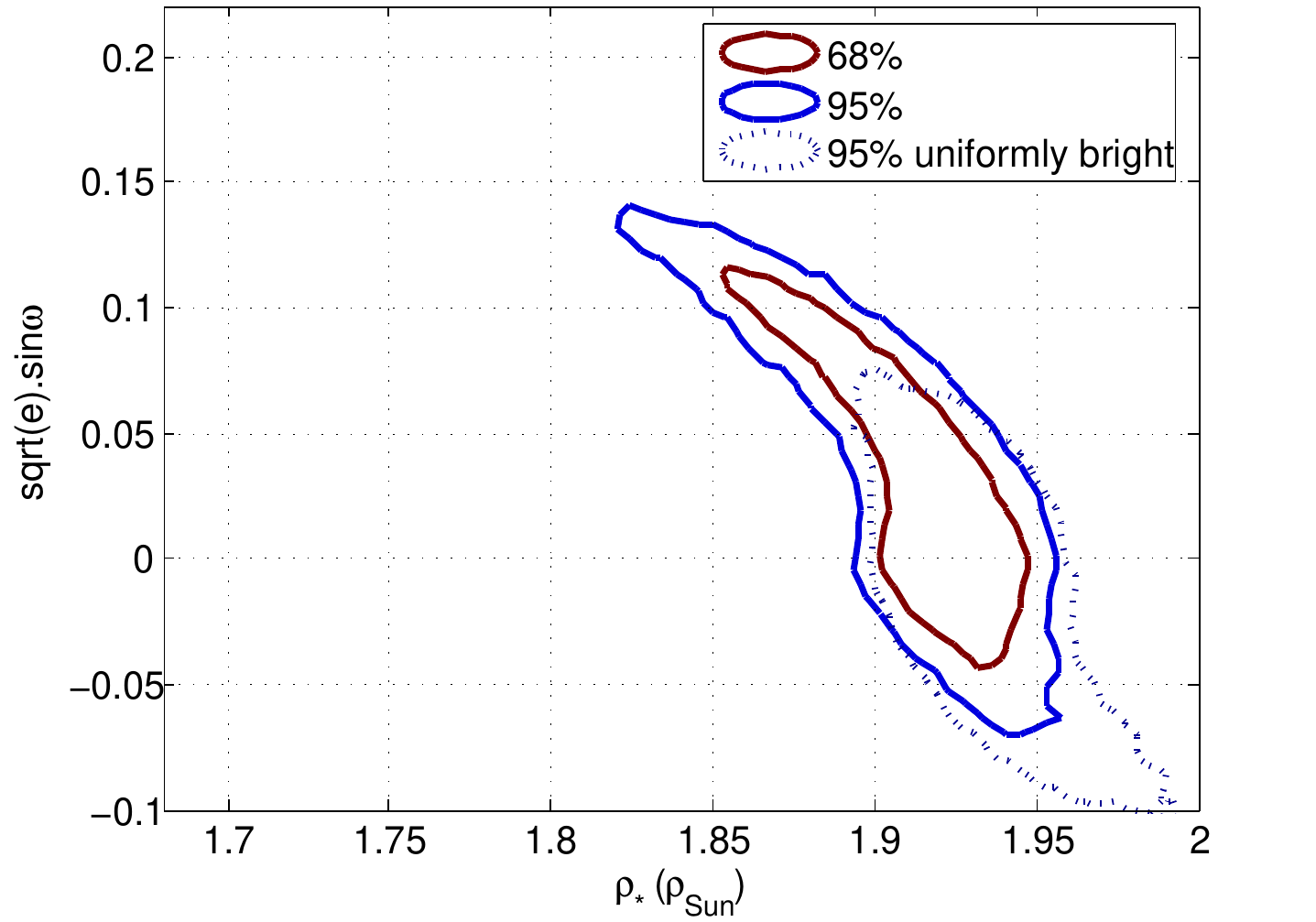}}
  ~  
  \subfloat[]{\label{fig:3_erho_ddps}\includegraphics[trim = 35mm 85mm 35mm 92mm,clip,width=0.55\textwidth,height=!]{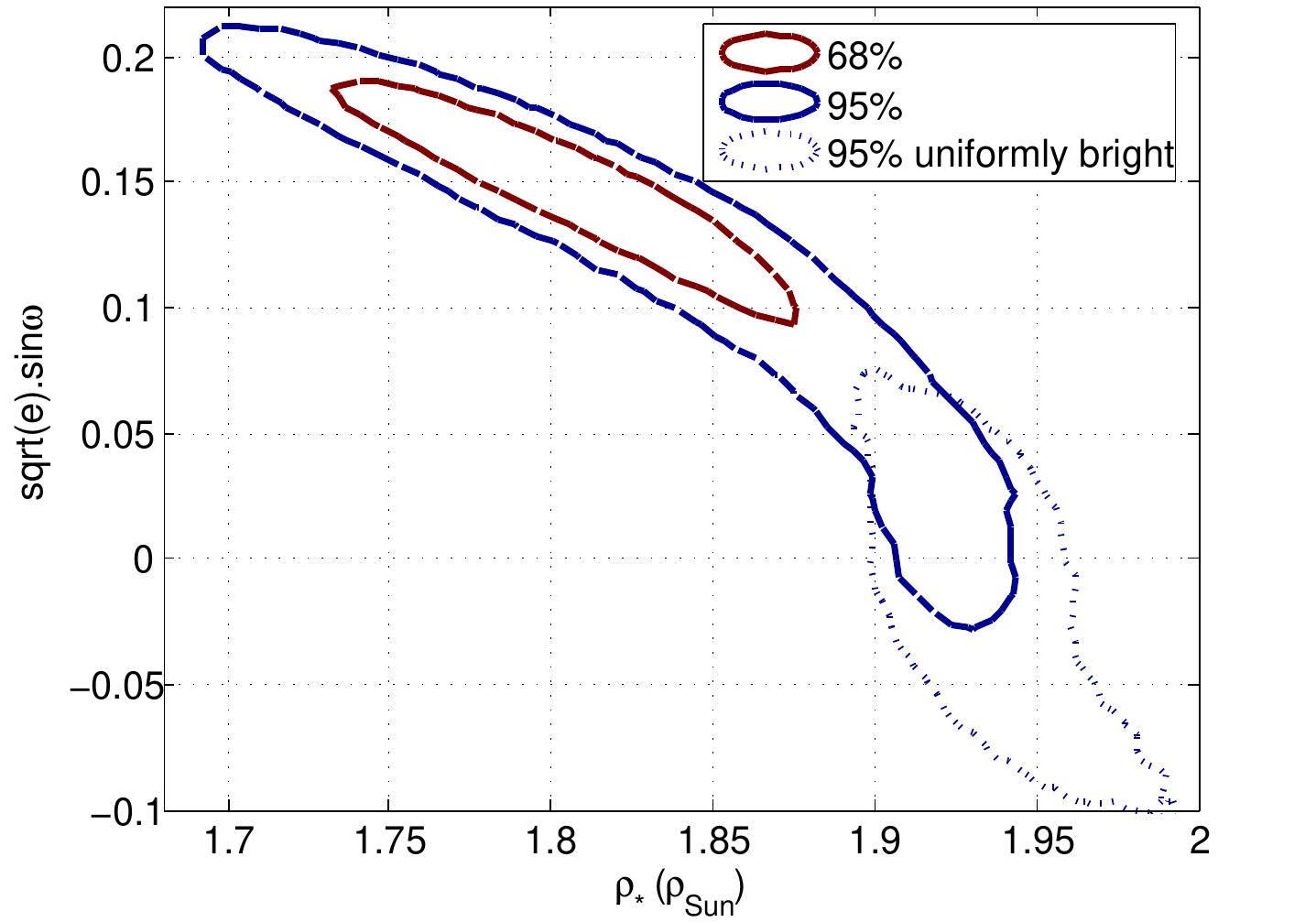}}

  \caption[Influence of the brightness model complexity on the system-parameter posterior probability distribution (PPD), using unipolar models.]{Influence of the brightness model complexity on the system-parameter posterior probability distribution (PPD), using unipolar models. \subref{fig:101_e_ddp} \&  \subref{fig:3_e_ddp} Marginal PPDs ($68\%$- and $95\%$-confidence intervals) of $\sqrt{e}\cos\omega$ and $\sqrt{e}\sin\omega$ for two unipolar brightness models, respectively, with a fixed and large structure ($\Gamma_{SH,1}$) and with free-confinement structure ($\Gamma_{2}$). A comparison with Figure\,\ref{fig:uniform_e_ddp} shows that $\sqrt{e}\cos\omega$ is constrained closer to zero. The reason is that non-uniform brightness models enable the exploration of additional dimensions of the parameter space and, therefore, provide additional adequate combinations of the other contributing factors to compensate an anomalous occultation. In particular, the uniform time offset is now mainly compensated by a non-uniformly-bright model, rather than by $\sqrt{e}\cos\omega$ as in conventional analysis. It shows also the evolution of the marginal PPD toward larger $\sqrt{e}\sin\omega$ when using more complex brightness models---which enable more localized brightness structure. \subref{fig:101_erho_ddps} \&  \subref{fig:3_erho_ddps} Marginal PPDs of $\rho_{\star}$ and $\sqrt{e}\sin\omega$ for, respectively, $\Gamma_{SH,1}$ and $\Gamma_{2}$. These show the effect of relaxing unassessed assumptions to probe consistently a correlated parameter space; in particular, it outlines the possible overestimation of $\rho_{\star}$ by 5\% (i.e., at 6\,$\sigma$ of the conventional estimate) when relaxing the conventional assumption of uniformly-bright disk to model occultations.}
  \label{fig:correl_mono}
\end{figure}

	We first present the effect of relaxing the uniformly-bright exoplanet assumption on the system parameters. For that purpose, we show in Figures\,\ref{fig:101_e_ddp} and \ref{fig:101_erho_ddps} respectively the marginal PPDs of $\left\lbrace\sqrt{e}\cos\omega,\sqrt{e}\sin\omega\right\rbrace$ and  $\left\lbrace\rho_{\star},\sqrt{e}\sin\omega\right\rbrace$ for the $\Gamma_{SH,1}$ model, and the ones for the $\Gamma_{2}$ model in Figures\,\ref{fig:3_e_ddp} and \ref{fig:3_erho_ddps}. We superimpose in Figures\,\ref{fig:101_erho_ddps} and \ref{fig:3_erho_ddps} the 95\% uncertainty interval obtained for the uniformly-bright model to extend our previous observations regarding the effect of relaxing unassessed assumptions to probe a correlated parameter space, here the circularized-orbit assumption (see Section\,\ref{sec:uniform}; Figure\,\ref{fig:uniform_brho_ddps}). We observe the increases of $\sqrt{e}\sin\omega$ and $b$ and the decrease of $\rho_{\star}$ while $\sqrt{e}\cos\omega$ is constrained closer to zero (see Table\,\ref{tab:BFP}). 
	
\begin{figure}
  \centering
  
   \subfloat[]{\hspace{-08mm}\label{fig:101_e_phi_ddp}\includegraphics[trim = 35mm 85mm 35mm 92mm,clip,width=0.55\textwidth,height=!]{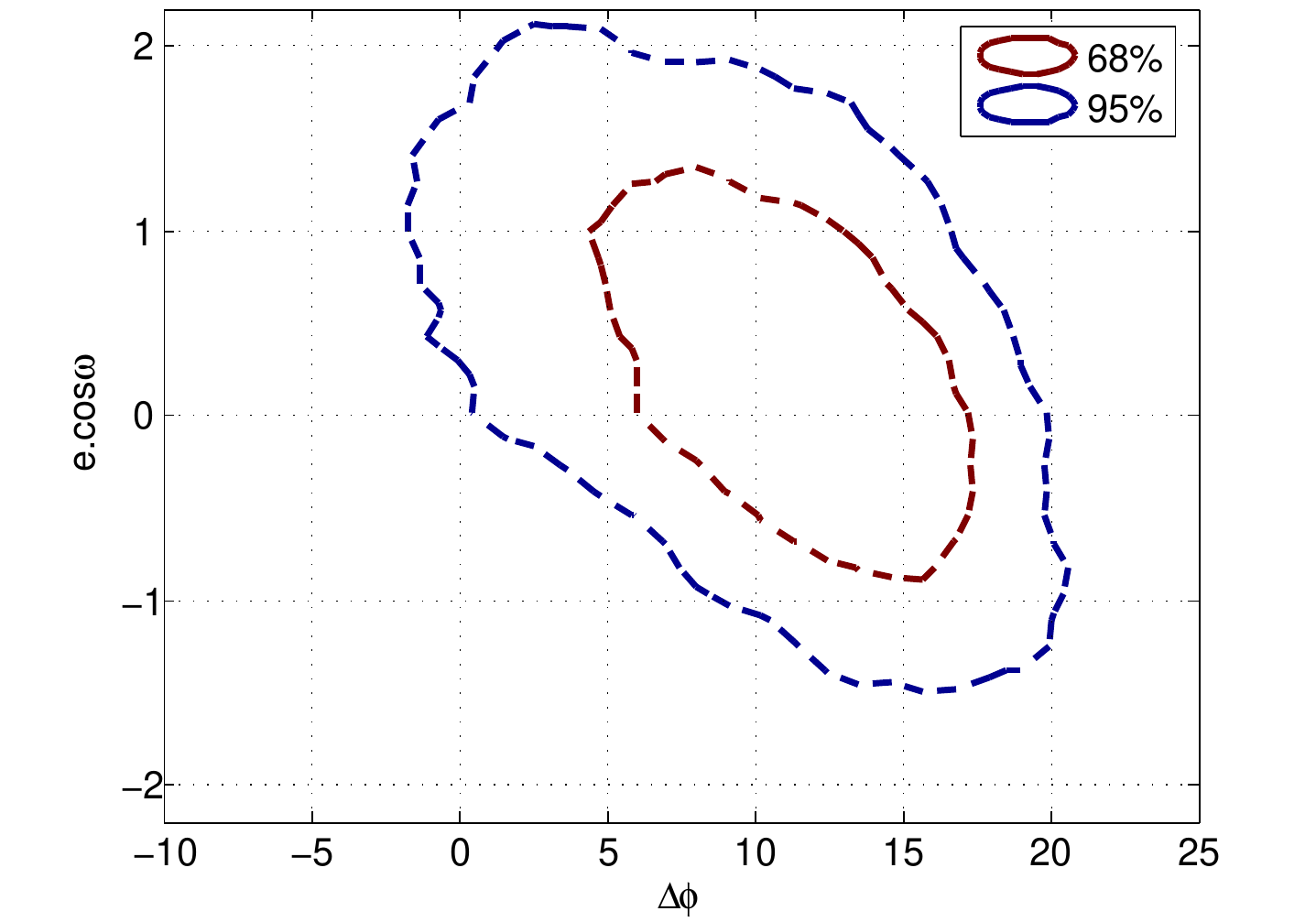}}	
  ~  
  \subfloat[]
  {\hspace{+00mm}\label{fig:101_peak_ddps}\includegraphics[trim = 35mm 85mm 35mm 92mm,clip,width=0.55\textwidth,height=!]{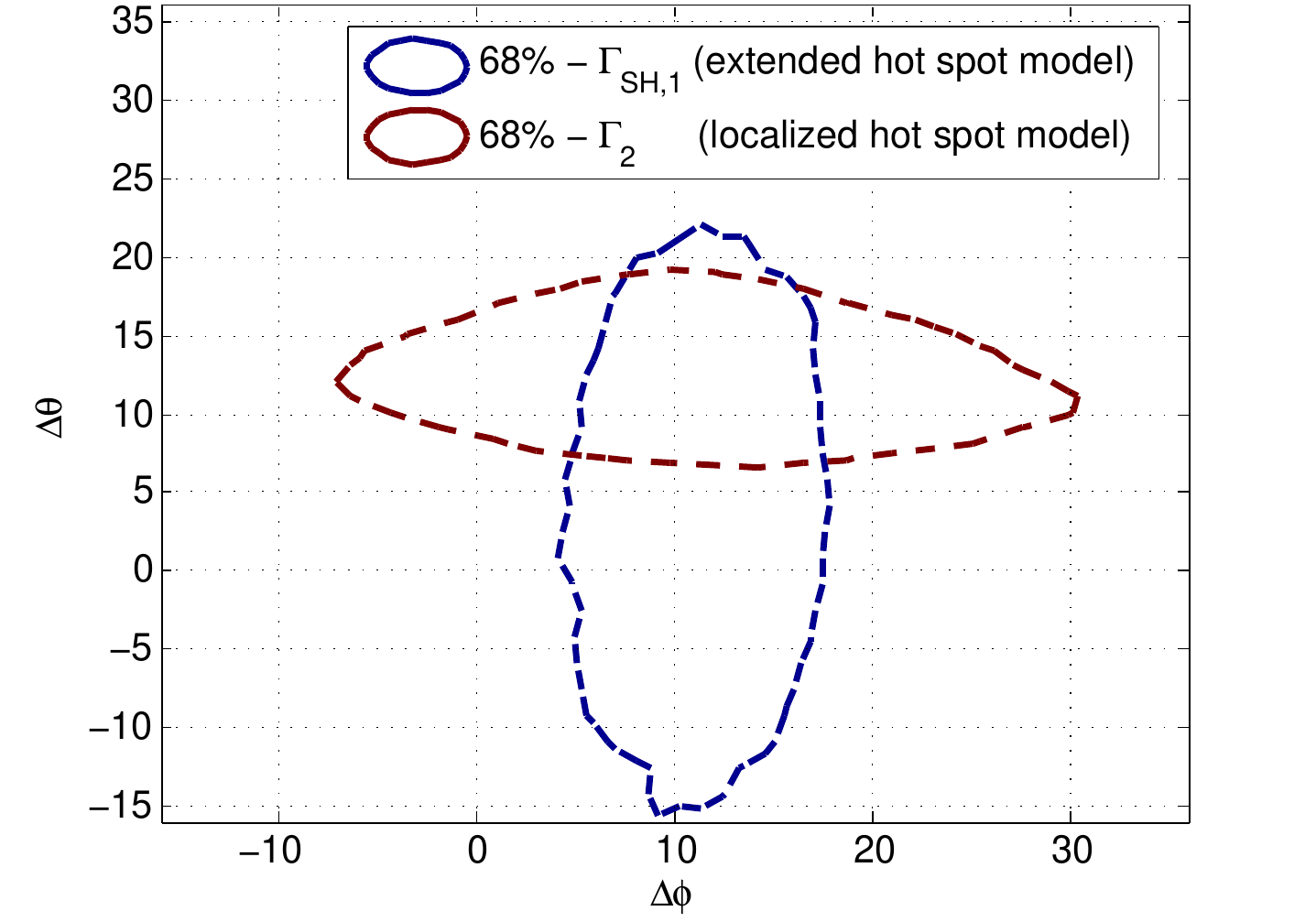}}
  
  \caption[Dependence and significance of the brightness peak localization.]{Dependence and significance of the brightness peak localization. \subref{fig:101_e_phi_ddp} Marginal PPD ($68\%$- and $95\%$-confidence intervals) of the brightness peak longitude, $\Delta\phi$, and $e\cos\omega$ for the $\Gamma_{SH,1}$ brightness model---the simplest non-uniform brightness model used in this study. \subref{fig:101_e_phi_ddp} shows the correlation between the brightness model and the uniform time offset. This correlation emerges from enabling compensation of HD\,189733b's anomalous occultation with a larger set of contributing factors, here by relaxing the uniformly-bright planet assumption. In particular, the occultation shape is now mainly compensated by the brightness distribution rather than by $\sqrt{e}\cos\omega$ as in the conventional analysis (see Figure\,\ref{fig:shape_brightness}, Equation\,\ref{DeltaToc} and Figure\,\ref{fig:Impact_syst_param}).   \subref{fig:101_peak_ddps} Marginal PPDs ($68\%$-confidence intervals) of the brightness peak localization for the $\Gamma_{SH,1}$ and $\Gamma_{2}$ brightness models. It shows that the brightness peak localization is model-dependent---mainly as a result of the fact that the mapping problem has degenerate solutions that are favored differently by each model. For example, the longitudinal $\Gamma_{SH,1}$ peak localization is constrained by the phase curve because of its large and constant extension; while the free extension of the $\Gamma_2$ model relaxes this longitudinal constraint (see Section\,\ref{sec:results}).}
  \label{fig:peak_101}
\end{figure}

	The reason is that the compensation of HD\,189773b's anomalous occultation is now also enabled by the non-uniform brightness models; which provide a better compensation than $e$ solely---with $\sqrt{e}\cos\omega$ for conventional analysis. Therefore, numerous combinations of $e$- and brightness-based compensations are adequate, in agreement with the correlation introduced in Section\,\ref{sec:degeneracies} as highlighted by the PPD in Figure\,\ref{fig:101_e_phi_ddp}. Finally, we note a progressive evolution of the system-parameter PPD with the brightness model complexity (from uniform to $\Gamma_2$). We assess further the validity of these observations, using spherical harmonics of higher degree.

\begin{figure}[!p]

    \hspace{-1.8cm}\includegraphics[angle = -90, trim = 70mm 10mm 70mm 10mm,clip,width=19cm,height=!]{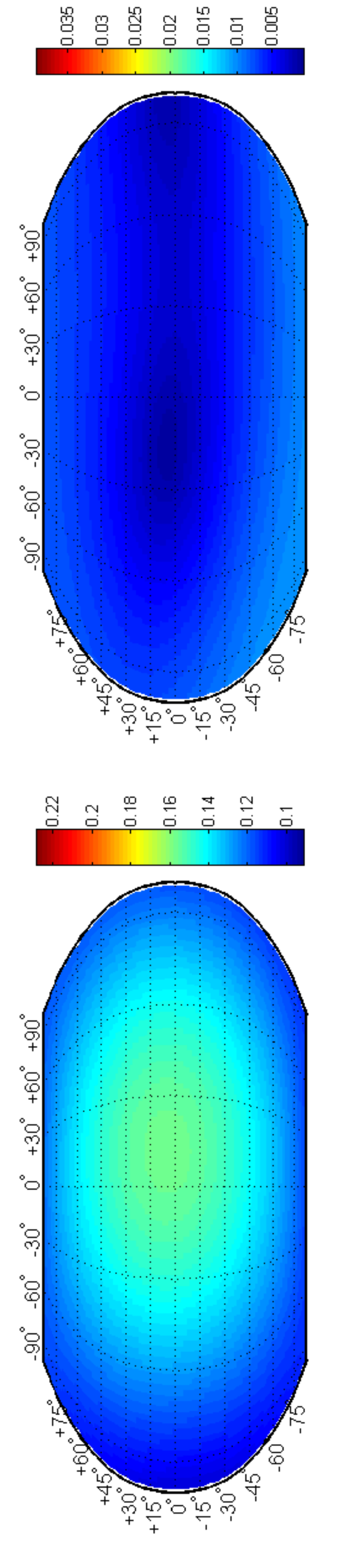}
  \vspace{-0.0cm}
  \caption[Estimate of HD\,189733b's brightness distribution in the IRAC 8-$\mu$m channel using the $\Gamma_{SH,1}$ brightness model.]{Estimate of HD\,189733b's brightness distribution in the IRAC 8-$\mu$m channel using the $\Gamma_{SH,1}$ brightness model. \textit{Left:} Relative brightness distribution of HD\,189733b's dayside. \textit{Right:} Dayside standard deviation. Because of its fixed and large structure, the $\Gamma_{SH,1}$ brightness model is well-constrained in amplitude (by the occultation depth) and in longitudinal localization (by the phase curve). However, it is less constrained in latitude (by the secondary eclipse scanning) than more confined model such as $\Gamma_{2}$ (see Figure\,\ref{fig:3_brightness_ddps}). These model-induced constraints are observable on the dayside standard deviation; which is significantly lower than for more complex brightness models (see Figures\,\ref{fig:3_brightness_ddps},\,\ref{fig:102_brightness_ddps} and \,\ref{fig:103_brightness_ddps}). In addition, the standard deviation distribution for the $\Gamma_{SH,1}$ model is related to its gradient (with a larger variation from the brightness peak localization along the latitude axis than along the longitude axis), because the brightness distributions accepted along the MCMC simulations differ from each other mainly in latitudinal orientation (see Figure\,\ref{fig:ddps_101}).}
  \label{fig:101_brightness_ddps}

\end{figure}

\begin{figure}[!p]
   
    \hspace{-1.8cm}\includegraphics[angle = -90, trim = 70mm 10mm 70mm 10mm,clip,width=19cm,height=!]{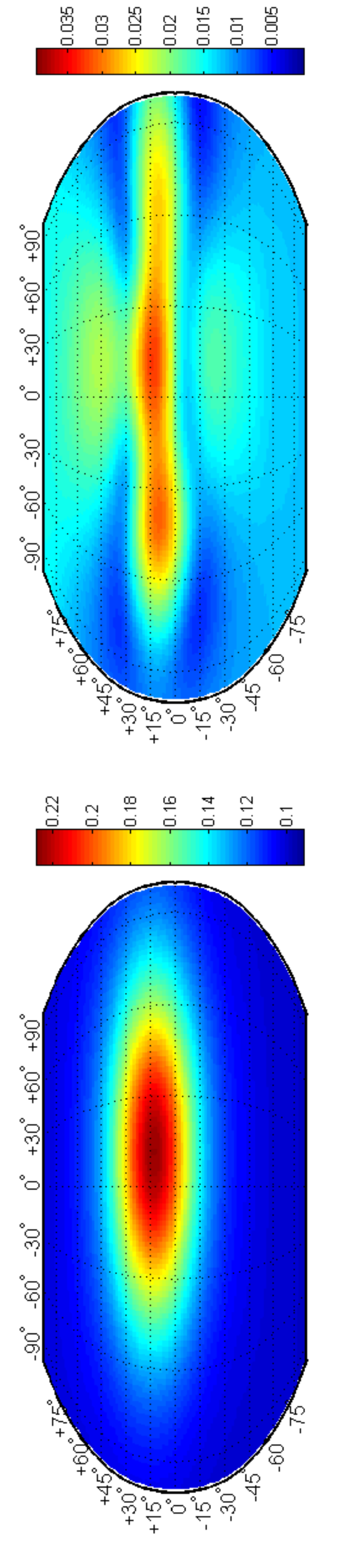}

  \vspace{-0.0cm}
    \caption[Estimate of HD\,189733b's brightness distribution in the IRAC 8-$\mu$m channel using the $\Gamma_{2}$ brightness model.]{Estimate of HD\,189733b's brightness distribution in the IRAC 8-$\mu$m channel using the $\Gamma_{2}$ brightness model. \textit{Left:} Relative brightness distribution of HD\,189733b's dayside. \textit{Right:} Dayside standard deviation. Because of its increased complexity, the $\Gamma_{2}$ brightness model enables more localized structure that are less constrained in amplitude (by the occultation depth) and in longitudinal localization (by the phase curve) than large-and-fixed-structure models such as, e.g., the $\Gamma_{SH,1}$ model (see Figure\,\ref{fig:101_brightness_ddps}). However, it is well-constrained in latitude by the secondary eclipse scanning that is sensitive to confined brightness structure. These model-induced constraints are observable on the dayside standard deviation; which shows a maximum at the brightness peak localization and extended wings towards west and east along the planetary equator. The reason is that the brightness distributions accepted along the MCMC simulations mainly affect the former by their amplitude changes and the latter by their structure changes (see Figure\,\ref{fig:ddps_3}).}
  \label{fig:3_brightness_ddps}

\end{figure}		
		
 	We now turn to HD\,189733b's brightness distribution. We show the dayside estimates for the $\Gamma_{SH,1}$ and $\Gamma_{2}$ models with their corresponding uncertainties in Figures\,\ref{fig:101_brightness_ddps} and \ref{fig:3_brightness_ddps} respectively; we focus on HD\,189733b's dayside as it is effectively constrained by the combination of the phase curve and the secondary eclipse scanning. In particular, note that Figures\,\ref{fig:101_brightness_ddps} and \ref{fig:3_brightness_ddps} present HD\,189733b's brightness relative to HD\,189733's hemisphere-averaged brightness in the IRAC 8-$\mu$m channel (i.e., $I_p(\phi,\theta)/<I_\star>|_{8\mu m}$). In addition, the figures are time-averaged; our estimates aim to approach the global pattern of HD\,189733b's BD based on eight snapshots taken from November 2005 to June 2008.  Finally, these estimates correspond to the median and standard deviation of the map trials accepted along the MCMC simulations, similarly to our approach for the corrected and phase-folded light curves (see Section\,\ref{sec:results&discussI}).

 	Both models retrieve a spatial feature in HD\,189733b's BD that corresponds to a hot spot. The $\Gamma_{SH,1}$ model retrieves a hot spot shifted to the east of the substellar point, see Figure\,\ref{fig:101_brightness_ddps}. The $\Gamma_{2}$ model retrieves a hot spot shifted to the east of the substellar point but also away from the equator, see Figure\,\ref{fig:3_brightness_ddps}. However, we cannot discuss the direction of this latitudinal shift due to a North-South ambiguity (E. Agol, private communication). 
 	
 	The BD estimates shown in Figures\,\ref{fig:101_brightness_ddps} and \ref{fig:3_brightness_ddps} are significantly different both in pattern and in intensity. These differences are due to the intrinsic degeneracy of the mapping problem (see Section\,\ref{sec:degeneracies}); which requires the use of different mapping models to enable a thorough discussion. For example, brightness models with non-constant structure (``complex'', i.e., in opposition to a dipole) are less constrained by a phase curve that is only dependent on the hemisphere-integrated brightness. To emphasize these model-induced constraints, we present in Figure\,\ref{fig:ddps_video} animations showing the compilations of dayside BDs accepted along the MCMC simulations for the $\Gamma_{SH,1}$ and $\Gamma_{2}$ models. These compilations show that \textbf{(1)} the amplitude and \textbf{(2)} the longitudinal localization for the $\Gamma_{SH,1}$ model are more constrained than for the $\Gamma_{2}$ model (by the occultation depth and by the phase curve, respectively) because of its fixed and large structure. However, \textbf{(3)} the $\Gamma_{SH,1}$ model is less constrained in latitude (by the secondary eclipse scanning) than the more complex $\Gamma_{2}$ model which enables more confined structures that induce larger deviations in occultation ingress/egress. For that reason, the brightness peak localization for the $\Gamma_{2}$ model is well-constrained in latitude (see Figure\,\ref{fig:101_peak_ddps}), while for the $\Gamma_{SH,1}$ model it is well-constrained in longitude.
 	
\begin{figure}[!p]
  \centering
  
   \subfloat[]{\label{fig:ddps_101}\includemovie[poster,toolbar,palindrome = false]{.8\textwidth}{.5\textwidth}{images/ddps_101.mp4}}	
    
  \subfloat[]
  {\hspace{+00mm}\label{fig:ddps_3}\includemovie[poster,toolbar,palindrome = false]{.8\textwidth}{.5\textwidth}{images/ddps_3.mp4}}
  
  \vspace{+0.5cm}
  \caption[Insight into the model-dependence of brightness distribution estimates.]{Insight into the model-dependence brightness distribution estimates. Animations showing the compilations of HD\,189733b's dayside brightness distributions accepted along MCMC chains for the $\Gamma_{SH,1}$ and $\Gamma_{2}$ models, \subref{fig:ddps_101} and \subref{fig:ddps_3} respectively. These animations show that \textbf{(1)} the amplitude and \textbf{(2)} the longitudinal localization for the $\Gamma_{SH,1}$ brightness model is more constrained than for the $\Gamma_{2}$ model (by the occultation depth and by the phase curve, respectively) because of its fixed and large structure. However, \textbf{(3)} the $\Gamma_{SH,1}$ model is less constrained in latitude (by the secondary eclipse scanning) than the more complex $\Gamma_{2}$ model, which enables more confined structures that induce larger deviation in occultation ingress/egress (schematic description in Figure\,\ref{fig:shape_brightness}).\\ \\(\textit{Videos available in the electronic version.})}
  \label{fig:ddps_video}
\end{figure} 	
 	
 	These observations recall that the problem of exoplanet mapping has degenerate solutions, hence, the brightness peak localization cannot be constrained uniquely without \textit{a priori} assumption (e.g., assuming a dipolar BD). Therefore, we will further refer to our brightness-distribution estimates instead of the brightness peak localization; which is not representative of complex BDs, in addition to being model-dependent. Nevertheless, we propose in Section\,\ref{sec:bpl} another unidimensional parameter to replace the brightness peak localization.

 	Finally, note that these model-induced constraints are also observable on the dayside standard deviation; which is significantly lower for $\Gamma_{SH,1}$ model than for more complex models. In particular, the standard deviation distribution for the $\Gamma_{SH,1}$ model is related to its gradient---with a larger variation from the brightness peak localization along the latitude axis than along the longitude axis, because the BDs accepted along MCMC chains differ from each other mainly in (latitudinal) orientation, see Figure\,\ref{fig:ddps_101}. This is in contrast with the standard deviation distribution for the $\Gamma_{2}$ model that shows a maximum at the brightness peak localization and extended wings towards west and east along the equator; because the BDs accepted along the MCMC simulations mainly affect the former by their amplitude change and the latter by their structure change (see Figure\,\ref{fig:ddps_3}).
\clearpage
 \paragraph*{Multipolar Brightness Distribution:}
 
 We observe an evolution of our inferences when increasing the complexity of our fitting model. To assess the validity of this observation, we present here the results obtained when using spherical harmonics up to the degrees 2 (quadrupole) and 3 (octupole).
 
\begin{figure}[!p]
  \centering
  
  \subfloat[]{\hspace{-08mm}\label{fig:102_e_ddp}\includegraphics[trim = 35mm 85mm 35mm 92mm,clip,width=0.55\textwidth,height=!]{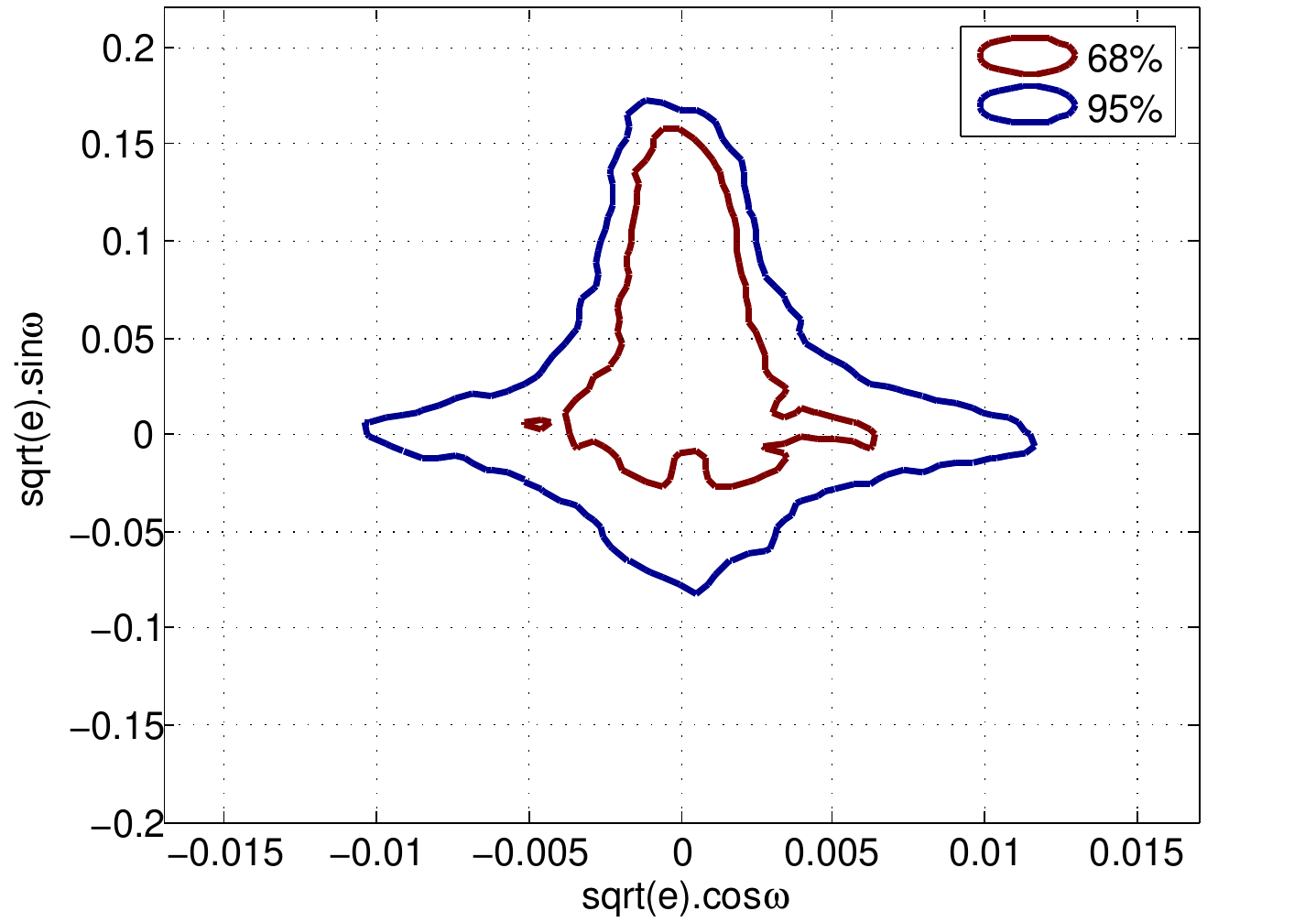}}
  ~ 
  \subfloat[]{\label{fig:103_e_ddp}\includegraphics[trim = 35mm 85mm 35mm 92mm,clip,width=0.55\textwidth,height=!]{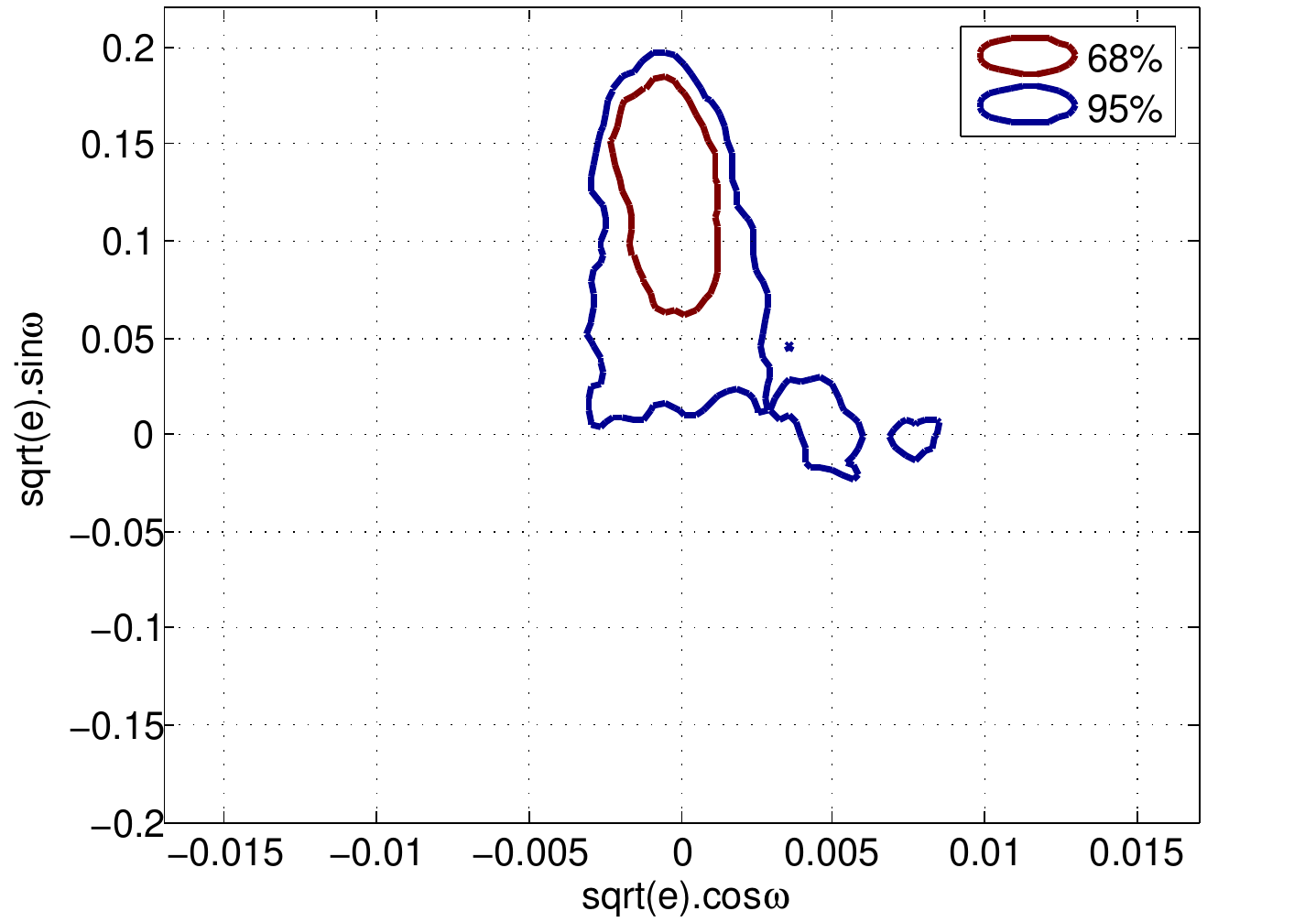}}

   \subfloat[]{\hspace{-08mm}\label{fig:102_erho_ddps}\includegraphics[trim = 35mm 85mm 35mm 92mm,clip,width=0.55\textwidth,height=!]{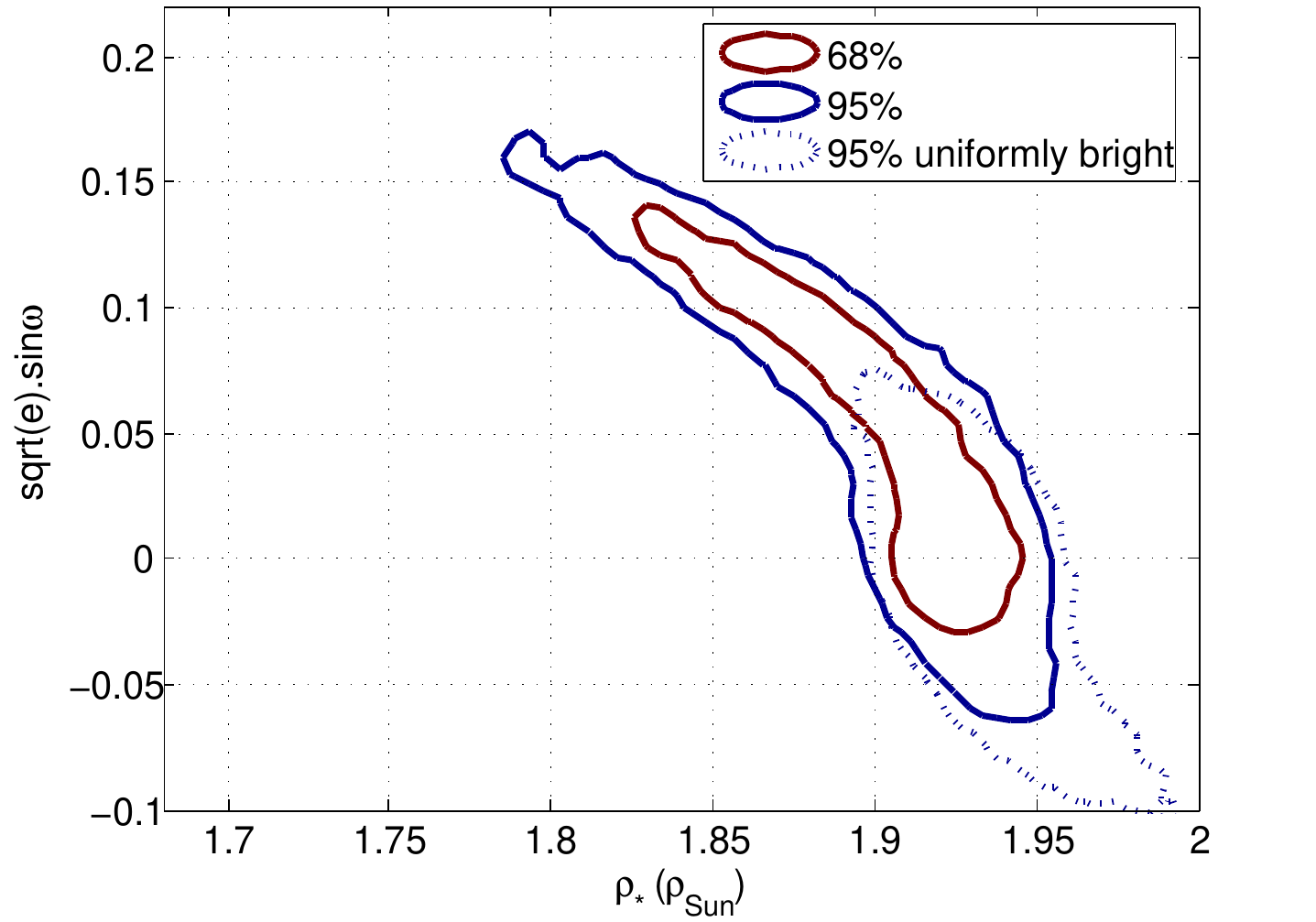}}
  ~  
  \subfloat[]{\label{fig:103_erho_ddps}\includegraphics[trim = 35mm 85mm 35mm 92mm,clip,width=0.55\textwidth,height=!]{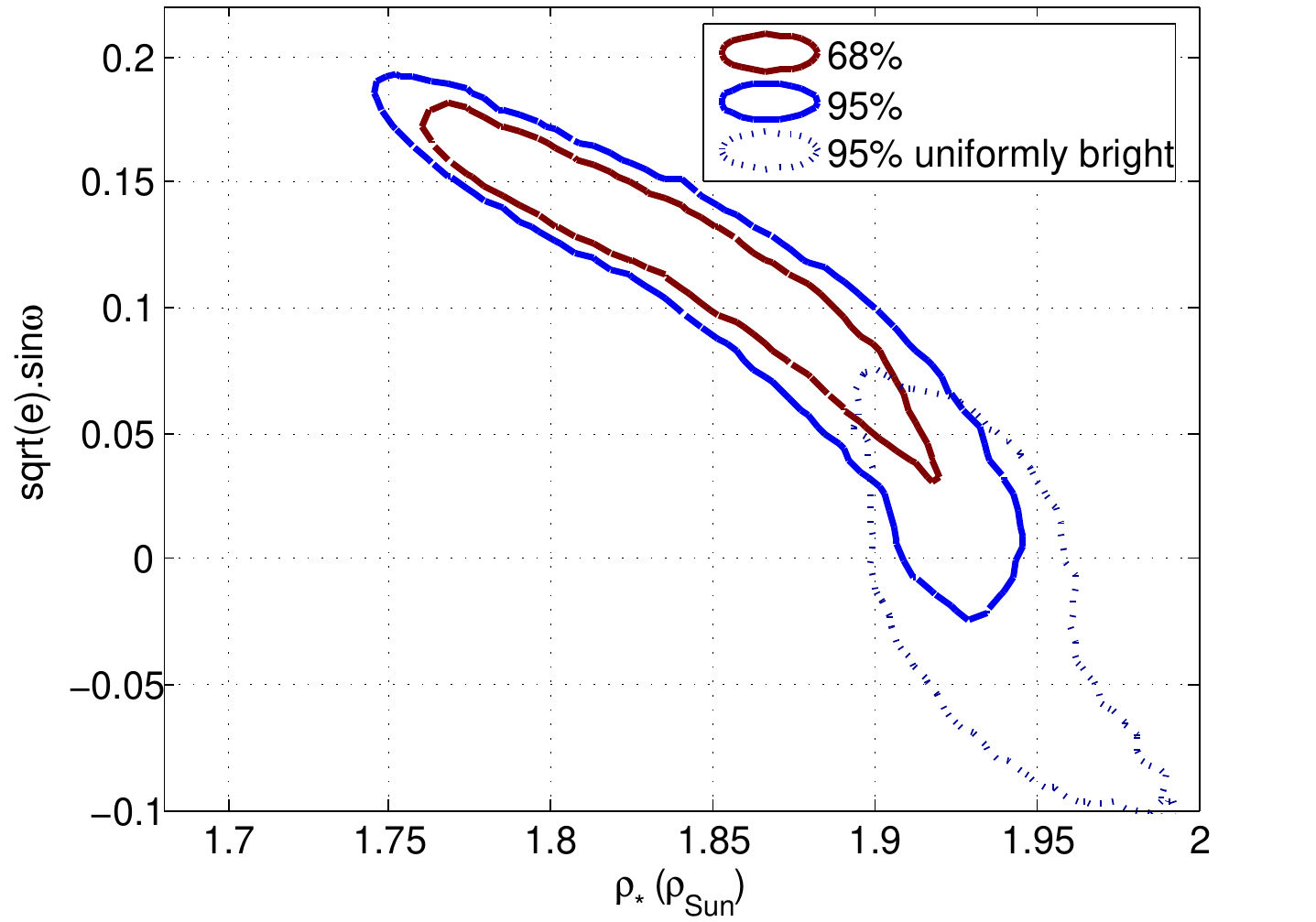}}

  \caption[Influence of the brightness model complexity on the system parameter posterior probability distribution (PPD), using multipolar models.]{Influence of the brightness model complexity on the system parameter posterior probability distribution (PPD), using multipolar models. \subref{fig:102_e_ddp} \&  \subref{fig:103_e_ddp} Marginal PPDs ($68\%$- and $95\%$-confidence intervals) of $\sqrt{e}\cos\omega$ and $\sqrt{e}\sin\omega$ for the quadrupolar and octupolar brightness models, respectively. \subref{fig:102_erho_ddps} \&  \subref{fig:103_erho_ddps} Marginal PPDs of $\rho_{\star}$ and $\sqrt{e}\sin\omega$ for the quadripolar and octopolar brightness models, respectively. These confirm the trend observed in Section\,\ref{sec:uniform} (Figure\,\ref{fig:correl_mono}) towards larger $\sqrt{e}\sin\omega$ and lower stellar density when increasing the complexity of the brightness distribution.}
  \label{fig:correl_multi}
\end{figure}

 We present in Figures\,\ref{fig:102_e_ddp} and \ref{fig:102_erho_ddps} respectively the marginal PPDs of $\left\lbrace\sqrt{e}\cos\omega,\sqrt{e}\sin\omega\right\rbrace$ and  $\left\lbrace\rho_{\star},\sqrt{e}\sin\omega\right\rbrace$ for the $\Gamma_{SH,2}$ model, and in Figures\,\ref{fig:103_e_ddp} and \ref{fig:103_erho_ddps} for the $\Gamma_{SH,3}$ model. These PPDs appear as intermediate steps between the results  obtained with the $\Gamma_{SH,1}$ and $\Gamma_{2}$ models (see Figure\,\ref{fig:correl_mono}). This gradual evolution reemphasizes the strong influence of underlying model assumptions in the context of correlated parameter space.
 
\begin{figure}[p]
      
      \subfloat[]{\hspace{-1.8cm}\label{fig:102_brightness_ddps}\includegraphics[angle = -90, trim = 70mm 10mm 70mm 10mm,clip,width=19cm,height=!]{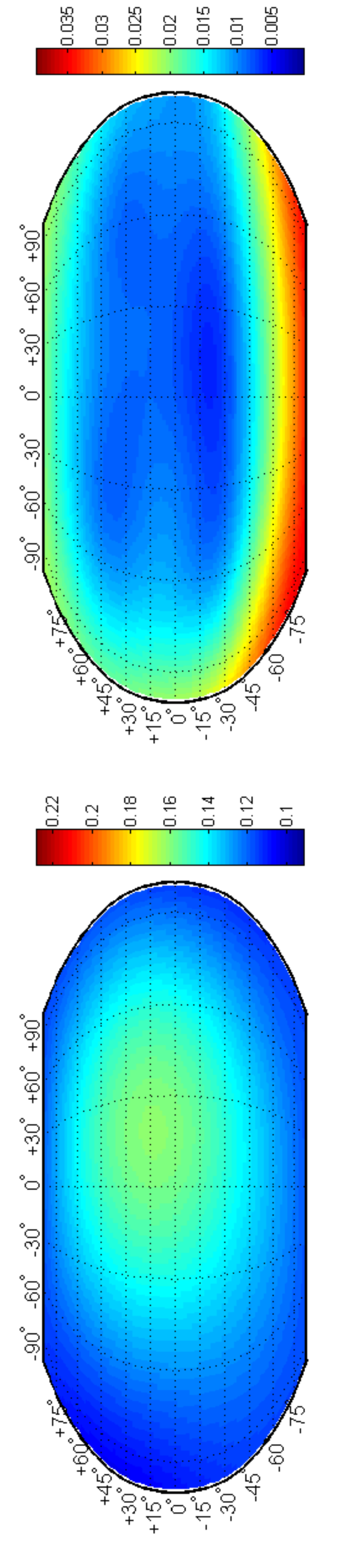}}
      
            \subfloat[]{\hspace{-18mm}\label{fig:103_brightness_ddps}\includegraphics[angle = -90, trim = 70mm 10mm 70mm 10mm,clip,width=19cm,height=!]{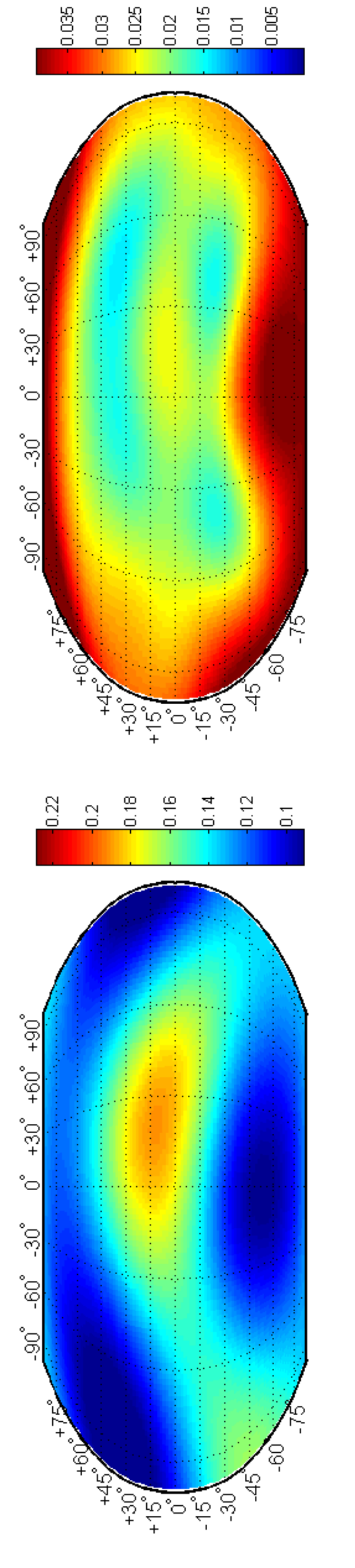}}

  \vspace{+0.5cm}
    \caption[Estimates of HD\,189733b's brightness distribution in the IRAC 8-$\mu$m channel using multipolar brightness models.]{Estimates of HD\,189733b's brightness distribution in the IRAC 8-$\mu$m channel using multipolar brightness models. \textit{Left:} Relative brightness distribution of HD\,189733b's dayside. \textit{Right:} Dayside standard deviation. \subref{fig:102_brightness_ddps} Estimate using the $\Gamma_{SH,2}$ brightness model. \subref{fig:103_brightness_ddps} Estimate using the $\Gamma_{SH,3}$ brightness model. These estimates confirm the trend toward a more localized and latitudinally-shifted hot spot when increasing the brightness-model complexity. Together with Figure\,\ref{fig:correl_multi}, these estimates outline that the more complex HD\,189733b's brightness model, the larger the eccentricity, the lower the densities, the larger the impact parameter and the more localized and latitudinally-shifted the hot spot estimated.}
  \label{fig:complex_brightness}

\end{figure}
 
 We show the dayside brightness estimates for the $\Gamma_{SH,2}$ and $\Gamma_{SH,3}$ models with their uncertainty in Figures\,\ref{fig:102_brightness_ddps} and \ref{fig:103_brightness_ddps}. As observed for the system-parameter PPDs, the brigthness distributions also appear as intermediate steps. The major evolutions are the shrinking of the structure retrieved and its shift away from the equator.

 These progressive evolutions of both the system-parameter PPD and the retrieved brightness structure show that, for HD\,189733's system, relaxing the eccentricity constraint and using more complex BDs lead to a lower stellar density and more localized and latitudinally-shifted hot spot. In particular, we find that the more complex HD\,189733b's brightness model, the larger the eccentricity, the lower the densities, the larger the impact parameter and the more localized and latitudinally-shifted the hot spot estimated. We discuss the significance of these findings in the next section.

 \clearpage
	\subsubsection{``The Most Adequate Model''}
	
	We present in Section\,\ref{sec:results} the results obtained using several fitting-models for assessing the model-dependence of our inferences in the context of underdetermined problem---i.e., degenerate solution (Section\,\ref{sec:degeneracies}). As a result, we quantitatively outline the correlation between the system parameters and the BD of an exoplanet introduced in Section\,\ref{sec:degeneracies}. In particular, we show a progressive evolution of the system-parameter PPD and the BD estimate when increasing the complexity of the fitting model. We discuss below the relevance of our model-complexity increase, as it may ultimately lead to data overfitting.
	
	We take advantage of our Bayesian framework using the BIC and the AIC. Both information criteria are in favor of models that relax the assumptions of a circularized orbit and a uniformly bright HD\,189733b. In particular, the $\Gamma_{SH,1}$ and $\Gamma_{SH,3}$ models are favored; the BIC insignificantly favors $\Gamma_{SH,3}$ (odds ratio $\sim$1.02) while the AIC significantly favors $\Gamma_{SH,3}$ (odds ratio $\sim$3.5). Because not decisive\footnote{Note that theoretical studies favor the AIC, e.g., \cite{Burnham2002,Yang2005}.}, the information criteria only suggest that the $\Gamma_{SH,3}$ model provides the most adequate constraints on
HD\,189733's system. 
			
	In particular, the information criteria suggest that HD\,189733b's hot spot is shifted both east of the substellar point and away from the equator and HD\,189733b's density has been overestimated by 3.6\%. Furthermore, it suggests that HD\,189733b's orbit is possibly not fully circularized ($ e = 0.015\pm^{0.09}_{0.012} $), although its eccentricity is consistent with zero. This recalls that the assumption of circularized orbit has to be continuously assessed with constantly-increasing data quality; even for old hot Jupiters that may show a hint of eccentricity \citep[e.g., CoRoT-16b, see][]{Ollivier2012}. Finally, for data-quality reason, the interpretation of HD\,189733b's BD has to focus on global trends: the presence of an asymmetrical hot spot.

	HD\,189733b's dayside presents a shifted hot spot. The eastward shift is in agreement with the literature: (1) with previous derivations, from HD\,189733b's phase curve \citep{Knutson2007} and an eclipse timing constraint \citep{Agol2010}; and (2) atmospheric models suggesting a super-rotating equatorial jet \citep[e.g.,][]{Showman2009}. In opposition, the suggested shift away from the equator is new. The small-scale origin of this latitudinal asymmetry remains unconstrained because we use large-scale brightness models to be consistent with the data quality. For that reason, additional observations would be required to improve our understanding of HD\,189733b's atmosphere (see Chapter\,\ref{chap:perspectives}); in particular, its interactions with HD\,189733 could induce unexpected thermal patterns, e.g., asymmetric patterns in its BD \citep{Lecavelier2012}. For example, magnetic star-planet interactions may lead to energy dissipation due to the stellar field penetration into the exoplanet envelope \citep[e.g.,][]{Laine2008} and to extensive energy injections into the auroral zones of the exoplanet from magnetic reconnections \citep[e.g.,][]{Ip2004}---similarly to the Jupiter-Io flux tube \citep[e.g.,][]{Bigg1964}. However, such magnetic reconnections have so far been only observed at the stellar surface, in the form of chromospheric hot spots rotating synchronously with the companions  \citep[e.g.,][]{Shkolnik2005,Lanza2009}.

\subsection{Global Analysis: Complementary Insights from RV Data}
 \label{sec:compl_analysis}
 
\begin{figure}
  \centering
  
  \subfloat[]{\vspace{-0.7cm}\hspace{+00mm}\label{fig:RVfit}\includegraphics[trim = 15mm 67mm 10mm 69mm,clip,width=12cm,height=!]{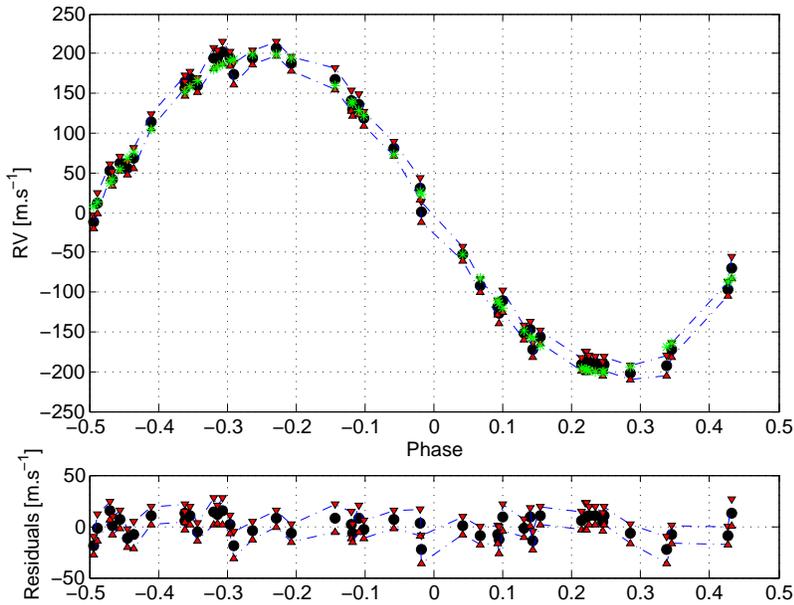}}
   
  \subfloat[]{\vspace{-1.cm}\hspace{-05mm}\label{fig:ddp_e_unif}\includegraphics[trim = 30mm 71mm 25mm 69mm,clip,width=9cm,height=!]{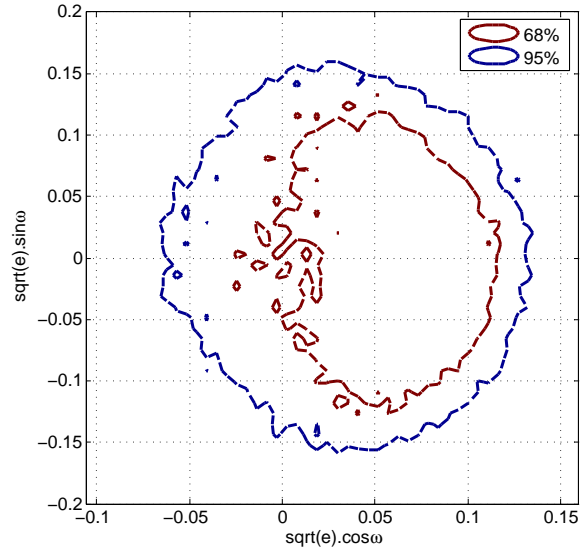}}

  \caption[Further insights gained using HD\,189733's RV measurements together with \textit{Spitzer}/IRAC 8-$\mu$m photometry.]{Influence of HD\,189733's RV measurements on inferences obtained from the \textit{Spitzer}/IRAC 8-$\mu$m photometry. \subref{fig:RVfit} Overall Keplerian fit (green) of the out-of-transit RV data (black dots) with their 1\,$\sigma$ error bars (red triangles) from \cite{Winn2006} and \cite{Boisse2009}. \subref{fig:ddp_e_unif} Marginal PPD (68\%- and 95\%-confidence intervals) of $\sqrt{e}\cos\omega$ and $\sqrt{e}\sin\omega$ obtained solely from the RV data. The RV data does not constrain further HD\,189733b's eccentricity than the \textit{Spitzer}/IRAC 8-$\mu$m photometry for low-complexity brightness models (see Figures\,\ref{fig:uniform_e_ddp}, \ref{fig:101_e_ddp}, and \ref{fig:102_e_ddp}). However, for more complex models that favor a localized hot spot and larger $\sqrt{e}\sin\omega$, HD\,189733's RV measurements may affect our inferences by rejecting the solutions involving $\sqrt{e}\sin\omega$ $ \gtrsim $ 0.15 in the context of the $ e $-$\omega$-$ b $-$\rho_{\star}$-BD correlation (see Section\,\ref{sec:degeneracies}).}
  \label{fig:RV_inputs}
\end{figure}

 We emphasize in our study the necessity of global analyses to fit high-SNR occultation (see Section\,\ref{sec:degeneracies}). For that reason, we purposely focus in the previous Section on the photometry (transits, occultations, and phase curve) in order to emphasize here the further insights gained while performing a global analysis including the RV measurements. RV measurements may provide a complementary insight into the $ e $-$\omega$-$ b $-$\rho_{\star}$-BD correlation. Therefore, we analyse first HD\,189733's out-of-transit RV data \citep{Winn2006,Boisse2009} separately to assess the constraint derived solely from the RV data on $e$ and $\omega$. We present in Figure\,\ref{fig:RV_inputs} our overall Keplerian fit and the marginal PPD of the parameters $\sqrt{e}\cos\omega$ and $\sqrt{e}\sin\omega$. This shows that the RV data does not constrain HD\,189733b's eccentricity further than the \textit{Spitzer}/IRAC 8-$\mu$m photometry does for low-complexity brightness models (see Figures\,\ref{fig:uniform_e_ddp}, \ref{fig:101_e_ddp} and \ref{fig:102_e_ddp}). However HD\,189733's RV measurements may affect our inferences for more complex models that favor a localized hot spot and larger eccentricity (see Figures\,\ref{fig:103_e_ddp} and \ref{fig:3_e_ddp}), by rejecting the solutions involving $\sqrt{e}\sin\omega$ $ \gtrsim $ 0.15. 

\begin{figure}
  \centering
  
  \subfloat[]{\hspace{-08mm}\label{fig:103_e_ddp_RV}\includegraphics[trim = 35mm 85mm 35mm 92mm,clip,width=0.55\textwidth,height=!]{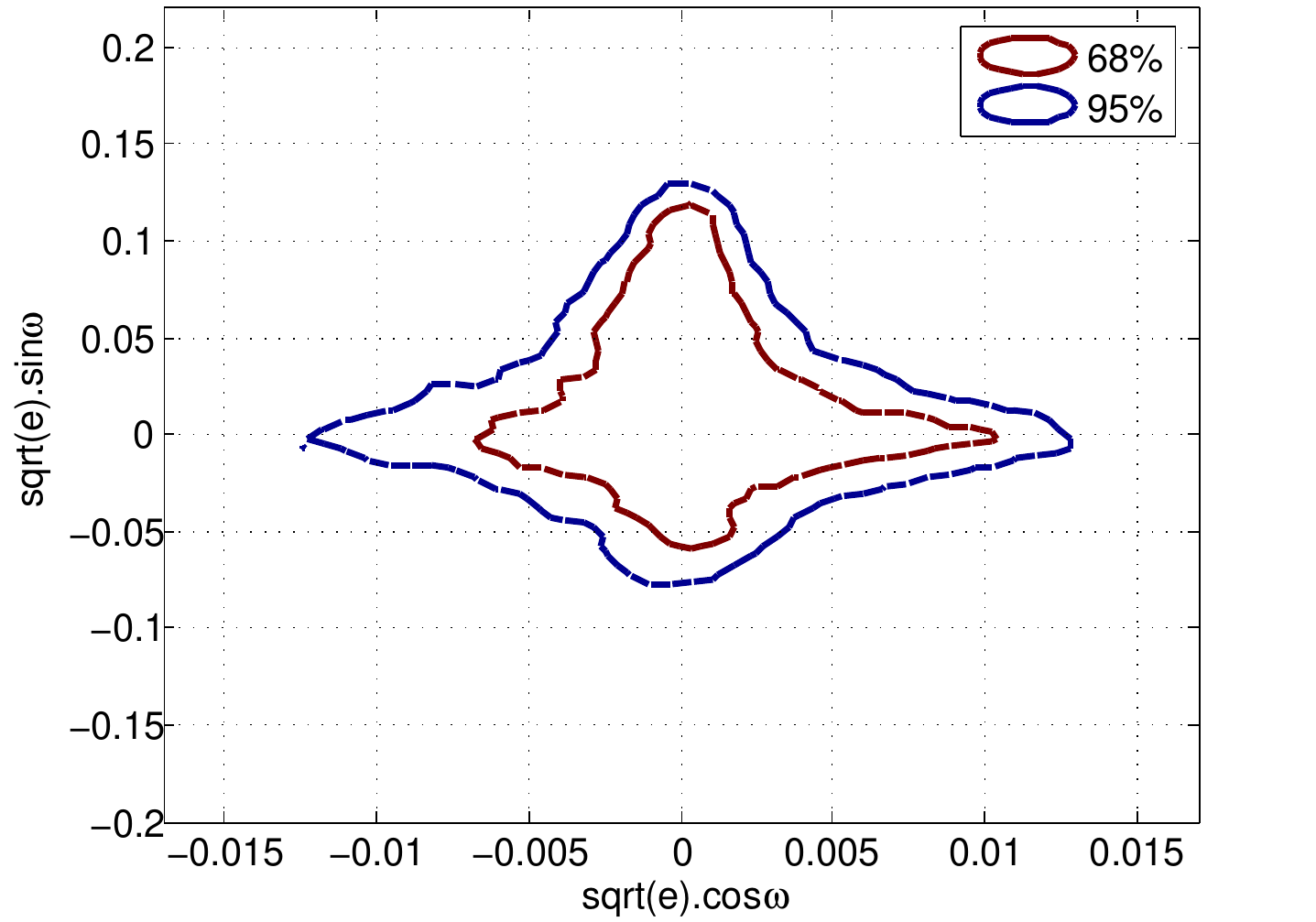}}
  ~ 
  \subfloat[]{\label{fig:3_e_ddp_RV}\includegraphics[trim = 35mm 85mm 35mm 92mm,clip,width=0.55\textwidth,height=!]{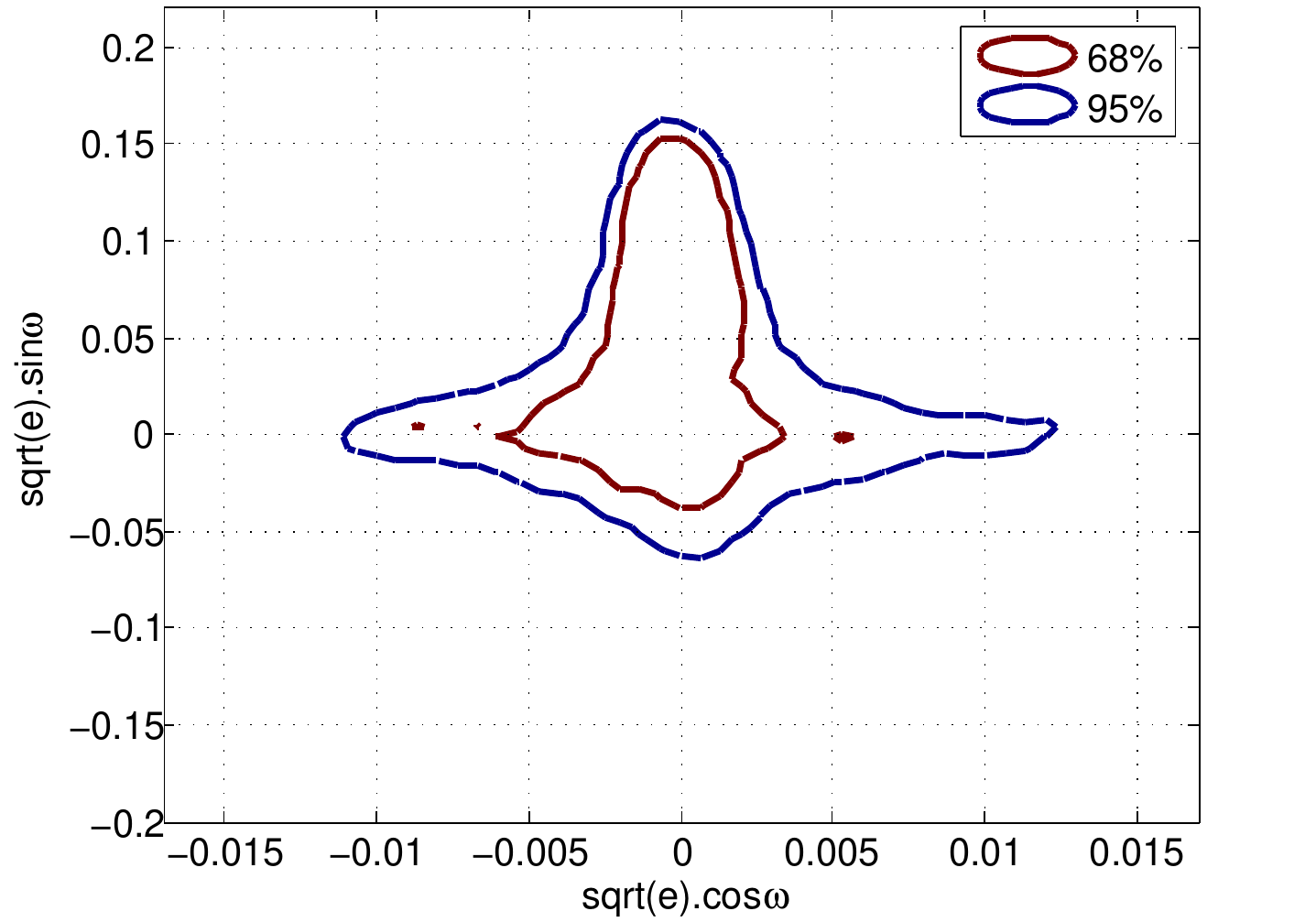}}

   \subfloat[]{\hspace{-08mm}\label{fig:103_erho_ddps_RV}\includegraphics[trim = 35mm 85mm 35mm 92mm,clip,width=0.55\textwidth,height=!]{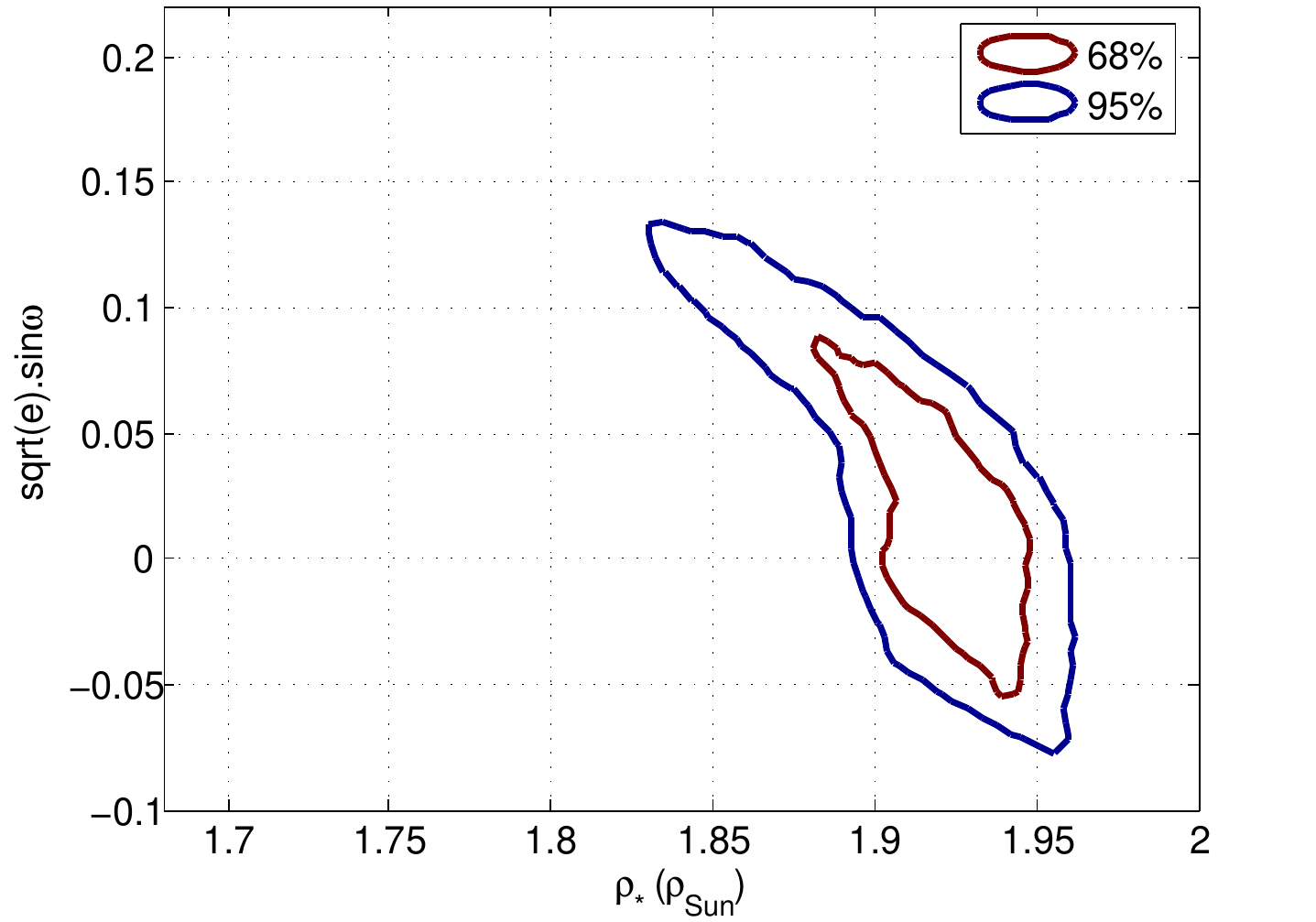}}
  ~  
  \subfloat[]{\label{fig:3_erho_ddps_RV}\includegraphics[trim = 35mm 85mm 35mm 92mm,clip,width=0.55\textwidth,height=!]{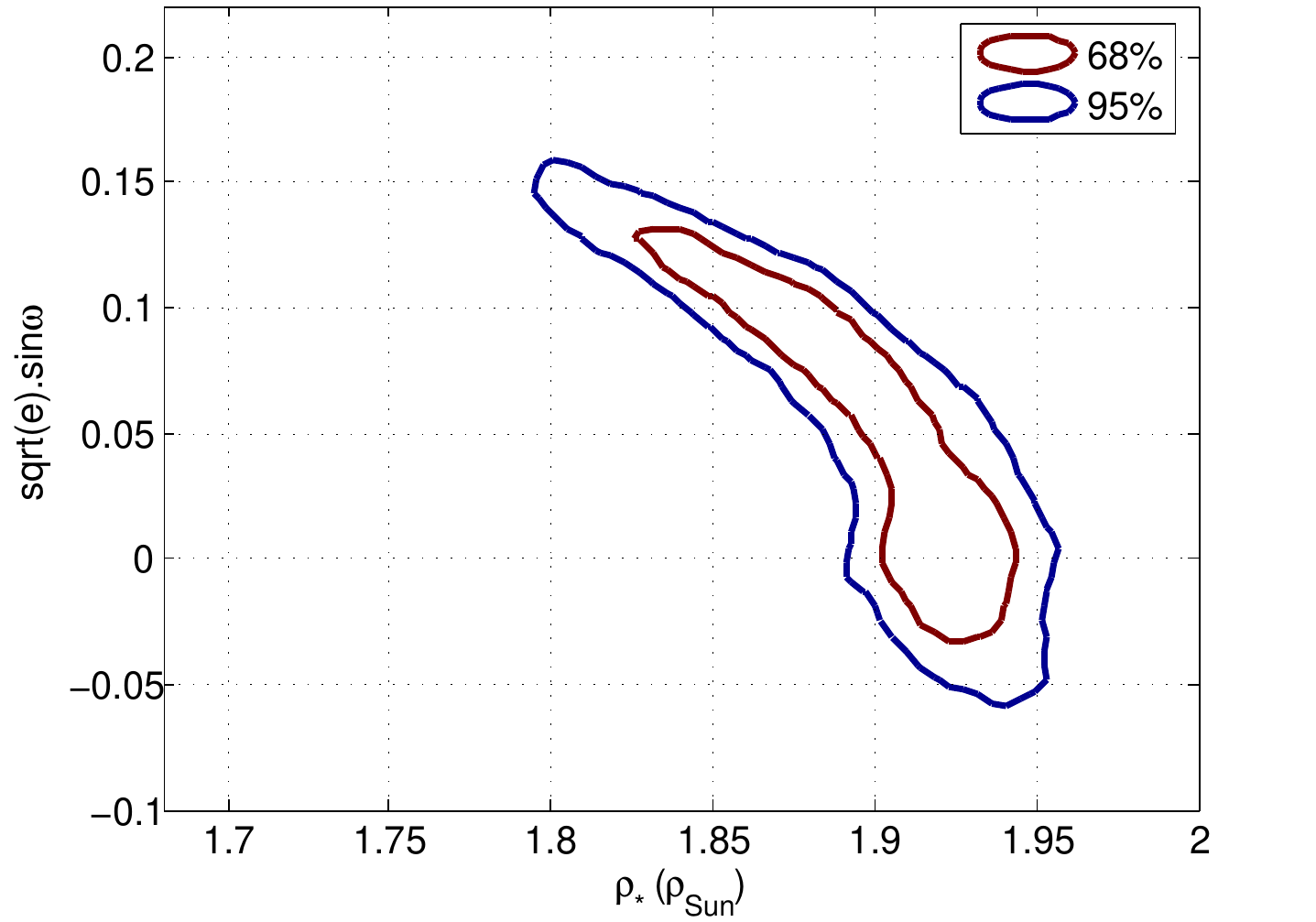}}

  \caption[Influence of the RV measurements on the system-parameter PPD.]{Influence of the RV measurements on the system-parameter PPD estimated using complex brightness models that suggest a non-zero eccentricity from the \textit{Spitzer}/IRAC 8-$\mu$m photometry  separately (see Figures\,\ref{fig:103_e_ddp} and \,\ref{fig:3_e_ddp}). \subref{fig:103_e_ddp_RV} \&  \subref{fig:3_e_ddp_RV} Marginal PPDs (68\%- and 95\%-confidence intervals) of $\sqrt{e}\cos\omega$ and $\sqrt{e}\sin\omega$ for the $\Gamma_{SH,3}$ and the $\Gamma_2$ brightness models, respectively. Conceptually these correspond to the marginalized product of the PPDs estimated using solely the photometry (see Figures\,\ref{fig:103_e_ddp} and \,\ref{fig:3_e_ddp}) and using solely the RV data (Figure\,\ref{fig:RVfit}). In particular, these highlight that the solutions involving $\sqrt{e}\sin\omega$ $ \gtrsim $ 0.15 are strongly rejected by HD\,189733's RV data, leading to a redistribution of the probability density. \subref{fig:103_erho_ddps_RV} \&  \subref{fig:3_erho_ddps_RV} Marginal PPDs of $\rho_{\star}$ and $\sqrt{e}\sin\omega$ for the $\Gamma_{SH,3}$ and $\Gamma_2$ brightness models, respectively. These show the impact of RV data in the context of the correlation introduced in Section\,\ref{sec:degeneracies}. RV data favors solutions with $\sqrt{e}\sin\omega$ $\lesssim $ 0.15 and, hence, solutions consistent with the inferences obtained with less complex brightness models (see Fig\,\ref{fig:correl_mono}).}
  \label{fig:correl_multi_RV}
\end{figure}

\begin{figure}
   \centering
   
      \subfloat[]{\hspace{-18mm}\label{fig:103_brightness_ddps_RV}\includegraphics[angle = -90, trim = 70mm 10mm 70mm 10mm,clip,width=19cm,height=!]{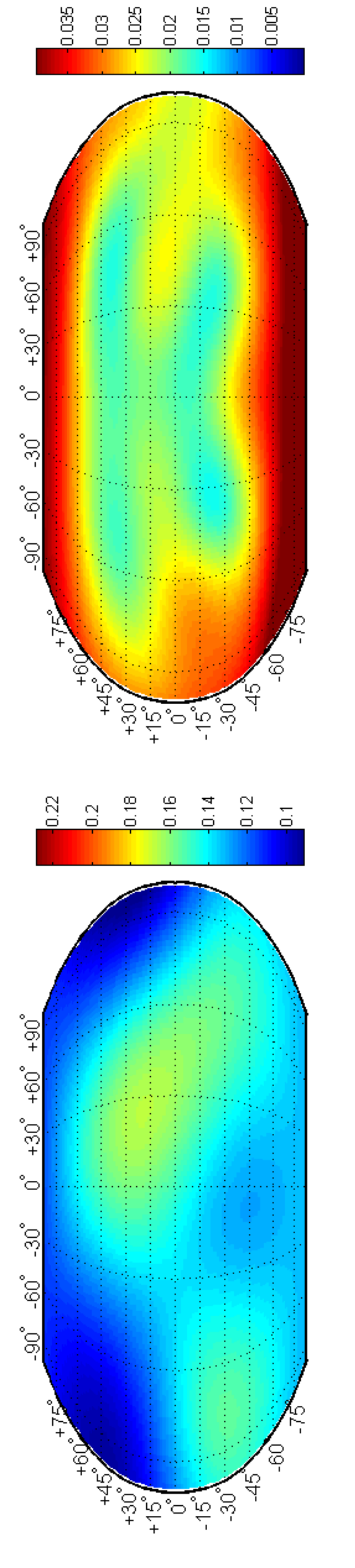}}
      
            \subfloat[]{\hspace{-18mm}\label{fig:3_brightness_ddps_RV}\includegraphics[angle = -90, trim = 70mm 10mm 70mm 10mm,clip,width=19cm,height=!]{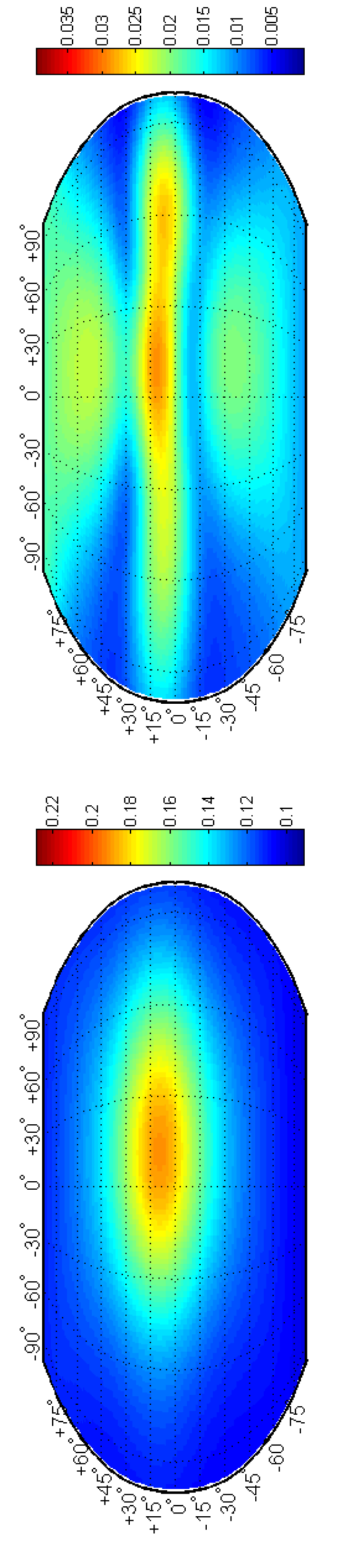}}

  \vspace{+0.5cm}
    \caption[Influence of HD\,189733's RV measurements on HD\,189733b's retrieved dayside brightness distribution.]{Influence of HD\,189733's RV measurements on HD\,189733b's dayside brightness distribution, estimated using complex brightness models that suggest non-zero eccentricity from the \textit{Spitzer}/IRAC 8-$\mu$m photometry (see Figures\,\ref{fig:103_e_ddp} and \,\ref{fig:3_e_ddp}). \textit{Left:} Relative brightness distribution of HD\,189733b's dayside. \textit{Right:} Dayside standard deviation. \subref{fig:103_brightness_ddps_RV} Estimate using the $\Gamma_{SH,3}$ brightness model. \subref{fig:3_brightness_ddps_RV} Estimate using the $\Gamma_2$ brightness model. The influence of HD\,189733's RV data is an extension of the brightness
patterns (compare with Figures\,\ref{fig:103_brightness_ddps} and \ref{fig:3_brightness_ddps}, respectively) associated with a decrease of the brightness peak---conservation of the hemisphere integrated flux. These show that localized brightness pattern that are favored by the photometry are rejected by the RV data, because these are associated with $\sqrt{e}\sin\omega$ $ \gtrsim $ 0.15.}
  \label{fig:complex_brightness_RV}

\end{figure}

	We observe no change of the system parameters PPDs for simple brightness models, as expected from the comparison of Figure\,\ref{fig:ddp_e_unif} and Figures\,\ref{fig:101_e_ddp} and \ref{fig:102_e_ddp}. However, we observe the effect of rejecting the solutions involving $\sqrt{e}\sin\omega$ $ \gtrsim $ 0.15---in the context of the $ e $-$\omega$-$ b $-$\rho_{\star}$-BD correlation---for complex brightness models ($\Gamma_{SH,3}$ and $\Gamma_2$). In particular, the evolution of the parameter estimates---toward lower densities, larger impact parameter and a more localized and latitudinally-shifted hot spot---providing larger eccentricity are mitigated. We present the influence on the system-parameter PPD in Figure\,\ref{fig:correl_multi_RV}. The $\left\lbrace\sqrt{e}\cos\omega,\sqrt{e}\sin\omega\right\rbrace$-PPD (Figures\,\ref{fig:103_e_ddp_RV} and \ref{fig:3_e_ddp_RV}) conceptually corresponds to the marginalized product of the PPDs estimated using solely the photometry (see Figures\,\ref{fig:103_e_ddp} and \,\ref{fig:3_e_ddp}) and using solely the RV data (Figure\,\ref{fig:RVfit}). These highlight the redistribution of the probability density that results from the rejection of solutions involving $\sqrt{e}\sin\omega$ $ \gtrsim $ 0.15 by HD\,189733's RV data. As a consequence of this probability redistribution in $\lbrace\sqrt{e}\cos\omega,\sqrt{e}\sin\omega\rbrace$, the set of acceptable combinations to compensate HD\,189733b's anomalous occultation is reduced too. In other words, the $ b $, $\rho_{\star}$ and BD estimates are affected. On the one hand, the system parameter estimates are consistent with those obtained for less complex brightness models (see Table\,\ref{tab:BFP_RV} and compare Figures\,\ref{fig:103_erho_ddps_RV} and \ref{fig:3_erho_ddps_RV} to Figures\,\ref{fig:101_erho_ddps} and \ref{fig:102_erho_ddps}). On the other hand, HD\,189733b's dayside brightness estimates present less confined patterns (compare Figures\,\ref{fig:103_brightness_ddps} and \ref{fig:3_brightness_ddps} with Figures\,\ref{fig:103_brightness_ddps_RV} and \ref{fig:3_brightness_ddps_RV}, respectively).

\begin{table}
\caption{Parameter estimates obtained from the photometry and the RV measurements. \label{tab:BFP_RV}}
	\centering
	\setlength{\extrarowheight}{4pt}

	\footnotesize{\begin{tabular}{c|c c|c c}
	
	\hline\hline
	\multirow{2}{*}{\textbf{Parameters (units)}}  & \multicolumn{2}{c}{\textbf{Unipolar brightness}} \vline & \multicolumn{2}{c}{\textbf{Multipolar brightness}}\\
	  & $\Gamma_{SH,1}$ & $\Gamma_{2}$ & $\Gamma_{SH,2}$ & $\Gamma_{SH,3}$\\
	\hline

			$b (R_\star)$	& $0.6591\pm^{0.0028}_{0.0023}$ & $0.6604\pm^{0.0063}_{0.0031}$ & $0.6590\pm^{0.0029}_{0.0025}$ & $0.6592\pm^{0.0036}_{0.0025}$
			  \\
			$\sqrt{e}\cos\omega$ & $0.0012\pm^{0.0039}_{0.0022}$ & $-0.0002\pm^{0.0027}_{0.0028}$ & $0.0003\pm^{0.0043}_{0.0038}$ & $0.0003\pm^{0.0046}_{0.0038}$
			 \\
			$\sqrt{e}\sin\omega$ &  $0.007\pm^{0.050}_{0.030}$  &  $0.031\pm^{0.079}_{0.043}$  &  $0.006\pm^{0.052}_{0.031}$  & $0.009\pm^{0.056}_{0.034}$
			\\
			$\rho_\star (\rho_\odot)$ &			$1.922\pm^{0.017}_{0.022}$ &			$1.912\pm^{0.023}_{0.051}$ &			$1.922\pm^{0.018}_{0.021}$ &			$1.922\pm^{0.019}_{0.025}$

	\end{tabular}}
	\vspace{-0.3cm}
\end{table}	

\subsubsection{Conclusion}	
	
	We estimate that HD\,189733b's brightness distribution in the IRAC 8-$\mu$m channel is best represented by Figure\,\ref{fig:101_brightness_ddps}, based on our global analysis of the \textit{Spitzer}/IRAC 8-$\mu$m photometry and the RV measurements of HD\,189733. In addition, we refine the constraint on HD\,189733b's orbital eccentricity to $ e\leq 0.011$ ($95\%$  confidence).

 \subsection{Discussion}
\label{sec:discussionII}
	\subsubsection{Adequacy of Conventional Analyses}
	\label{sec:impact1}

	We show based on high-SNR data the limitation of conventional analyses to interpret light curves because of they assume the planet to be uniformly bright despite the $ e $-$\omega$-$ b $-$\rho_{\star}$-BD correlation. The significance of this limitation is related to an occultation SNR---which is particularly high for HD\,189733b's occultation in the \textit{Spitzer}/IRAC 8-$\mu $m channel. A ``sufficient'' SNR on the occultation requires resolving the occultation ingress/egress; therefore, it has to be about one order of magnitude less than the occultation depth, for a time bin about one order of magnitude less than the occultation ingress/egress duration. In other words, conventional analyses are typically adequate when the ratio eclipse depth to photometric precision on a $\sim$1-min bin is $\ll10$. For occultations with significantly lower SNR, the underlying assumption of conventional analyses have a negligible effect on the system parameter PPD, and hence conventional analysis are adequate.
	
	We briefly outline that the conventional assumption of a uniformly-bright exoplanet could also affect the inferred planetary interior models; because $ \rho_{\star} $ is possibly affected and, therefore, $ \rho_{p} $ is too (e.g., suggested 3.6\%-overestimation for HD\,189733b's).

	 \subsubsection{Reducing Brightness Distributions to Unidimensional Parameters}
 \label{sec:bpl}

We show in Section\,\ref{sec:unipolar} that  the light curve of an exoplanet does not constrain uniquely its brightness peak localization. We discuss here the reasons why we strongly advocate discussing the BD estimates and, if necessary, using with care the dayside barycenter as an alternative possibility for representative 1-D parameter. The reason is that the dayside brightness barycenter weights the BD according to the geometrical configuration at superior conjunction (i.e., it contains partial 2D information).

\begin{figure}
  \centering
   \begin{center}
    \hspace{-0.cm}\includegraphics[trim = 35mm 85mm 35mm 92mm,clip,width=0.7\textwidth,height=!]{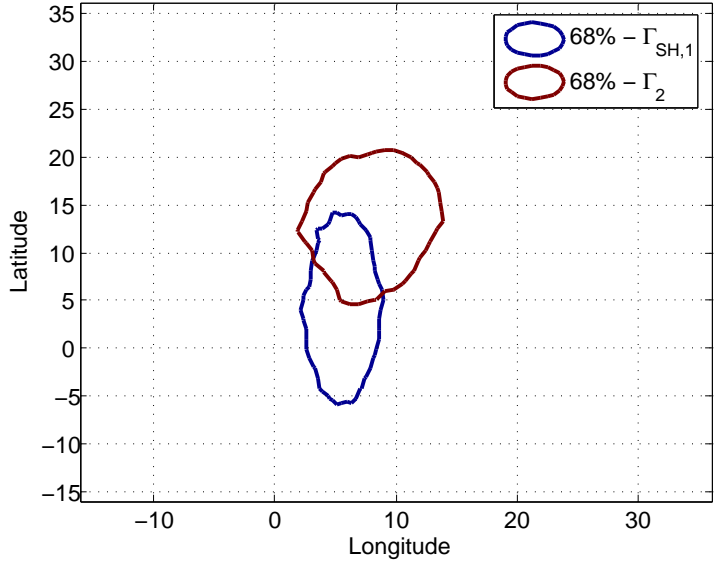}
  \end{center}
  \vspace{-0.5cm}  
  \caption[Reducing brightness distributions to unidimensional parameters.]{Reducing brightness distributions to unidimensional parameters. Marginal PPDs ($68\%$-confidence intervals) of the dayside barycenter brightness peak localization for the $\Gamma_{SH,1}$ and $\Gamma_{2}$ brightness models. A comparison with the marginal PPDs of the brightness peak localization (Figure\,\ref{fig:101_peak_ddps}) shows the reduced model-dependence of the dayside barycenter. In particular, it shows a less-extended PPD for the $\Gamma_{2}$-model barycenter; because this extension for the brightness-peak-localization PPD emerges from the model wings, weighted by the dayside barycenter. In addition, it shows the shift and slight shrinking of the $\Gamma_{SH,1}$ PPD that reflect the barycenter weighting according to the geometrical configuration at superior conjunction; map cells closer to the substellar point have more weight.}
  \label{fig:barycenter}
\end{figure}

We show in Figure\,\ref{fig:barycenter} the marginal PPDs ($68\%$-confidence intervals) of the brightness peak localization for the $\Gamma_{SH,1}$ and $\Gamma_{2}$ brightness models. A comparison with the marginal PPDs of the brightness peak localization (Figure\,\ref{fig:101_peak_ddps}) shows the reduced model-dependence of the dayside barycenter. In particular, it shows a less-extended PPD for the $\Gamma_{2}$-model barycenter; because this extension for the PPD of the brightness-peak-localization emerges from the model wings---weighted by the dayside barycenter. In addition, it shows the shift and slight shrinking of the $\Gamma_{SH,1}$ PPD that reflects the barycenter weighting according to the geometrical configuration at superior conjunction; map cells closer to the substellar point have more weight. This emphasizes the primary drawback of the dayside barycenter that is to attenuate the offset of BDs. This recalls that unidimensional parameters cannot stand adequately for complex BDs and, therefore, have to be used complementary to BD estimates.

\subsubsection{Effect of HD\,189733's activity}

\begin{figure}
   \centering

  \begin{center}
    \hspace{-0.cm}\includegraphics[trim = 10mm 70mm 5mm 70mm,clip,width=0.8\textwidth,height=!]{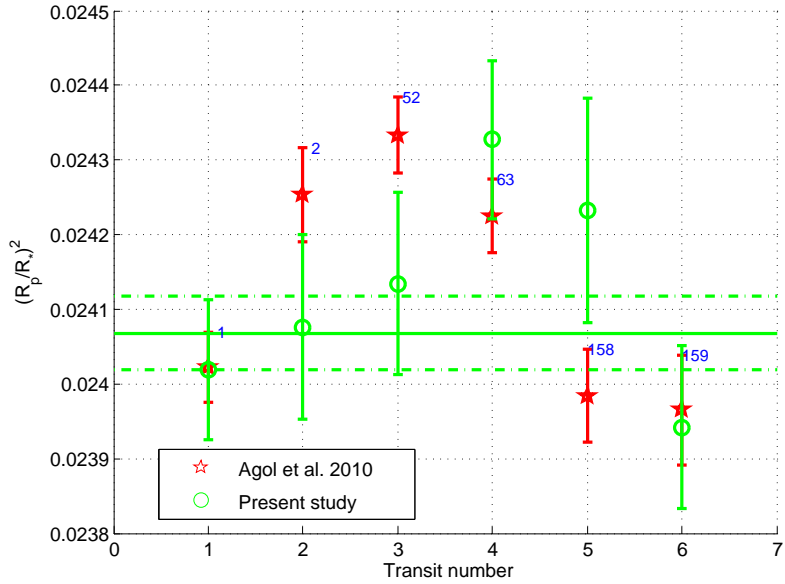}
  \end{center}
  \vspace{-0.5cm}
  \caption[Influence of HD\,189733's variability on individual transit-depth estimates in the \textit{Spitzer}/IRAC 8-$\mu$m channel.]{Influence of HD\,189733's variability on individual transit-depth estimates in the \textit{Spitzer}/IRAC 8-$\mu$m channel. Transit-depth estimated for six transits in individual (markers) and global (solid line, the dashed lines refer to the 1-$\sigma$ error bar) analyses. Our estimates (green) show no significant transit-depth variation with the relative eclipse phase (blue numbers) in opposition to \cite{Agol2010} (red).}
  \label{fig:DdF}

\end{figure}

HD\,189733 presents high-activity levels that may affect the transit parameters---incl., $ R_p^2/R_\star^2 $, due to occulted/unocculted star spots \citep[][]{Pont2007,Sing2011}. Therefore, treating coherently spots of active stars is key. However, while important in the optical, the stellar activity may be negligible at 8\,$\mu $m with the current data quality. We assess this statement performing individual analysis of the 6 transits used in this study (see Table\,\ref{tab:AOR}). We present in Figure\,\ref{fig:DdF} our individual transit-depth estimates. These estimates show no significant temporal variation \citep[in opposition to][who attributed these to stellar activity]{Agol2010}. Similarly, we observe no significant variation of the transit parameters from one individual analysis to another. In addition, we observe no pattern specific to the occultation of a star spot \citep[i.e., similar to, e.g., ``Features A and B'' in the Figure 1 of][]{Pont2007}. Therefore, we consider that our time-averaged  inferences (see Section\,\ref{sec:results}) are not biased by HD\,189733's activity.

\subsubsection{Comparison with \cite{Majeau2012}}
 \label{sec:compare_2_M12}
We obtain qualitatively similar results as M12: an offset hot spot. Nevertheless, significant differences exist between both studies. \textbf{(1)} The starting point of our mapping analysis is our detection of HD\,189733b's eclipse scanning at the 6$\sigma$ level (in contrast to their $\sim$3.5\,$\sigma$). Our study provides consistent estimates of the contributions of the multiple possible factors to HD\,189733b's occultation shape. \textbf{(2)} In addition, and related to the second point, we do not constrain \textit{a priori} the system parameters to the best-fit of a conventional analysis nor the orbital eccentricity to zero; instead, we estimate the system parameters simultaneously with the BD. \textbf{(3)} We also investigate the model-dependence of our inferences because the mapping problem has degenerate solutions, in contrast to M12 who focused on a dipolar brightness model (see related discussion in Section\,\ref{sec:unipolar}). In particular, we show their impact on the system-parameter PPD (see Section\,\ref{sec:results}). 

As a consequence, M12 estimate the brightness peak localization with narrow error bars ($21.8\pm1.5^\circ$east and $3.1\pm9.4^\circ$ away from the equator) for a circularized HD\,189733b's orbit; while we show that the brightness peak localization, as well as the system-parameter PPD, are model-dependent because of the $ e $-$\omega$-$ b $-$\rho_{\star}$-BD correlation. Nevertheless, for a direct comparison to M12's estimate of the brightness peak localization for a dipolar brightness model, we estimate it to $11.5\pm4.3^\circ$east and $3.1\pm11.4^\circ$ away from the equator, see Section\,\ref{sec:discussionII}, Figure\,\ref{fig:101_peak_ddps}.

\newpage
\section{Application II: Kepler-7 b}
\label{sec:K7b}
The mapping work introduced in this Section was performed for and published in \cite{Demory2013}. The overall analysis presented in  \cite{Demory2013} not only presents the first evidence of optically thick clouds in an exoplanetary atmosphere but also the first cloud map of an exoplanet (see Figure\,\ref{fig:k7bmaps}). Kepler-7 b's cloud map provides unprecedented insights into its climate: our current understanding is that its equatorial jet brings hot gas from its dayside to its nightside where the temperature drops enough to form clouds---possibly silicate-dominated clouds---that are then brought to the dayside where they reflect the host-star light until their properties change as they encounter higher temperature around the substellar point.

\vspace{-0.3cm}
\subsection*{Introduction}

Kepler-7b is a hot Jupiter orbiting a subgiant G star in 4.89 days  \citep{Latham2010}. Its relatively low mass $M_p = 0.44 \pm 0.04M_{Jup}$ and large radius $R_p = 1.61 \pm 0.02 R_{Jup}$ result in a very low density $\rho_p = 0.14 g.cm^{−3}$ \citep{Demory2011}. Remarkably, Kepler-7b has a significant geometric albedo
$A_g \sim 0.35$ and exhibits a clear phase-curve modulation in the Kepler bandpass \citep{Demory2011,Kipping2011,Coughlin2012}. Kepler-7b's effective temperature places it in an exceptionally rich region of condensation
phase space. Because of the extreme difference between its equilibrium temperature and the brightness temperature as derived from its occultation in the Kepler bandpass, the origin of Kepler-7b’s albedo has been attributed to the presence of a cloud or haze layer in its atmosphere or to Rayleigh scattering \citep{Demory2011}.

\cite{Demory2013} presents a detailed study of Kepler-7 b's observation using \textit{Spitzer} and \textit{Kepler}: \textbf{(1)} Description, reduction, and analysis of \textit{Spitzer}'s 3.6 and 4.5 $\mu$m photometry and \textbf{(2)} Description, reduction, and analysis of \textit{Kepler}'s photometry spanning over three years of quasi-continuous monitoring---which includes the mapping work based on the phase-folded light curve. Finally, combining the insight gained from infrared and visible dataset, \cite{Demory2013} constrain the origin of Kepler-7 b's visible flux as scattered light from optically thick clouds and discussed the possible properties of those clouds. Only the work concerning Kepler-7 b's mapping is part of this thesis and is introduced hereafter. For further details regarding the overall and multidisciplinary analysis yielding to the first map of cloud in an exoplanet atmosphere, please refer to \cite{Demory2013}.

\vspace{-0.3cm}
\subsection*{Mapping using \textit{Kepler}'s Dataset}

\begin{figure}
   
     \vspace{-1.5cm}\hspace{-2.2cm}\includegraphics[trim = 00mm 00mm 00mm 00mm,clip,width=19cm,height=!]{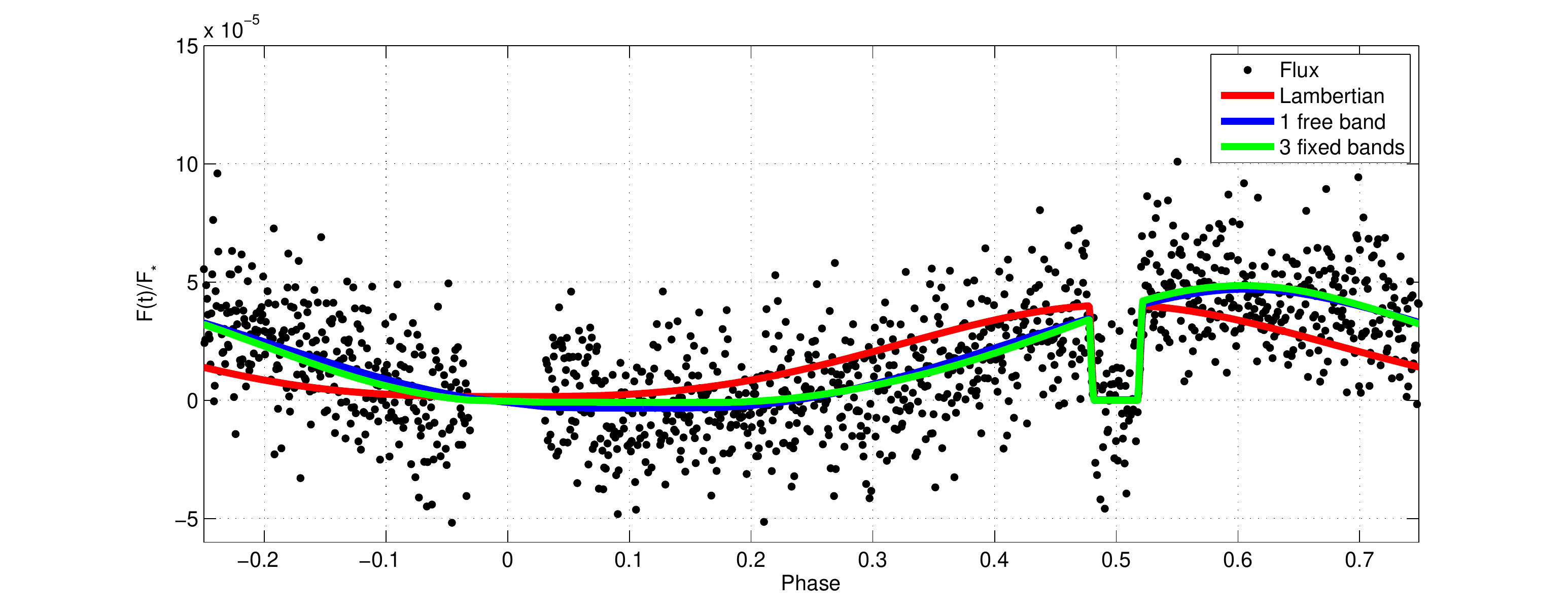}

  \vspace{-0.3cm}
  \caption[Kepler-7b's phase curve based on \textit{Kepler} Q1-Q14 data.]{Kepler-7b's phase curve based on \textit{Kepler} Q1-Q14 data. Data are binned per 5 minutes. The Lambertian sphere (red), 1-free-band (blue), and 3-fixed-band (green) best-fit models are superimposed.}
  \vspace{-0.3cm}
  \label{fig:k7bphasecurve}
\end{figure}

We map Kepler-7b using its corrected and phase-folded phase curve obtained in the visible using the \textit{Kepler Space Telescope} over more than three years (Figure\,\ref{fig:k7bphasecurve}). The data reduction and correction is detailed in \cite{Demory2013}, together with a discussion of the robustness of Kepler-7b's phase curve over the different quarters.

Kepler-7b's phase curve deviates from a pure Lambert-law phase-dependent behavior \citep[e.g.,][]{Sobolev1975} expected for isotropic scattering alone (Figure\,\ref{fig:k7bphasecurve}, red). The main feature of Kepler-7b's phase curve is a delay of $13 \pm 3.5$ hr of the phase-curve's peak from the occultation center. This delay implies that the hemisphere-integrated flux is maximum to the west of Kepler-7b's substellar point. We further measure a phase-curve amplitude of $50\pm2$ ppm and an occultation depth of $48\pm3$ ppm, corresponding to a geometric albedo $A_g = 0.35 \pm 0.02$. This occultation depth translates to a brightness temperature of $2645\pm^{20}_{30}$ K in the \textit{Kepler} bandpass, which is 1000 K and 800 K larger than the infrared brightness temperatures upper limits measured at 3.6 and 4.5$\mu$m respectively \citep[see Section 2 of ][]{Demory2013}. The phase-curve amplitude and occultation depth are in agreement with previous analyses \citep{Demory2011,Kipping2011,Coughlin2012}.
The key features of Kepler-7b's phase-curve translate directly into constraints on maps \citep{Cowan2008} assuming a tidally locked planet on a circular orbit. Indeed, a planetary phase-curve $F_p/F_\star$ measures the planetary hemisphere-averaged relative brightness $<I_p>/<I_\star>$ as follows:
\begin{eqnarray}
	\frac{F_p}{F_\star}(\alpha) & = & \frac{<I_p>(\alpha)}{<I_\star >} \left(\frac{R_p}{R_\star}\right)^2
	,\label{eq:phasecurve} 
\end{eqnarray}
where $\alpha$ is the orbital phase.

We first notice that Kepler-7b's planetary flux contribution starts from phase $0.18 \pm 0.03$, when the meridian centered $25\pm12^\circ$ east of the substellar point appears. Second, the phase-curve's maximum is located at phase $0.61\pm0.03$, implying that the brightest hemisphere is centered on the meridian located $41 \pm 12^\circ$ west of the substellar point. Third, the planetary flux contribution vanishes around the transit, implying that the ``bright'' area extends up to the western terminator, while its extension to the east of the substellar point is nominal. We finally note that the phase-curve's amplitude of $50 \pm 2$ ppm converts into an hemisphere-averaged relative brightness $74 \pm 2 \times 10^{−4}$ (Equation\,\ref{eq:phasecurve}).

\begin{figure}
   \subfloat[]{\label{fig:k7bmap1band}\hspace{-0.cm}\includegraphics[trim = 00mm 00mm 191mm 00mm,clip,width=15cm,height=!]{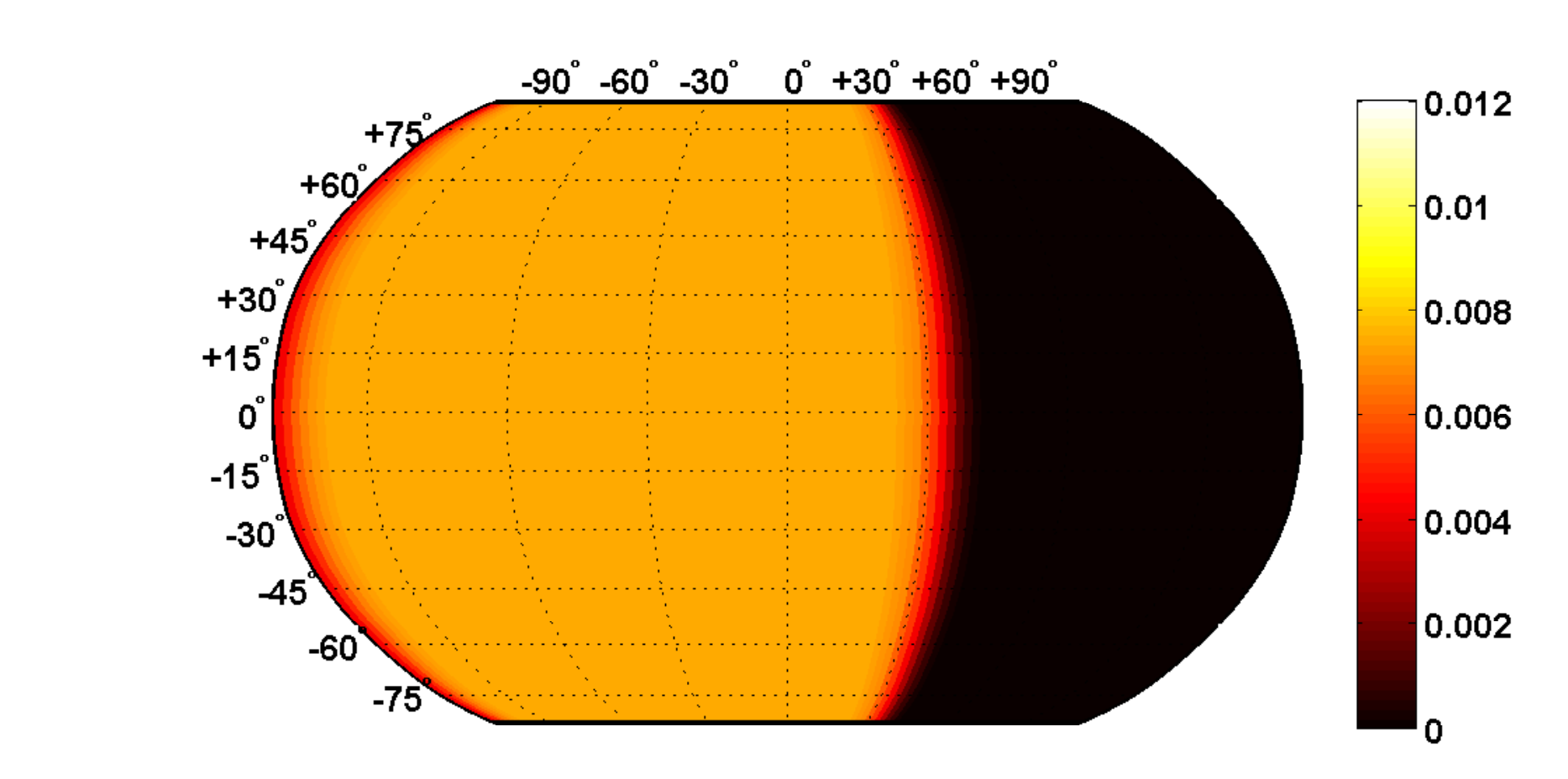}}
      
      \subfloat[]{\label{fig:k7bmap3bands}\hspace{+0.cm}\includegraphics[trim = 191mm 00mm 00mm 00mm,clip,width=15cm,height=!]{brice_map_HOT.pdf}}

  \caption[Longitudinal brightness maps of Kepler-7b.]{Longitudinal brightness maps of Kepler-7b. Kepler-7b's longitudinal brightness distributions $I_p(\alpha)/I_\star$ as retrieved in Kepler's bandpass using the 1-free-band
model \subref{fig:k7bmap1band} and the 3-fixed-band model \subref{fig:k7bmap3bands}. It shows that Kepler-7b's is bright in \textit{Kepler}'s band in an area that extends from its dayside western terminator to $\sim 30^{\circ}$ east of its substellar point.}
  \label{fig:k7bmaps}
\end{figure}

We longitudinally map Kepler-7b using the MCMC implementation presented in Section\,\ref{sec:analysismethod}. We use the 1D models as Kepler-7b's brightness distribution is mainly constrained longitudinally. We choose the two simplest models from these families: a 3-fixed-band model and 1-free-band model so as to extract Kepler-7b's longitudinal dependence of the dayside brightness as well as the extent of the ``bright'' area. For both models, we compute each band's amplitude from their simulated light curve by using a perturbed singular value decomposition method. The corresponding median brightness maps are shown on Figure\,\ref{fig:k7bmaps}. The 1-free-band model (Figure\,\ref{fig:k7bphasecurve}, blue) finds a uniformly bright longitudinal area extending from $105\pm12^{\circ}$ west to $30 \pm 12^{\circ}$ east with a relative brightness $78\pm4\times10^{−4}$ (Figure\,\ref{fig:k7bmap1band}). The 3-fixed-band model (Figure\,\ref{fig:k7bphasecurve}, green) finds bands of relative brightness decreasing from the west to the east with the following values: 100 to 68 and $3\pm6\times10^{−4}$ (Figure\,\ref{fig:k7bmap3bands}). We finally note that the 1-freeband model finds a bright sector extending to the night side, due
to the sharp flux increase observed around transit (Figure\,\ref{fig:k7bphasecurve}).

%% file: Chapter3/chap3.tex
\renewcommand\thefootnote{\fnsymbol{footnote}}

\chapter[Constraining Exoplanet Mass from Transmission Spectroscopy]{Constraining Exoplanet Mass from Transmission Spectroscopy\footnote{Work published in part in \textit{Science}, Volume 342, Issue 6165, pp. 1473-1477 (2013), see \cite{deWit2013}.}}
\label{chap:mass}

\vspace{-0.7cm}
\renewcommand\thefootnote{\arabic{footnote}}
\addtocounter{footnote}{-1}

The mass of a planet is a fundamental parameter because it is connected to its internal and atmospheric structure. A planet's mass affects basic planetary processes such as interior cooling and plate tectonics \citep{Stamenkovic2012,Stamenkovic2014}, magnetic field generation \citep{Stevenson2003,Stamenkovic2011}, outgassing \citep{Elkins2008}, and atmospheric escape \citep{Lammer2009}. Measurement of a planetary mass can in many cases reveal the planet bulk composition, allowing to determine whether the planet is a gas giant or is rocky and suitable for life as we know it. 

Planetary mass is traditionally constrained with the radial velocity (RV) technique, using single-purpose dedicated instruments. The RV technique measures the Doppler shift of the stellar spectrum to derive the planet-to-star (minimum) mass ratio as the star orbits the planet-star common center of mass. Although the RV technique has a pioneering history of success laying the foundation of the field of exoplanet detection, it is mainly effective for massive planets around relatively bright and quiet stars. Most transiting planets have host stars that are too faint for precise RV measurements. For sufficiently bright host stars, stellar perturbations may be larger than the companion-induced signal, possibly preventing a determination of the planet mass with RV measurements even for hot Jupiters \citep[e.g.,][]{Collier2010}. In the long term, the limitation due to the faintness of targets will be reduced with technological improvements. However, host star perturbations may be a fundamental limit that cannot be overcome, meaning that the masses of small planets orbiting relatively quiet stars would remain out of reach.

Current alternative mass measurements to RV are based on modulations of planetary-system light curves \citep{Mislis2012} or transit-timing variations \citep{Fabrycky2010}. The former works for massive planets on short period orbits and involves detection of both beaming and ellipsoidal modulations \citep{vanKerkwijk2010, Faigler2011, Shporer2011}. The latter relies on gravitational perturbations of a companion on the transiting planet's orbit \citep{Miralda2002,Agol2005,Holman2005}. This method is most successful for companions that are themselves transiting and in orbital resonance with the planet of interest \citep{Carter2012}. For unseen companions the mass of the transiting planet is not constrained, but an upper limit on the mass of the unseen companion can be obtained to within 15 to 50\% \citep{Steffen2013}.

Transiting exoplanets are of special interest because their densities can be estimated providing insights into their internal structures. Furthermore, the atmospheric properties of a transiting exoplanet can be retrieved from the host-star light passing through its atmosphere when it transits. However, the quality of atmospheric retrieval is reduced if the planet's mass is inadequately constrained \citep{Barstow2013}. 

In this Chapter, we show that a planet's mass is a key parameter of its transmission spectrum that has to be accounted for by atmospheric retrieval methods to avoid the deterioration of the retrieval quality introduced by \cite{Barstow2013}. Therefore, we develop \textit{MassSpec}, a method for simultaneously and self-consistently constraining the mass and the atmospheric properties of an exoplanet based solely on transit observations, thereby enabling mass measurements for transiting planets for which the RV method fails.

\section{\textit{MassSpec}'s Concept and Feasibility}
\label{sec:spectrum2mass}

	The mass of a planet affects its transmission spectrum through the pressure profile of its atmosphere (i.e., $p(z)$ where $z$ is the altitude), and hence its atmospheric absorption profile. For an ideal-gas atmosphere in hydrostatic equilibrium, the pressure varies with the altitude as $\text{d}\ln(p) = -\frac{1}{H} \text{d} z$, where $H$ is the local atmospheric scale height defined as 		
\begin{eqnarray}
H & = & \frac{kT}{\mu g},
\label{sh1}
\end{eqnarray}
where $k$ is Boltzmann's constant and $T$, $\mu$ and $g$ are the local (i.e., altitude dependent) temperature, mean molecular mass, and gravity. By expressing the local gravity in terms of the planet's sizes (mass, $M_p$, and radius, $R_p$), Equation\,\ref{sh1} can be rewritten as
\begin{equation}
M_p=\frac{kTR_p^2}{\mu GH}.
\label{M2}
\end{equation} 
Thus, our method conceptually requires constraining the radius of the target as well as its atmospheric temperature, mean molecular mass, and scale height. Can these independent constraints be provided solely by a transmission spectrum?

\vspace{-0.0cm}
\begin{wrapfigure}{r}{0.4\textwidth}
   \begin{center}
   \begin{minipage}[c]{0.4\textwidth}
    \vspace{-1.0cm}\includegraphics[width=7.0cm,height=!]{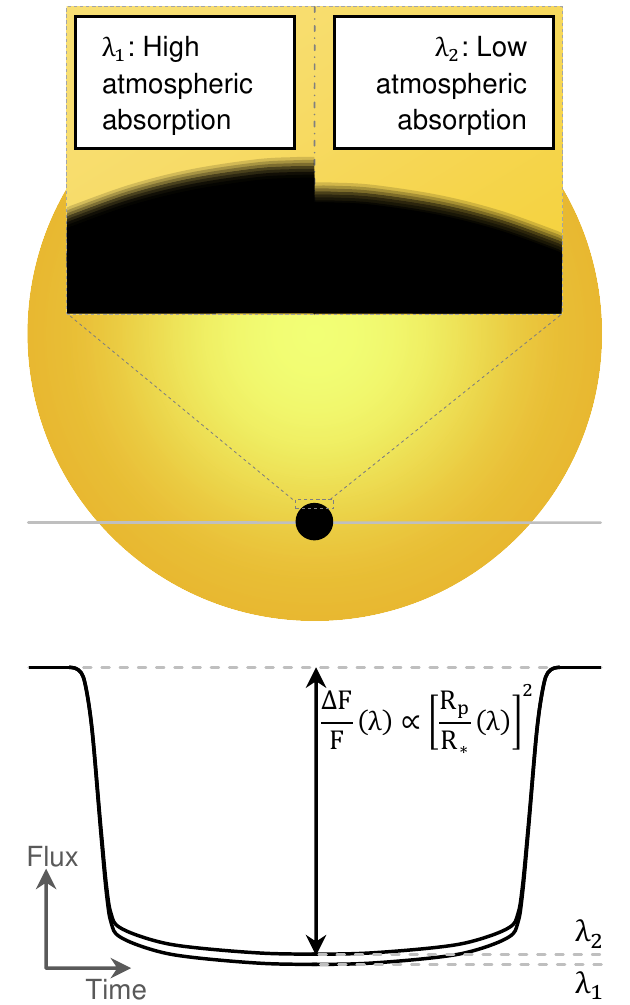}  
  \end{minipage}\hfill
  \end{center}
  \vspace{-1.8cm}
  \hspace{-0.6\textwidth}\begin{minipage}[c]{\textwidth}
  \caption[Transit-depth variations, $ \frac{\Delta F}{F} (\lambda)$, induced by the wavelength-dependent opacity of a transiting planet atmosphere.]{Transit-depth variations, $ \frac{\Delta F}{F} (\lambda)$, induced by the wavelength-dependent opacity of a transiting planet's atmosphere. The stellar disk and the planet are not resolved; the flux variation of a point source is observed.}
  \vspace{-5cm}
  \label{fig:transmission_spectrum}
  \end{minipage}
\end{wrapfigure}

 \vspace{-.0cm}A planet transmission spectrum can be seen as a wavelength-dependent drop in the apparent brightness of the host star when the planet transits (Figure\,\ref{fig:transmission_spectrum}). At a wavelength with high atmospheric absorption, $\lambda_1$, the planet appears larger than at a wavelength with lower atmospheric absorption, $\lambda_2$---because of the deeper transit depth due to the larger opaque atmospheric annulus. In particular, a relative flux-drop, $ \frac{\Delta F}{F} (\lambda)$, is associated with an effective planetary radius, $R_p(\lambda) = R_{\star}\sqrt{\frac{\Delta F}{F} (\lambda)}$. Transmission spectroscopy mainly probes low-pressure levels; therefore, the mass encompassed in the sphere of radius $R_p(\lambda)$ (Equation\,\ref{M2}) is a good proxy for the planetary mass. 
\newline
\newline
\newline

\begin{figure}[!ht]
  
    \vspace{-0.6cm}\hspace{-1.8cm}{\includegraphics[trim = 00mm 170mm 00mm 35mm,clip,width=19cm,height=!]{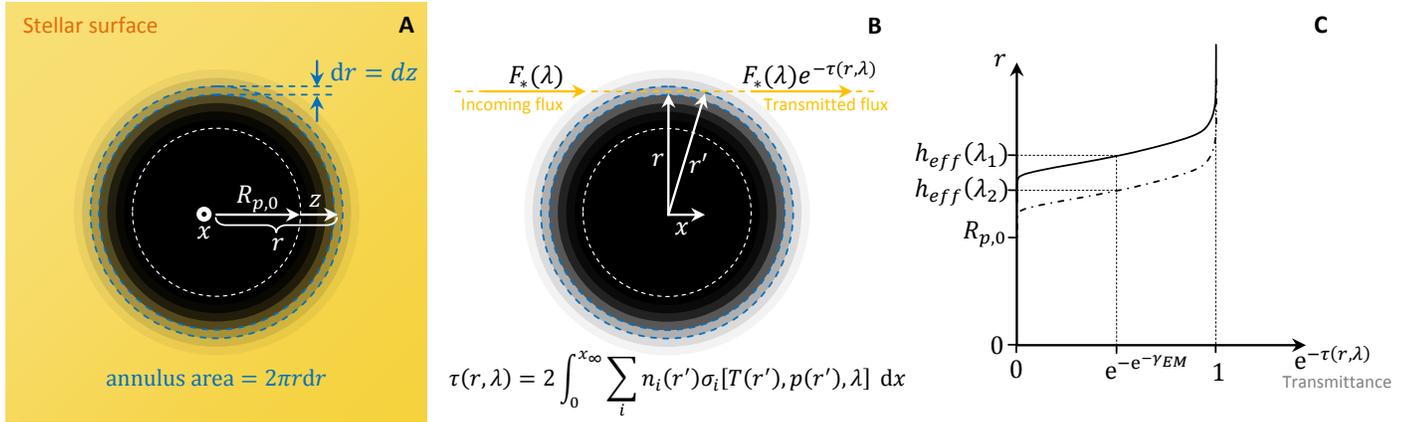}}
  
  \vspace{-0.0cm}
  \caption[Basics of a planet's transmission spectrum]{Basics of a planet's transmission spectrum (planetary atmosphere scaled up to enhance visibility). (\textbf{A}) In-transit geometry as viewed by an observer presenting the areas of the atmospheric annulii affecting the transmission spectrum. (\textbf{B}) Side-view showing the flux transmitted through an atmospheric annulus of radius $r$. (\textbf{C}) Transmittance as a function of the radius at wavelengths with high and low atmospheric absorption---$\lambda_1$ (solid lines) and $\lambda_2$ (dash-dotted lines), respectively. Due to higher atmospheric absorption at $\lambda_1$, the planet will appear larger than it does at $\lambda_2$, because of the more-extended opaque atmospheric annulus $\left[h_{eff}(\lambda_1)) > h_{eff}(\lambda_2)\right]$ that translates into an additional flux drop.}
  \label{fig:transmission_spectrum_basics}
\end{figure}

 The atmospheric properties of a planet directly determine its effective radius in the following manner:
	\begin{eqnarray}
\pi R_p^2(\lambda) = \pi \left[R_{p,0} + h_{eff}(\lambda)\right]^2 & = & \int_{0}^{\infty} 2\pi r \left(1-e^{-\tau(r,\lambda)}\right) \text{ d}r,
\label{eq:transmission_spectrum_h}
\end{eqnarray}
where $R_{p,0}$, $h_{eff}(\lambda)$, and $e^{-\tau(r,\lambda)}$ are respectively a planetary radius of reference\footnote{A planetary radius of reference is any radial distance at which the body is optically thick in limb-looking over all the spectral band of interest, i.e., $\tau(R_{p,0},\lambda) \gg 1\text{ }\forall \lambda \in$ the band.\vspace{-0.7cm}}, the effective atmosphere height, and the planet's transmittance at radius $r$ (Figure\,\ref{fig:transmission_spectrum_basics}). $\tau(r,\lambda)$ is the slant-path optical depth defined as
\begin{eqnarray}
\tau(r,\lambda) & = & 2\int_{0}^{x_{\infty}} \sum_in_{i}(r') \times \sigma_{i}\left[T(r'),p(r'),\lambda\right] \text{ d}x,
 \label{eq:optical_depth}
\end{eqnarray} 
where $r' = \sqrt{r^2+x^2}$ (Figure\,\ref{fig:transmission_spectrum_basics}.B), and $n_{i}(r')$ and $\sigma_{i}\left[T(r'),p(r'),\lambda\right]$ are the number density and the extinction cross section of the $i^{th}$ atmospheric component at the radial distance $r'$ \citep{Seager2010}. A planet's atmospheric properties $\left[n_{i}(z),T(z),\mbox{ and }p(z)\right]$ are embedded in its transmission spectrum through $\tau(r,\lambda)$ (Equations\,\ref{eq:transmission_spectrum_h} and\,\ref{eq:optical_depth}).

The integral in Equation\,\ref{eq:transmission_spectrum_h} can be split at the radius of reference (because the planet is opaque at all $\lambda$ at smaller radii), and thus Equation\,\ref{eq:transmission_spectrum_h} becomes
	\begin{eqnarray}
\left[R_{p,0} + h_{eff}(\lambda)\right]^2 & = & R_{p,0}^2 + R_{p,0} c \text{, }\label{eq:transmission_spectrum_h2}\\
 c &\triangleq& 2 \int_{0}^{\infty}  (1+y) (1-e^{-\tau(y,\lambda)}) \text{ d}y, \footnotemark \nonumber \\
 y & = & z/R_{p,0},  \nonumber
\end{eqnarray}
{\footnotetext{Edited from the version published in \cite{deWit2013} to correct Equation\,\ref{eq:transmission_spectrum_h3} and have consistent definitions of $c$ in Equations\,\ref{eq:transmission_spectrum_h2} and \ref{eq:def_yeff}, based on Stephen Messenger's comments.}
leading directly to the expression of the effective atmosphere height
	\begin{eqnarray}
h_{eff}(\lambda)  = R_{p,0} \left(-1 + \sqrt{1+c}\right).
\label{eq:transmission_spectrum_h3}
\end{eqnarray}
The embedded atmospheric information can be straightforwardly accessed for most optically active wavelength ranges using
\begin{eqnarray}
 h_{eff}(\lambda) & = & R_{p,0}B(\gamma_{EM}+\ln{A_{\lambda}}),
 \label{eq:h_eff_in_text}
\end{eqnarray}
where $\gamma_{EM}$ is the Euler-Mascheroni constant \citep{Euler1740}---$\gamma_{EM}= \lim_{n \to +\infty} \sum_{k = 1}^n \frac{1}{k} - \ln{n} \approx 0.57722$ (see Section\,\ref{app:dependency}). In the above equation, $B$ is a multiple of the dimensionless scale height and $A_\lambda$ is an extended slant-path optical depth at reference radius. The exact formulation of $B$ and $A_\lambda$ depends on the extinction processes affecting the transmission spectrum at $\lambda$ (Table\,\ref{tab:A_B}). As an example, for Rayleigh scattering
\begin{eqnarray}
 B & = & \frac{H}{R_{p,0}} \text{ and }\\
 \label{eq:B_RS_in_text}  
A_{\lambda} & = & \sqrt{2 \pi R_{p,0} H} n_{sc,0} \sigma_{sc}(\lambda),
 \label{eq:Al_RS_in_text}
\end{eqnarray}
where $n_{sc,0}$ and $\sigma_{sc}(\lambda)$ are the number density at $R_{p,0}$ and the cross-section of the scatterers. Conceptually, Equation\,\ref{eq:h_eff_in_text} tells us the height where the atmosphere can be considered transparent in limb-looking. As an example, if $A_\lambda$ is $10^4$ then the atmosphere can be considered transparent at $9\left(\approx\ln(10^4)\right)$ scale heights above the reference radius.

Interestingly, we also show that Equation\,\ref{eq:h_eff_in_text} can be rewritten as 
\begin{eqnarray}
h_{eff}(\lambda) = \left\lbrace z:\tau(z,\lambda) = e^{-\gamma_{EM}}\right\rbrace,
 \label{eq:h_eff_tau_eq}
\end{eqnarray}
meaning that the slant-path optical depth at the effective height is the constant $e^{-\gamma_{EM}}$ independently of $\lambda$ (see Figure\,\ref{fig:transmission_spectrum_basics}.C). In other words, $\tau(h_{eff}(\lambda),\lambda) \triangleq \tau_{eq} = e^{-\gamma_{EM}}$. Therefore 
\begin{eqnarray}
\tau_{eq}  =  \lim_{n \to +\infty} n\prod_{k = 1}^n e^{-1/k} \approx 0.56146, 
\end{eqnarray}
which analytically extends previous numerical observations that $\tau_{eq}\approx0.56$ in some cases \citep{LecavelierDesEtangs2008}. Equations\,\ref{eq:h_eff_in_text} and\,\ref{eq:h_eff_tau_eq} can be rewritten as 
\begin{eqnarray}
 R_p(\lambda) & = & R_{p,0}\left[1+B\left(\gamma_{EM}+\ln{A_{\lambda}}\right)\right] \text{ and }\label{eq:R_eff_tau_eq}\\
 R_p(\lambda) & = & \left\lbrace r:\tau(r,\lambda) = e^{-\gamma_{EM}}\right\rbrace,
\end{eqnarray}
providing a simple analytical way to implement and validate transmission spectrum models (see respectively, Section\,\ref{app:generalization} and Appendix\,\ref{sec:generalizedtransmissionspectra}, and Appendix\,\ref{sec:testtransmissionspectrummodels}).

Most importantly, Equation\,\ref{eq:h_eff_in_text} shows the dependency of a transmission spectrum on its key parameters: in particular, $A_\lambda$ is dependent in unique ways on the scale height, the reference pressure, the temperature, and the number densities of the main atmospheric constituents (for a summary of these dependencies, see Section\,\ref{app:generalization}), which lead to the mean molecular mass. The uniqueness of these dependencies enables, in theory, the independent retrieval of each of these key parameters. Therefore, a planet's mass can, in theory, be constrained uniquely by transmission spectroscopy (Equation\,\ref{M2}). In practice, the determination of an exoplanet's mass and atmospheric properties using solely transmission spectroscopy requires a high signal-to-noise ratio (SNR) over an extended spectral coverage (e.g., 0.2 to 5 $\mu$m) adequately sampled ($\lambda/\Delta\lambda \geqq 300$)---see Sections\,\ref{sec:masspechd189733b} and\,\ref{app:results}.

The next section introduces the analytical derivations that identify the key parameters of an exoplanetary transmission spectrum and its dependencies on these. By identifying those dependencies, we aim to find out if independent constraints on an atmosphere's temperature, composition, and pressure can be obtained solely from transmission spectroscopy.

\newpage
\section[Demonstration]{Demonstration: Key Parameters of and their Effects on a Transmission Spectrum}
\label{app:dependency}}

Here, we derive analytically the dependencies of a transmission spectrum on its main parameters for different extinction processes. In particular, we show that the effective height takes the form $R_{p,0}B(\gamma_{EM}+\ln{A_{\lambda}})$ for extinction processes such as Rayleigh scattering, collision-induced absorption (CIA), and molecular absorption for most optically active wavelength ranges. $R_{p,0}B$ is a multiple of the scale height and $A_{\lambda}$ is an extended slant-path optical depth at $R_{p,0}$. $B$ and $A_{\lambda}$ summarize how a planet's atmospheric properties are embedded in its transmission spectrum. In particular, the formulations of $B$ and $A_{\lambda}$ identify the key parameters behind a planet's transmission spectrum. The formulations of $B$ and $A_{\lambda}$ (i.e., the way the atmospheric properties are embedded by transmission spectroscopy) depend on the extinction processes. Therefore, we first approach in detail the case of Rayleigh scattering---or any processes with an extinction cross-section independent of $p$. Then we extend our demonstration to other processes such as CIA and molecular absorption.

For the coming demonstrations, we will use the following assumptions: (a1) the extent of the optically active atmosphere is small compared to the planetary radius $\left(z \ll R_p(\lambda)\right)$, (a2) the atmosphere can be assumed isothermal $\left(d_zT(R_p(\lambda))\simeq0\right)$, and (a3) the atmosphere can be assumed isocompositional, $d_zX_i(R_p(\lambda))\simeq0$ (where $X_i$ is the mixing ratio of the $i^{th}$ atmospheric constituent). For a later generalization (see Appendix\,\ref{sec:generalizedtransmissionspectra}), we specify for each case at which step these assumptions are used.  
	
\subsection{Dependency for Rayleigh Scattering}
\label{app:ray}
For extinction processes like Rayleigh scattering, the cross section is independent of the pressure---i.e.,$\sigma_{\lambda} \neq f_{\lambda}(p)$. Using the assumptions a1, a2, and a3, the slant-path optical depth (Equation\,\ref{eq:optical_depth}) can be formulated as
\begin{eqnarray}
\tau(z,\lambda) & = & \sum_i \sigma_{i}(\lambda) n_{i,0} e^{-z/H} \sqrt{2\pi (R_{p,0}+z) H},
 \label{eq:tau_fortney}
\end{eqnarray}
where the last term comes from the integral over d$x$  \citep{Fortney2005}, or as,
\begin{eqnarray}	
    \tau(y,\lambda) \simeq A_{\lambda} e^{-y/B},
 & \mbox{where } & 
	\left\{{
   \begin{array}{c c c}
     y & = & z/R_{p,0}\\
 A_{\lambda} & = & \sqrt{2 \pi R_{p,0} H} \sum_i n_{i,0} \sigma_{i}(\lambda) \\
 B & = & H/R_{p,0}
  \end{array}}
  	\right.
  	,
  	 \label{eq:tau_var1}
\end{eqnarray}
i.e., $y$ and $B$ are the dimensionless altitude and atmospheric scale height, respectively, and $A_{\lambda}$ is the slant-path optical depth at $R_{p,0}$---we recall that the reference radius is any radial distance at which the body is optically thick in limb-looking over all the spectral band of interest. Therefore, Equation\,\ref{eq:transmission_spectrum_h2} can be rewritten as
\begin{eqnarray}
y_{eff}^2(\lambda)+2y_{eff}(\lambda) & = c = & 2 \int_{0}^{\infty}  (1+y) \left(1-e^{-A_{\lambda} e^{-y/B}}\right) \text{ d}y.
\label{eq:def_yeff}
\end{eqnarray}
By solving Equation,\,\ref{eq:def_yeff}, one finds $y_{eff}(\lambda) = -1 + \sqrt{1+c}$. The integral in Equation\,\ref{eq:def_yeff} evaluated analytically over $\tau$ is
\begin{eqnarray}
\frac{c}{2} & = & \int_{A_{\lambda}}^{0}  \left(1-B\ln{\frac{\tau}{A_{\lambda}}}\right) \left(1-e^{-\tau}\right) \left(-\frac{B}{\tau}\right) \text{ d}\tau \label{cvA2_0}\\
& = & \int_{A_{\lambda}}^{0} -\frac{B}{\tau} + \frac{B e^{-\tau}}{\tau} + \frac{B^2\ln{\frac{\tau}{A_{\lambda}}}}{\tau} - \frac{B^2\ln{\frac{\tau}{A_{\lambda}}}e^{-\tau}}{\tau} \text{ d}\tau.
 \label{cvA2}
\end{eqnarray}
An evaluation of the integral of each term of Equation\,\ref{cvA2} leads to
\begin{eqnarray}
\int_{A_{\lambda}}^{0} -\frac{B}{\tau} \text{ d}\tau & = & -B\ln{\tau}|_{A_{\lambda}}^{0}, \label{intA1} \\
\int_{A_{\lambda}}^{0} \frac{B e^{-\tau}}{\tau} \text{ d}\tau & = & B E_i(\tau)|_{A_{\lambda}}^{0}, \label{intA2} \\
\int_{A_{\lambda}}^{0} \frac{B^2\ln{\frac{\tau}{A_{\lambda}}}}{\tau} \text{ d}\tau & = & -B^2\ln{A_{\lambda}}\ln{\tau}|_{A_{\lambda}}^{0} + 0.5B^2\ln^2{\tau}|_{A_{\lambda}}^{0}, \text{ and} \label{intA3} \\
\int_{A_{\lambda}}^{0} - \frac{B^2\ln{\frac{\tau}{A_{\lambda}}}e^{-\tau}}{\tau} \text{ d}\tau & = & B^2\ln{A_{\lambda}}E_i(\tau)|_{A_{\lambda}}^{0} - B^2 \int_{A_{\lambda}}^{0} \frac{\ln{\tau}e^{-\tau}}{\tau} \text{ d}\tau,
 \label{intA4}
\end{eqnarray}
where $E_i(x)$ is the exponential integral. The integral remaining in Equation\,\ref{intA4} is equal to  $ \left[\tau\text{ }_3F_3(1,1,1;2,2,2;-\tau) - 0.5\ln(\tau)\left(\ln(\tau) + 2\Gamma(0,\tau) + 2\gamma_{EM}\right)\right]|_{A_{\lambda}}^{0}$ \newline where $_pF_q(a_1,...,a_p;b_1,...,b_q;z)$ is the generalized hypergeometric function and $\gamma_{EM}$ is the Euler-Mascheroni constant \citep{Euler1740}. This new insight for the transmission spectrum equations (Equations\,\ref{cvA2}-\ref{intA4}) allows further developments of Equation\,\ref{cvA2_0} using the following series expansions:
\begin{itemize}
\item $E_i(x)\big|_{x=0} = \ln{x}+\gamma_{EM}-x+O(x^2)$.
\item $\left[\tau\text{ }_3F_3(1,1,1;2,2,2;-x) - 0.5\ln(x)\left(\ln(x) + 2\Gamma(0,x) + 2\gamma_{EM}\right)\right]\big|_{x=0} =\frac{\ln^2{x}}{2}+O(x)$.
\item $\left[\tau\text{ }_3F_3(1,1,1;2,2,2;-x) - 0.5\ln(x)\left(\ln(x) + 2\Gamma(0,x) + 2\gamma_{EM}\right)\right]\big|_{x\gg1} =\frac{6\gamma_{EM}^2+\pi^2}{12}+O(x^{-5})$.    
\end{itemize}
The use of this last series expansion is appropriate if $ x \gtrsim 5$, i.e., in optically-active spectral bands where $ A_{\lambda} \gtrsim 5$ because absorbers/diffusers prevent the light transmission at $R_{p,0}$.

In active spectral bands, we can rewrite Equation\,\ref{cvA2} as 
 \begin{eqnarray}
\frac{c}{2} & = & -B\ln{0} + B\ln{A_{\lambda}} + B(\ln{0} + \gamma_{EM}) - B E_i(A_{\lambda}) - B^2\ln{A_{\lambda}}\ln{0} + B^2\ln^2{A_{\lambda}} + 0.5B^2\ln^2{0}\nonumber \\&\text{  }& -0.5B^2\ln^2{A_{\lambda}}+B^2\ln{A_{\lambda}}(\ln{0} + \gamma_{EM})-B^2\ln{A_{\lambda}}E_i(A_{\lambda})-0.5B^2\ln^2{0}+B^2\frac{6\gamma_{EM}^2+\pi^2}{12}, \nonumber \label{simplify_negl_} \\
& = & \left(\gamma_{EM} + \ln{A_{\lambda}} - E_i(A_{\lambda})\right)(B+B^2\ln{A_{\lambda}}) + B^2\left(-\frac{\ln^2{A_{\lambda}}}{2}+ \frac{6\gamma_{EM}^2+\pi^2}{12}\right) \label{simplify_negl_0}
\end{eqnarray}
Because $ E_i(A_{\lambda}) \ll 1$ and $ \gamma_{EM}\gg B\frac{6\gamma_{EM}^2+\pi^2}{12}$ ($B \ll 1$ because of assumption a1), we can finally rewrite Equation\,\ref{simplify_negl_0} as
 \begin{eqnarray}
c & = &  (B\ln{A_{\lambda}})^2+2(1+B\gamma_{EM})B\ln{A_{\lambda}}+(2B\gamma_{EM}) .\label{simplify_negl_1}
\end{eqnarray}
 The last step is to use $B\gamma_{EM} \ll 1$ to write 
 \begin{eqnarray}
1 + c & \simeq &  (B\gamma_{EM} + B\ln{A_{\lambda}} + 1)^2. \label{simplify_negl_c1}
\end{eqnarray} 
 Therefore, we obtain the following solution to Equation\,\ref{eq:def_yeff}
 \begin{eqnarray}
y_{eff}(\lambda) & \simeq & B(\gamma_{EM}+\ln{A_{\lambda}}) \text{ and} \label{eq:y_eff_ana}\\
\tau(y_{eff}) & \simeq & e^{-\gamma_{EM}} \approx 0.5615
\label{eq:tau_y_eff_ana}
\end{eqnarray}
---note that a first order approximation on $c$ leads to $y_{eff}(\lambda) = B\ln{A_{\lambda}}$ and $\tau(y_{eff}) = 1 $.

Equations\,\ref{eq:y_eff_ana} and \,\ref{eq:tau_y_eff_ana} summarize the way in which a planet's atmospheric properties ($n_{i},T,\mbox{ and }p$) are embedded in its transmission spectrum (Equation\,\ref{eq:transmission_spectrum_h}). Conceptually, Equation\,\ref{eq:y_eff_ana} tells us the height (expressed in planetary radius) where the atmosphere can be considered transparent. As an example, if $A_\lambda$ is $10^4$ then the atmosphere can be considered transparent at $9\left(\approx\ln(10^4)\right)$ scale heights above the reference radius. Equation\,\ref{eq:tau_y_eff_ana} shows that the slant-path optical depth at the effective height is the constant $e^{-\gamma_{EM}}$ (Figure\,\ref{fig:transmission_spectrum_basics}, panel C)---this extends previous numerical observations that $\tau_{eq}\approx0.56$ in some cases \citep{LecavelierDesEtangs2008}.

\begin{figure}
 \centering
  \begin{center}
    \includegraphics[trim = 00mm 00mm 00mm 10mm,clip,width=15cm,height=!]{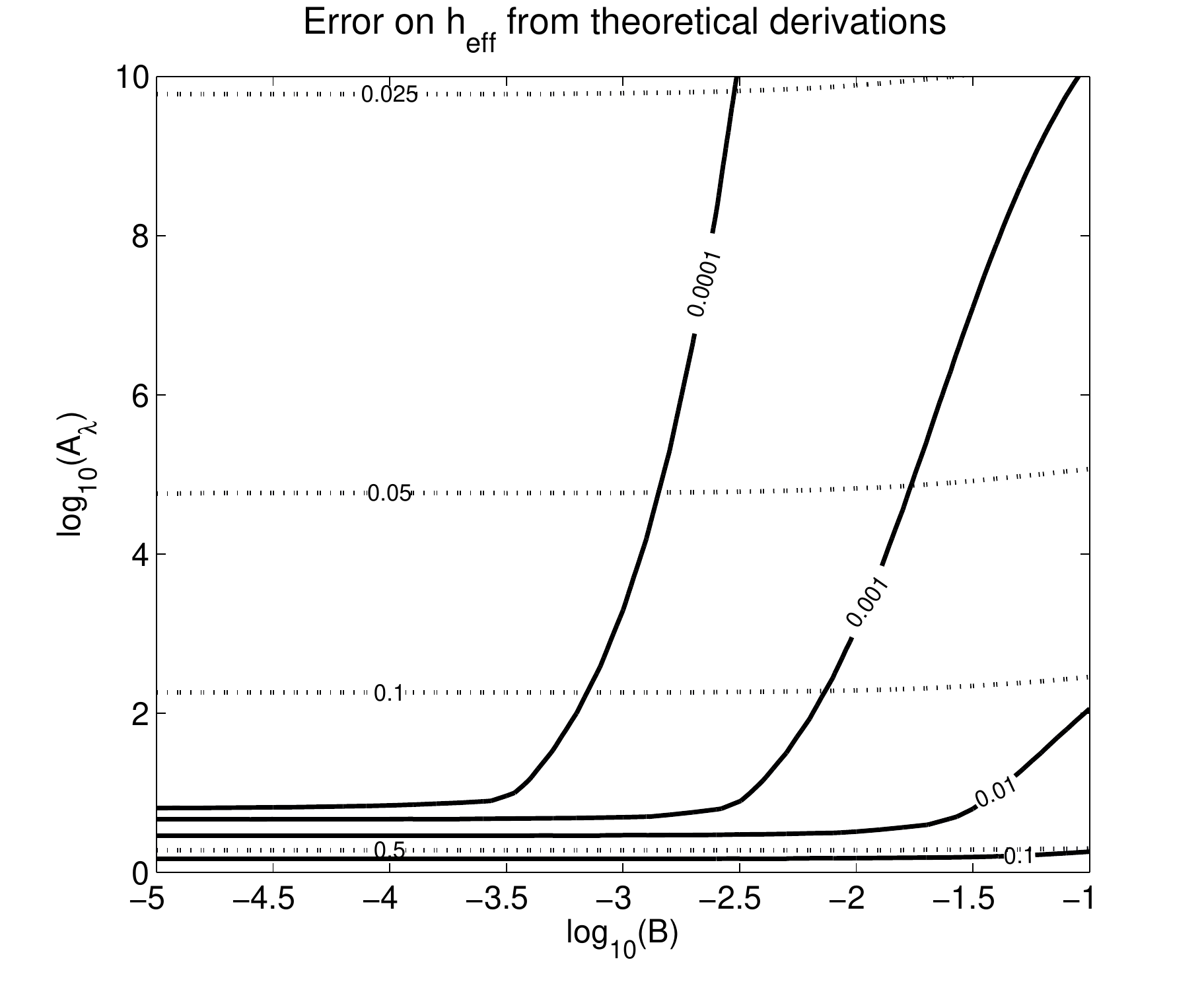}
  \end{center}
  \vspace{-0.7cm}
  \caption[Relative errors on the effective atmosphere height ($h_{eff}(\lambda)$) using the analytic solution to Equation\,\ref{eq:transmission_spectrum_h} (or Equation\,\ref{eq:def_yeff}) for extinction processes with a cross-section independent of $T$ and $p$.]{Relative errors on the effective atmosphere height ($h_{eff}(\lambda)$) using the analytic solution to Equation\,\ref{eq:transmission_spectrum_h} (or Equation\,\ref{eq:def_yeff}) for extinction processes with a cross-section independent of $T$ and $p$ (such as Rayleigh scattering). Relative errors for the zeroth-order derivation $h_{eff}(\lambda) = \left\lbrace z:\tau(z,\lambda) = 1\right\rbrace$ correspond to the dotted lines, and the ones for our derivation $h_{eff}(\lambda) = \left\lbrace z:\tau(z,\lambda) = e^{-\gamma_{EM}}\right\rbrace$ are the thick lines.}
  \vspace{-0.0cm}
  \label{fig:Error_on_cst_cross_sect_derivation}
\end{figure}

The appropriateness of Equations\,\ref{eq:y_eff_ana} and\,\ref{eq:tau_y_eff_ana} is emphasized in Figure\,\ref{fig:Error_on_cst_cross_sect_derivation} that shows the relative deviation on the effective height between numerical integration of Equation\,\ref{eq:def_yeff} and our analytical solution (Equation\,\ref{eq:y_eff_ana}). For Earth ($B \approx 0.1\%$), the relative errors in the active spectral bands will be below 0.1\% which corresponds to an uncertainty on $h_{eff}(\lambda)$ and $R_p(\lambda)$ below 10 meters---which is well below the observational precision we ever hope to achieve.

\subsection{Dependency for Collision-Induced Absorption}
\label{app:cia}
We extend here the results of Section\,\ref{app:ray} to extinction processes with $\sigma\propto P^l $ like CIA. For such processes, Equation\,\ref{eq:tau_var1} can be rewritten as
\begin{eqnarray}	
    \tau(y,\lambda) \simeq A_{\lambda} e^{-y/B},
 & \mbox{where } & 
	\left\{{
   \begin{array}{c c c}
     y & = & z/R_{p,0}\\
 A_{\lambda} & = & \sqrt{\frac{2 \pi R_{p,0} H}{l+1}} \sum_i \alpha_{l,i,T_0} \\
 B & = & \frac{H}{(l+1)R_{p,0}}
  \end{array}}
  	\right.
  	\mbox{and }
  	 \label{eq:tau_var1cia}
\end{eqnarray}
$\alpha_{l,i,T,0}$ is the temperature- and species-dependent absorption coefficient. For example, for CIA $l = 1$ and the $\alpha_{l,i,T} = K_i(T) n_i^2$ where $K_i(T)$ depends solely on the temperature \citep{Borysow2002}. Now that we obtain the same form for $\tau$ as in Section\,\ref{app:ray} (Equation\,\ref{eq:tau_var1}) we can apply the same derivation leading to Equations\,\ref{eq:y_eff_ana} and\,\ref{eq:tau_y_eff_ana}. Note that $B$ and $A_{\lambda}$ have different formulations from the ones derived in Section\,\ref{app:ray}---although the general formulation of Equation\,\ref{eq:tau_var1cia} for processes with $\sigma\propto P^l $ encompasses the case of Rayleigh scattering ($l = 0$).

\subsection{Dependency for Molecular Absorption}
\label{app:molecules}

\begin{figure}

    \vspace{-0.7cm}\hspace{-2.5cm}\includegraphics[trim = 15mm 00mm 20mm 0mm,clip,width=19cm,height=!]{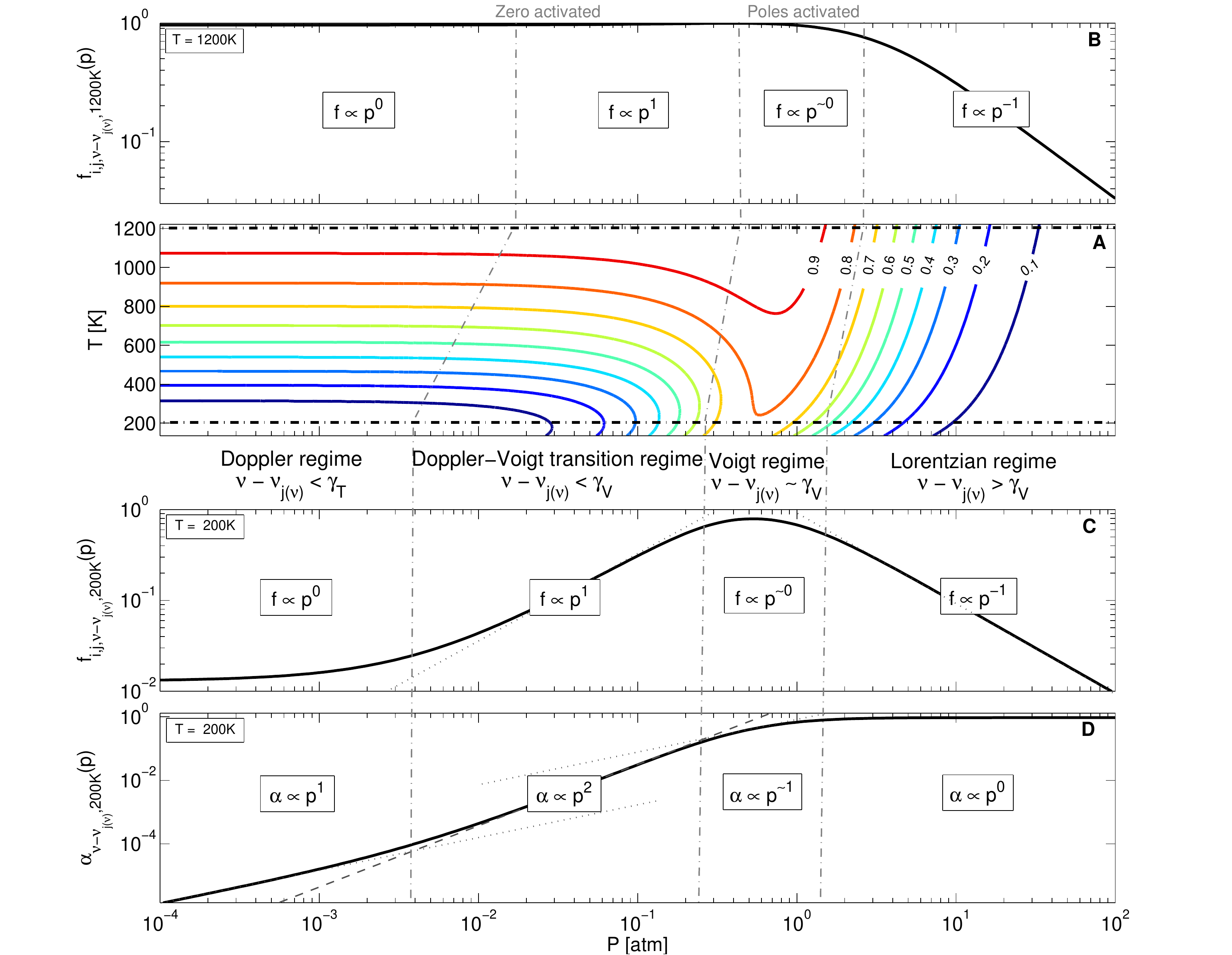}\hspace{-0cm}

  \vspace{-0.7cm}
  \caption[Rational-function-like dependence of a line profile on the pressure at fixed temperature and frequency.]{A line profile ($f_\nu$) depends on the pressure ($p$) as a rational function at fixed temperature ($T$) and frequency ($\nu$). (\textbf{A}) Dependency of $f_\nu$ on $T$ and $p$, at fixed $\nu$, that shows the four domains of different dependency regimes of $f_\nu$ on $p$ whose boundaries are $T$-dependent (gray dot-dash lines). The black dot-dash lines represents the position of the slices in the $\left\lbrace T-p-f\right\rbrace$ space used to highlight that $f_\nu$ behaves as a rational function of $p$, at $\{T,\nu\}$ fixed. Planels \textbf{B} and \textbf{C} present the slices at $T=1200$K and $T=200$K, respectively. These slices show that $f_\nu$ behaves as a rational function of $p$ with a zero and a pair of conjugated zeros (Figure\,\ref{fig:rational_function}). In particular, the absolute value of the zero is less than the poles', as underscored by the sequential transition from the following dependency regimes $\propto p^{0}$, $\propto p^{1}$, $\propto p^{\sim0}$, and $\propto p^{-1}$ with increasing $p$---the dotted and the dashed lines represent a slope of 1 and 2, respectively. (\textbf{D}) Dependency of the absorption coefficient ($\alpha_\nu$) on $p$, at $T=200$K. The exponent of the $\alpha_\nu$ dependency on $p$ increases by one compared to $f_\nu$ dependency on $p$. The exponent increase by one because of $\alpha_\nu$'s additional zero at $p=0$ that originates from the number density.}
  \vspace{-0.0cm}
  \label{fig:T_p_dependence_of_line_profile}
\end{figure}

We extend here the results of Section\,\ref{app:ray} to molecular absorption. We apply the same strategy as for CIA by showing that the slant-path optical depth (Equation\,\ref{eq:optical_depth}) can be formulated as $\tau(y,\lambda) = A_\lambda e^{-y/B}$---around $h_{eff}(\lambda)$ and for most $\lambda$---and then relate to the derivation in Section\,\ref{app:ray}. For molecular absorption, the cross-section can be expressed as
\begin{eqnarray}
\sigma_i(\lambda,T,p) = \sum_j S_{i,j}(T) f_{i,j}(\lambda-\lambda_{i,j},T,p),
 \label{lines}
\end{eqnarray}
where $S_{i,j}$ and $f_{i,j}$ are the intensity and the line profile of the $j^{th}$ line of the $i^{th}$ atmospheric species. Each quantity can be approximated by
\begin{eqnarray}
S_{i,j}(T) & \approx & S_{i,j}(T_{ref}) \sum_{m = 0}^{n_T} a_{T,i,m}T^{m} \text{ and} \label{line_intensity_approx} \\
f_{i,j}(\lambda-\lambda_{i,j},T,p) & \approx & A_{i,j}(\lambda,T) \frac{p+a_{i,j}(\lambda,T)}{p^2+b_{i,j}(\lambda,T)}, \label{line_profile_approx}
\end{eqnarray}
where $n_T = 3$ is sufficient to interpolate the line intensity dependency to $T$ \citep{Gamache1990}. $A_{i,j}, a_{i,j},$ and $b_{i,j}$ are the new parameters we introduce to model the variation of the line with $p$ at fixed $\{\lambda,T\}$ (Figure\,\ref{fig:T_p_dependence_of_line_profile}, panel B)---as an example, at low pressure $f_{i,j}(\lambda-\lambda_{i,j},T,p) \approx  A_{i,j}(\lambda,T) a_{i,j}(\lambda,T)/b_{i,j}(\lambda,T)$ where $ A_{i,j}(\lambda,T) a_{i,j}(\lambda,T)/b_{i,j}(\lambda,T)$ is the amplitude of the Doppler profile of at $\{\lambda,T\}$. The second term of Equation\,\ref{line_profile_approx} is a dimensionless rational function with a zero, $-a_{i,j}$, and a pair of complex conjugate poles, $\pm\sqrt{b_{i,j}}$---details on rational functions and their properties are provided in Appendix\,\ref{app:TF}. The positions of the zero and the poles in $\mathbb{C}$ induce four regimes of specific dependency of $f_{i,j}$ on $T$ and $p$ (Figure\,\ref{fig:T_p_dependence_of_line_profile}.C).
\begin{enumerate}
\item \textbf{Doppler regime:} while $p < a_{i,j}$, $f_{i,j}$ is independent of $p$ (i.e., $f_{i,j} \approx A_{i,j}a_{i,j}/b_{i,j}$) because neither the zero nor the poles are activated. In terms of distance to the line center ($\nu_j$), the Doppler regime dominates when $(\nu-\nu_j) < \gamma_T$, where $\gamma_T \triangleq \left\lbrace \nu : \mbox{d}^2_{\nu} \ln f_V|_\nu = 0 \right\rbrace$ ($f_V$ and $\gamma_V$ are the Voigt profile and its FWMH, respectively). 

\item \textbf{Voigt-to-Doppler transition regime:} while $p^2 < b_{i,j}$, $f_{i,j} \propto p^1$ because only the zero is activated (i.e., $p \geq a_{i,j}$). In particular, $f_{i,j} = A_{i,j}(p+a_{i,j})/b_{i,j}$. In terms of distance to the line center, this regime dominates when $(\nu-\nu_j) < \gamma_V$. 

\item \textbf{Voigt regime:} while $p^2 \sim b_{i,j}$ and $p \geq a_{i,j}$, $f_{i,j}$ behaves as the rational fraction introduced in Equation\,\ref{line_profile_approx} because the zero and the poles are activated. In terms of distance to the line center, this regime dominates when $(\nu-\nu_j) \sim \gamma_V$. 

\item \textbf{Lorentzian regime:} while $p^2 \geq b_{i,j}$ and $p \gg a_{i,j}$, $f_{i,j} \propto p^{-1}$ because one zero and two poles are activated. In terms of distance to the line center, this regime dominates when $(\nu-\nu_j) > \gamma_V$. 

\end{enumerate}

While $a_{i,j}$ and $b_{i,j}$ govern the regime of the line profile, $A_{i,j}$ models its variation with temperature; $A_{i,j} \propto T^w$, where $w$ is the broadening exponent. In the Lorentzian regime, $w$ ranges from 0.4 to 0.75, while it is -0.5 in the Doppler regime \citep{Seager2010}.

Using Equations\,\ref{line_intensity_approx} and\,\ref{line_profile_approx} and the ideal gas law, the absorption coefficient can be formulated as
\begin{eqnarray}
\alpha(\lambda,T,p) & = & \sum_i n_i(T,p) \sigma_i(\lambda,T,p) \\
 & = & \frac{Vp}{R} \sum_i X_i \sum_{m = 0}^{3} a_{T,i,m}T^{m-1}  \sum_j S_{i,j}(T_{ref}) A_{i,j}(\lambda,T) \frac{p+a_{i,j}(\lambda,T)}{p^2+b_{i,j}(\lambda,T)}.
 \label{line_alpha}
 \end{eqnarray}

 For most $\lambda$, the extinction process is dominated by one line. Therefore, the overall dependency of the absorption coefficient can be formulated as
 \begin{eqnarray}
 \alpha(\lambda,T,p) = \Lambda_{\kappa} p \frac{p+a_{\kappa}}{p^2+b_{\kappa}},
 \label{line_alpha2}
\end{eqnarray}
where $\kappa=\left\lbrace\lambda,T,X_i\right\rbrace$ and $\Lambda_{\kappa}, a_{\kappa},$ and $b_{\kappa}$ are model parameters introduced to fit the absorption coefficient variation in the $T-p-X$ space with $\lambda$ being fixed. All of these parameters are known \textit{a priori} from quantum physics and/or lab measurements \citep{Rothman2009}. \textbf{In particular, for most wavelengths, $\Lambda_{\kappa}, a_{\kappa},$ and $b_{\kappa}$ are $X_i S_{i,j(\lambda)}(T) A_{i,j(\lambda)}(\lambda)\mbox{, }a_{i,j(\lambda)}(\lambda),$ and $b_{i,j(\lambda)}(\lambda)$, respectively, where $j(\lambda)$ refers to the line that dominates at $\lambda$.} We show in Figure\,\ref{fig:T_p_dependence_of_line_profile}.D that the absorption coefficient behaves like $f_{i,j}$ (Equation\,\ref{line_profile_approx}) but with an additional zero at $p=0$, which originates from the number density.

Using the assumptions a1, a2, and a3, Equation\,\ref{eq:optical_depth} becomes
\begin{eqnarray}
\tau(z,\lambda)= 2 \Lambda_{\kappa}\int_{0}^{\infty} \frac{p^2+a_{\kappa}p}{p^2+b_{\kappa}}\text{ d}x,
 & \mbox{where } & 
	\left\{{
   \begin{array}{c c c}
     p& = & p_0 \exp(-z'/H)\\
     z'& = & z'(z,x)\approx \frac{x^2}{2R_{p,0}}+z
  \end{array}}
  	\right.
  	.
 \label{tau_lines_T_cst}
\end{eqnarray}

Transmission spectroscopy probes a limited range of atmospheric layers at each wavelength \citep[see, e.g., Figure 9-12(b) from][]{Elachi2006}.
Therefore, only a limited part of the dependency on $T-p$ of the dominant line at $\lambda$ is recorded. We use this property to extend our demonstration assuming that each wavelength records only one regime of dependency (Figure\,\ref{fig:T_p_dependence_of_line_profile}). By doing so, we show that the slant-path optical depth (Equation\,\ref{eq:optical_depth}) can be approached by $\tau(y,\lambda) = A_\lambda e^{-y/B}$ in the probed atmospheric layers (i.e., around $h_{eff}(\lambda)$)---and for most $\lambda$. This approach extends the use of Equations\,\ref{eq:y_eff_ana} and \,\ref{eq:tau_y_eff_ana} to molecular absorption based on the derivation performed in Section\,\ref{app:ray}.

\subsubsection{Doppler regime}
 When the dominant line at $\lambda$ behaves as a Doppler profile ($p \ll a_{\kappa}$) in the atmospheric layers probed around $h_{eff}(\lambda)$, the absorption coefficient depends on the pressure only through the number density (i.e., alike the Rayleigh-scattering case, Section\,\ref{app:ray}). Therefore, Equation\,\ref{tau_lines_T_cst} can be rewritten as 
\begin{equation}
\tau(z,\lambda) = \sqrt{2 \pi R_{p,0} H} \Lambda_{\kappa} \frac{a_{\kappa}}{b_{\kappa}} p_0 e^{-z/H}.
 \label{tau_lines_T_cst_Dop}
\end{equation}
 
\subsubsection{Voigt-to-Doppler transition regime}

In this regime, $p^2 < b_{\kappa}$, therefore, Equation\,\ref{tau_lines_T_cst} can be rewritten as 
\begin{equation}
\tau(z,\lambda) = \sqrt{2 \pi R_{p,0} H} \Lambda_{\kappa} \frac{a_{\kappa}}{b_{\kappa}} p_0 e^{-z/H} \left(1+  \frac{p_0}{\sqrt{2}a_{\kappa}} e^{-z/H}\right).
 \label{tau_lines_T_cst_V2D_0}
\end{equation}
This regime encompasses the three following subregimes: $p \ll a_{\kappa}$, $p \sim a_{\kappa}$, and $p \gg a_{\kappa}$. The first subregime corresponds to the Doppler regime (Equation\,\ref{tau_lines_T_cst_Dop}). Equation\,\ref{tau_lines_T_cst} is rewritten for the second and third subregimes, respectively, as 
 \begin{eqnarray}
 \tau(z,\lambda) & = & \sqrt{\frac{4}{3} \pi R_{p,0} H} \Lambda_{\kappa} \frac{2\sqrt{a_{\kappa}}}{b_{\kappa}} \left(p_0e^{-z/H}\right)^{3/2}\text{ and}\\
 \label{tau_lines_T_cst_trans}
 \tau(z,\lambda) & = & \sqrt{2 \pi R_{p,0} H} \frac{\Lambda_{\kappa}}{\sqrt{2}b_{\kappa}} \left(p_0 e^{-z/H}\right)^2.
 \label{tau_lines_T_cst_btw}
\end{eqnarray}

\subsubsection{Voigt regime}

This regime refers to the general formulation of the problem, i.e., when the Doppler and the Lorentzian behaviours affect the line profile with comparable magnitudes. This formulation does not simplify the integral in Equation\,\ref{tau_lines_T_cst}. Therefore, we rewrite Equation\,\ref{tau_lines_T_cst} for the transition case---i.e.,$p^2 \sim b_{\kappa} \gg a_{\kappa}^2$;
 \begin{eqnarray}
\tau(z,\lambda) & = & \sqrt{2 \pi R_{p,0} H} \Lambda_{\kappa}\frac{2}{\sqrt{b_{\kappa}}} p_0 e^{-z/H}.
 \label{tau_lines_T_cst_infl}
\end{eqnarray}

\subsubsection{Lorentzian regime}

In this regime, $p^2 \geq -b_{i,j}$ and $p \gg -a_{i,j}$; therefore, the absorption coefficient is mostly pressure-independent (Figure\,\ref{fig:T_p_dependence_of_line_profile}, panel D). As a result, such a regime is not expected to be recorded, under the assumptions a2 and a3.

\subsection{Summary and Discussion}
\label{app:generalization}

\begin{figure}
 \centering
  \begin{center}
    \includegraphics[trim = 00mm 00mm 00mm 00mm,clip,width=15cm,height=!]{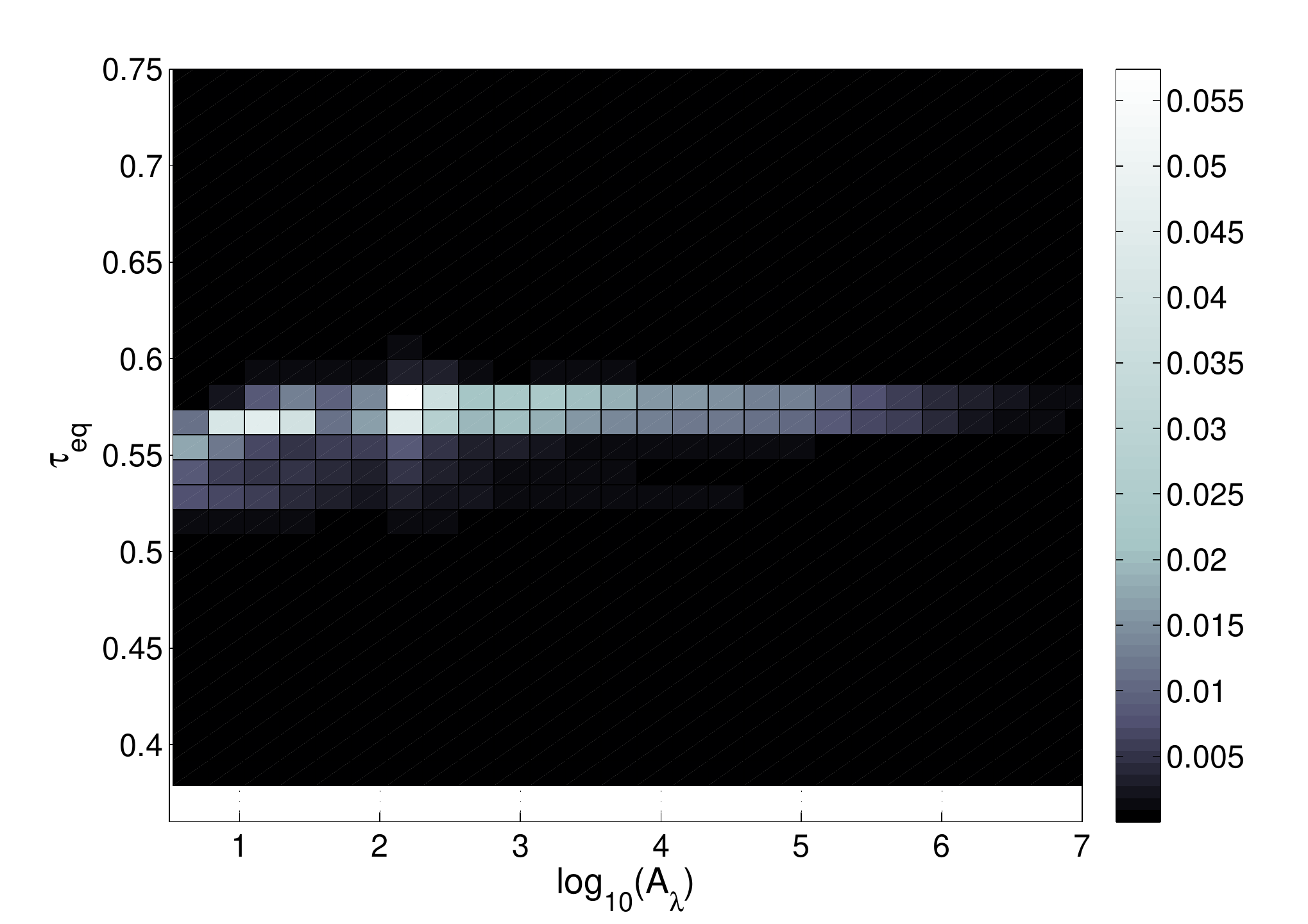}
  \end{center}
  \vspace{-0.7cm}
  \caption[Validity of $\tau_{eq} \approx 0.56$ for an isothermal-isocomposition Earth.]{Two-dimensional histogram of $\tau_{eq}(\lambda)$ as a function of $\log_{10}(A_{\lambda})$ for an isothermal-isocomposition Earth. The color scale is linearly proportional to the count in each bin. $\sim$99$\%$ of the active spectral bins have a $\tau_{eq}\approx \exp^{-\gamma_{EM}}$.}
  \vspace{-0.0cm}
  \label{fig:tau_distribution_isothermal_and_real_Earth}
\end{figure}

Now that we have gone through all the different extinction processes, we can find out what formulation of the effective height is generally true and identify the key parameters behind a planet's transmission spectrum. We demonstrate that the slant-path optical depth (Equation\,\ref{eq:optical_depth}) is of the form
 \begin{eqnarray}
 \tau(y,\lambda) = A_\lambda e^{-y/B},
 \label{eq:tau_summ}
\end{eqnarray}
for most $\lambda$ under the assumptions a1, a2, and a3. As a result, the effective atmospheric height can be expressed as
 \begin{eqnarray}
 h_{eff} = R_{p,0}B(\gamma_{EM}+\ln{A_{\lambda}}).
 \label{eq:h_eff_summ}
\end{eqnarray}
We show  the appropriateness of our demonstration in Figure\,\ref{fig:tau_distribution_isothermal_and_real_Earth} using the numerical simulation of the transmission spectrum\footnote{For details on our transmission spectrum model please refer to Section\,\ref{app:transmission}.} of an Earth-sized planet with a isothermal and isocompositional atmosphere---same abundances as at Earth's surface. It confirms the pivotal role of $\gamma_{EM}$ in transmission spectroscopy as
\begin{eqnarray}
\tau_{eq} \approx e^{-\gamma_{EM}}
\label{eq:tau_eq_summ}
\end{eqnarray}
for a significant fraction of the active spectral bins ($\sim$99$\%$). In addition, we note that a large fraction of the active bins ($\sim$70$\%$) recorded a $\left\lbrace \propto P^2 \right\rbrace$-dependency of $\tau$. This means that \textbf{transmission spectroscopy preferably records transitions from the Voigt regime to the Doppler regime}; because $\gamma_V$ is small, the spectral bins are more likely to be on a line wing, rather than close to the line center. The fact that transmission spectroscopy records preferably the Voigt-to-Doppler transition regime is important because this regime embeds independent information about the pressure (Equations\,\ref{tau_lines_T_cst_trans} and\,\ref{tau_lines_T_cst_btw}). 

\begin{figure}[!p]
 \centering
  \begin{center}
    \includegraphics[trim = 00mm 00mm 00mm 00mm,clip,width=15cm,height=!]{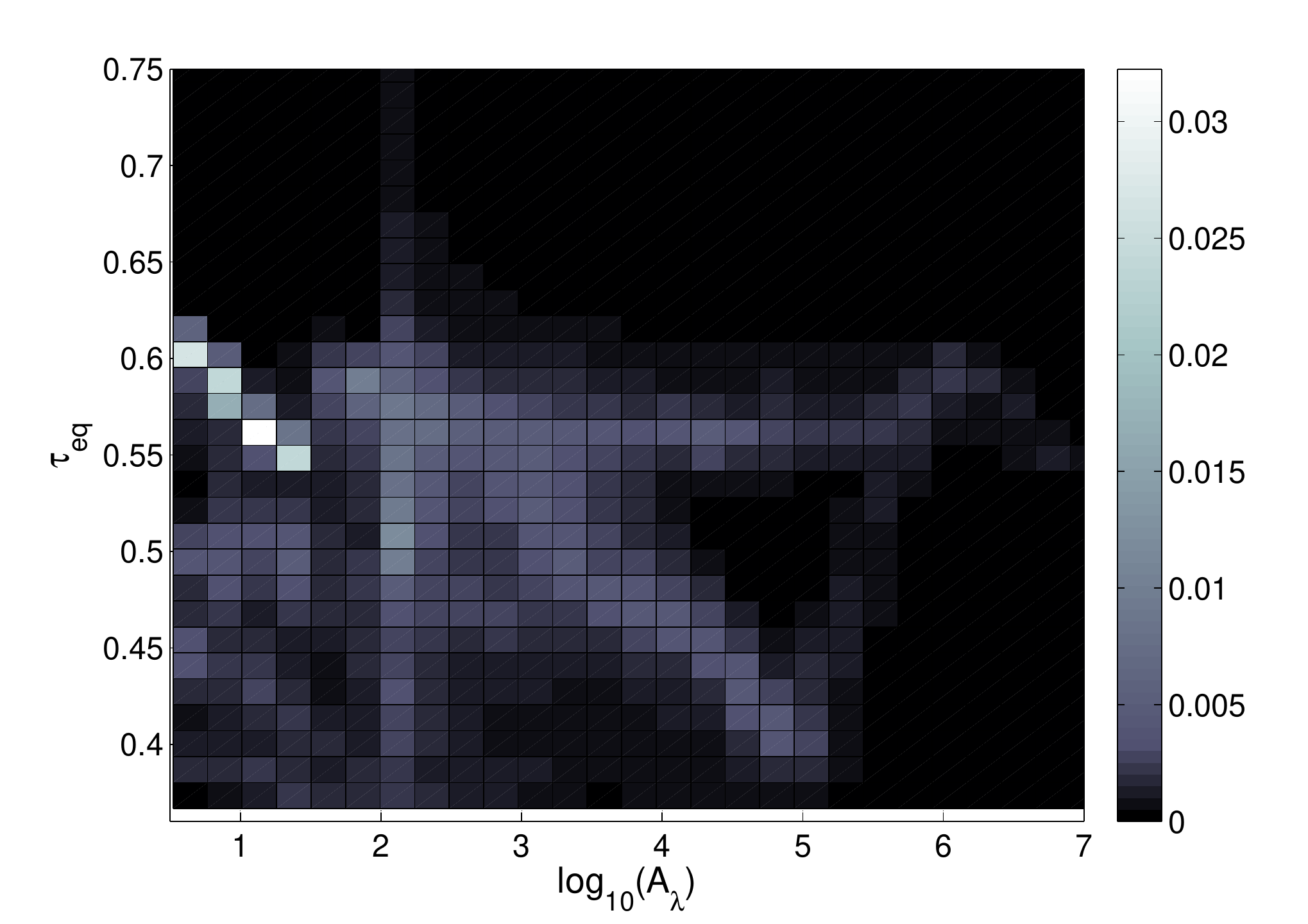}
  \end{center}
  \vspace{-0.7cm}
  \caption[Validity of $\tau_{eq} \approx 0.56$ for Earth.]{Two-dimensional histogram of $\tau_{eq}(\lambda)$ as a function of $\log_{10}(A_{\lambda})$ for Earth---temperature-pressure-mixing ratio profiles from \cite{Cox2000}. The color scale is linearly proportional to the count in each bin. $\sim$50$\%$ of the active spectral bins have a $\tau_{eq}\approx \exp^{-\gamma_{EM}}$.}
  \vspace{-0.0cm}
  \label{fig:tau_eq_vs_a_lambda_Earth}
\end{figure}

\begin{figure}
 \centering
  \begin{center}
    \includegraphics[trim = 00mm 00mm 00mm 00mm,clip,width=15cm,height=!]{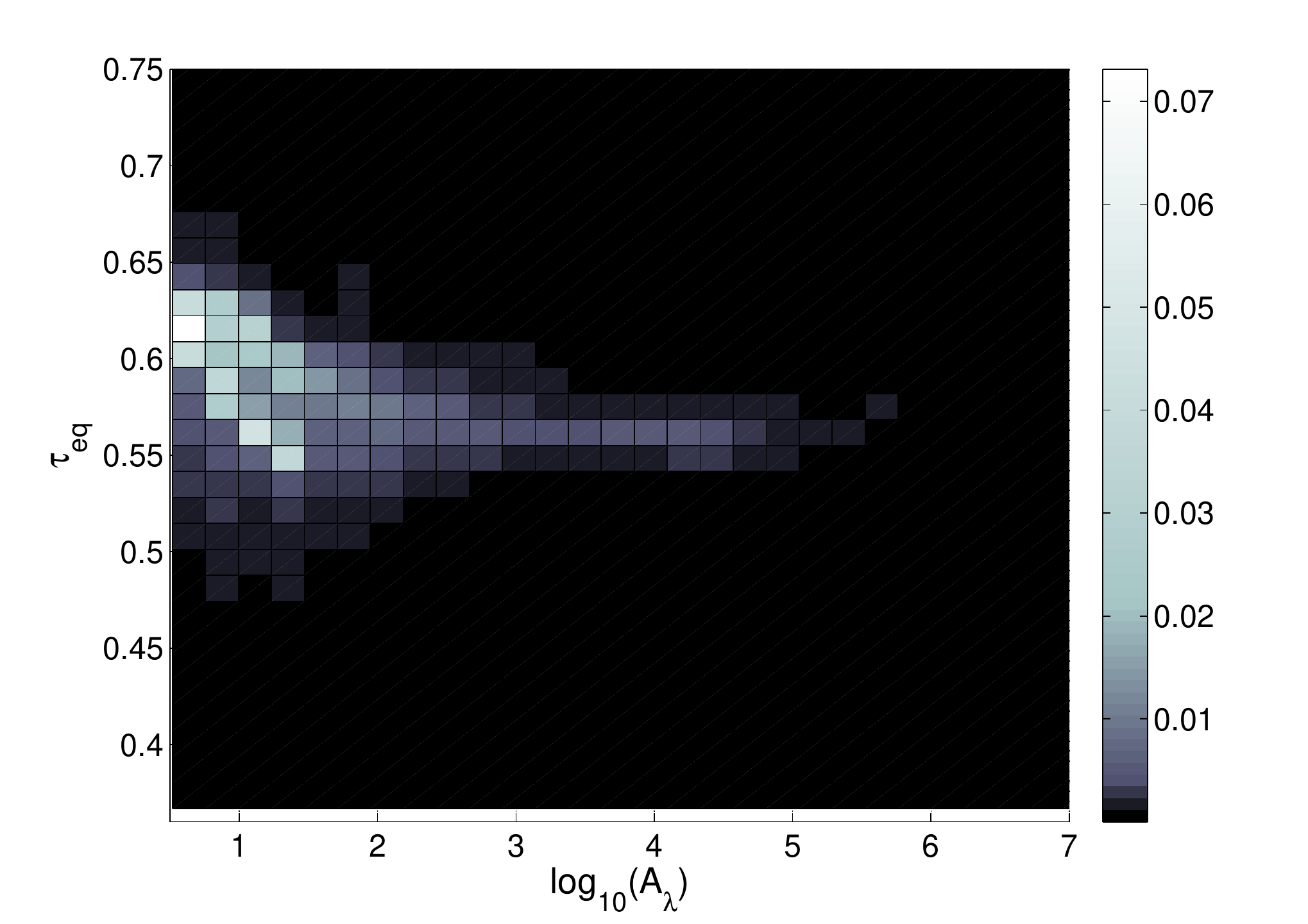}
  \end{center}
  \vspace{-0.7cm}
  \caption[Validity of $\tau_{eq} \approx 0.56$ for a dry Earth.]{Two-dimensional histogram of $\tau_{eq}(\lambda)$ as a function of $\log_{10}(A_{\lambda})$ for Earth with no water in its atmosphere---temperature-pressure-mixing ratio profiles from \cite{Cox2000}. The color scale is linearly proportional to the count in each bin. $\sim$80$\%$ of the active spectral bins have a $\tau_{eq}\approx \exp^{-\gamma_{EM}}$.}
  \vspace{-0.0cm}
  \label{fig:tau_eq_vs_a_lambda_Earth_without_water}
\end{figure}

We show in Figure\,\ref{fig:tau_eq_vs_a_lambda_Earth} that for $50\%$ of the optically active bins, our formulation for $h_{eff}$ is still adequate for Earth, i.e., for a planet with atmospheric temperature and mixing ratios strongly dependent on the altitude\footnote{Steps towards a generalization of the derivations performed in Section\,\ref{app:dependency} are introduced in Appendix\,\ref{sec:generalizedtransmissionspectra}.}. We emphasize in Figure\,\ref{fig:tau_eq_vs_a_lambda_Earth_without_water} that water is the main cause for the deviation of $\tau_{eq}$'s distribution from $\sim0.56$ shown in Figure\,\ref{fig:tau_eq_vs_a_lambda_Earth}. The removal of water from Earth's atmosphere leads to $\tau_{eq} \approx 0.56$ for 80\% of the active bins. The effect of water on $\tau_{eq}$'s distribution is due to the strong variation of the water mixing ratio with the altitude close to Earth's surface. In particular, the water scale height is smaller than the local atmospheric scale height, which invalidates the demonstration in Section\,\ref{app:molecules}. Therefore, $h_{eff}$ deviates from the formulation $R_{p,0}B(\gamma_{EM}+\ln{A_{\lambda}})$. As discussed above, it is not because $h_{eff}$ deviates from $R_{p,0}B(\gamma_{EM}+\ln{A_{\lambda}})$ that the specificity of the dependencies of a transmission spectrum to its parameters is lost---we are just currently unable to provide the general derivation in such a general atmosphere. Note that the significant effect of the scale height of a specie---such as water---on a transmission spectrum is favorable for habitability assessment (see the paragraph at the section end).

\paragraph*{Insights into the Dependencies}
$A_\lambda$ and $B$ record the dependency of a planet's transmission spectrum on atmospheric properties in ways that vary based on the extinction process and the regime recorded at $\lambda$ (Table\,\ref{tab:A_B}). While $B$ is solely affected by $H$, $A_\lambda$ is affected in independent ways by the following parameters:
\begin{itemize}
\item \textbf{The scale height}, $H$, affects $A_\lambda$ via $\sqrt{H}$ for all extinction processes, which is a geometry factor from the light path.

\item \textbf{The number densities} of the main atmospheric absorbents, $n_i$, affects $A_\lambda$ proportionally for molecular absorption---recall $\Lambda_{\kappa}p_0 = n_{i,0} S_{i,j(\lambda)}(T) A_{i,j(\lambda)}(\lambda)$---and for Rayleigh scattering, while their squares affect $A_\lambda$ for CIA. These different dependencies on $n_i$ result in stronger constraints on $n_i$ when assessed primarily from CIA signal.

\item \textbf{The reference pressure}, $p_0$, affects $A_\lambda$ for molecular absorption in the Voigt-to-Doppler regime ($A_\lambda \propto p_0^{0.5}$ if $p \sim a_{\kappa}$ and $A_\lambda \propto p_0^1$ if $p \gg a_{\kappa}$). Fortunately, this regime is recorded by most spectral bins and, hence, allows constraining the atmospheric pressure.

\item \textbf{The temperature}, $T$, affects $A_\lambda$ in two ways: (i) through $\Lambda_{\kappa}$ due to the line-intensity dependence on $T$ (via a third-order polynomial) and (ii) through the line-profile dependence on $T$ ($T^w$-dependency of the line broadening). In addition, processes like CIA are also dependent on $T$, in ways that are known \textit{a priori} from quantum mechanics \citep[e.g.,][]{Borysow2002}. 

\end{itemize}
Although the temperature and the number densities are both embedded in $\Lambda_{\kappa}$, fortunately they affect $\Lambda_{\kappa}$ in different ways. $\Lambda_{\kappa} \propto n_i$ independently of $\lambda$, while it depends on $T$ in a $\lambda$-dependent way through the line intensity and profile. Note again that those dependency are known---or can be known---\textit{a priori} from quantum mechanics or lab measurements.

In summary, the key parameters of a transmission spectrum---the atmospheric scale height, the number densities of the main absorbents, the temperature, and the pressure---affect it in unique ways. Therefore, it is theoretically possible to retrieve each of these parameters independently.

\begin{table}[!h]
\caption{Dependency of $A_{\lambda}$ and $B$ to the regimes recorded \label{tab:A_B}}
	\centering
	\setlength{\extrarowheight}{4pt}
	\setlength{\tabcolsep}{2pt}

	\begin{tabular}{c|r l|c}
	
	\hline\hline
	\textbf{Extinction process}---Recorded regime & \multicolumn{2}{c}{$\mathbf{A_{\lambda}}$} \vline & $\mathbf{B}$\\
	\hline
	{\textbf{Rayleigh scattering}} &  $\sqrt{2 \pi R_{p,0} H} $&$ \sum_i n_{i,0} \sigma_{i}(\lambda)$ & $\frac{H}{R_{p,0}}$\\
	
	{\textbf{Collision Induced Absorption}} &  $ \sqrt{\text{ } \pi R_{p,0} H} $&$ \sum_i K_i(T) n_{i,0}^2$ & $\frac{H}{2R_{p,0}} $\\
	
	{\textbf{Molecular absorption}} & &  \\
	{Doppler regime} ($p < a_{\kappa}$) & $\sqrt{2 \pi R_{p,0} H} $&$ \Lambda_{\kappa} \frac{a_{\kappa}}{b_{\kappa}} p_0$ & $\frac{H}{R_{p,0}}$\\
	{Voigt-to-Doppler transition regime} ($p^2 < b_{\kappa}$)  & & \\
	 {{if $p \sim a_{\kappa}$}} & $\sqrt{\frac{4}{3} \pi R_{p,0} H} $&$ \Lambda_{\kappa} \frac{2\sqrt{a_{\kappa}}}{b_{\kappa}} p_0^{1.5} $ & $\frac{H}{1.5R_{p,0}}$ \\
	  {{if $p \gg a_{\kappa}$}} & $\sqrt{\text{ } \pi R_{p,0} H} $&$ \Lambda_{\kappa} \frac{1}{b_{\kappa}} p_0^{2} $ & $\frac{H}{2R_{p,0}}$ \\
	{Voigt regime} & & \\
	  {{if $p^2 \sim b_{\kappa} \gg a_{\kappa}^2$}} & $\sqrt{2 \pi R_{p,0} H} $&$ \Lambda_{\kappa} \frac{2}{\sqrt{b_{\kappa}}} p_0 $ & $\frac{H}{R_{p,0}}$ \\

	\hline

	\end{tabular}
	\vspace{-0.3cm}
\end{table}

\paragraph*{Discussion}
The analytical derivations performed in this section required using the three following assumptions: \textbf{(a1)} the extent of the optically active atmosphere is small compared to the planetary radius ($z \ll R_p$), \textbf{(a2)} the atmosphere can be assumed isothermal $[d_zT(R_p(\lambda))\simeq0]$, and \textbf{(a3)} the atmosphere can be assumed isocompositional $[d_zX_i(R_p(\lambda))\simeq0]$. We discuss below how Section\,\ref{app:dependency}'s conclusions are mostly unaffected by the relaxation of the assumptions a2 and a3---a1 being justified. In particular, we explain conceptually that the effect of each key parameter remains unique while relaxing these assumptions, meaning that \textit{MassSpec} can be applied to any exoplanet atmosphere, theoretically. In practice, \textit{MassSpec}'s application requires the transmission spectra to be of high-SNR, sufficient spectral resolution, and with an extended spectral coverage.

The analytical derivations become more complex if assumptions a2 and a3 are relaxed. \textbf{We provide in Appendix\,\ref{sec:generalizedtransmissionspectra} the first steps towards a generalization of the previous derivations.} Below, we describe conceptually how these relaxations affect the previous results.

\textbf{(i) A non-isothermal atmosphere} translates primarily into an altitude-dependent scale height [$H=H(z)$]---therefore, it is required to model the planet's atmosphere with a $z$-dependent scale height in atmospheric retrieval methods (see, e.g., Section\,\ref{app:retrieval}). However, the temperature can still be self-consistently retrieved from a planet's transmission spectrum because it affects in its own way the slant-path optical depth profile, $\tau(r,\lambda)$, through the extinction cross section profile, $\sigma_{i}\left(T(r'),p(r'),\lambda\right)$.

For molecular absorption, $T$ affects $\sigma_{i}\left(T(r'),p(r'),\lambda\right)$ through the line intensities and profiles in ways that are known a priori from quantum mechanics and/or lab measurements. $T$ affects mainly the line intensities through the total internal partition sum (TIPS) and the Boltzmann populations for molecular absorption. The TIPS describes the overall population of the molecule's quantum states, and is solely dependent on the molecular structure (i.e., the species) and the local temperature. The effect of temperature on the TIPS can be appropriately approximated by a third order polynomial \citep{Gamache1990}. On the other hand, the temperature effect on the population of a molecule's individual state depends solely on the temperature and the energy of the state. Therefore, the extinction cross-section depends on the temperature in ways that vary with the wavelength, in opposition to the TIPS. Finally, the temperature affects the line broadening as $\propto T^w$, where $w$ is the broadening exponent and ranges from 0.4 to 0.75 in the Lorentzian regime (classical value: 0.5),  while it is -0.5 in the Doppler regime \citep{Seager2010}.

The overall effect of a change in $T$ cannot be compensated/mimicked by other atmospheric parameters because it is specific---and \textit{a priori} known. For example, although a local decrease in temperature could be compensated at the zeroth order by an increase of the local number densities:
\begin{itemize}
\item The increase in number densities required to compensate the change in the line intensities will be inconsistent with the increase required to mitigate the change in local pressure. While a molecule's local number density will have to scale as a third order polynomial to compensate line intensity changes, it should scale as $1/T$ to compensate the change in pressure.
\item The compensation enabled by an increase in number densities is wavelength-independent while the effect of $T$ on the absorption coefficient is strongly wavelength-dependent.
\end{itemize}

\textbf{(ii) A non-isocompositional atmosphere} translates into different number-density scale height for each component. Alike in the case of temperature changes, changes in composition affect the slant-path in specific ways, preserving \textit{MassSpec}'s capability to constrain independently the parameters of the mass equation (Equation\,\ref{M2}). A change in number density is species specific and, hence, cannot be compensated by global atmospheric quantities such as the temperature, scale height, or reference pressure. 

In some case, molecules such as water may require the use of an individual scale height for proper retrieval. \textbf{The use of species-specific scale heights will likely be of primary importance to assess the habitability of planets}, as a significantly smaller scale height for the water number density could indicate the presence of water surface reservoir. For such applications, comparisons with the scale height of other molecules will be required. Molecules that are expected to present constant mixing ratio throughout the atmosphere could then be considered as independent markers to extract the pressure scale height. The best marker candidate is carbon dioxide because (i) it is a chemically-stable molecule that is usually well-mixed in a planet's atmosphere and (ii) it presents numerous strong absorption bands that enable CO$_2$ detection at low abundance (down to $\sim0.1$ppm). In the extreme cases where water and carbon dioxide are depleted from a planet's atmosphere due to condensation, nitrogen and/or hydrogen---which are known to be chemically stable at temperatures below which CO$_2$ condensates---would be the dominant species and, hence, the primary marker for the pressure scale height.

\section{Proof of Concept with HD\,189733b}
\label{sec:masspechd189733b}

We demonstrate in Section\,\ref{app:dependency} that, in theory, the transmission spectrum of an exoplanet constrains independently its atmospheric properties (temperature, pressure, and composition) and, hence, its mass. In practice, the determination of an exoplanet's mass and atmospheric properties using \textit{solely} transmission spectroscopy requires a high SNR over an adequately-sampled extended spectral coverage (e.g., from 0.2 to 5 $\mu$m with $\lambda/\Delta\lambda \geqq 300$), which will be available with future facilities (see Section\,\ref{app:results}). This means that with current facilities, \textit{MassSpec} cannot yield independent constraints on the temperature, composition, and pressure of a planet's atmosphere \textit{solely} from its transmission spectrum. Therefore, complementary information is required for immediate applications---see Equation\,\ref{M2}. 

We show here that \textit{MassSpec} can be applied immediately to some hot Jupiters. Their temperature can be determined by emission spectroscopy. Their mean molecular mass is known a priori (H/He-dominated atmosphere: $\mu \approx 2.3$). And, their scale height can be directly derived from transmission spectroscopy via their ``Rayleigh-scattering slope''. In spectral bands where Rayleigh scattering dominates, the effective planetary radius (Equation\,\ref{eq:R_eff_tau_eq}) can be rewritten using Equations\,\ref{eq:Al_RS_in_text} and\,\ref{eq:B_RS_in_text} as
\begin{equation}
	\alpha H=\frac{\text{d}R_p(\lambda)}{\text{d}\ln \lambda},
	\label{eq:scaleheightlecavray}
\end{equation} 
because the cross section is of the form $\sigma_{sc}(\lambda) = \sigma_0(\lambda/\lambda_0)^\alpha$ with $\alpha =$ -4 \citep{LecavelierDesEtangs2008}. In other words, the slope of the effective planetary radius in the $\ln \lambda$-space relates to the local scale height\footnote{We refer to Figures\,\ref{fig:weighing_WASP33b} and\,\ref{fig:rayleigh_slope_when_non_isothermal} for examples of Rayleigh-scattering slopes for isothermal and non-isothermal atmospheres, respectively.}. 

Therefore, using Equations\,\ref{M2} and\,\ref{eq:scaleheightlecavray} the planet mass can be derived from 
\begin{equation}
	M_p=-\frac{4kT[R_p(\lambda)]^2}{\mu G\frac{\text{d}R_p(\lambda)}{\text{d}\ln \lambda}}
	\label{eq:M2scattering}.
\end{equation} 
\textit{MassSpec}'s estimate of HD\,189733b's mass is 1.15 $M_{Jup}$, based on estimates of $T$ ($\approx$ 1300 K), $\text{d}R_p(\sim 0.8\,\mu\text{m})/\text{d}\ln \lambda$ ($\approx$ -920 km), and $R_p(\sim 0.8\,\mu\text{m})$ ($\approx$ 1.21$R_{Jup}$) derived from its emission and transmission spectra \citep{Madhusudhan2009,Pont2008,Pont2013}. \textit{MassSpec}'s estimate of HD\,189733b's mass is in excellent agreement with the mass derived from RV measurements \citep[1.14$\pm$0.056$M_{Jup}$][]{Wright2011} for this extensively observed Jovian exoplanet. 

\section{Preliminary Discussions}

\subsection{The Importance of Accurate Extinction Cross Section Databases}
\label{app:cross_section_database}

Atmospheric retrieval methods based on spectroscopy---like \textit{MassSpec}---require accurate extinction cross sections (Equation\,\ref{eq:optical_depth}). Conceptually, atmospheric retrieval methods solve the inverse problem of determining the conditions (i.e., temperature, pressure, composition) of the medium probed from its transmission spectrum knowing how the transmission of light is affected by a medium's conditions.  Hence, our capability for characterizing exoplanetary atmospheres dependents on the accuracy of our knowledge concerning the dependency of a medium's optical properties on its conditions. That is why we advocate for devoting significant efforts to generate accurate extinction cross section databases that cover various atmospheric conditions (i.e., temperature, pressure, composition).

\subsection{Complementarity of \textit{MassSpec} and RV}
\label{app:scaling_laws}

Here we introduce \textit{MassSpec}'s sensitivity to the planetary system parameters and show \textit{MassSpec}'s complementarity with the RV method to yield planetary mass measurements. For that purpose, we derive scaling laws to identify the main dependency of each method's signal on the properties of the observed planetary system, such as the planet's semi-major axis. 

\subsubsection*{Intensity of the RV and Transmission Signals}
\label{app:scaling_laws_signals}

 The signal targeted by the RV method, the RV shift ($K_{\star}$), can be expressed as
\begin{eqnarray}
	K_{\star} = M_p\sqrt{\frac{G}{(M_p+M_{\star})a}}\frac{\sin i}{\sqrt{1-e}},
	\label{eq:RV_shift}
\end{eqnarray}
where $M_{\star}$, $M_p$, $a$, $i$, and $e$ are the host star's mass and the planet's mass, orbital semi-major axis, inclination, and eccentricity \citep{Murray2010}. (We do not discuss here the effects of $i$ and $e$.)
 On the other hand, the signal in transmission depends solely on the area of the opaque atmospheric annulus, $2\pi R_{p,0}h_{eff}(\lambda)$, and the host star spectral radiance $B_{\lambda}(T_{\star})$ (where $B_{\lambda}$ is the Planck function and $T_{\star}$ the star's effective temperature). Using Equation\,\ref{eq:h_eff_in_text}, we can write that $h_{eff}(\lambda)\propto H$---for active molecular bands $A_\lambda \geq 10^3$ therefore $h_{eff}(\lambda) = nH$ with $n \geq 6$. $H=kT/\mu g$ where T can be approached by the planet's equilibrium temperature at first order,
 \begin{eqnarray}
	T_{eq} = T_{\star}(R_{\star}/a)^{0.5}\left(f'(1-A_B)\right)^{0.25},
	\label{eq:T_eq}
\end{eqnarray}
where $R_{\star}$ is the star's radius and $a$, $f'$, and $A_B$ are respectively the semi-major axis, a parameter for the heat redistribution in the planet's atmosphere\footnote{$f'=1/4$ if the heat deposited by the stellar radiation is uniformly distributed. $f'=2/3$ for a tidally-locked planet without atmospheric advection.} and the Bond albedo \citep{Seager2010}. By rewriting the planet's surface gravity as $g = 4\pi G \rho_p R_p/3$ ($\rho_p$ is the planetary density), we can summarize how the system parameters affect the transmission spectrum and the RV signals (Table\,\ref{tab:Tr_RV_app}).

\begin{table}[!h]
\caption{Dependency of a signal intensity in transmission spectroscopy and RV to the system parameters. \label{tab:Tr_RV_app}}
	\centering
	\setlength{\extrarowheight}{4pt}
	\setlength{\tabcolsep}{2pt}

	\begin{tabular}{c|c|c}
	
	\hline\hline
	\textbf{Signal}& \textbf{Planetary parameters} & \textbf{Stellar parameters}\\
	\hline
Transmission spectrum & $\rho_p^{-1} \mu^{-1} a^{-0.5}$ & $B_{\lambda}(T_{\star})T_{\star}R_{\star}^{0.5}$\\
RV shift & $M_p a^{-0.5}$ & $M_{\star}^{-0.5}$\\
	\hline

	\end{tabular}
	\vspace{-0cm}
\end{table}

Table\,\ref{tab:Tr_RV_app} highlights that signals in transmission are more intense for low-density planets and atmospheres and bright or large stars, while the RV shift is amplified by massive planets and low-mass stars. In particular, bright and large stars increase the atmospheric temperature, hence the atmospheric scale height, for a fixed planetary density, atmospheric composition, and semi-major axis.
Note that \textit{MassSpec} will also be relevant for gas giants, especially for those whose star's activity prevents a mass measurement with RV \citep[e.g., the hottest known planet, WASP-33b][]{Collier2010}. 

\subsubsection*{Sensitivity of a Transmission Spectrum SNR}
\label{app:scaling_laws_snr}

In order to derive the actual scaling law of transmission spectrum SNR---and, hence, \textit{MassSpec}'s sensitivity---one has to account for the observational parameters. The overall significance of a signal in transmission scales as
\begin{eqnarray}
	SNR_{ST} \propto \frac{\frac{2\pi R_{p,0}h_{eff}(\lambda)}{d^2} t A \eta B_{\lambda}(T_{\star})}{\sqrt{\frac{\pi R_{\star}^2}{d^2}t A \eta B_{\lambda}(T_{\star})} },
	\label{eq:SNRt_general_scalinglaw_0}
\end{eqnarray}
where the numerator and the denominator relate, respectively, to the number of photons blocked by the planet atmosphere and the uncertainty on the baseline. The latter corresponds to the square root of the total number of photon emitted by the host star and collected out-of-transit in the same spectral band (Poisson process), respectively. $t$ is the observation time, $A$ and $\eta$ are the telescope's collecting area and total optical throughput, and $d$ is the host star's distance to Earth. Equation\,\ref{eq:SNRt_general_scalinglaw_0} can be rewritten as
\begin{eqnarray}
	SNR_{ST} \propto T_{\star} \left(t A \eta B_{\lambda}(T_{\star})\right)^{0.5} \left(R_{\star}^{0.5}d \rho_p \mu a^{0.5}\right)^{-1}.
	\label{eq:SNRt_general_scalinglaw}
\end{eqnarray}
For a given planet ($R_{p,0}$, $M_{p}$, $\mu$, $a$, and $T$ fixed), Equation\,\ref{eq:SNRt_general_scalinglaw} can be rewritten as
\begin{eqnarray}
	SNR_{ST} \propto \frac{\sqrt{t A \eta B_{\lambda}(T_{\star})}}{R_{\star} d}.
	\label{eq:SNRt_general_scalinglaw_fixed_planet}
\end{eqnarray}

\begin{figure}[!h]
 \centering
  \begin{center}
    \includegraphics[trim = 00mm 00mm 00mm 00mm,clip,width=12cm,height=!]{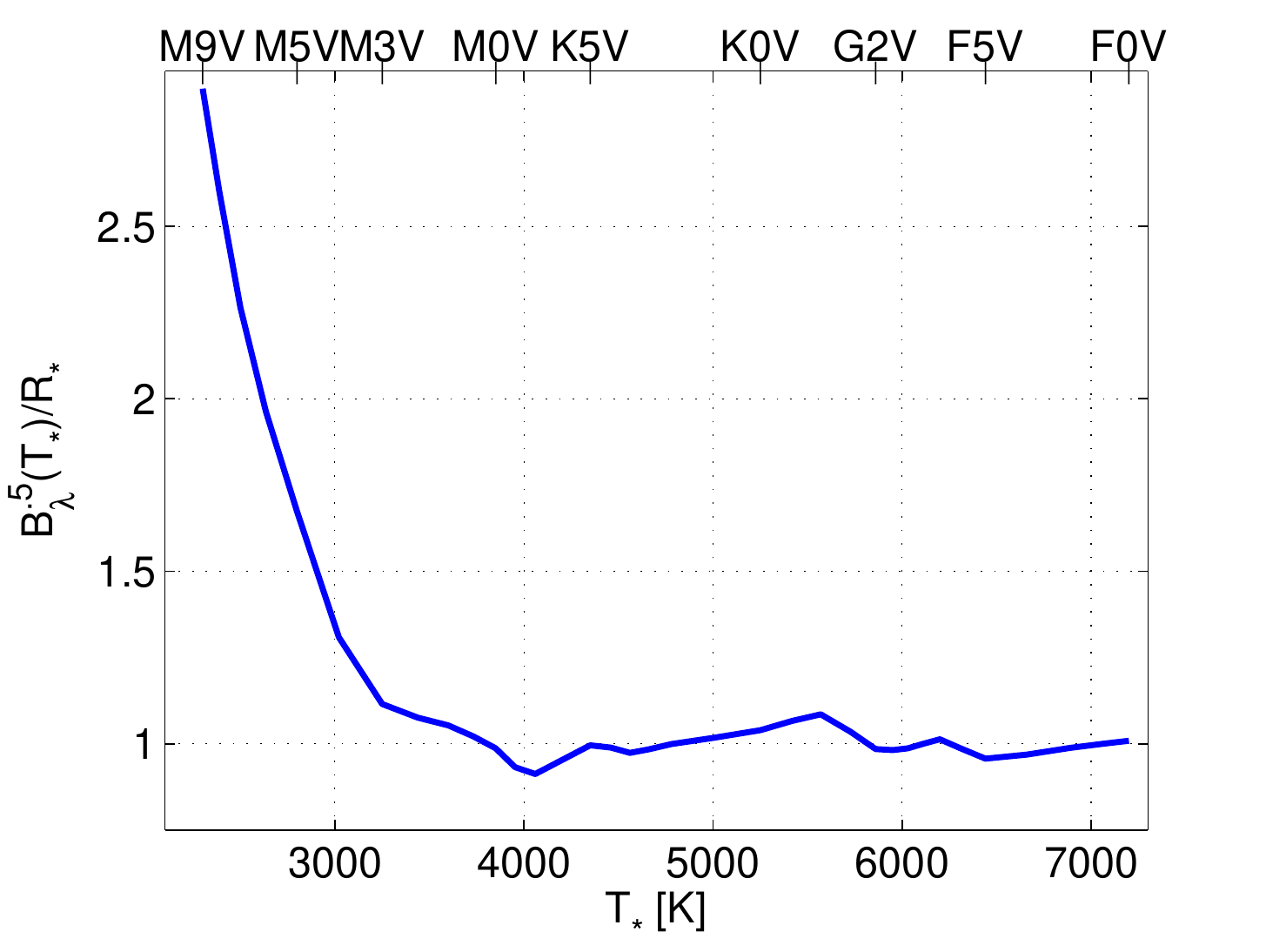}
  \end{center}
  \vspace{-0.7cm}
  \caption[Dependence of in-transit planetary signals on the host-star type.]{The significance of in-transit signals increases for planets transiting M dwarfs. The figure shows the ratio $\sqrt{B_{\lambda}(T_{\star})}/R_{\star}$---normalized for a Sun-like star---as a function of the stellar effective temperature \citep[$T_{\star}-R_{\star}$ relation based on][]{Reid2005,Torres2010}. $\sqrt{B_{\lambda}(T_{\star})}/R_{\star}$ scales as the overall significance of an in-transit signal such as a transmission spectrum, for fixed planetary properties (Equation\,\ref{eq:SNRt_general_scalinglaw_fixed_planet}). For stars with earlier spectral types than M2V, the significance is independent of the host-star type. However, the significance increase significantly towards late M dwarfs.}
  \vspace{-0.0cm}
  \label{fig:key_ratio_fixed_planet}
\end{figure}

Equation\,\ref{eq:SNRt_general_scalinglaw_fixed_planet} highlights the interest of M dwarfs for obtaining high-SNR transmission spectrum (or transit light-curve) for fixed planet properties (e.g., $T_{eq} \sim 300$ K when searching for habitable planets). Equation\,\ref{eq:SNRt_general_scalinglaw_fixed_planet} shows that for a given planet, the signal significance scales as $\sqrt{B_{\lambda}(T_{\star})}/(R_{\star} d)$. Figure\,\ref{fig:key_ratio_fixed_planet} shows the ratio $\sqrt{B_{\lambda}(T_{\star})}/R_{\star}$---normalized for a Sun-like star---as a function of the stellar effective temperature \citep[$T_{\star}-R_{\star}$ relation based on][]{Reid2005,Torres2010}. For stars with earlier spectral types than M2V, the significance is independent of the host-star type. However, the significance increases substantially for M dwarfs. We take advantage of this favorable properties of late M dwarfs to show in Chapter\,\ref{chap:perspectives} that \textit{MassSpec} will enable the characterization of Earth-sized planets within the next decade, possibly leading to the identification of the first habitable planets.

\subsection{Possible Insights Into Planetary Interiors}

Mass and radius are not always sufficient to obtain insights into a planet's interior. \textit{MassSpec}'s simultaneous constraints on a planet's atmospheric properties and average density may help to break this degeneracy, in some cases. A precision on a planet mass of 3 to 15\%, combined with the planetary radius can yield the planetary average density and hence bulk composition. Even with a relatively low precision of 10 to 15\%, it is possible to infer whether or not a planet is predominantly rocky or predominantly composed of H/He \citep{Seager2007,Fortney2007}. With a higher planet mass precision, large ranges of planetary compositions can be ruled out for high- and low-mass planets, possibly revealing classes of planets with densities intermediate between terrestrial and giant planets with no Solar System counterpart \citep{Rogers2010,Kipping2013}. Typically the bulk density alone cannot break the planet interior composition degeneracy, especially for planets of intermediate density. However, measurement of atmospheric species may add enough information to reduce some of the planet interior composition degeneracies---e.g., the rejection of H/He as the dominant atmospheric species constrains significantly the bulk composition, independently of the mass uncertainty.

\subsection{Enhancing \textit{MassSpec}'s Capabilities}

We show that transmission spectroscopy can constrain the mass and the atmospheric properties of an exoplanet. Therefore, any dataset that can independently enhance the constraints on the mass and/or the atmospheric properties of an exoplanet can enhance \textit{MassSpec}'s capabilities via a joint analysis. In particular, RV measurements and emission spectra would be great assets to provide complementary constraints on a target's mass and its atmospheric properties, respectively. High-SNR emission spectra will be acquired in alternation with transmission spectra and are particularly sensitive to temperature.

%% file: Chapter4/chap4.tex
\renewcommand\thefootnote{\fnsymbol{footnote}}

\chapter[Future Prospects]{Future Prospects\footnote{Work published in part in \textit{Science}, Volume 342, Issue 6165, pp. 1473-1477 (2013), see \cite{deWit2013} and in \textit{Experimental Astronomy}, see \cite{Parmentier2014}. Work performed in part to support observation proposals: two for the \textit{Spitzer Space Telescope}---PI: de Wit, rejected, and PI: Lewis, accepted, see \cite{Lewis2013prop}---and one for the \textit{Hubble Space Telescope}---PI: Crossfield, under review.}}
\label{chap:perspectives}

\vspace{-0.7cm}
\renewcommand\thefootnote{\arabic{footnote}}
\addtocounter{footnote}{-1}

In this Chapter, we assess the potential of the methods introduced in Chapters\,\ref{chap:mapping} and\,\ref{chap:mass} when applied to observations from space-based facilities\footnote{Both methods can be applied to ground-based observations.}. We introduce short-term goals such as mapping for the first time an exoplanet's atmosphere in 3D and weighting the hottest exoplanet known with the \textit{Spitzer Space Telescope} \citep[\textit{Spitzer},][]{Werner2004} and the \textit{Hubble Space Telescope} (\textit{HST}), respectively. We discuss long-term goals such as obtaining time-dependent 3D maps of exoplanetary atmospheres and assessing the habitability of Earth-sized planets with the \textit{James Webb Space Telescope} \citep[\textit{JWST}; launch date 2018,][]{Clampin2010} and \textit{EChO}-class missions\footnote{\textit{Exoplanet Characterisation Observatory} (\textit{EChO}) was a M3 mission candidate of the European Space Agency \citep{Tinetti2012}. While \textit{EChO} was not funded, such a mission could still revolutionize our understanding of exoplanet atmospheres. Hence, we point out here the potential of our methods when applied to the observations of an \textit{EChO}-class mission---``\textit{EChO}'', for brevity.}, i.e. within the next decade.


\vspace{-0.3cm}
\section{Telescopes Noise Models}
	
	We introduce here the optical performance models used to assess the potential of our methods with current and future telescopes. Conceptually, such models are mainly effected by \textbf{(1)} the collecting area, \textbf{(2)} the total optical throughput\footnote{The total optical throughput of a telescope is the ratio of photons collected by its primary mirror to electrons read over a spectral resolution element of its detector.}, and \textbf{(3)} the instrument's duty cycle for a given target.
	
\subsubsection{\textit{HST}/WFC3}
\label{app:wfc3}
	We model the optical efficiency of \textit{HST} and the \textit{Wide Field Camera 3} (WFC3) as follow. We use a telescope effective area of $4.5$ square meter, a global optical efficiency of 0.14 $0.25$ to $0.95$ $\mu$m and of 0.20 from $1.00$ to $1.75$ $\mu$m---including optical throughput and detector quantum efficiency. We derive the duty cycle on a target-by-target basis to account adequately for the detector saturation---e.g., $\sim 40\%$ for the brightest target of our sample, WASP-33b. We validate our model with real data. In particular, our model predicts a RMS of 110 ppm for 103-sec bin of WASP-43, which is in agreement with the RMS of actual observation is $\sim100$ ppm (data from GO 13467, PI Bean). 
	
	We recall here that \textit{HST} is in a low Earth orbit (LEO), hence it does not allow continuous monitoring for most targets, but rather $\sim48$-min viewing windows.

\subsubsection{\textit{Spitzer}/IRAC}

	To date, no model of \textit{Spitzer}/IRAC's optical efficiency has been developed in the context of this thesis. The mapping capabilities introduced in Section\,\ref{sec:mappingpotential} are based on scaling the RMS on previous observations with the square root of number of observation to be performed.

\subsubsection{\textit{JWST/NIRSpec}}
\label{app:nirspec}

	\textit{NIRSpec} \citep{RauscherB2007} is the Near-Infrared Spectrograph for the \textit{James Webb Space Telescope} \citep[\textit{JWST}, see][]{Clampin2010}. The purpose of \textit{NIRSpec} is to provide low (R = 100), medium (R = 1000), and high-resolution (R = 2700) spectroscopic observations from 0.6 to 5 $\mu$m. We focus here on the medium resolution mode because \textit{MassSpec} requires a sufficient spectral resolution. Furthermore, a larger spectral dispersion enables the observation of brighter stars. However, the medium and high resolution modes require the necessity to use individually three grisms to obtain a full spectrum, hence, decreasing the effective observation time by three.

\paragraph*{\textit{JWST/NIRSpec} Total Optical Throughput}

\begin{figure}
  \begin{center}
    \includegraphics[trim = 00mm 00mm 00mm 05mm,clip,width=12cm,height=!]{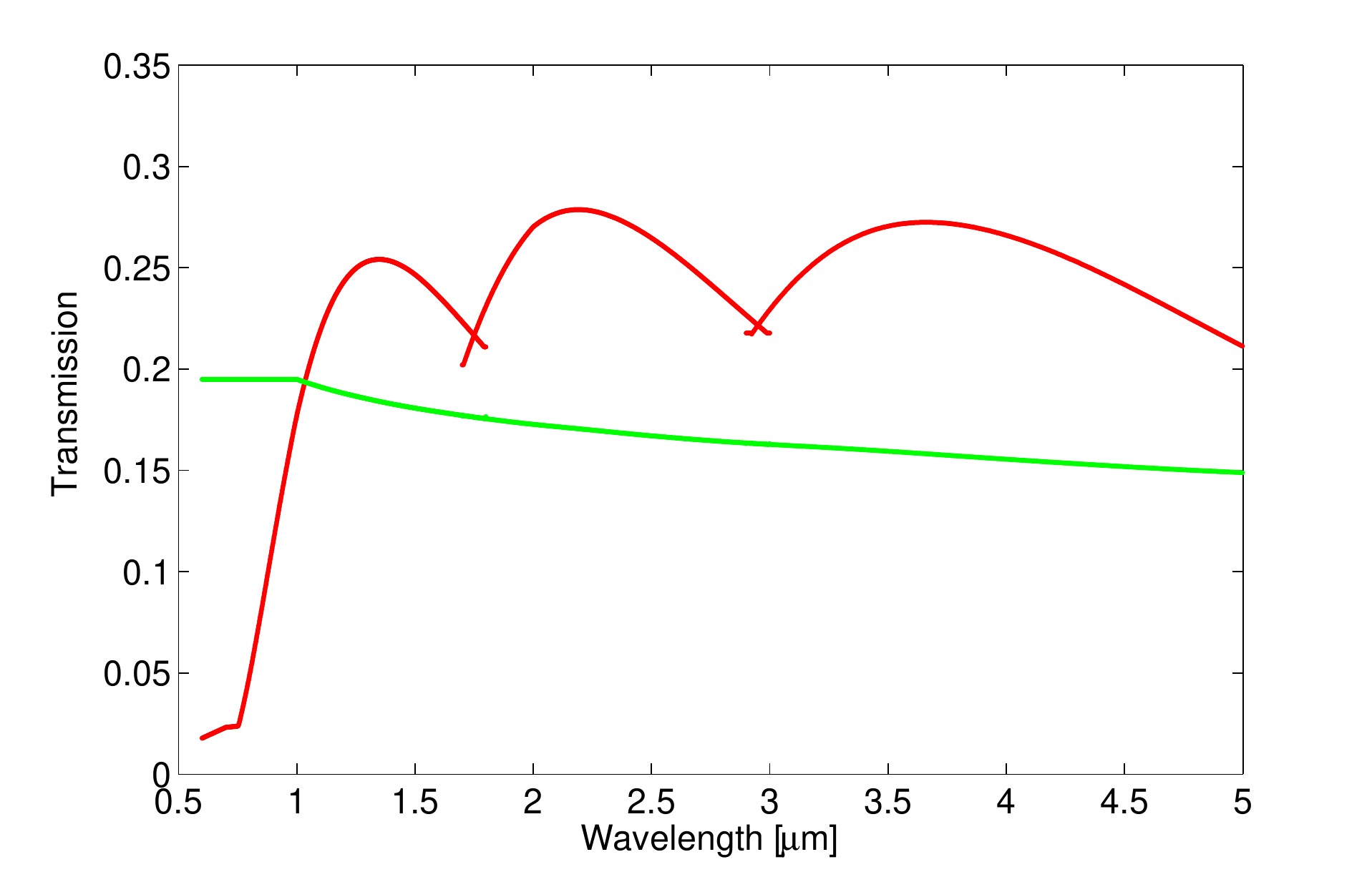}
  \end{center}
  \vspace{-0.6cm}
  \caption[\textit{JWST}/\textit{NIRSpec} optical performance for the medium spectral resolution mode (R = 1000)]{\textit{JWST}/\textit{NIRSpec} optical performance for the medium spectral resolution mode (R = 1000). The red line shows the total optical throughput for each spectral resolution element. The green line shows the flux fraction going to the brightest pixel of each spectral resolution element.}
  \vspace{-0.2cm}
  \label{fig:nirspec_perfo}
\end{figure}

	 We determine the total optical throughput of \textit{JWST/NIRSpec} according to \cite{Boker2010}. We model the detector pixel efficiency as a plateau at the 75\% level with a drop of 3\% at the pixel edges (G. Cataldo, private communication). We integrate the spectrograph $\lambda$-dependent PSF over the pixel grid to estimate the resolution element sensitivity. By doing so, we account for distortions for wavelengths under 1 $\mu$m using the 1 $\mu$m PSF-size. We present the throughput budget summary for the \textit{NIRSpec} medium-resolution mode in Figure\,\ref{fig:nirspec_perfo}. Figure\,\ref{fig:nirspec_perfo} also shows our estimate of the flux fraction going to the brightest pixel of each resolution element which we will use further to derive the saturation time of the detector. 

\vspace{-0.6cm}

	\paragraph*{JWST/NIRSpec Noise Budget}
	
	We derive the budget noise from the instrumental performance shown in Figure\,\ref{fig:nirspec_perfo}. First, we estimate the electron flux produced on the brightest pixel of the detector for a given star. Then, we derive the saturation time from the electron flux and a pixel well capacity of 60,000 e$^-$ to ensure linear response. We derive the variance of a frame, $\sigma_{1F}(\lambda)$, including a readout noise of 6 e$^-$\,px$^{-1}$ rms and a dark current of 0.03 e$^-$\,(px\,s)$^{-1}$ and assuming the read mode to be MULTIACCUM-2x1 \citep{Boker2010}. We derive the duty cycle based on a readtime of $\sim$0.53 sec, which, together with $\sigma_{1F}(\lambda)$, leads to the signal-to-noise ratio (SNR) for a given observation time, $t$.

	\paragraph*{\textit{JWST/NIRSpec} Noise Model Comparison}

	We find \textit{JWST/NIRSpec} observations will be photo-noise dominated, unlike \cite{Deming2009} that suggested that the intrapixel-sensitivity induced flux variations (see Section\,\ref{sec:systematics}) would be the main source of noise. Recent estimates of \textit{NIRSpec}'s detector performance show that its pixel-phase deviates mainly from a plateau at the pixel edge with a relative drop below 2\% due mainly by cross-talk (G. Cataldo, private communication)---i.e., the information is mainly transferred, not lost. These variations are more than an order of magnitude less than presented in \cite{Deming2009}. Furthermore, even for current facilities significantly affected by pixel-phase like the \textit{Spitzer Space Telescope}, the observations are within 10 to 20\% of the photon noise after systematic correction. In addition, we take into account the necessity to change the grisms to obtain a full spectrum \citep{Boker2010}. For practical purposes including stability and baseline follow-up, we consider that one transit is observed in a unique grism. For that reason we further scaled the SNR by $1/\sqrt{3}$ to include the inherent sharing of the integration time between the three channels of \textit{NIRSpec}'s medium-resolution module.
	
\subsubsection{\textit{EChO-Class Mission}}
\label{sec:echo}
While \textit{EChO} was not funded, such a mission could still revolutionize our understanding of exoplanet atmospheres. For that reason, we chose to point out the potential of our methods when applied to the observations of an \textit{EChO}-class mission.
In order to model a future \textit{EChO}-class mission, we use \textit{EChO}'s noise model introduced in \cite{Barstow2013}.  In particular, we use a telescope effective area of $1.13$ square meter, a detector quantum efficiency of $0.7$, a duty cycle of $0.8$, and an optical throughput of 0.191 from 0.4 to 0.8 $\mu$m, 0.284 from 0.8 to 1.5
$\mu$m, 0.278 from 1.5 to 2.5 $\mu$m, 0.378 from 2.5  to 5 $\mu$m, 0.418 from 5 to 8.5 $\mu$m,
0.418 from 8.5 to 11 $\mu$m, 0.326 from 11 to 16 $\mu$m.

\subsubsection{Future-generation telescope}

We use a scaled-up version of \textit{EChO} to model the 20-meter space-based telescope. In particular, we use a spectral resolution of $1000$, a detector quantum efficiency of $0.7$, a duty cycle of $0.8$, and an optical throughput of 0.4 from 0.4 to 16 $\mu$m.

\section{Mapping Potential}
\label{sec:mappingpotential}

\subsection{On the Benefits of Multi-Wavelength Observations}
	
		The spectral range used for light-curve observations  determines the layers probed in the target's atmosphere, as its optical depth is $\lambda$-dependent. Hence, if the SNR is high enough it is possible to map exoplanets' atmospheres in 3D with multi-wavelength observations. Such observations also mitigate the correlation between the system parameters and a planet's brightness distribution. The system parameters are not wavelength-dependent, hence their effects on the occultation shape can better be disentangled from the effect of the wavelength-dependent brightness distribution. 
			
		Applications of our method to observations in the visible can also provide us with insights into a planet's cloud properties (e.g., Section\,\ref{sec:K7b}). Recently, \cite{Madhusudhan2012} provide a theoretical framework for interpreting geometrical albedos from phase curves, which is is indicative of the scattering and absorptive properties of the atmosphere. Our method could, hence, complement such frameworks with 2D constraints on the geometrical albedo. 
	
\subsection{Towards the First 3D Map of an Exoplanet's Atmosphere}

	As a contribution to the accepted \textit{Spitzer} proposal ID $\#$10103 \citep{Lewis2013prop} we demonstrate that complementary observations of the hot Jupiters HD\,189733b and HD\,209458b can yield to the first 3D maps of exoplanetary atmospheres. In particular, we estimate that a total of nine and seven eclipses respectively in \textit{Spitzer}/IRAC 3.6 and 4.5-$\mu$m channels would enable to map HD\,189733b's atmospheric layers probed at those wavelengths and twenty-one and sixteen eclipses of HD\,209458b respectively at 3.6 and 4.5 $\mu$m would yield the same results. 
	
	The end of the complementary observations is scheduled before 2015, hence we expect the first 3D maps of exoplanet atmospheres within a year. Note that their resolution will be limited both in longitude/latitude and in altitude because of a limited SNR and spectral resolution, respectively.

	We demonstrate the feasibility of mapping in 3D HD\,189733b's and HD\,209458b's atmospheres as follow: 
	\begin{enumerate}
	\item We determine the current RMS achievable on archived occultation observations of our targets in the channels of interest---200 ppm and 175 ppm per 1-min bin in IRAC's 3.6 and 4.5-$\mu$m channels respectively for HD\,189733b and 300 ppm and 250 ppm for HD\,209458b. These RMS estimates account also for the correlated noise.

	\item We estimate the amplitude of the deviation in occultation ingress/egress due to the targets' brightness patterns. We expect deviation with an amplitude of $\sim180$ ppm (see Figure\,\ref{fig:3d_map_spitzer_hd189733b_lc} and compare with Figure\,\ref{fig:in_eg_structures}).

	\item We estimate the number of eclipses required to secure a 10$\sigma$ detection of these deviations in ingress/egress. We choose to aim for a high significance in order to mitigate the correlation between the brightness distribution and the system parameters (see Section\,\ref{sec:degeneracies}). We estimate the significance of the deviation as follow. The ingress/egress duration of our targets is 25 minutes, hence a total of $N_b\sim$50 bins capture an average deviation ($D_a$) of half the deviation amplitude ($D_a\sim$90 ppm). We thus need to achieve a RMS of $\sim$70 ppm per 1-min bin---i.e., $70 \approx D_a/\sigma*\sqrt{N_b}$ with $\sigma = 10$ the significance level chosen. To do so, we need a total of nine, $\sim (200/70)^2$, and seven eclipses, $\sim (175/70)^2$, at 3.6 and 4.5 $\mu$m for HD\,189733b's and twenty-one and sixteen---$\sim (300/70)^2$ and $\sim (250/70)^2$, respectively---at 3.6 and 4.5 $\mu$m for HD\,209458b. 
	
	\vspace{-0.5cm}
	\begin{figure}[!ht]
\begin{center}
\hspace{-0.0cm}\includegraphics[trim = 00mm 00mm 00mm 05mm,clip,width=12cm,height=!]{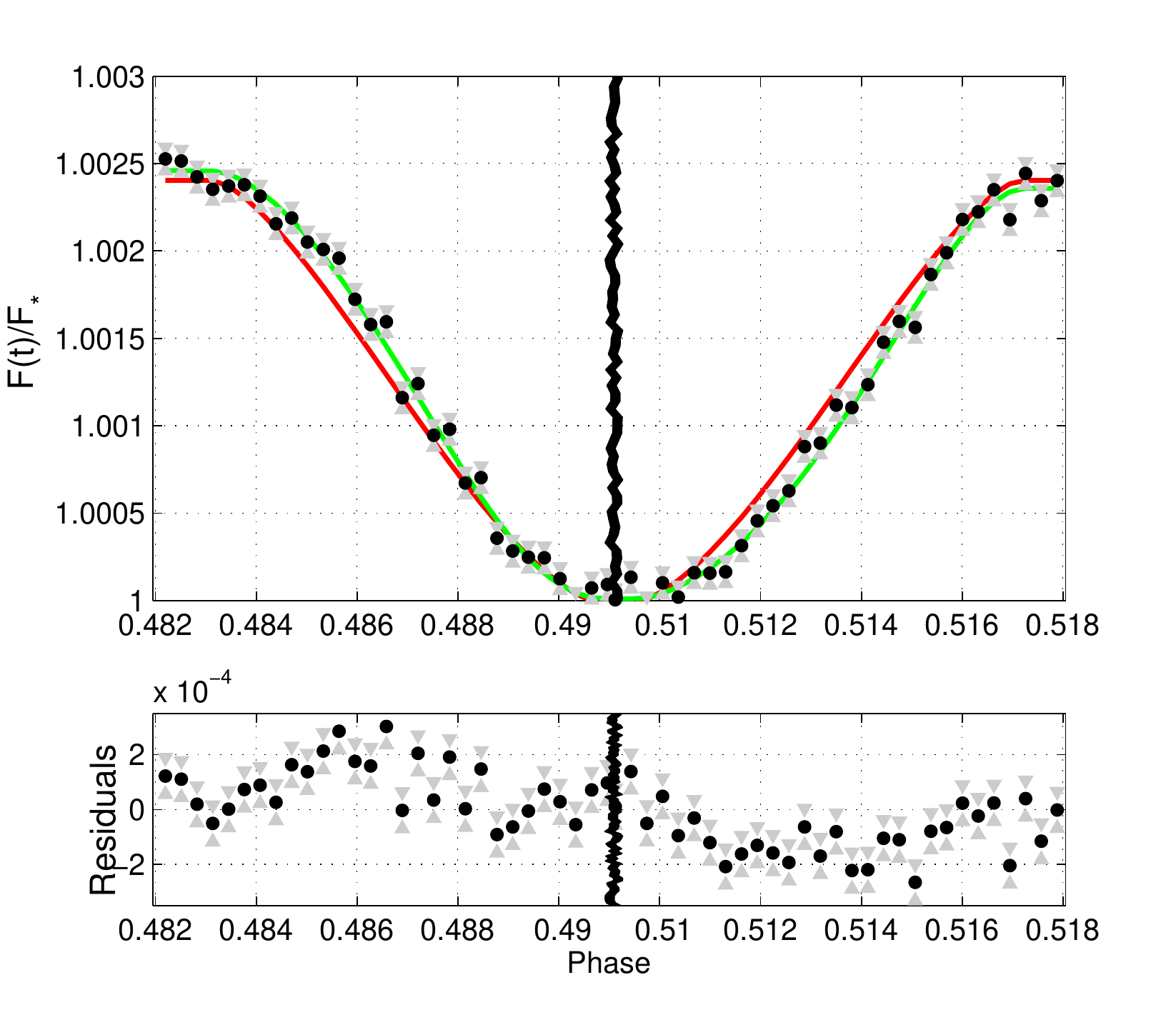}
\end{center}    
  \vspace{-0.8cm}
  \caption[Synthetic observations of HD\,189733b in \textit{Spitzer}/IRAC 4.5-$\mu$m channel.]{Synthetic observations of HD\,189733b in \textit{Spitzer}/IRAC 4.5-$\mu$m channel for the brightness map shown in Figure\,\ref{fig:3d_map_spitzer_hd189733b_map_simu}. The red line represents the best fit using a conventional model (i.e., assuming HD\,189733b to be a uniformly-bright disk) and the green
line represents the best fit using a non-uniform dayside brightness model. The deviation of the occultation shape from the occultation of a uniformly-bright disk is shown in the bottom pannel.}
  \vspace{-.2cm}
  \label{fig:3d_map_spitzer_hd189733b_lc}
\end{figure}	

\begin{figure}

   \subfloat[]{\label{fig:3d_map_spitzer_hd189733b_map_simu}\hspace{1.2cm}\includegraphics[trim = 00mm 00mm 191mm 00mm,clip,width=13cm,height=!]{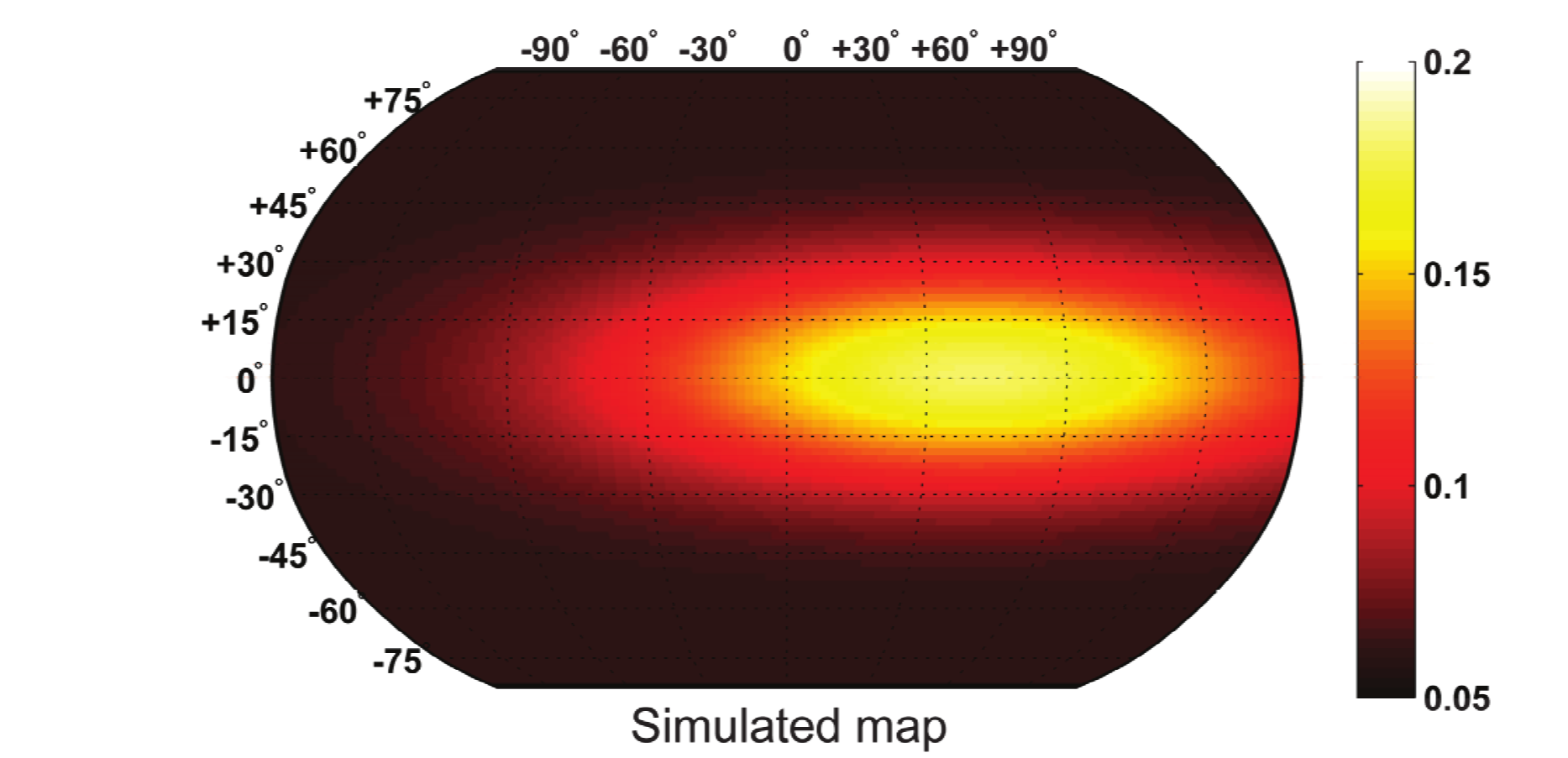}}
      
      \subfloat[]{\label{fig:3d_map_spitzer_hd189733b_map_retrieved}\hspace{1.2cm}\includegraphics[trim = 191mm 00mm 00mm 00mm,clip,width=13cm,height=!]{3d_map_spitzer_hd189733b_maps.pdf}}

  \caption[Feasibility of mapping HD\,189733b and HD\,209458b using \textit{Spitzer}/IRAC 3.6 and 4.5$\mu$m channels.]{Feasibility of mapping HD\,189733b and HD\,209458b using \textit{Spitzer}/IRAC 3.6 and 4.5$\mu$m channels. \subref{fig:3d_map_spitzer_hd189733b_map_simu} shows one of the synthetic map used to simulate synthetic \textit{Spizter}/IRAC's datasets of HD\,189733b (see Figure\,\ref{fig:3d_map_spitzer_hd189733b_lc}) and to estimate the mapping capabilities with seven eclispes. \subref{fig:3d_map_spitzer_hd189733b_map_retrieved} retrieved brightness map that is consistent with \subref{fig:3d_map_spitzer_hd189733b_map_simu} and hence demonstrate the feasibility of our proposal---i.e., of mapping in 3D HD\,189733b's and HD\,209458b's atmospheres.}
  \label{fig:3d_map_spitzer_hd189733b_maps}
\end{figure}
	
	\item Finally, we demonstrate that the RMS requested above would be sufficient to retrieve meaningful constraints on the brightness distributions of our targets. For that purpose, we simulate synthetic light curves and show that we could retrieve the brightness maps used to produce the synthetic light curve. As an example, Figure\,\ref{fig:3d_map_spitzer_hd189733b_lc} shows the significance of deviations from a uniformly-bright disk that we expect to detect for both targets at 3.6 and 4.5~$\mu$m. Figures\,\ref{fig:3d_map_spitzer_hd189733b_map_simu} and\,\ref{fig:3d_map_spitzer_hd189733b_map_retrieved} show the simulated and retrieved brightness maps, respectively. Figure\,\ref{fig:3d_map_spitzer_hd189733b_maps} shows that the numbers of eclipses requested yield a SNR sufficient to retrieve a brightness map consistent with the synthetic map used to simulate the synthetic data. This demonstrates the feasibility of our proposal. We also investigate what constraints on the location and extent of bright ``hot-spots''  would be achievable, as those constraints provide insights into the underlying physical processes shaping these features. We show that the complementary observations will enable:
\begin{enumerate}
\item Localization of the peak in the dayside brightness distribution with an uncertainty $\sim$2-3$^{\circ}$ in longitude and $\sim$6-10$^{\circ}$ in latitude. 
         If the hottest regions of the planet were substantially shifted from the equatorial region (e.g. 45$^{\circ}$ in latitude) the precision with which we could localize the peak in the dayside brightness would be decreased to $\sim$3.5-6.0$^{\circ}$ in longitude and $\sim$12-20$^{\circ}$ in latitude.
         
\item A precision better than 17$^{\circ}$ on the latitudinal extent of the hot-spots and better than 8$^{\circ}$ on the longitudinal one, i.e., a precision relevant for comparisons with three-dimensional hot-Jupiter GCMs.   
\end{enumerate}

	\end{enumerate}

	\subsection{3D Map of the Hottest Jupiters in the Visible}
	
	As a contribution to Dr. I. Crossfield's \textit{HST} proposal to map the hottest planetary atmospheres, we applied a methodology analogous to the one described at the previous Section. The main differences with the methodology adopted for the \textit{Spitzer} proposal are that we investigate the entire list of planets known to determine which would be the best targets to perform mapping and we use the \textit{HST}/WFC3 model introduced in Section\,\ref{app:wfc3}. Note that \textit{HST}'s orbital configuration implies that the observation of an occultation ingress or egress requires at least two \textit{HST} orbits, one to perform the pointing and a second to perform the science---in some cases, it might be necessary to perform an observation before and after the science in order to get the light-curve baseline.
	
	\subsubsection*{Finding the Most Suitable Planets to Map in the Visible}
	
\begin{figure}
\begin{center}
\hspace{-0.0cm}\includegraphics[trim = 00mm 00mm 00mm 00mm,clip,width=12cm,height=!]{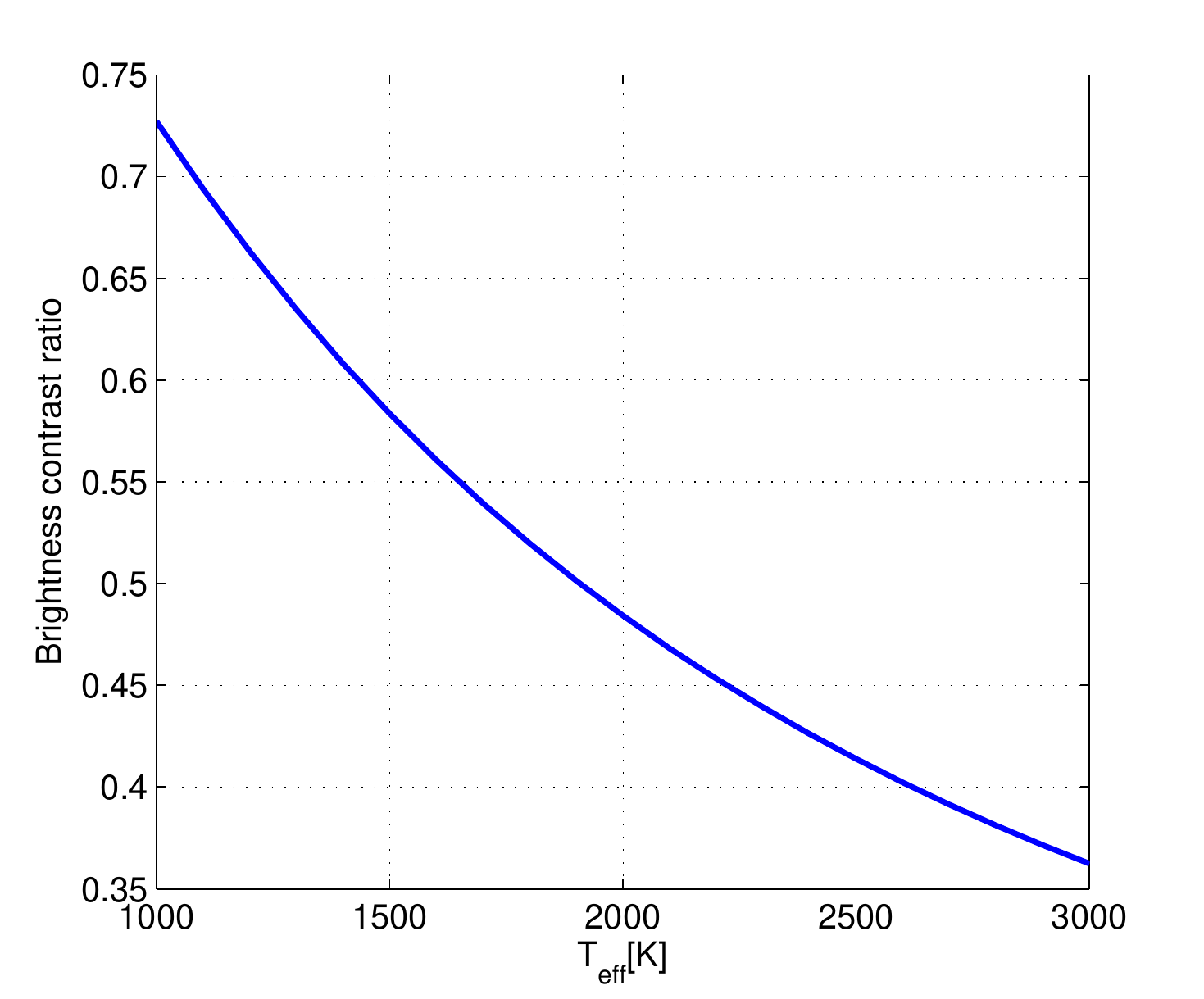}
\end{center}    
  \vspace{-0.6cm}
  \caption[Dependence on the brightness temperature of the brightness contrast ratio between a planetary hot spot and the background atmosphere in \textit{HST}/WFC3 spectral band.]{Dependence on the brightness temperature of the brightness contrast ratio between a planetary hot spot and the background atmosphere in \textit{HST}/WFC3 spectral band. The brightness contrast is shown for a modeled 10$\%$ temperature difference between the hot spot and the background.}
  \label{fig:brigthness_contrast_vs_T_eff}
\end{figure}

	Finding the best targets to map with a given telescope requires determining the number of occultations necessary to reach a sufficient SNR on the occultation ingress/egress. We considered the exoplanets listed in \cite{Wright2011} and \cite{Schneider2011} and proceed  for each of them as follow:
	\begin{enumerate}
	\item Determine the RMS per 1-min bin over the \textit{HST}/WFC3 spectral range based on their host-star properties.  
	\item Based on the target's brightness temperature, we derive its eclipse depth. If brightness temperature is not yet constrained, we derive its equilibrium temperature (Equation\,\ref{eq:T_eq}).  
	\item We derived the number of eclipses required to yield a ratio eclipse depth to RMS per 1-min bin of 30, which ensure a sufficient precision of the ingress/egress. We note that we use an extra factor to account for the dependence on the planet's brightness temperature of the brightness contrast ratio between a hot spot and the background temperature. Figure\,\ref{fig:brigthness_contrast_vs_T_eff} shows that the signal in occultation ingress/egress of colder planets is enhanced by an increased contrast ratio, and hence partially compensate for their lower eclipse depth in \textit{HST}/WFC3 spectral band.
	\end{enumerate}
	
	We present in Table\,\ref{tab:HSTWFC3Targets} the ten best targets to perform exoplanet mapping with \textit{HST}/WFC3. We sort these targets based on the number of eclipses required to secure a sufficient SNR to perform eclipse mapping with \textit{HST}/WFC3. In Dr. I. Crossfield's proposal, we proposed to target WASP-18b and WASP-19b for scientific considerations that are not part of the present work but also because: \textbf{(1)} WASP-33b orbits a Delta Scuti variable, hence its pulsations would mitigate the mapping performance---particularly in the context of \textit{HST}'s non-continuous monitoring---and \textbf{(2)} we have currently very view observational constraints concerning WASP-76b. For future references, we also note that mapping WASP-12b would require a particular attention as it is distorted by its host \citep{Li2010}---see Section\,\ref{sec:degeneracies}.
	
	\begin{table}
\caption{Ten most favorable exoplanets to map with \textit{HST}/WFC3 \label{tab:HSTWFC3Targets}}
	\centering
	\setlength{\extrarowheight}{3pt}
	\footnotesize{\begin{tabular}{c c}
	
	\hline\hline
	{\normalsize\textbf{Target}} & {\normalsize\textbf{Number of eclipses}}\\
	\hline
	 WASP-33b & 8\\
	 WASP-18b & 18\\
	 WASP-76b & 18\\
	 WASP-19b & 36\\
	 WASP-12b & 40\\
	 WASP-103b & 58\\
	 WASP-77 Ab & 74\\
	 WASP-82b & 93\\
	 WASP-3b & 102\\
	 WASP-14b & 120
			
	\end{tabular}}
	
\end{table}

\subsubsection*{3D Maps of WASP-18b and WASP-19b with \textit{HST}/WFC3}

	As for \cite{Lewis2013prop}, we demonstrate that the number of eclipses mentioned in Table\,\ref{tab:HSTWFC3Targets} for WASP-18b and WASP-19b would be sufficient to retrieve meaningful constraints on the brightness distributions of our targets. For that purpose, we simulate synthetic light curve---including partial phase curve---and show that we could retrieve the brightness map used to produce the synthetic light curve. We also show that these observations would enable:
\begin{enumerate}
\item The localization of the peak in the dayside brightness distribution with an uncertainty $\sim$10$^{\circ}$ in longitude and latitude. 
\item Constraints on the extent and the intensity distribution of the hot-spots with a precision better than 20$^{\circ}$ on the extent in longitude and in latitude.  
\end{enumerate}

	The \textit{HST} proposal partially described above is currently under review for \textit{HST} cycle 22.

	\subsection{Time Variability of Exoplanets' Amospheres 3D Structures}
	\label{sec:map_echo_jwst}
	
	As a contribution to \cite{Parmentier2014}, we investigate the mapping capability of an \textit{EChO}-class mission based on the \textit{EChO}'s model introduced in Section\,\ref{sec:echo}. We also discuss here \textit{JWST}'s mapping capabilities.
	
\begin{figure}[!p]

\hspace{-4.0cm}\includegraphics[trim = 00mm 00mm 00mm 00mm,clip,width=22cm,height=!]{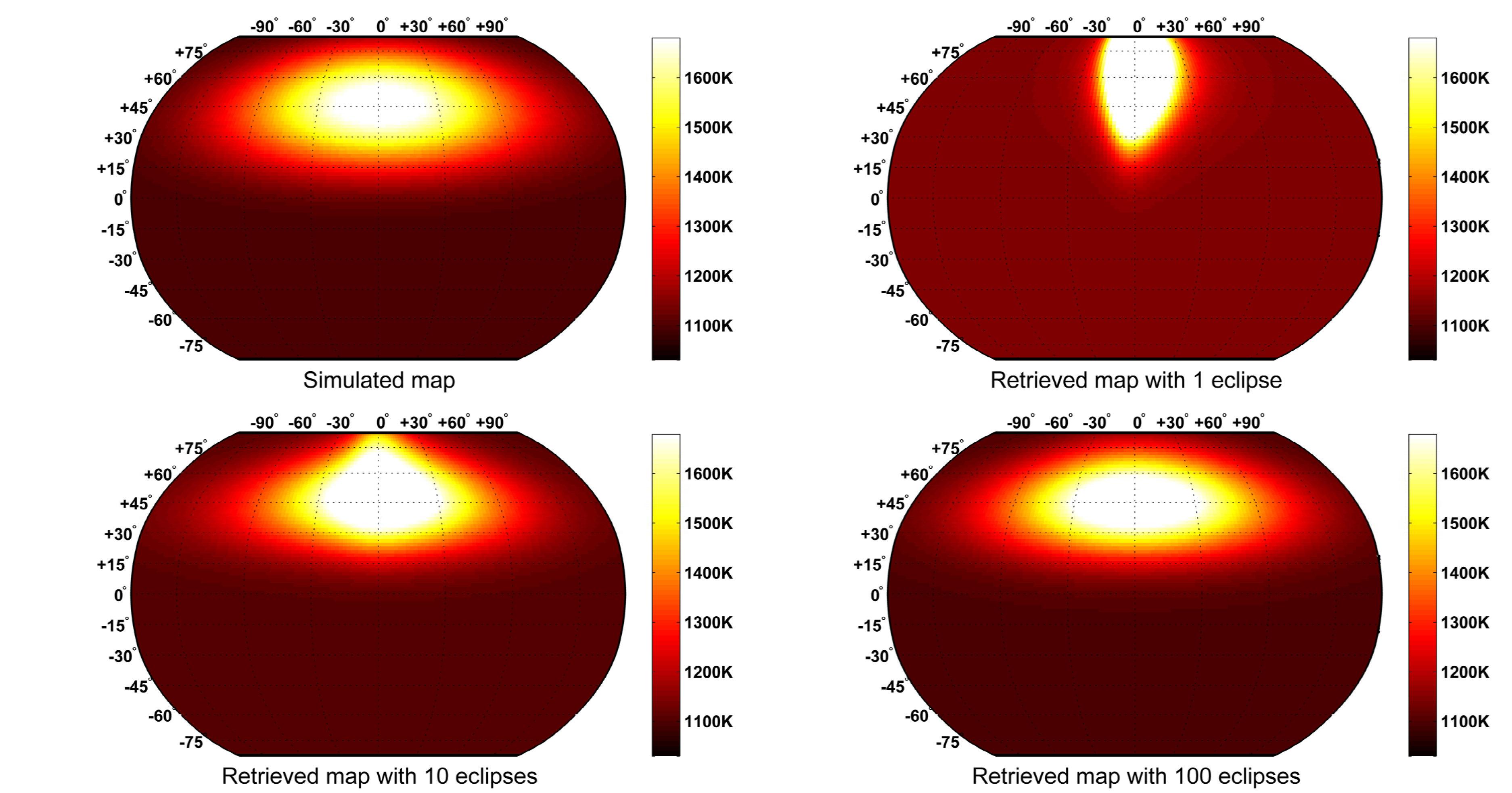}
  \vspace{-0.6cm}
  \caption[Mapping potential of an \textit{EChO}-class mission.]{Mapping potential of an \textit{EChO}-class mission. Synthetic distribution of brightness temperature for the hot Jupiter HD\,189733b (``Simulated map'') and the retrieved map as a function of the number of occulation (1, 10, and 100) observed in a spectral bin of resolution 20---here, in the 4.3$\mu$m band of carbon dioxide.}
  \label{fig:retrieved_maps_EChO}
\end{figure}

	We show in Figure\,\ref{fig:retrieved_maps_EChO} the mapping potential of an \textit{EChO}-class mission. In particular, Figure\,\ref{fig:retrieved_maps_EChO} shows the map retrieval of a synthetic brightness map of the hot Jupiter HD\,189733b harboring a hot spot with a temperature contrast of 30$\%$ and located on the northern hemisphere. The retrieval assumes observations in a spectral bin of resolution $\sim20$---here, in the 4.3$\mu$m band of carbon dioxide. Such a hot spot could be formed by the presence of patchy clouds or chemical differences between the poles and the equator. With one secondary eclipse, \textit{EChO} would detect the presence of latitudinal asymmetry in the planet's brightness distribution. With $\sim 10$ and  (resp. $\sim 100$) secondary eclipses, the temperature contrast will be measured with a precision of 300 K (resp. 100 K) and the latitudinal location of the hot-spot will be determined with a precision of 10$^\circ$ (resp. 3.5$^\circ$). 

\textbf{An \textit{EChO}-class mission} would thus be able to yield 3D maps of exoplanets as such 2D maps would be available at different optical depths, for the most favorable targets. In addition, we find that for \textit{EChO}:
\begin{itemize}
\item Observation programs dedicated to 3D mapping in the infrared will require a minimum of 150 hours of observation.

\item Observation programs aiming to constrain cloud properties will only provide longitudinal constraints \citep[similarly to][]{Demory2013} and will require a minimum of 250 hours of phase curve observation. It will not be possible to constrain latitudinally the cloud coverage because the contrast of feature in reflection (clouds) is $\sim$10 times less than in emission (hot or dark spots), in terms of relative brightness. 

\item \textit{EChO} would be able to capture temporal variabilities in an planetary atmosphere if these variations have either large amplitudes (e.g., hot spot oscillation from one tropic to another) or large time scales (i.e., $>10$ orbital periods to enable capturing with a sufficient SNR the extreme brightness distribution of the variations).
\end{itemize}

	\textbf{JWST}'s large collecting area yields to a significant reduction of eclipses required to perform 3D mapping. \textit{JWST} would yield a 3D map with a single occultation for most targets for which an \textit{EChO}-class mission could yield 3D mapping. Therefore, the observations of consecutive occultations will provide unprecedented insights into the physics of those planet's atmosphere in the form of constraints on the time variability of their 3D atmospheric structures. However, \textit{JWST} will not yield cloud maps because its spectral band do not cover the visible. Still, \textit{JWST} could provide strong inference on cloud coverage via the detections of dark spots in the infrared---if those spots can be associated to large albedo using complementary facilities.

\subsection{From Maps to Insights into Atmospheric Physics}

The analysis method introduced in Chapter\,\ref{chap:mapping} constrains the brightness distribution of a target, not its underlying physics. A way around that would be to use brightness models (see Section\,\ref{sec:planet_contribution}) with parameters carrying a physical meaning. As an example, \cite{Heng2014} derived key analytical formulations for the global structure of planetary atmospheres---under the shallow-water assumption---which are governed by up to three dimensionless parameters related to the atmospheric conditions. Therefore, the implementation of such formulations in our mapping method could be used to map exoplanets while gaining additional insights into the physicals processes at play in the target's atmosphere via the determination of meaningful brightness-model parameters. \footnote{The discussion about such an implementation was initiated by Dr. K. Heng at the Exoclimes III conference.}

\newpage
\section{\textit{MassSpec}'s Potential}

	\textit{MassSpec}'s applications require a high SNR over an extended adequately-sampled spectral coverage (e.g., from 0.2 to 5 $\mu$m with $\lambda/\Delta\lambda \geqq 300$) to constrain independently an exoplanet's mass and atmospheric properties. We assess in this Section the full potential\footnote{We refer to ``\textit{MassSpec}'s full potential'' for its capability to retrieve the mass and the atmospheric properties of a planet solely from transmission spectroscopy.} of \textit{MassSpec} using future space-based facilities. But first, we show below that although \textit{MassSpec}'s applications are currently limited (see Section\,\ref{sec:masspechd189733b}) it could already yield significant science using \textit{HST}/WFC3, such as the mass of the hottest planet known. 
	
	\subsection{Short-Term: Weighing the Hottest Planet Known}
\begin{figure}[!ht]

\vspace{-0.7cm}\centering\includegraphics[trim = 00mm 00mm 00mm 00mm,clip,width=12cm,height=!]{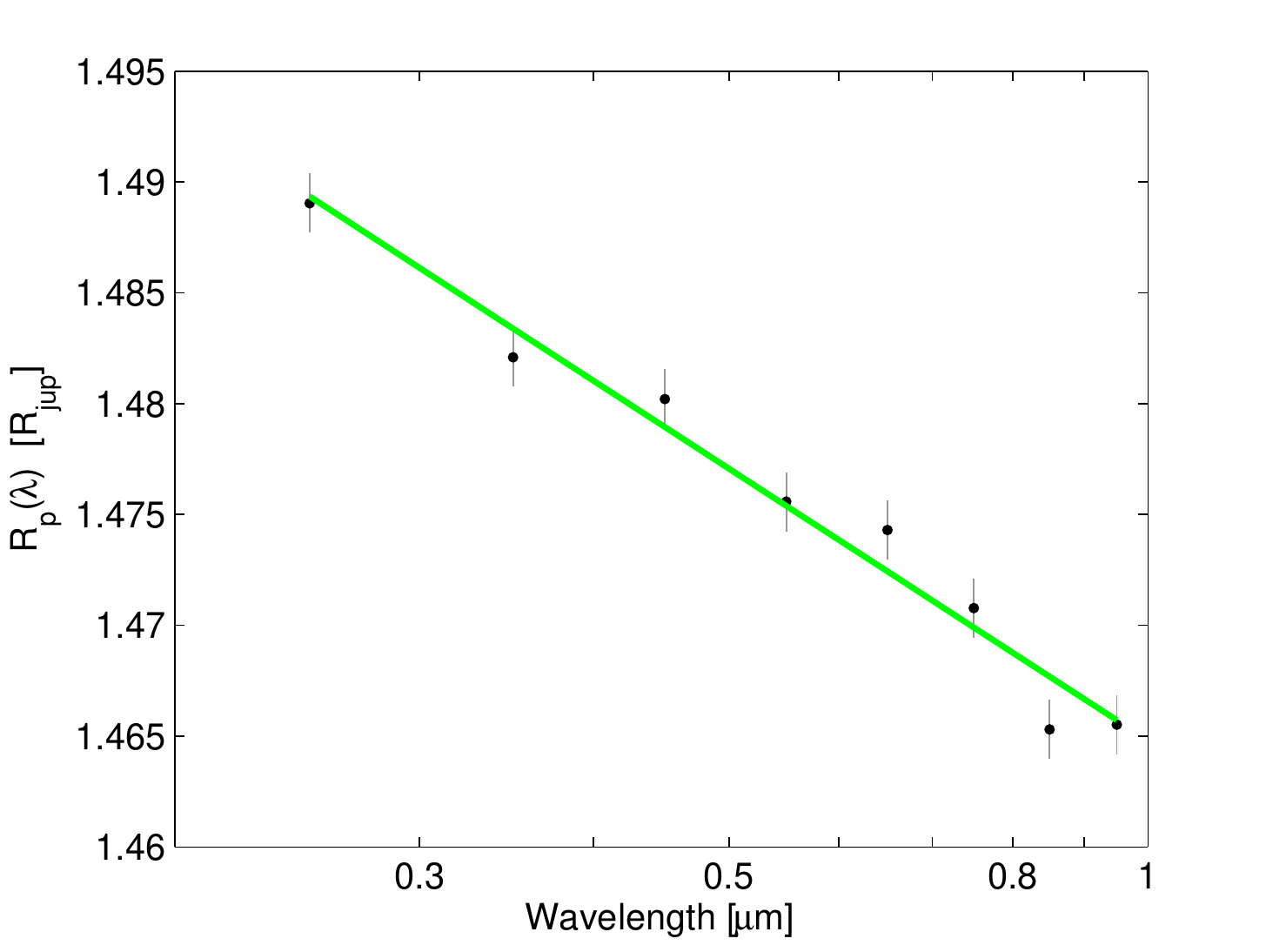}
  \vspace{-0.0cm}
  \caption[Weighing the hottest planet known with \textit{HST}.]{Weighing the hottest planet known with \textit{HST}. Synthetic transmission spectrum of WASP-33b (black dots) and their uncertainties (grey bars) in the \textit{HST}/WFC3 UV/Visible band together with the best fit (green).}
  \vspace{-.4cm}
  \label{fig:weighing_WASP33b}
\end{figure}

	\textit{MassSpec}'s application is currently limited to giant planets (Section\,\ref{sec:masspechd189733b}). Although most giant planets's masses can be determined from RV measurements, for some their host-stars' activities prevent a mass measurement \citep[e.g., the hottest  planet known, WASP-33b,][]{Collier2010}. We show here how a reasonable amount of \textit{HST}'s time can yield a mass measurement for WASP-33b. For that purpose, we proceed as follow:
	\begin{enumerate}
	 \item We estimate the RMS on the transit depth in 0.1$\mu$m-bins in the \textit{HST}/WFC3 UV/Visible channel using the noise model introduced in Section\,\ref{app:wfc3}. We assume a contribution of correlated noise post-correction for WASP-33's variability of 25$\%$\footnote{We consider that a simultaneous and continuous monitoring of WASP-33 using multi-channel ground-based instruments enable to correct for WASP-33 activity over a transit \citep[see, e.g.,][]{vonEssen2014}.} and estimate the precision on the transit depth to $\sim$45 ppm per bin. 
	 \item We simulate WASP-33b's synthetic transmission spectrum assuming a pure scattering slope (Figure\,\ref{fig:weighing_WASP33b}) because no molecular absorption is expected at the target's extreme temperature \citep[$T=3200$ K and $R\simeq1.48$ $R_{Jup}$,][]{Smith2011}. We model the spectrum based on Equation\,\ref{eq:M2scattering} for a synthetic WASP-33b of 3 Jupiter's mass .
	 \item We perturb the theoretical transmission spectrum using our estimate of the uncertainty on the transit depth in each spectral bin.
	 \item We fit the synthetic signal to retrieve the scattering slope and derive the relative uncertainty on the WASP-33b's mass that would originate from the uncertainty on the scattering slope.
	\end{enumerate}
	
	We find that the uncertainty on WASP-33b's mass due to the scattering slope uncertainty could be as low as 6$\%$ using five transits observed with \textit{HST}/WFC3. This precision is in strong contrast with the current constraints on WASP-33b's mass from RV measurements. However, we note that the uncertainty on WASP-33b's mass would then be primarily affected by our uncertainty on its temperature---i.e., on its temperature profile as a vast range of pressure levels are probed between 0.2 and 1$\mu$m\footnote{In practice, the Rayleight-scattering ``slope'' would also provide the target's atmospheric temperature profile (Appendix\,\ref{sec:tra_spec_equ} and Figure\,\ref{fig:rayleigh_slope_when_non_isothermal}).}. In particular, when deriving the mass of a gas giant solely from its scattering slope Equation\,\ref{eq:M2scattering}, the relative uncertainty on its mass, $\frac{\sigma_{M_p}}{M_p}$, can be approached at first order as follow
\begin{equation}
\frac{\sigma_{M_p}}{M_p}=\sqrt{{\frac{\sigma_{T}}{T}}^2+2{\frac{\sigma_{R_p}}{R_p}}^2+{\frac{\sigma_{\mu}}{\mu}}^2+{\frac{\sigma_{H}}{H}}^2} 
\label{Mstd3}
\end{equation} 
where $\sigma_{X}$ is the uncertainty on the parameter $X$. We note that for WASP-33b the relative uncertainty on $R_p$ and $\mu$ are negligible compare to the uncertainty on the temperature profile $\frac{\sigma_{T}}{T} \lesssim 0.15$. Hence, we are confident in \textit{MassSpec}'s capacity to yield the mass of the hottest planet known with a relative uncertainty of $\lesssim 20\%$ together with refined constraints on its temperature profile with less than 20 \textit{HST} orbits.

\subsection{Transmission Spectrum Model}
\label{app:transmission}
	We model high resolution\footnote{It is necessary to model the radiative transfer process at high resolution to approach adequately the absorption lines and their effects on the stellar light. Once the simulated transmission spectrum is modeled it can be binned down to the facility's spectral resolution for comparison.} ($R = \lambda/\Delta \lambda = 10^5$) transmission spectra following Equation\,\ref{eq:transmission_spectrum_h}: (i) we calculate the slant-path optical depth for different altitudes and (ii) integrated the contribution of each projected atmospheric annulus to the overall flux drop---$2\pi r \mbox{d}r (1-\exp[-\tau(r,\lambda)]) $ (Figure\,\ref{fig:transmission_spectrum_basics}). The extinction cross section accounts for molecular absorption, collision-induced absorption (CIA), and Rayleigh scattering. We compute the monochromatic molecular absorption cross sections from the HITRAN 2008 database \citep{Rothman2009} and model the line profile according to \cite{Liu2001} to increase the computational speed. We use opacity tables from \cite{Borysow2002} for H$_2-$H$_2$ CIA. We determine the Rayleigh-scattering cross section, $\sigma_{R,i}$ in cgs units, from
	\begin{equation}
\sigma_{R,i}(\lambda) = \frac{24\pi ^3}{n_{i}^2\lambda^{4}} \left(\frac{N_{i}(\lambda) ^2-1}{N_{i}(\lambda) ^2+2}\right)^2 F_{i}(\lambda),
\label{eq:rayleight}
\end{equation} 
	where $\lambda$ is the wavelength and $n_{i}$, $N_{i}(\lambda)$, and $F_{i}(\lambda)$ are the number density, the refractive index, and the King correction factor for the depolarization of the i$^{th}$ atmospheric species. In particular, we use the refractive indices of N$_2$, CO$_2$, CO, CH$_4$, N$_2$O from \cite{Sneep2005}\footnote{We observe discrepancies between the measured data and the functional forms proposed in \cite{Sneep2005} for the refractive indices of CO$_2$ and CO---their equations (13) and (17). Therefore, we use the following corrected forms:
	\begin{equation}
\begin{split}
\frac{n_{CO_2}-1}{1.1427 \times 10^{\boldsymbol{3}}}  &= \frac{5799.25}{(128908.9)^2-\lambda^{-2}} + \frac{120.05}{(89223.8)^2-\lambda^{-2}}\\ 
&+ \frac{5.3334}{(75037.5)^2-\lambda^{-2}} + \frac{4.3244}{(67837.7)^2-\lambda^{-2}}+\frac{\boldsymbol{0.1218145}}{(2418.136)^2-\lambda^{-2}}, 
\end{split}
\label{eq:n_sneepco2}
\end{equation}
	\begin{equation}
\frac{n_{CO}-1}{1 \times 10^{-8}} = 22851 + \frac{0.456\times 10^{\boldsymbol{14}}}{(71427)^2-\lambda^{-2}}.
\label{eq:n_sneepco}
\end{equation}
	}.

	A transmission spectrum simulation requires the computation of millions of absorption lines for each atmospheric species and numerous $T-p$ conditions. In addition, we develop our retrieval method in a Bayesian framework which requires a large number of transmission model runs to converge. Therefore, we compute the extinction cross section for each component of HITRAN as a function of $\lambda$, $T$ and $p$ to interpolate later for the required conditions. In particular, we generate the extinction cross section 4-D array 
	\begin{itemize}
	 \item at a spectral resolution of $10^5$ for 0.4$\mu$m to 250$\mu$m,
	 \item for 17 pressure values spread in the $\log_{10}p[\text{Pa}]$ space from 7 to -3 with a higher density around 4---because most of the information is recorded around 1 mbar,
	 \item for 12 temperature values homogeneously spread from 150 K to 700 K.
	 \end{itemize} 
	 
	 We validate our extinction cross section model with \cite{Benneke2012}. In addition, we validate our transmission spectrum model comparing a synthetic Earth transmission spectrum with \cite{Kaltenegger2009} and \cite{Hu2013}. Except for a 3.2$\mu$m-water signature\footnote{We observe an additional water band at 3.2$\mu$m in \cite{Kaltenegger2009}; it could be due to a complementary list used in this study. The HITRAN water absorption lines are uniformly spaced in terms of wavenumber, as expected from quantum mechanics. Furthermore, HITRAN's aim is to provide the spectroscopic parameters required to simulate the transmission and emission light for Earth-like conditions. Therefore, it is unlikely that such a significant water band would not be included in HITRAN.}, our simulation is consistent with \cite{Kaltenegger2009} and \cite{Hu2013}; for that reason, we consider our transmission model appropriate for the present study.

\subsection{Atmospheric Retrieval Method}
\label{app:retrieval}
We use an adaptive Markov Chain Monte Carlo (MCMC, see Section\,\ref{sec:analysismethod} for an introduction) algorithm to retrieve the properties of an exoplanet's atmosphere embedded in its transmission spectrum (Equation\,\ref{eq:transmission_spectrum_h}). The key improvement compared to previous studies \citep[e.g.,][]{Madhusudhan2009,Benneke2012,Lee2012} is the use of a self-consistent set of parameters derived from first principle of transmission spectroscopy (Section\,\ref{app:dependency}) that uniquely constrains the planetary mass. In other words, we use as jump parameters: the temperature, the pressure scale height and the species number density at a given planetary radius---whose choice does not affect the retrieval method, similarly to the ``reference radius''.    We assume a uniform prior distribution for all these jump parameters and draw at each step a random stellar radius based on a Gaussian prior assuming here a 1$\%$ relative uncertainty on the stellar radius.

\subsubsection{Synthetic exoplanet scenarios}
\label{app:scenarios}
	We assess \textit{MassSpec}'s capabilities for different super-Earths and Earth-sized planets. We use GJ\,1214b's \citep{Charbonneau2009} and Earth's sizes, respectively. We use the ``hot mini-Neptune'', ``Hot Halley world'', and ``nitrogen-rich world'' atmospheric scenarios introduced in \cite{Benneke2012} to provide a representative overview of \textit{MassSpec}'s potential. We assume a well-mixed atmospheres and use temperature profiles similar to \cite{Miller-Ricci2009}---in particular, $T(p\gtrsim1\text{ mbar})\approx300$ K.
	
	We assume a total in-transit observation time of 200 hrs \citep{Deming2009} and a M1V\footnote{A M1V star is a representative host in terms of transmission-spectrum significance (Section\,\ref{app:scaling_laws_snr}).} host star located at 15 pc, except for the Earth-like planet observed with \textit{JWST} for 200 hrs around a M7V at 15 pc. We note that for Earth-sized planets, we investigate in greater details the nitrogen-world scenario (``Earth-like" planet) because (1) the first exoplanets to be confirmed habitable are likely to be hot desert world \citep{Zsom2013}, (2) in most cases hydrogen should have escaped from an Earth-sized planet atmosphere, and (3) nitrogen-dominated atmospheres are less favorable for transmission spectroscopy due to their larger mean molecular mass (Equation\,\ref{eq:SNRt_general_scalinglaw}) hence \textit{MassSpec}'s capabilities derived from those are more conservative.

\newpage
\subsection{Long-Term: Characterizing Planets as Small as Earth}
\label{app:results}

We find that future space based facilities designed for exoplanet atmosphere characterization will also be capable of mass measurements for super-Earths and Earth-sized planets with a relative uncertainty as low as $\sim2\%$---a precision that has not yet been reached using RV measurements, even for the most favorable cases of hot Jupiters.

\begin{figure}[!ht]
 \centering
  \begin{center}
    \includegraphics[trim = 30mm 20mm 40mm 23mm,clip,width=12.3cm,height=!]{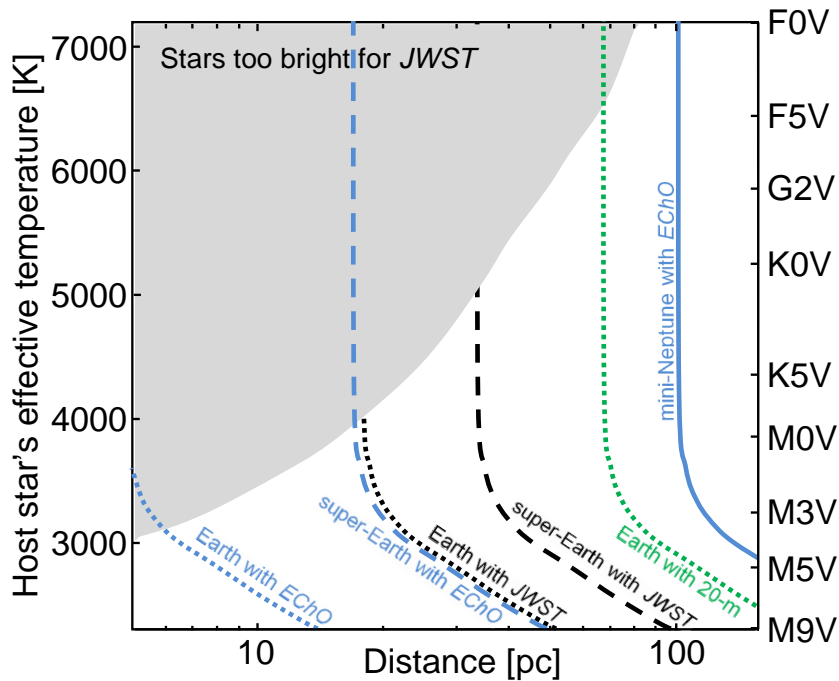}
  \end{center}
  \vspace{-1.2cm}
  \caption[\textit{MassSpec}'s potential using future space-based facilities.]{Boundaries of \textit{MassSpec}'s application domain for 200 hours of in-transit observations. Using \textit{JWST}, \textit{MassSpec} could yield the mass and atmospheric properties of super-Earth and Earth-sized planets up to the distances shown by the black dashed, and dotted lines, respectively. Similarly, the maximum distances to a mini-Neptune, a super-Earth, and an Earth-sized planets using \textit{EChO}'s observations are represented by the blue solid, dashed, and dotted lines, respectively. The green dotted line refers to the case of an Earth-sized planet observed with a 20-meter space telescope. The grey area shows the stars too bright for \textit{JWST}/\textit{NIRSpec} in the R=1000 mode (J-band magnitude $\lesssim$ 7).}
  \vspace{-1.0cm}
  \label{fig:MassSpec_app_domain_final_in_text}
\end{figure}

The pool of planets accessible to \textit{MassSpec} will extend down to Earth-sized planets with the future observatories. We estimate that with data from \textit{JWST}, \textit{MassSpec} could yield the masses and atmospheric properties of mini-Neptunes, super-Earths,
and Earth-sized planets up to distances of 500 pc, 100 pc, and 50 pc, respectively, for M9V stars and 200 pc, 40 pc and 20 pc for M1V stars or stars with earlier spectral types (Figure\,\ref{fig:MassSpec_app_domain_final_in_text}). For an \textit{EChO}-class mission, the numbers would be 250 pc, 50 pc and 13 pc, and 100 pc, 20 pc and 6 pc, respectively.

\vspace{-0.7cm}

\subsubsection{Earth-sized Planets}

\begin{figure}
    \hspace{-2.0cm}\includegraphics[trim = 00mm 00mm 00mm 00mm,clip,width=19cm,height=!]{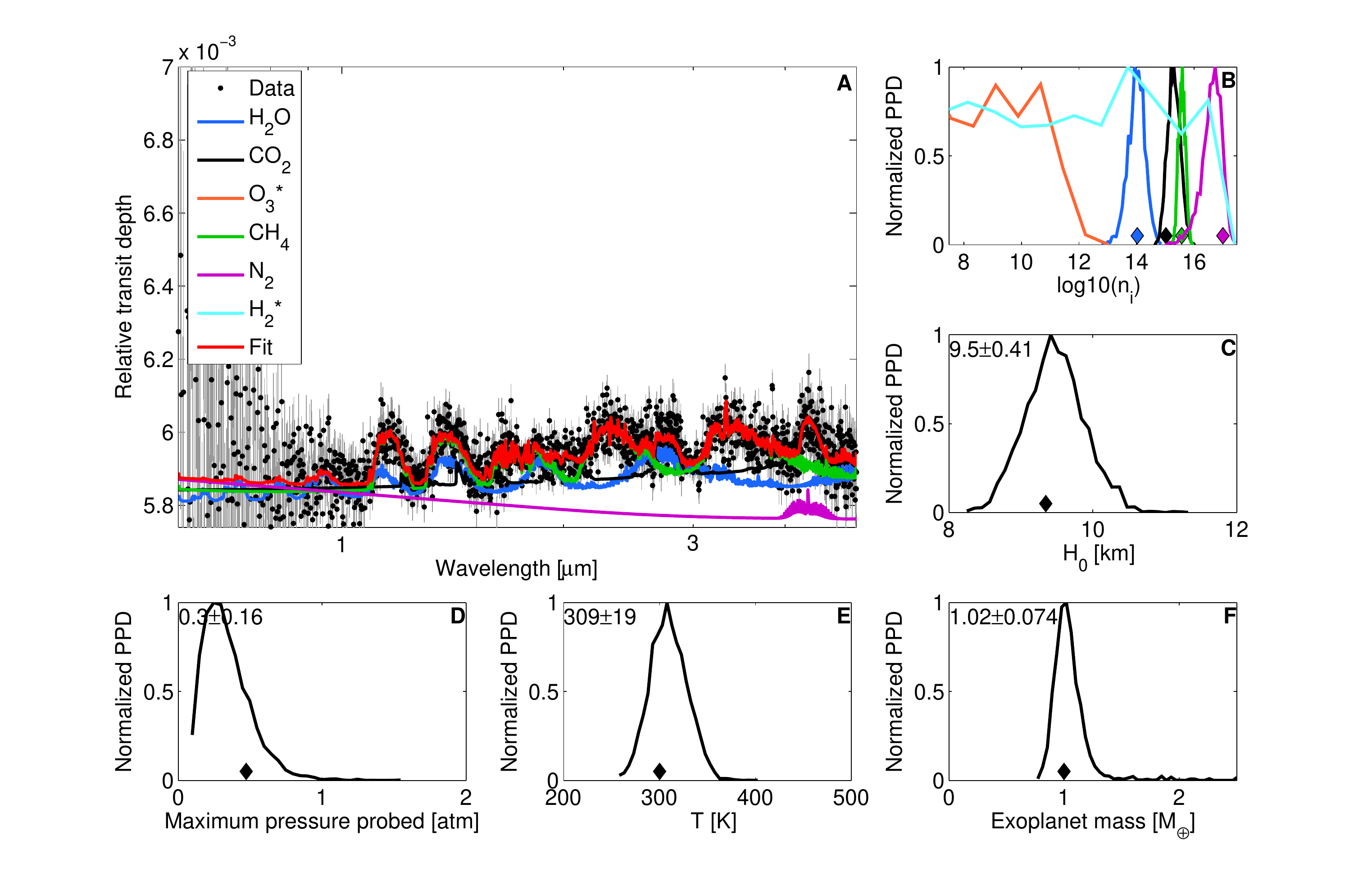}
  \vspace{-0.6cm}
  \caption[\textit{MassSpec}'s application to the synthetic transmission spectrum of an Earth-like exoplanet transiting a M7V star at 15 pc as observed with \textit{JWST}]{\textit{MassSpec}'s application to the synthetic transmission spectrum of an Earth-like exoplanet transiting a M7V star at 15 pc as observed with \textit{JWST} for a total of 200 hrs in-transit. (\textbf{A}) Synthetic data and the best fit together with the individual contributions of the atmospheric species. (\textbf{B}) Normalized posterior probability distribution (PPD) of the atmospheric species number densities at the reference radius. (\textbf{C}) Normalized PPD for the scale height. (\textbf{D}) Normalized PPD for the pressure at deepest atmospheric level probed by transmission spectroscopy. (\textbf{E}) Normalized PPD for the temperature. (\textbf{F}) Normalized PPD for the exoplanet mass. The diamonds indicate the values of atmospheric parameters used to simulate the input spectrum and the asterisks in the panel A legend indicate molecules that are not used to simulate the input spectrum. 
The atmospheric properties (number densities, scale height, and temperature) are retrieved with significance yielding to a mass measurement with a relative uncertainty of $\sim 8\%$.}
  \label{fig:MassSpec_results_in_text_nw}
\end{figure}

\begin{figure}
  \hspace{-2.0cm}\includegraphics[trim = 00mm 00mm 00mm 00mm,clip,width=19cm,height=!]{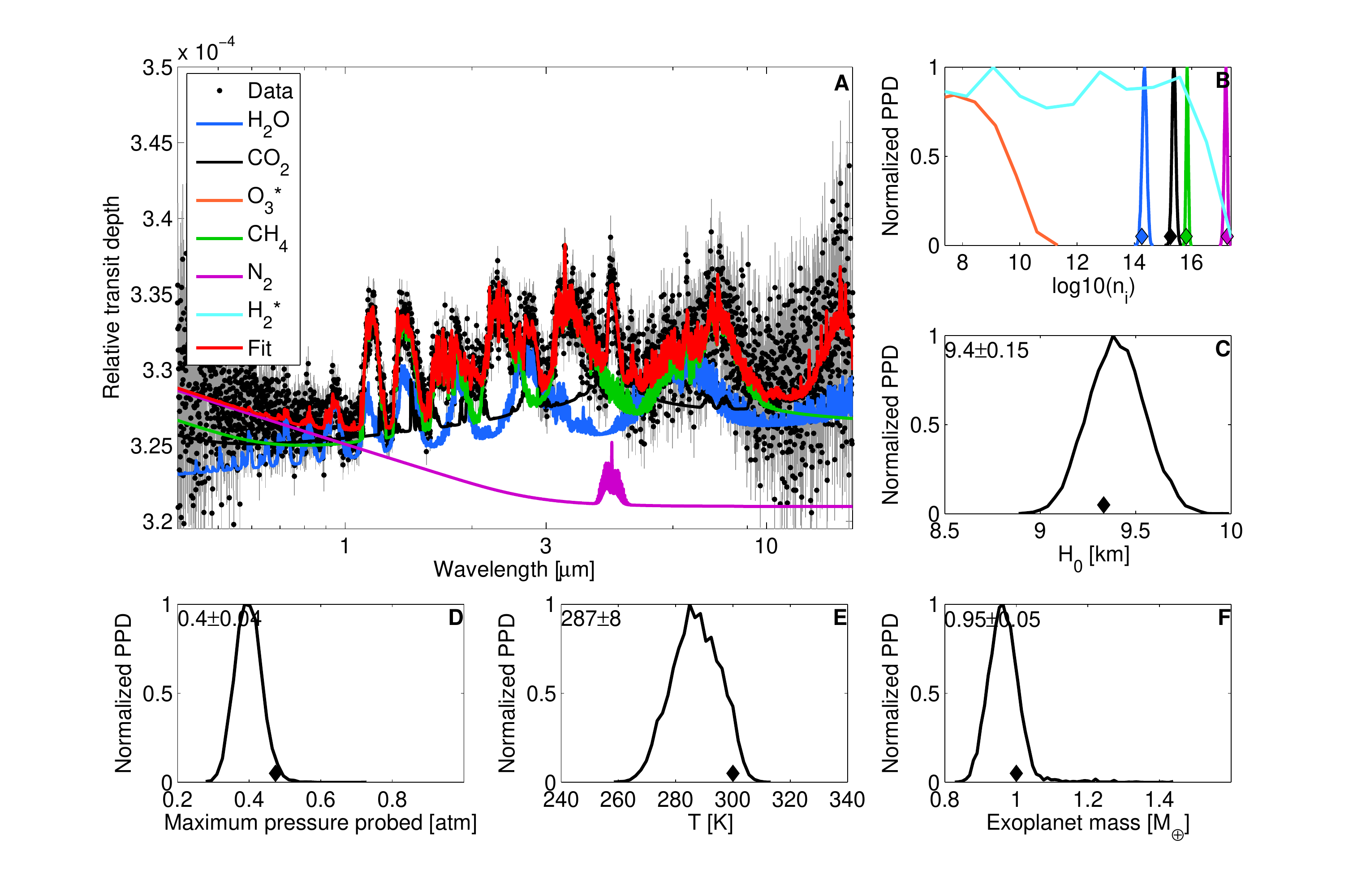}
  \vspace{-0.6cm}
  \caption[\textit{MassSpec}'s application to the synthetic transmission spectrum of an Earth-sized planet with a nitrogen-dominated atmosphere transiting a M1V star at 15 pc as observed with a future-generation 20-meter space telescope.]{\textit{MassSpec}'s application to the synthetic transmission spectrum of an Earth-sized planet with a nitrogen-dominated atmosphere transiting a M1V star at 15 pc as observed with a future-generation 20-meter space telescope for a total of 200 hrs in-transit. The panels show the same quantities as on Figure\,\ref{fig:MassSpec_results_in_text_nw}.  
The atmospheric properties are retrieved with high significance yielding to a mass measurement with a relative uncertainty of $\sim5\%$. Note that the significant observation of the Rayleigh-scattering slope combined with the lack of $H_2-H_2$ CIA feature at 3 microns (Figure\,\ref{fig:mini_neptune_JWST}) yields to the retrieval of nitrogen as the dominant atmospheric species.}
  \label{fig:nitrogen_world_20m}
\end{figure}

We find that \textit{MassSpec} could determine the masses and atmospheric properties of Earth-sized planets using \textit{JWST} or an \textit{EChO}-class mission (Figure\,\ref{fig:MassSpec_app_domain_final_in_text}). For example, even for case of Earth-like planets, \textit{MassSpec} could constrain their masses with a relative uncertainty of $\leq8\%$ using \textit{JWST} (Figure\,\ref{fig:MassSpec_results_in_text_nw})---if these are transiting late M dwarfs within 15 pc of Earth. In the future era of 20-meter space telescopes, sufficiently high quality transmission spectra of Earth-sized planets will be available \citep{Ehrenreich2006,Kaltenegger2009}. By using \textit{MassSpec}, such facilities could yield the mass of Earth-sized planets transiting a M1V star (or stars with earlier spectral types) at 15 pc with a relative uncertainty of $\sim 5\%$ (Figure\,\ref{fig:nitrogen_world_20m}). For M9V stars, such an observatory enable the characterization of Earth-sized planets up to 200 pc and for M1V stars or stars with earlier spectral types, up to 80 pc (Figure\,\ref{fig:MassSpec_app_domain_final_in_text}).

\subsubsection{Super-Earths}

\begin{figure}
    \hspace{-2.0cm}\includegraphics[trim = 00mm 00mm 00mm 00mm,clip,width=19cm,height=!]{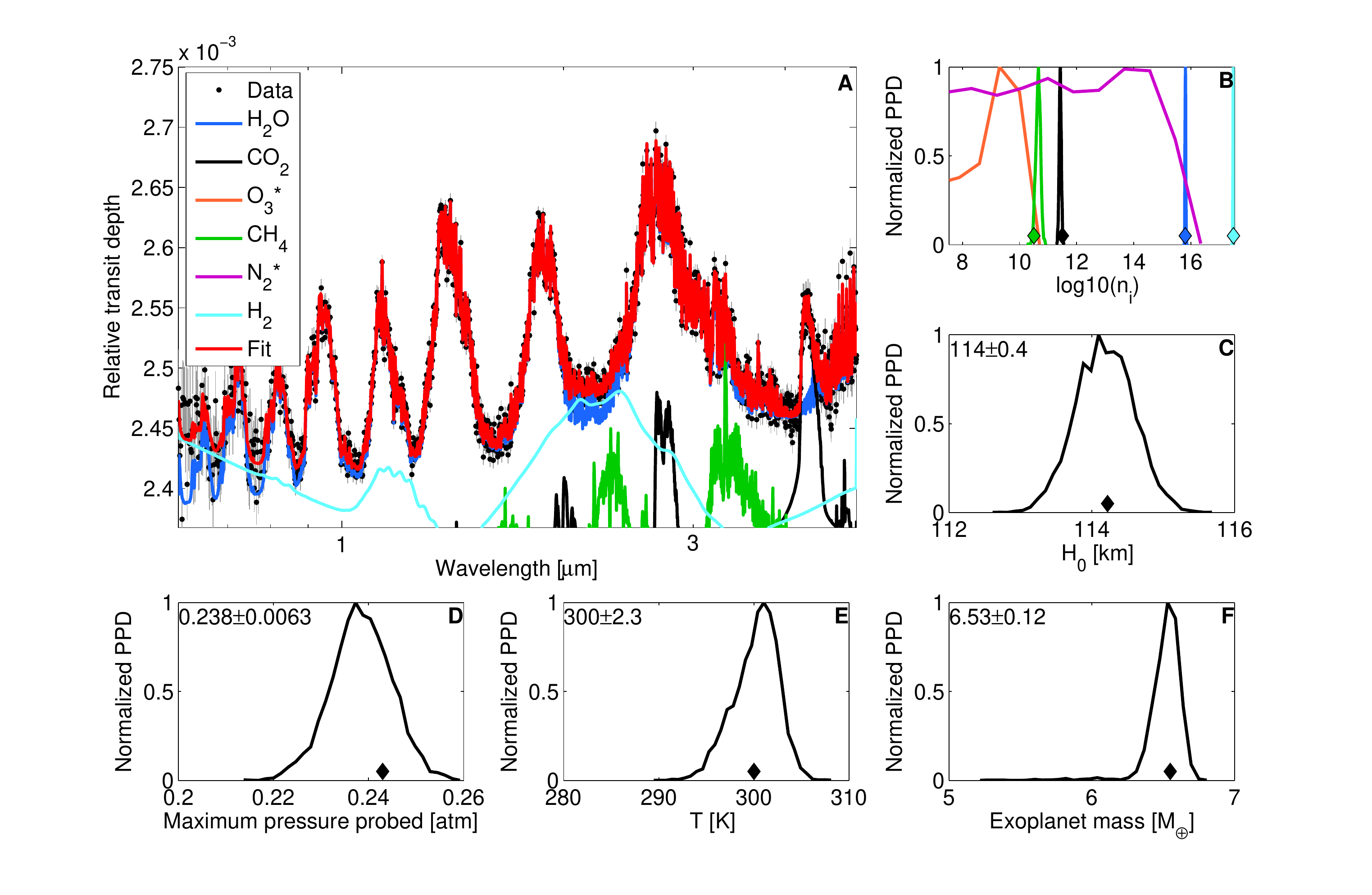}
  \vspace{-0.6cm}
  \caption[\textit{MassSpec}'s application to the synthetic transmission spectrum of a hydrogen-dominated super-Earth transiting a M1V star at 15 pc as observed with \textit{JWST}.]{\textit{MassSpec}'s application to the synthetic transmission spectrum of a hydrogen-dominated super-Earth transiting a M1V star at 15 pc as observed with \textit{JWST} for a total of 200 hrs in-transit. The atmospheric properties (number densities, scale height and temperature) are retrieved with high significance yielding to a mass measurement with a relative uncertainty of $\sim2\%$. Note the significant difference between the number density PPD’s of hydrogen, water, carbon dioxide, and methane and those of ozone and nitrogen (\textbf{B}). The latter two gases were not part of the synthetic atmosphere. Ozone and nitrogen are not detected, because no constraints on their mixing ratios can be made. Hydrogen-dominated planets are targets that are particularly favorable for \textit{MassSpec} because their extended atmosphere leads to high-SNR transmission spectra.}
  \label{fig:mini_neptune_JWST}
\end{figure}

\begin{figure}
    \hspace{-2.0cm}\includegraphics[trim = 00mm 00mm 00mm 00mm,clip,width=19cm,height=!]{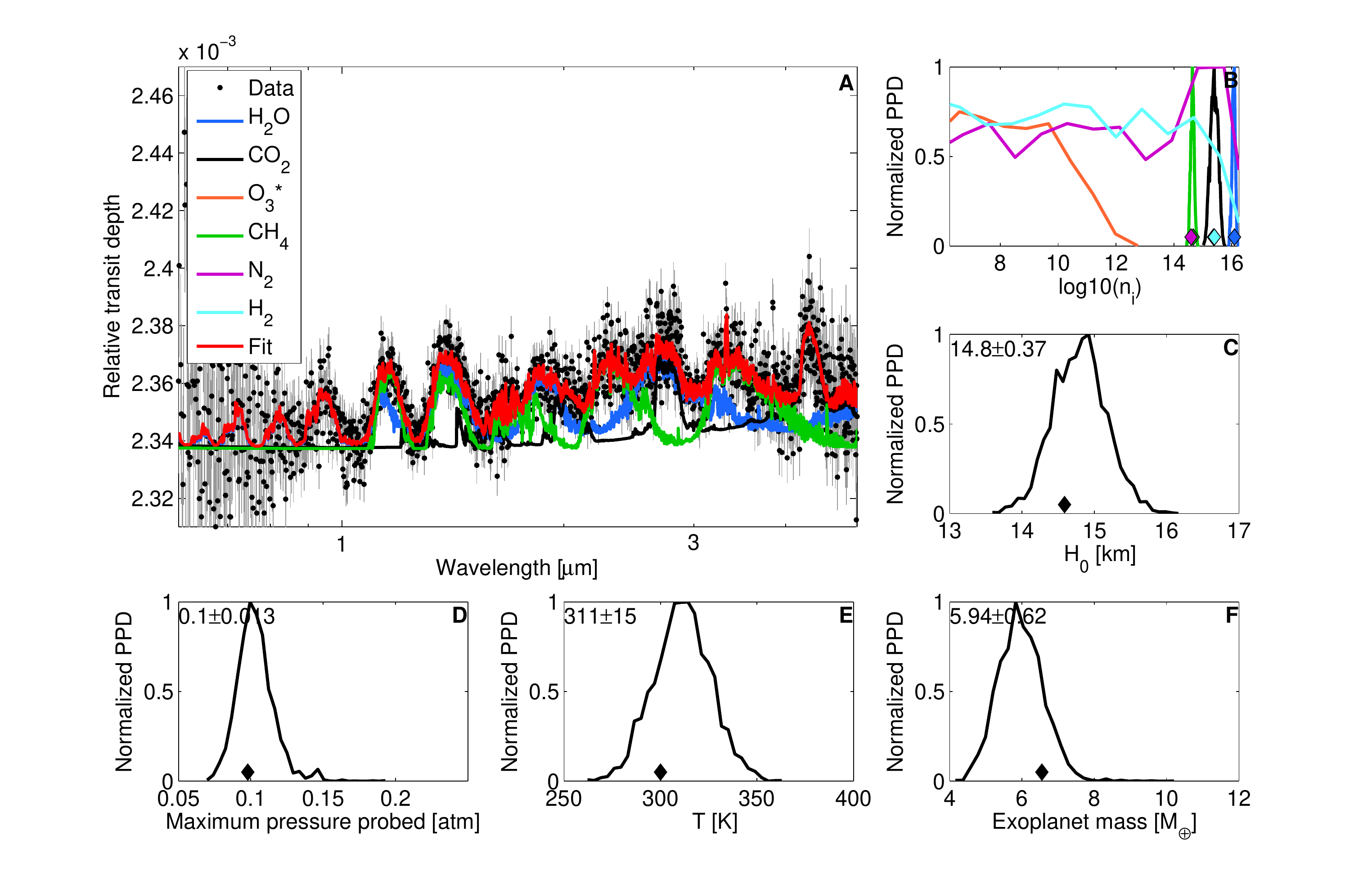}
  \vspace{-0.6cm}
  \caption[\textit{MassSpec}'s application to the synthetic transmission spectrum of a water-dominated super-Earth transiting a M1V star at 15 pc as observed with \textit{JWST}.]{\textit{MassSpec}'s application to the synthetic transmission spectrum of a water-dominated super-Earth transiting a M1V star at 15 pc as observed with \textit{JWST} for a total of 200 hrs in-transit. The panels show the same quantities as on Figure\,\ref{fig:MassSpec_results_in_text_nw}.  
The atmospheric properties (number densities, scale height, and temperature) are retrieved with significance yielding to a mass measurement with a relative uncertainty of $\sim10\%$.}
  \label{fig:MassSpec_results_in_text_ww}
\end{figure}

\begin{figure}
    \hspace{-2.0cm}\includegraphics[trim = 00mm 00mm 00mm 00mm,clip,width=19cm,height=!]{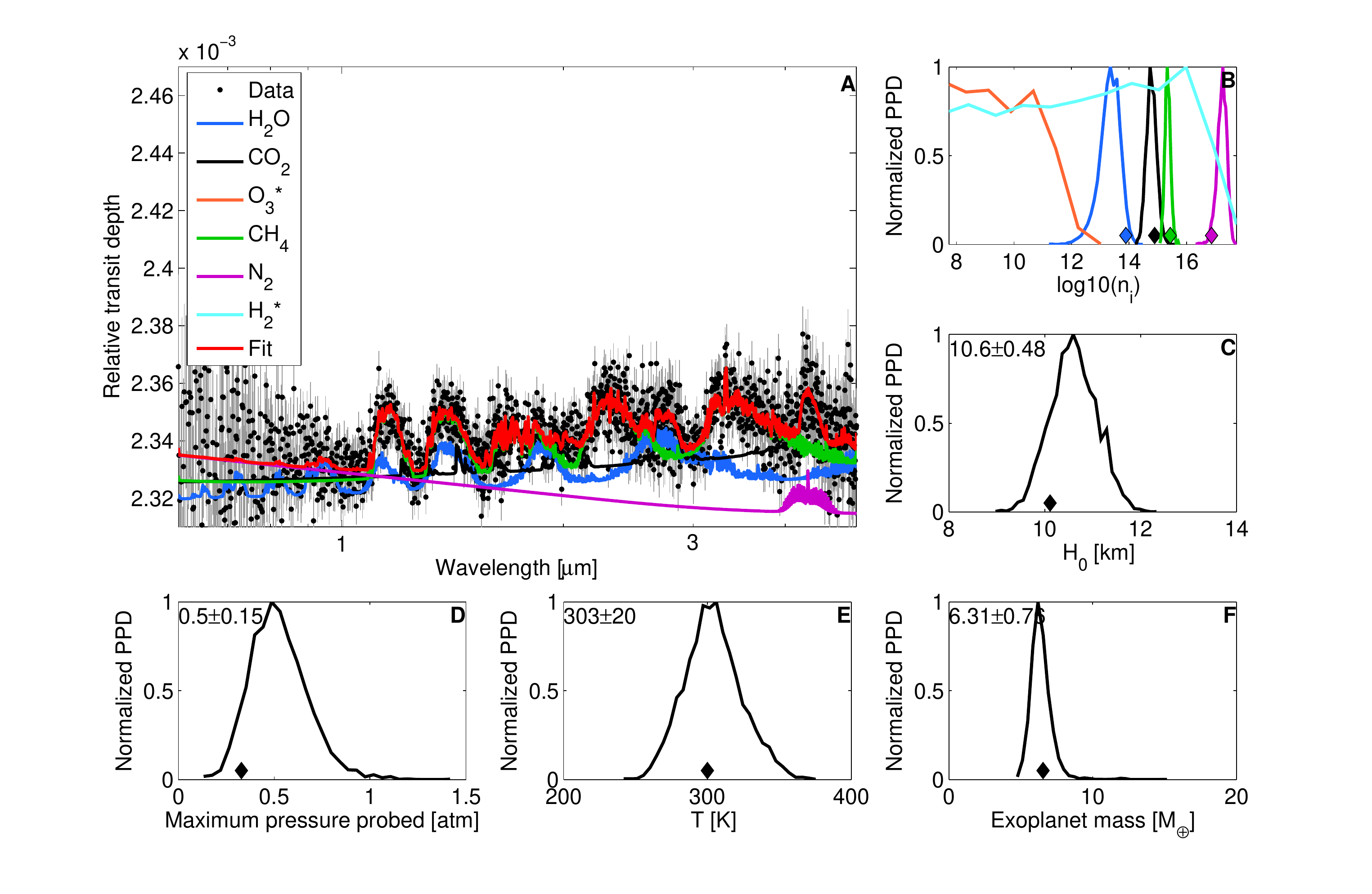}
  \vspace{-0.6cm}
  \caption[\textit{MassSpec}'s application to the synthetic transmission spectrum of a super-Earth with a nitrogen-dominated atmosphere transiting a M1V star at 15 pc as observed with \textit{JWST}.]{\textit{MassSpec}'s application to the synthetic transmission spectrum of a super-Earth with a nitrogen-dominated atmosphere transiting a M1V star at 15 pc as observed with \textit{JWST} for a total of 200 hrs in-transit. The panels show the same quantities as on Figure\,\ref{fig:MassSpec_results_in_text_nw}. 
The atmospheric properties are retrieved with high significance yielding to a mass measurement with a relative uncertainty of $\sim15\%$.}
  \label{fig:nitrogen_world_JWST}
\end{figure}

\begin{figure}
    \hspace{-2.0cm}\includegraphics[trim = 00mm 00mm 00mm 00mm,clip,width=19cm,height=!]{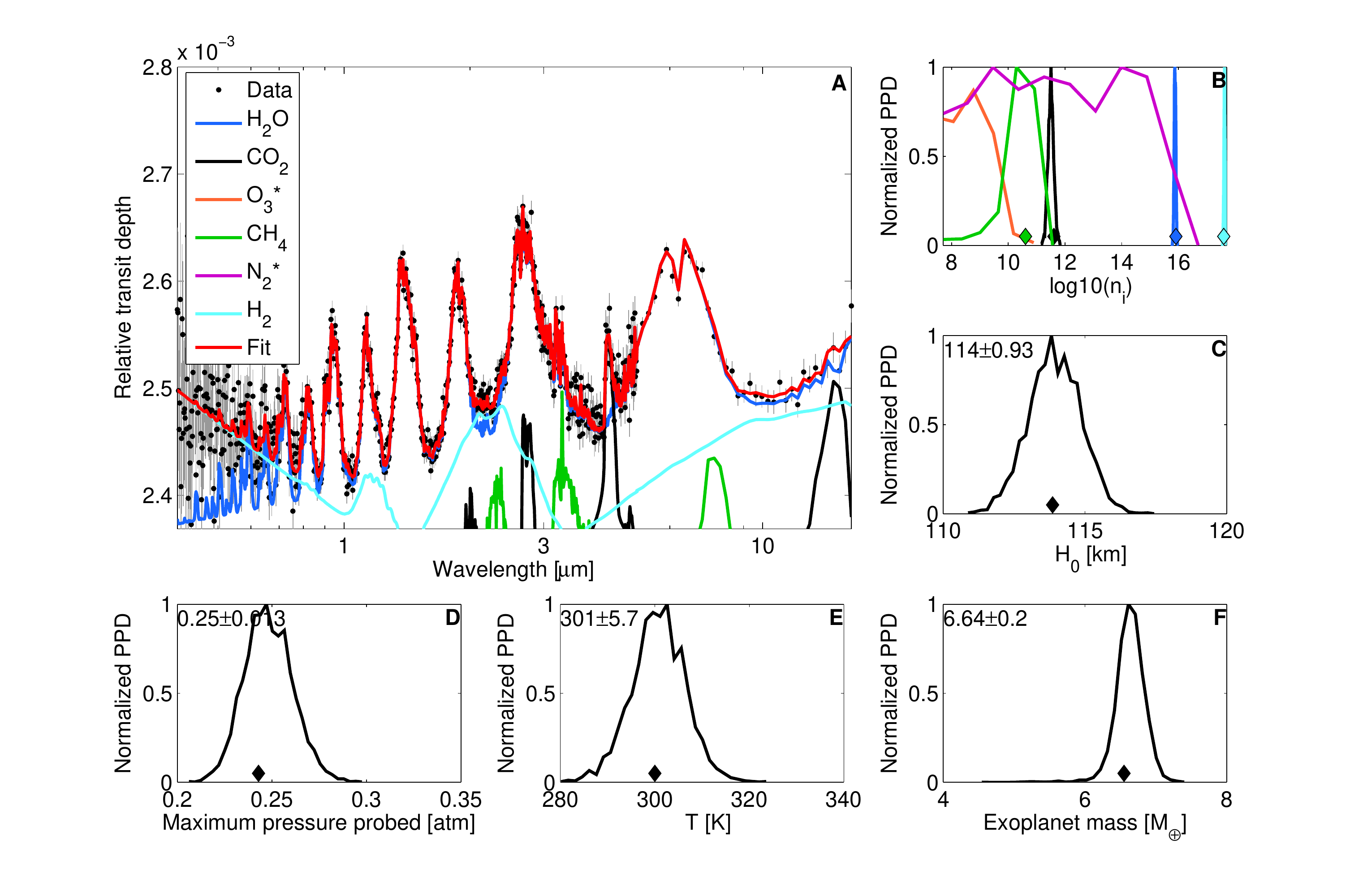}
  \vspace{-0.6cm}
  \caption[\textit{MassSpec}'s application to the synthetic transmission spectrum of a super-Earth with a hydrogen-dominated atmosphere transiting a M1V star at 15 pc as observed with \textit{EChO}.]{\textit{MassSpec}'s application to the synthetic transmission spectrum of a super-Earth with a hydrogen-dominated atmosphere transiting a M1V star at 15 pc as observed with \textit{EChO} for a total of 200 hrs in-transit. The panels show the same quantities as on Figure\,\ref{fig:MassSpec_results_in_text_nw}. 
The atmospheric properties are retrieved with high significance yielding to a mass measurement with a relative uncertainty of $\sim3\%$. Note that \textit{EChO}'s capabilities for mass measurements are enhanced by its large spectral coverage that yields the Rayleigh-scattering slope, which is particularly valuable for mass and atmospheric retrieval.}
  \label{fig:mini_neptune_EChO}
\end{figure}

\begin{figure}
    \hspace{-2.0cm}\includegraphics[trim = 00mm 00mm 00mm 00mm,clip,width=19cm,height=!]{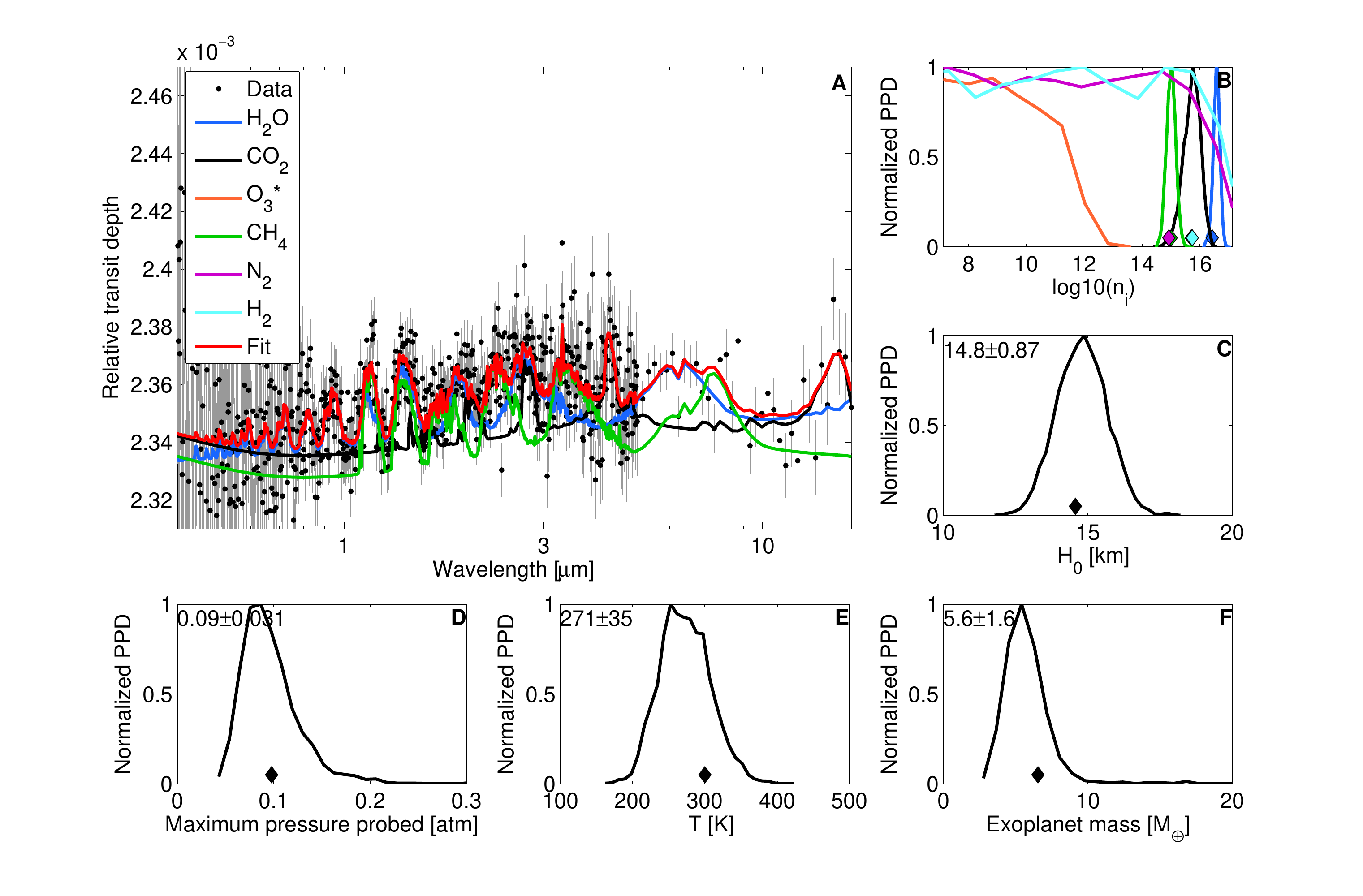}
  \vspace{-0.6cm}
  \caption[\textit{MassSpec}'s application to the synthetic transmission spectrum of a super-Earth with a water-dominated atmosphere transiting a M1V star at 15 pc as observed with \textit{EChO}.]{\textit{MassSpec}'s application to the synthetic transmission spectrum of a super-Earth with a water-dominated atmosphere transiting a M1V star at 15 pc as observed with \textit{EChO} for a total of 200 hrs in-transit. The panels show the same quantities as on Figure\,\ref{fig:MassSpec_results_in_text_nw}.  
The atmospheric properties are retrieved with high significance yielding to a mass measurement with a relative uncertainty of $\sim25\%$.}
  \label{fig:water_world_EChO}
\end{figure}

\begin{figure}
    \hspace{-2.0cm}\includegraphics[trim = 00mm 00mm 00mm 00mm,clip,width=19cm,height=!]{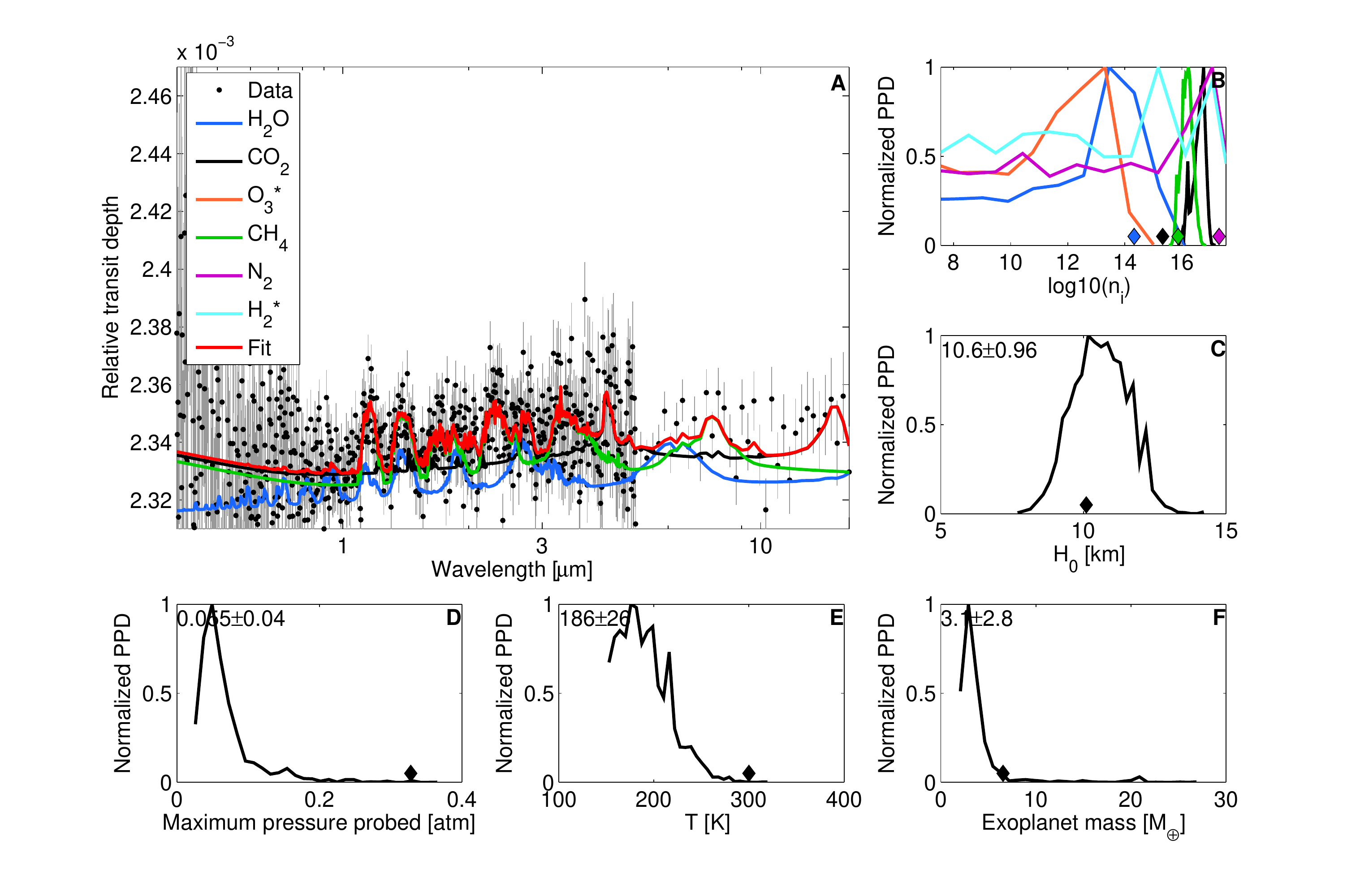}
  \vspace{-0.6cm}
  \caption[\textit{MassSpec}'s application to the synthetic transmission spectrum of a super-Earth with a nitrogen-dominated atmosphere transiting a M1V star at 15 pc as observed with \textit{EChO}.]{\textit{MassSpec}'s application to the synthetic transmission spectrum of a super-Earth with a nitrogen-dominated atmosphere transiting a M1V star at 15 pc as observed with \textit{EChO} for a total of 200 hrs in-transit. The panels show the same quantities as on Figure\,\ref{fig:MassSpec_results_in_text_nw}. The planet mass is not retrieved because the data quality does not yield the atmosphere composition and temperature---although the scale height and the signatures of water, methane, and carbon dioxide are retrieved.}
  \label{fig:nitrogen_world_EChO}
\end{figure}

We find that \textit{MassSpec} would yield the mass measurements of super-Earths transiting an M1V star at 15 pc with a relative uncertainty of $\sim2\%$($\sim10\%$)[$\sim15\%$] for \\hydrogen(water)[nitrogen]-dominated atmospheres (Figures\,\ref{fig:mini_neptune_JWST}, \ref{fig:MassSpec_results_in_text_ww}, and \ref{fig:nitrogen_world_JWST}), if applied to 200 hrs of such a target. The larger significance of the mass measurements obtained for hydrogen-dominated super-Earths results from higher SNR of their transmission spectra (Equation\,\ref{eq:SNRt_general_scalinglaw})---the small mean molecular masses of their atmospheres reduce their extent. For the same super-Earths with hydrogen(water)-dominated atmospheres, \textit{EChO}'s data would yield mass measurements with a relative uncertainty of $\sim3\%$($\sim25\%$) (Figures\,\ref{fig:mini_neptune_EChO} and \ref{fig:water_world_EChO}), respectively---for a nitrogen world in the same configuration it will not be possible to constrain its mass (Figure\,\ref{fig:nitrogen_world_EChO}).

\subsubsection{Giant Planets}

\textit{JWST}'s and \textit{EChO} 's observations of planets larger than super-Earths could provide high-significance mass and atmospheric constraints with a limited number of transits. Using the scaling law (Equation\,\ref{eq:SNRt_general_scalinglaw}), we estimate that for giant planets within 100 pc half a dozen of transits would secure mass measurements with a relative uncertainty $\leq 5\%$. An \textit{EChO}-class mission would be particularly useful to characterize giant planets as its wide spectral coverage would allow to measure their Rayleigh-scattering slope at short wavelengths. \textit{MassSpec}'s applications to giant planets will be particularly important for giant planets whose star's activity prevents a mass measurement with RV \citep[e.g., the hottest known planet, WASP-33b][]{Collier2010}. In addition, independent high-significance mass measurements of targets accessible by RV will also be beneficial to compare the capabilities of both techniques and to assess their future complementarity.

\subsection{Discussion}
\label{app:discussion_massspec}

\subsubsection{How Many (Habitable) Planets Are Within Reach?}

We present in Table\,\ref{tab:howmany} estimates of the number of exoplanets that would be characterizable with \textit{MassSpec}, or any other atmospheric retrieval methods based on transmission spectroscopy. These estimates assume a stellar density of 0.14 stars per pc$^3$ and consider that $80\%$ of those are M dwarfs. We derive the planet's transit probability ($\sim R_*/a$) assuming them to be at the center of their host's habitable zone $\left(a = (R_*/R_\odot)^2 (T_*/T_\odot)^4\right)$. We use a planet occurrence rate of 0.15 for Earth-sized planets in their host's habitable zone and 1 for other planet types \citep{Dressing2013}. The number within brackets refer to the estimate assuming the most optimistic occurrence rate of HZ planets from Zsom, in prep.

\begin{figure}[!ht]

\vspace{-0.0cm}\centering\includegraphics[trim = 00mm 00mm 00mm 00mm,clip,width=15cm,height=!]{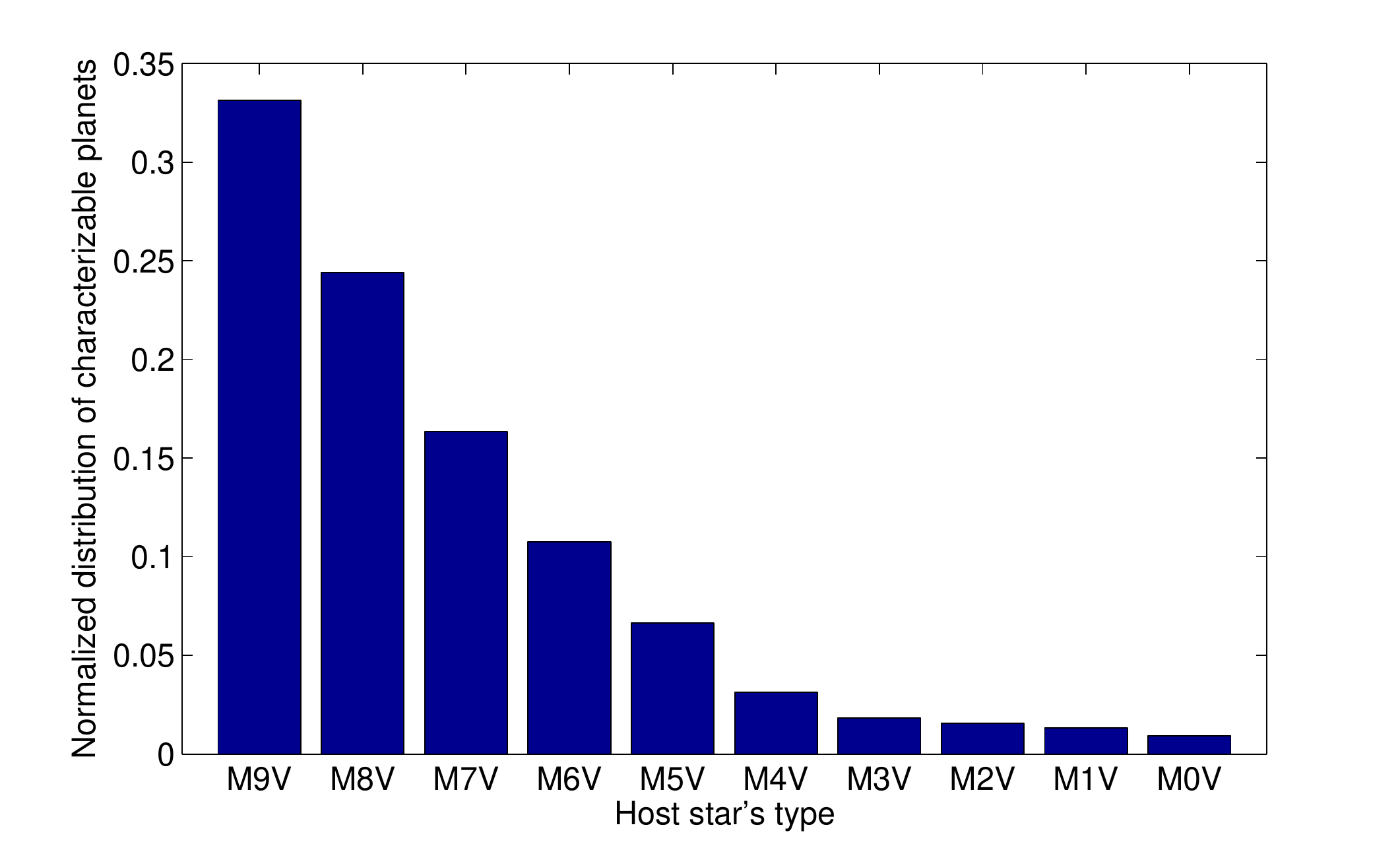}
  \vspace{-0.0cm}
  \caption{Estimated fraction of exoplanets characterizable via transmission spectroscopy as a function of the host star's type.}
  \vspace{-.4cm}
  \label{fig:fraction_of_planet_characterizable}
\end{figure}

We note that about $75\%$ of the planets to be accessible for in-depth characterization orbit a host with a spectral type later than M5V. We develop this point in Figure\,\ref{fig:fraction_of_planet_characterizable} and show the estimated fraction of characterizable planets as a function of the host star's type. As it was pointed out in Section\,\ref{app:scaling_laws_snr}, late M dwarfs are favorable for transmission spectroscopy because of their large ratio of radiance over projected area (Figure\,\ref{fig:key_ratio_fixed_planet}). For this reason, atmospheric retrieval methods are applicable to larger distances for late M dwarfs than for stars with earlier spectral types (Figure\,\ref{fig:MassSpec_app_domain_final_in_text}). Hence, the number of favorable hosts is considerably larger as it scales with the cube of the distance. In addition, the transit probability is larger for stars with later spectral types. 

\begin{table}[!h]
\caption{Estimated number of exoplanets within reach for atmosphere characterization. \label{tab:howmany}}
	\centering
	\setlength{\extrarowheight}{3pt}
	\setlength{\tabcolsep}{8pt}

	\hspace{-0cm}\normalsize{\begin{tabular}{c | c c c}
	
	\hline\hline
	\multirow{2}{*}{\textbf{Observatories}} & \multicolumn{3}{c}{\textbf{Exoplanet types}}\\
	  & Earth-sized in Habitable-Zone & Earth-sized & Super-Earth \\
	\hline

			\textit{EChO}	& 5 (10) & 40 & 600 
			  \\
			\textit{JWST} & 200 (400) & 2000  & 30000
			 \\
			20-m & 15000 (30000) & 130000 & 2$\times10^6$

	\end{tabular}}
	
\end{table}	

This new insight for the future exoplanet characterization highlight that:

\textbf{(i)} If Earth-sized planets are found around the habitable zone of M dwarfs before 2018, one could constrain their atmospheric properties and mass with future facilities such as \textit{JWST}. In other words, \textbf{the first habitable distant worlds could be identified within the next decade}. 

\textbf{(ii)} If found by 2018, there will be too many favorable planets to be characterized with the available facilities. As an example, the sole  characterization of each of the $\sim 200$ possibly-habitable Earth-sized planets would require a continuous use of \textit{JWST} over its lifetime (5 years). Not only will JWST have other scientific goals for the field of exoplanetology, but it is also not dedicated to planetary science. Hence, it will be key for the community to pre-select targets of interest---which requires to discover these well in advance of \textit{JWST}'s launch.

\subsubsection{(i) Detecting the Earth-Sized Planets to be Identify Habitable in the Next Decade}

\textit{JWST}'s optical performance could yield the characterization of about 200 Earth-sized planets in the habitable zone of their host stars, primarily late M dwarfs. All of them are yet to be detected. Their early detection (prior to 2018) is key to ensure that \textit{JWST} would monitor enough of their transits to secure a sufficient SNR. In practice, half a dozen Earth-sized planets in the habitable zone of their host stars could be properly characterized over \textit{JWST}'s lifetime. To date, one mission is dedicated to this aim: \textit{SPECULOOS} (\textit{Search for Habitable Planets Eclipsing Ultra-cool Stars}). \textit{SPECULOOS} is a European Research Council mission that will begin observing the coolest M dwarfs in 2016. The mass of the companions found will not be constrained by RV because of the faintness of their host stars. However, \textit{MassSpec}'s application to their \textit{JWST} spectra would yield both their masses and atmospheric properties (Figure\,\ref{fig:MassSpec_results_in_text_nw}). Hence, \textit{MassSpec} is likely to play a pivotal role in the assessment of their potential habitability.

\subsubsection{(ii) Optimizing the Scientific Return of \textit{JWST}}

Time prioritization will be key for the future of exoplanetology. The number of targets favorable for characterization and the time it would require will be too large to allow for a systematic characterization due to our limited capabilities. That is why we suggested to aim for community efforts (1) to discover targets of interest well in advance of \textit{JWST}'s launch and (2) pre-select the targets for which \textit{JWST}'s time will be dedicated over its life-time. To help mitigate the drawback of dedicating \textit{JWST}'s time to a limited amount of targets, we suggest to consider more modest missions such as \textit{EChO}. The smaller
aperture of \textit{EChO} would enable it to observe brighter stars (i.e., early-type and close-by stars), hence \textit{EChO}'s and \textit{JWST}'s time could be respectively prioritized on super-Earths and Earth-sized planets for M9V stars closer than 25 pc and for M1V stars (or stars with earlier spectral type) closer than 10 pc (Figure\,\ref{fig:MassSpec_app_domain_final_in_text}). Similarly, \textit{EChO}'s and \textit{JWST}'s time could be respectively prioritized on giant planets and super-Earths for M9V stars closer than 125 pc and for M1V stars (or stars with earlier spectral type) closer than 50 pc. \textit{EChO} would be particularly useful to determine the mass---and atmospheric properties---of giant planets because its wide spectral coverage would allow to measure their Rayleigh-scattering slope at short wavelengths.

The same argument concerning the synergy between \textit{JWST} and an \textit{EChO}-class mission in terms of mapping capabilities can be found in Section\,\ref{sec:map_echo_jwst}.

\subsubsection{Clouds Will Not Overshadow \textit{MassSpec}}

\begin{figure}[!p]
    \hspace{-2.0cm}\includegraphics[trim = 00mm 00mm 00mm 00mm,clip,width=19cm,height=!]{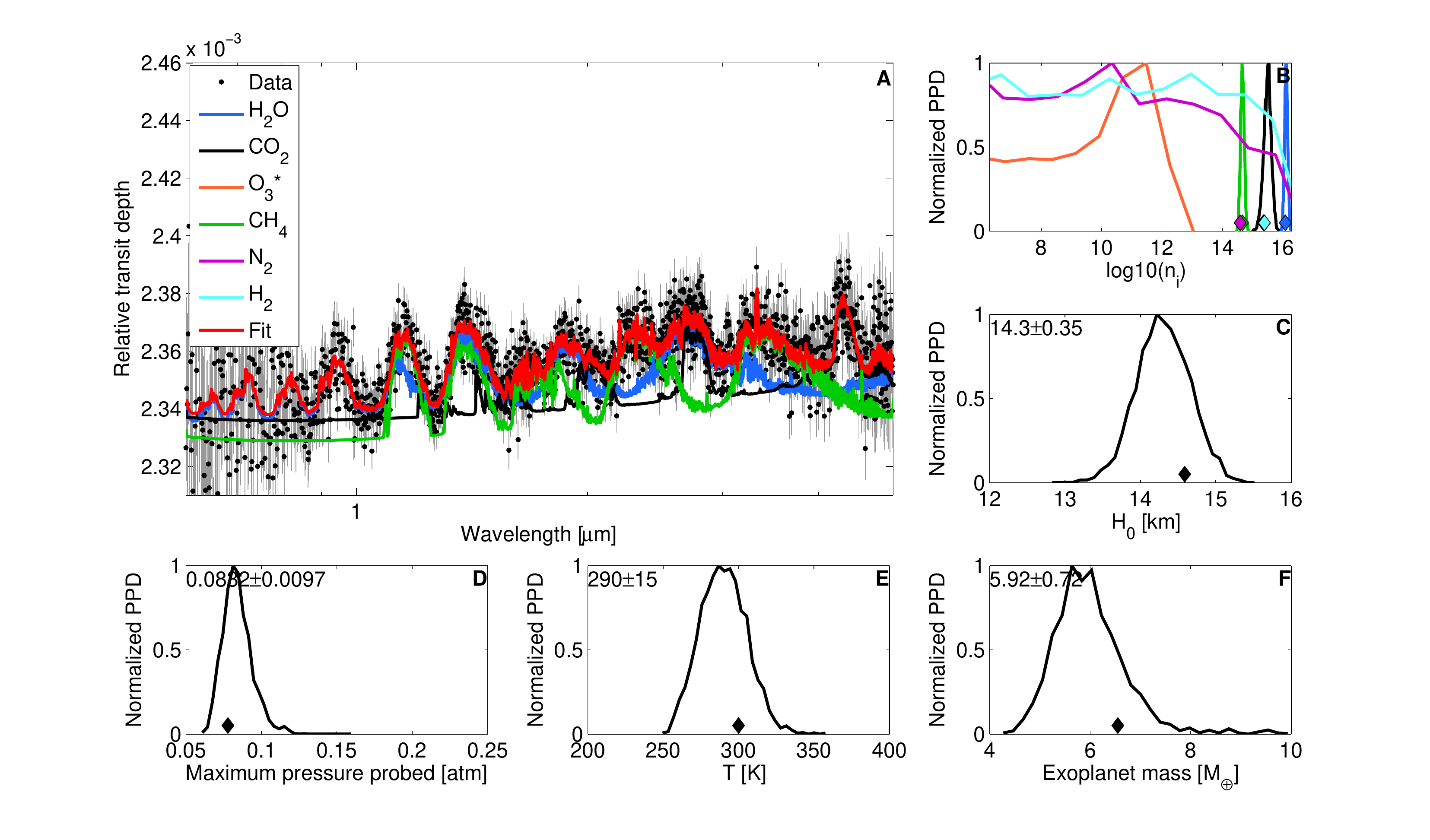}
  \vspace{-0.6cm}
  \caption[Effects of a cloud deck at 100 mbar on \textit{MassSpec}'s retrieval.]{\textit{MassSpec}'s application to the synthetic transmission spectrum of a super-Earth with a water-dominated atmosphere presenting a cloud deck at 100 mbar (such as Venus) transiting a M1V star at 15 pc as observed with \textit{JWST} for a total of 200 hrs in-transit. The panels show the same quantities as on Figure\,\ref{fig:MassSpec_results_in_text_nw}.
The atmospheric properties are retrieved with high significance yielding to a mass measurement with a relative uncertainty of $\sim10\%$. \textit{MassSpec}'s capabilities are not affected by the 100-mbar cloud deck because transmission spectroscopy does not probe deeper than 10 mbar (Figure\,\ref{fig:MassSpec_results_in_text_ww}, panel D).}
  \label{fig:water_world_JWST_C1}
\end{figure}

\begin{figure}
    \hspace{-2.0cm}\includegraphics[trim = 00mm 00mm 00mm 00mm,clip,width=19cm,height=!]{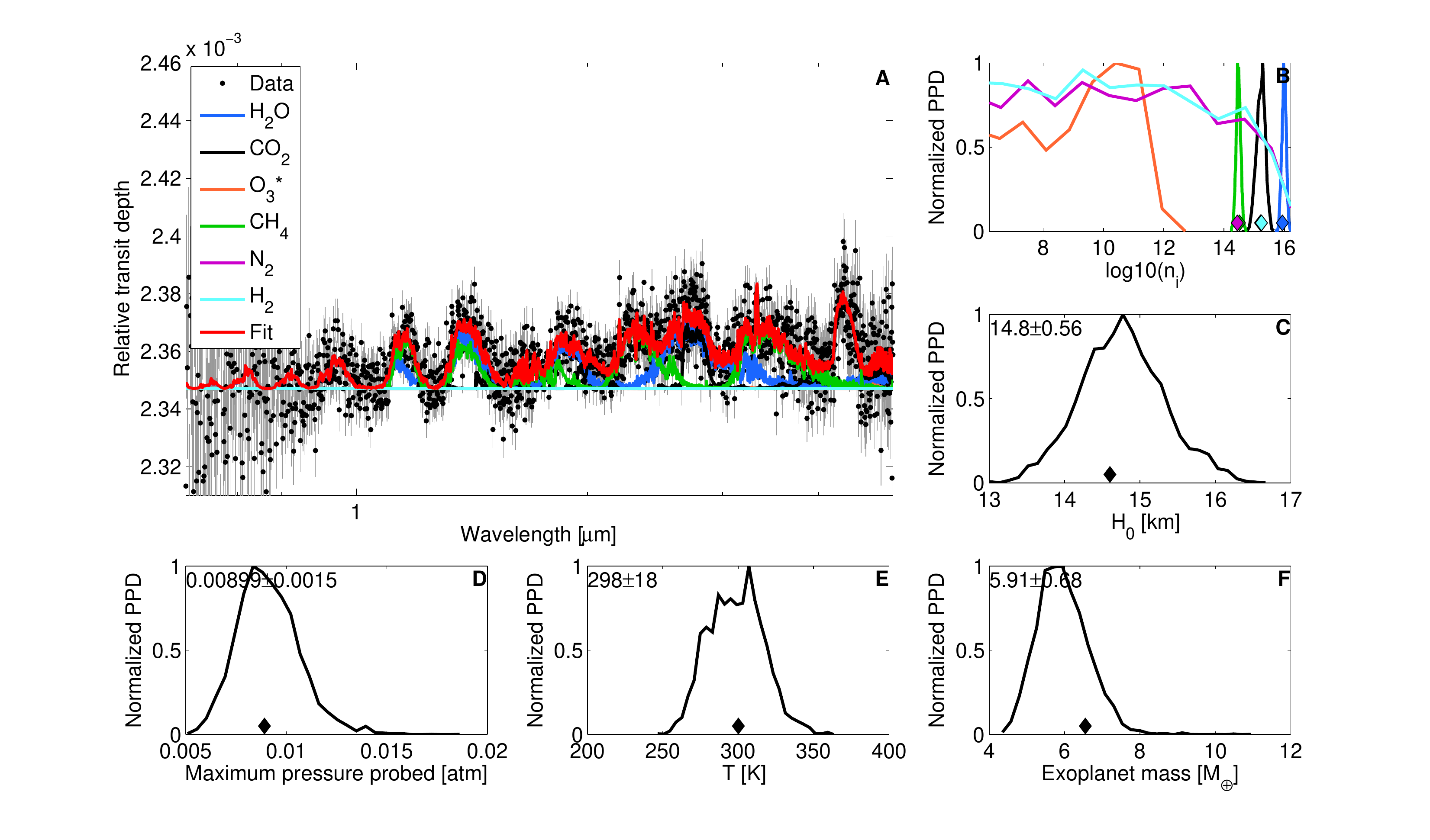}
  \vspace{-0.6cm}
  \caption[Effects of a cloud deck at 10 mbar on \textit{MassSpec}'s retrieval.]{\textit{MassSpec}'s application to the synthetic transmission spectrum of a super-Earth with a water-dominated atmosphere presenting a cloud deck at 10 mbar transiting a M1V star at 15 pc as observed with \textit{JWST} for a total of 200 hrs in-transit. The panels show the same quantities as on Figure\,\ref{fig:MassSpec_results_in_text_nw}. The atmospheric properties are retrieved with high significance yielding to a mass measurement with a relative uncertainty of $\sim10\%$. \textit{MassSpec}'s capabilities are marginally affected by a 10-mbar cloud deck because a limited fraction of the spectral bins probe deeper than 10 mbar.}
  \label{fig:water_world_JWST_C2}
\end{figure}

\begin{figure}
    \hspace{-2.0cm}\includegraphics[trim = 00mm 00mm 00mm 00mm,clip,width=19cm,height=!]{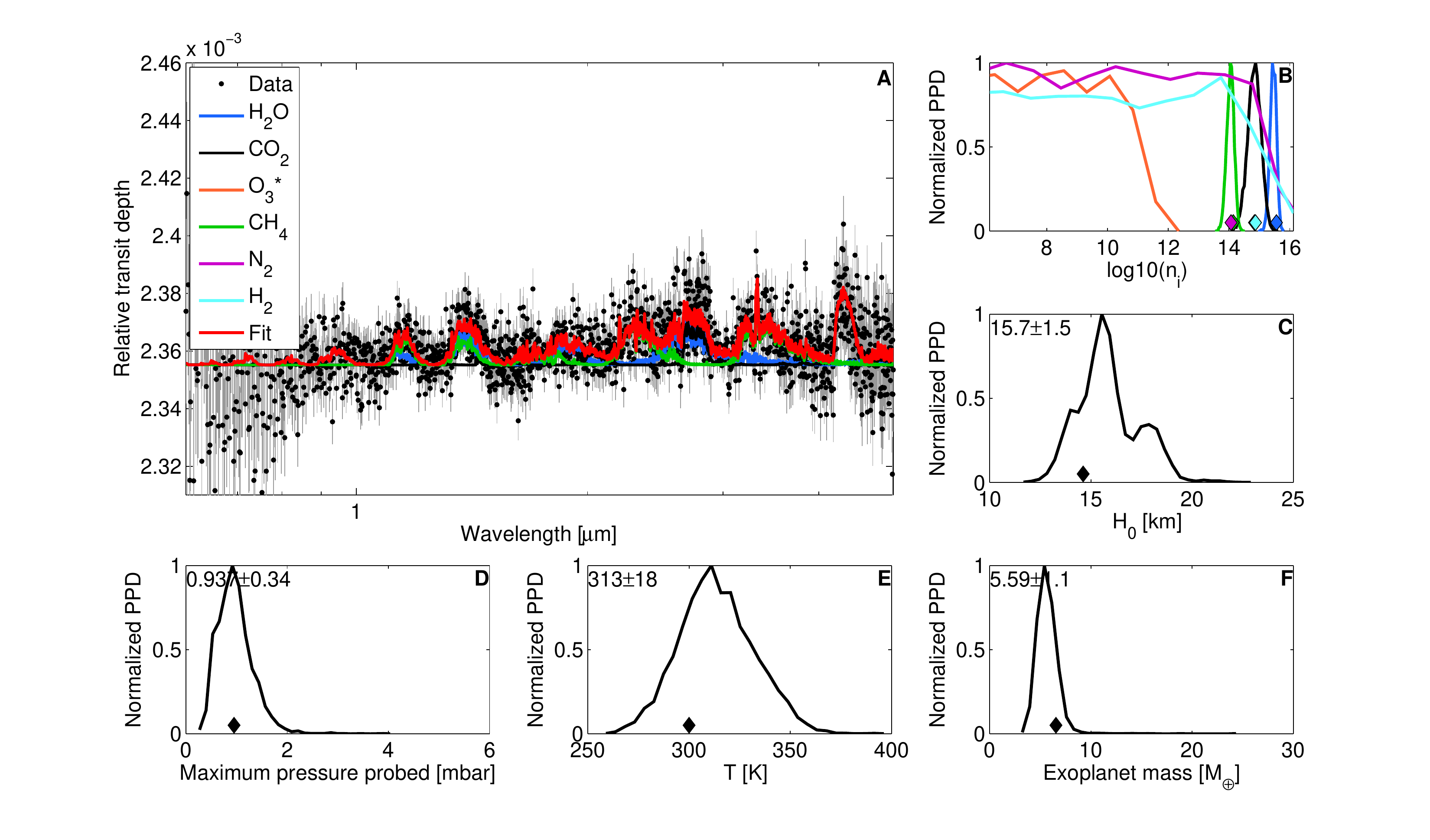}
  \vspace{-0.6cm}
  \caption[Effects of a cloud deck at 1mbar on \textit{MassSpec}'s retrieval.]{\textit{MassSpec}'s application to the synthetic transmission spectrum of a super-Earth with a water-dominated atmosphere presenting a cloud deck at 1 mbar \citep[lowest pressure level where clouds are expected][]{Howe2012} transiting a M1V star at 15 pc as observed with \textit{JWST} for a total of 200 hrs in-transit. The panels show the same quantities as on Figure\,\ref{fig:MassSpec_results_in_text_nw}. Despite the presence of high and thick clouds, the atmospheric properties are retrieved with sufficient significance to yield a mass measurement with a relative uncertainty of $\sim20\%$ (i.e., twice larger than in the cloud-free scenario).}
  \label{fig:water_world_JWST_C3}
\end{figure}

	Clouds are known to be present in exoplanet atmospheres \citep{Demory2013} and to affect transmission spectra because atmospheric layers below the cloud deck are not probed \citep{Benneke2013,Barstow2013b}. However, \textit{MassSpec} is not rendered ineffectual by clouds because clouds are deeper than the lowest pressure probed by transmission spectroscopy. In other words, there will always be atmospheric information available from transmission spectroscopy, in theory. In practice, spectral features would be detectable in the presence of clouds but with a reduced significance (compare panels A of Figures\,\ref{fig:water_world_JWST_C1} to\,\ref{fig:water_world_JWST_C3}): the higher the cloud deck, the larger is the uncertainty on the mass estimate due to the reduced amount of atmospheric information available. We show the effect of cloud decks at 100, 10, and 1 mbar in Figures\,\ref{fig:water_world_JWST_C1}, \ref{fig:water_world_JWST_C2}, and \ref{fig:water_world_JWST_C3}, respectively. Note that 100 mbar corresponds to Earth's and Venus' clouds's pressure level and 1 mbar is the lowest pressure where thick clouds are expected \citep{Howe2012}. Cloud decks at 100 and 10 mbar affect marginally \textit{MassSpec}'s capabilities because a limited fraction of the spectral bins probe as deep as than 10 mbar in transmission. However a 1-mbar cloud deck affects significantly atmospheric retrieval results, increasing the uncertainty on the mass estimate by a factor of $\sim2$ over the one derived in the cloud free scenario. The increased uncertainty in the mass estimate results mainly from an increased uncertainty on the scale height and the mean molecular mass estimates. The reason is that (1) the precision on scale-height estimates are reduced because the signatures of atmospheric species are truncated and (2) additional atmospheric scenarios are now possible because numerous combinations of molecular signatures could be masked by a high-cloud deck. Disentangling between such scenarios requires a sufficient spectral resolution.

\subsubsection{Number Densities vs Mixing Ratios in Atmospheric Retrieval Method}

\begin{figure*}[!ht]
 \centering
  \begin{center}
    \includegraphics[trim = 00mm 00mm 00mm 00mm,clip,width=14cm,height=!]{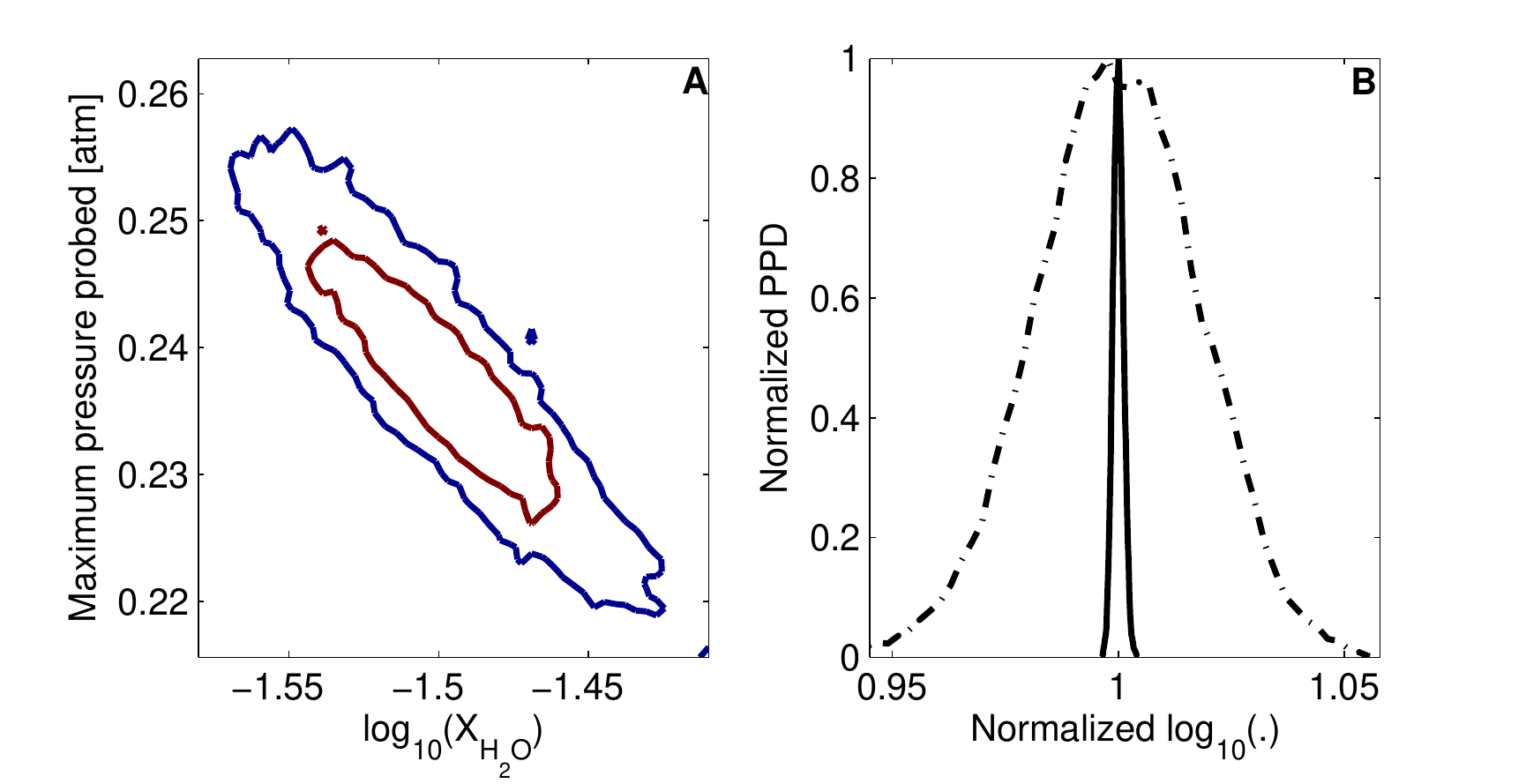}
  \end{center}
  \vspace{-0.7cm}
  \caption[On the use of species' mixing ratios in atmospheric retrieval method.]{The use of species' mixing ratios in atmospheric retrieval method. (\textbf{A}) Marginal posterior probability distribution (69$\%$ and 95$\%$ confidence intervals, respectively in red and blue) of maximum pressure probed and H$_2$O mixing ratio shows the correlation between both parameters. The correlation shown between the pressure and a species mixing ratio translates the fact that the key parameter of a planet's transmission spectrum are the number densities, not the mixing ratios. Therefore, the uncertainty on mixing ratios combines the uncertainty on the pressure and the number densities. (\textbf{B}) Water's mixing ratio PPD (dot-dash line) and water's number density PPD (solid line), shifted along the x-axis for comparison. The larger extent of the water mixing ratio PPD results from the combination of the uncertainty on the number density and pressure.}
  \vspace{-0.0cm}
  \label{fig:pressuremix}
\end{figure*}

Species' mixing ratios are not adequate parameters for atmospheric retrieval method. The reason is that the uncertainty on mixing ratios encompasses the uncertainty on the number densities and the pressure. We demonstrate that pressure and number densities are the atmospheric parameters embedded in a planet's transmission spectrum. This implies that pressure and mixing ratios are correlated in the context of transmission spectroscopy (Figure\,\ref{fig:pressuremix}.A). Hence, the posterior probability distributions of mixing ratios are the results of the convolution of the number densities' and pressure's PPDs (Figure\,\ref{fig:pressuremix}.B). Therefore, the significance of a molecular detection based on the mixing ratio will be less significant than if based on the number densities, in particular because number densities are more constrained than the pressure.

\subsubsection{No False Molecular Detections}

\textit{MassSpec}'s applications do not lead to false molecular detections---assuming that all lines and bands of a molecule are known. As emphasized previously, the application of \textit{MassSpec} requires a planet's transmission spectrum of sufficient quality (SNR, spectral resolution, and coverage) to identify the major molecular species. Once these criteria are met, the specificity of the species' signature in extinction (shown in panels A of Figures\,\ref{fig:MassSpec_results_in_text_nw} to\,\ref{fig:water_world_JWST_C3}) prevents from the false positive detection of atmospheric species. As an example, we try to retrieve ozone that was not part of the synthetic atmospheres and it is detected in none of our retrieval.

%% file: Chapter5/chap5.tex
\chapter{Summary and Conclusions}
\label{chap:conclusion}
\vspace{-0.7cm}

The detailed characterization of extra-solar planets is the next milestone for the field of exoplanetology. The quality of the data available to characterize exoplanets is constantly improving. In that context, we approached the question ``What new insights into the atmospheric and interior properties of exoplanets will future high-quality data bring within reach?''. For that purpose, this thesis focused on transiting exoplanets, which are of special interest due to the wealth and diversity of data made available by their favorable orbital configuration. We introduced two new characterization methods. The first method aims to map transiting exoplanets's atmospheres based on their light curves, with a particular focus on their secondary eclipses. The second method aims to determine transiting planet masses and atmospheric properties solely from transmission spectra, i.e. the starlight filtered by a planet's atmosphere during transits. 

For both methods, we introduced the underlying concepts and theory, emphasized the possible caveats, and presented the Bayesian frameworks developed to analyze available datasets and assess their potential with future observatories. Below we summarize our main results and conclusions for each method. In particular, we provide  a takeaway message for each method, a reminder of what is possible at the time this thesis was written, and what the future has to offer. We conclude with general recommendations for the field of exoplanet characterization.
\newpage
\section{Mapping Exoplanet Atmospheres}

\textbf{Takeaway message:} We found that a planet's shape, brightness distribution, and system parameters are underdetermined by its light curve (i.e., correlated). This means that assumptions on any of these properties affect the others and yield biased estimates and also underestimated error bars. For example, we show that assuming HD\,189733b to be uniformly bright in the \textit{Spitzer}/IRAC 8\,$\mu$m channel leads to an overestimation of the stellar density (and hence, the planetary density) by 5$\%$. For that reason, we developed a method to map exoplanets based on global analyses using datasets complimentary to transits, such as radial velocity measurements, to relax assumptions.

\textbf{Present:} We mapped the hot Jupiter HD\,189733b's emission in the infrared (8\,$\mu$m) in 2D and constrained the characteristic of its hot spot. We mapped Kepler-7b in the visible and contributed to the first map of cloud coverage on a planet outside of the Solar System.

\textbf{Future:} Exoplanet mapping will provide unprecedented insights into the physics of exoplanetary atmospheres via their temperature and composition patterns. By 2015, we will obtain the first 3D maps of exoplanet atmospheres. By 2020, we should be able to assess the time variability of 3D structures in exoplanet atmospheres.
\begin{center}
\rule{3cm}{0.4pt}
\end{center}
\textbf{Summary of key points:}
\vspace{-0.3cm}\begin{itemize}
\item Mapping exoplanets is an underdetermined problem because the target's brightness distribution is underdetermined by its light curve, hence multiple solution can fit a given dataset. Therefore, mapping methods have to ensure that all the possible solutions are revealed.

\item Mapping exoplanets also involves degenerate solutions because four of the systems parameters (namely, $e$, $\omega$, $b$, and $\rho_\star$) and the target's projected shape at conjunctions can also affect the occultation shape, hence those are correlated with its brightness distribution. To mitigate the effects of this degeneracy, it is crucial to avoid unassessed assumptions on one of the above properties as it could bias the estimates of the others. For example, we show that assuming HD\,189733b to be uniformly bright in \textit{Spitzer}/IRAC-8$\mu$m channel leads to an overestimation of the stellar density (and hence, the planetary density) by 5$\%$---which is at 6\,$\sigma$ of the actual estimate. We advocate for relaxing assumptions in global analyses of occultation, transits, phase curves, and RV measurements to consistently probe  the correlated parameter space of the light-curve fitting problem. We note that the effects of assumptions---such as ``the planet can be modeled as a uniformly-bright disk''---are adequate to fit light curves with low SNR, typically when the ratio eclipse of depth to photometric precision on a $\sim$1-min bin is $\ll10$.

\item We applied our mapping method to the hot Jupiter HD\,189733b. We detected the deviation of its occultation shape from the one of a uniformly bright disk at the 6\,$\sigma$ level in \textit{Spitzer}/IRAC's 8$\mu$m channel. We demonstrated that this deviation emerges mainly from a large-scale brightness structure in HD\,18733b's atmosphere---we rejected HD\,189733b's shape as a possible contributing factor based on its transit residuals. HD\,18733b's brightness distribution indicated a hot spot shifted east of the substellar point within the atmospheric layers probed at 8\,$\mu $m. In addition, we investigated the practical effects of underlying model assumptions on our inferences due to the correlation between the system parameters and the brightness distribution. Notably, we found that the more complex HD\,189733b's brightness model, the larger the eccentricity, the lower the densities, the larger the impact parameter and the more localized and latitudinally-shifted the hot spot estimated. We outlined the influence of HD\,189733's RV measurements on our inferences obtained solely from \textit{Spitzer}/IRAC 8-$\mu $m photometry. In particular, we observed redistributions of the probability density for the posterior probability distributions (PPDs) obtained with complex brightness models, which result from the rejection of solutions involving $\sqrt{e}\sin\omega$ $ \gtrsim $ 0.15 by HD\,189733's RV data---favored by the photometry for complex brightness models. This reemphasized the necessity of global analyses to consistently probe the correlated parameter space of exoplanets' light curves and results in a new upper limit of HD\,189733b's orbital eccentricity, $ e\leq 0.011$ ($95\%$  confidence).

\item We applied our mapping method to the hot-Jupiter Kepler-7b as a contribution to \cite{Demory2013}. We found that the large geometric albedo of Kepler-7b in \textit{Kepler}'s bandpass originates mainly from the western sector of its dayside. In particular, we found Kepler-7b has a bright sector extending from the western terminator of its dayside to $\sim30^{\circ}$ east of the substellar point  with a mean relative brightness of  $78\pm4\times10^{−4}$ ppm.

\item We highlighted that multi-wavelength observations enable 3D mapping of exoplanet atmospheres and mitigate the degeneracy between the brightness distributions and the system parameters. Multi-wavelength observations probe different optical depths under the same orbital configuration, hence yielding 3D mapping while recording the same information about the system parameters.

\item We showed how applying our method to complementary observations of HD\,189733b (and HD\,209458b) could yield to the first 3D map of exoplanet atmospheres within a year \citep[as a contribution to the \textit{Spitzer} proposal $\#$10103, PI Dr. Nikole Lewis, see][]{Lewis2013prop}. 

\item We determined the best known exoplanets to perform eclipse mapping in the visible using \textit{HST}/WFC3 are WASP-18b, WASP-19b, and WASP-76b (Table\,\ref{tab:HSTWFC3Targets}).
 
\item We showed that future spaced-based observatories like \textit{JWST} and \textit{EChO}-class missions could yield time-dependent 3D maps of distant worlds. 

\item We pointed out the synergy of \textit{JWST} and \textit{EChO}-class missions to constrain the cloud coverage of an exoplanet. 

\item Finally, we discussed how the implementation of atmospheric circulation models  \citep[e.g.,][]{Heng2014} in our mapping method could yield direct constraints on the atmosphere's circulation parameters  while mapping it.

\end{itemize}
\newpage
\section{Insights into Exoplanets from Transmission Spectroscopy}
\textbf{Takeaway message:} We demonstrated from first principles that a planet's mass is a key parameter behind its transmission spectrum, akin to the planet's atmospheric properties (temperature, pressure, and composition). Hence, for consistency, atmospheric retrieval methods have to account for the planetary mass. On this basis, we developed a new retrieval method called \textit{MassSpec}. 

\textbf{Present:} We applied a simplified version of \textit{MassSpec} to the hot-Jupiter HD\,189733b and retrieved a mass in excellent agreement with the mass estimated from RV measurements. We showed how the same procedure could determine the mass of the hottest planet known, WASP-33b, with less than 20 \textit{HST} orbits.

\textbf{Future:} We demonstrated \textit{MassSpec}'s capability to determine the masses and atmospheric properties of planets as small as Earth with future observatories. We estimated that \textit{JWST} could characterize the properties of $\sim 200$ Earth-sized planets in their hosts' habitable zones, if such planets are discovered. The vast majority of these hosts will be late M dwarfs, which may prevent a mass determination via RV measurements. For that reason, it is likely that \textit{MassSpec} will play a pivotal role in the identification of the first habitable exoplanet, which could occur within the next decade.
\begin{center}
\rule{3cm}{0.4pt}
\end{center}
\textbf{Summary of key points:}
\vspace{-0.3cm}\begin{itemize}
\item To date, mass constraints for exoplanets are predominantly based on radial velocity measurements, which are not suited for planets with low masses, large semi-major axes, or those orbiting faint or active stars. We presented an alternative method to the RV method, \textit{MassSpec}, to determine the mass of an exoplanet solely from transmission spectroscopy. \textit{MassSpec} also provided a solution to the problem of reduced atmospheric retrieval quality based on a planet's transmission spectrum when its mass is inadequately constrained \citep[see][]{Barstow2013} by consistently determining (i.e., simultaneously) a planet's mass and its atmospheric properties.

\item We prove \textit{MassSpec}'s feasibility analytically by identifying the key parameters of and their effects on an exoplanet's transmission spectrum. In particular, we demonstrated that a transmission spectrum depends in unique ways on the atmospheric scale height, reference pressure, temperature and number densities of the main atmospheric absorbents---which constrain the mean molecular mass. The uniqueness of these dependencies enables the independent retrieval of each of these key parameters that in turns leads to the planet's mass. 

\item We found that the Euler-Mascheroni constant \citep[$\gamma_{EM}$, see][]{Euler1740} plays a pivotal role in the equations of transmission spectroscopy. In particular, $\gamma_{EM}$ is central in the identification of the key parameters of transmission spectrum. Furthermore, we showed that it is also key to improving the computational efficiency for modeling transmission spectra via solving $h_{eff}(\lambda) = [z:\tau(z,\lambda) = e^{-\gamma_{EM}}]$. 

\item We tested \textit{MassSpec} on real data using HD\,189733b's transmission spectrum and showed an excellent agreement between the masses retrieved by \textit{MassSpec} and from RV mass measurements.

\item We discussed the complementary nature of RV measurements and transmission spectra to constrain planetary masses. To do so, we expressed the sensitivity of each signal's significance to the system parameters. 

\item We reemphasized the benefit of late M dwarfs for transit observations. We showed that the significance of in-transit signals scales as $\sqrt{B_{\lambda}(T_{\star})}/R_{\star}$ and is thus increased up to a factor of three when the host-star of a given planet is a late M dwarf, instead of, e.g., a Sun-like star. In that context, we also discussed the characterization of habitable earth-sized planets around late M dwarfs in the next decade.

\item We demonstrated \textit{MassSpec}'s capacity to yield the mass of the hottest planet known, WASP-33b, with a relative uncertainty of $\lesssim 20\%$ with less than 20 \textit{HST} orbits. 

\item We demonstrated numerically that future space-based facilities designed for exoplanet atmosphere characterization will also be capable of mass measurements for super-Earths and Earth-sized planets
with a relative uncertainty as low as $\sim 2\%$---a precision that has not yet been reached using RV measurements, even for the most favorable cases of hot Jupiters. In particular, we showed that \textit{JWST}'s observations of unfavorable cases, such as Earth-sized planets with nitrogen-dominated atmosphere (``Earth-like planets''), could yield mass measurements with a relative uncertainty of $\sim 8\%$---if the targets are transiting late M dwarfs within 15 pc of Earth. We estimated that with data from \textit{JWST}, \textit{MassSpec} could yield the mass of mini-Neptunes, super-Earths, and Earth-sized planets up to distances of 500 pc, 100 pc, and 50 pc, respectively, for M9V stars and 200 pc, 40 pc and 20 pc for M1V stars or stars with earlier spectral types. For \textit{EChO}-class missions, the numbers would be 250 pc, 50 pc and 13 pc for M9V stars and 100 pc, 20 pc and 6 pc for M1V stars or stars with earlier spectral types.

\item We estimated \textit{JWST} could determine the mass and the atmospheric properties of $\sim$2000 Earth-sized planets, 10$\%$ of which would be in their host's habitable zones. For an \textit{EChO}-class mission, this numbers would be 40. This highlights how close we are to identifying habitable exoplanets.

\item We discussed \textit{MassSpec}'s input to gain insight into planetary interiors. Although \textit{MassSpec} provides constraints on the mass and radius of a target, those are not always sufficient to obtain insights into a planet's interior. \textit{MassSpec}'s simultaneous constraints on a planet's atmosphere and bulk density may help to break this degeneracy, in some cases. 

\item We emphasized that joint analyses of transmission spectra with any dataset that can independently constrain a planet's mass and/or atmospheric properties can enhance \textit{MassSpec}'s capabilities. In particular, RV measurements and emission spectra would be great assets to provide complementary constraints on the mass and the atmospheric properties, respectively.

\item We showed that, in principle, \textit{MassSpec} is not rendered ineffectual by clouds because clouds are deeper than the lowest pressure probed by transmission spectroscopy. In other words, there will always be atmospheric information available from transmission spectroscopy, although high-altitude clouds may increase the uncertainty on the planetary mass estimates by a factor of $\sim 2$.

\end{itemize}

\section{General Recommendations for Future Progress}

\subsection{On the Need for Accurate Extinction Cross Section Databases}

Like most methods aimed at determining the properties of distant object, the methods introduced in this thesis require accurate extinction cross sections (Equation\,\ref{eq:optical_depth}). On the one hand, 3D mapping of exoplanetary atmospheres necessitates the determination the atmospheric layer from which the detected flux originates. On the other hand, atmospheric retrieval methods---like \textit{MassSpec}---solve the inverse problem of determining the conditions (i.e., temperature, pressure, composition) of the medium probed from its transmission spectrum. In both cases, the key to a data interpretation is to know \textit{a priori} how the transmission of light is affected by a medium's conditions (i.e., temperature, pressure, composition). That is why, we advocate for devoting significant efforts to generate accurate extinction cross section databases that cover various atmospheric conditions in order to prevent them from limiting our ability to characterize exoplanets.

\subsection{How to Optimize the Scientific Return of \textit{JWST}}

The number of targets favorable for characterization and the time it would require will exceed the time allotted to exoplanets with \textit{JWST}. This is why we suggested to aim for community efforts (1) to discover targets of interest well in advance of \textit{JWST}'s launch with missions like \textit{SPECULOOS} and (2) pre-select the targets for which \textit{JWST}'s time will be dedicated over its lifetime. To help mitigate the drawbacks of dedicating \textit{JWST}'s time to a limited number of targets, we suggest to consider more modest missions such as \textit{EChO}. We pointed out the synergy between \textit{JWST} and an \textit{EChO}-class mission to map exoplanets' atmospheres, constrain their cloud coverage, and determine their atmospheric properties and masses.

%% file: AppendixA/appa.tex
\chapter{Introduction to Rational Functions}
\label{app:TF}

\vspace{-0.7cm}

A rational function, $f(.)$, is a function that can be expressed as the ratio of polynomials, $P(.)/Q(.)$. The zeros and poles of $f(.)$ are the zeros of $P(.)$ and the zeros of $Q(.)$, respectively. Rational functions are used in various fields of science and engineering (signal processing, acoustics, aerodynamics, structural dynamics, electronic circuitry, control theory, etc.) as mathematical representation of the relation of the inputs ($i(.)$) and outputs ($o(.)$) of a system, called the transfer function of the system. The transfer function of a system ($H(s)$) is the linear mapping of the Laplace/Fourier transform of its inputs ($I(s) = \mathcal{L}(i(.))$) to the Laplace/Fourier transform of its outputs ($O(s) = \mathcal{L}(o(.))$), i.e., $O(s) = H(s)I(s)$ where $s$ is a spatial or temporal frequency if $i$ and $o$ are functions of space or time, respectively. As an example, the transfer function of imaging devices is the Fourier transform of the point spread function (PSF)---because the theoretical input leading to the PSF is a spatial impulse (i.e., a Dirac function) in the field of view and the Fourier transform of an impulse is one, therefore $\mathcal{L}(PSF) = H(s)$. 

\begin{figure}
     \hspace{-2.1cm}\includegraphics[trim = 20mm 00mm 20mm 00mm,clip,width=19cm,height=!]{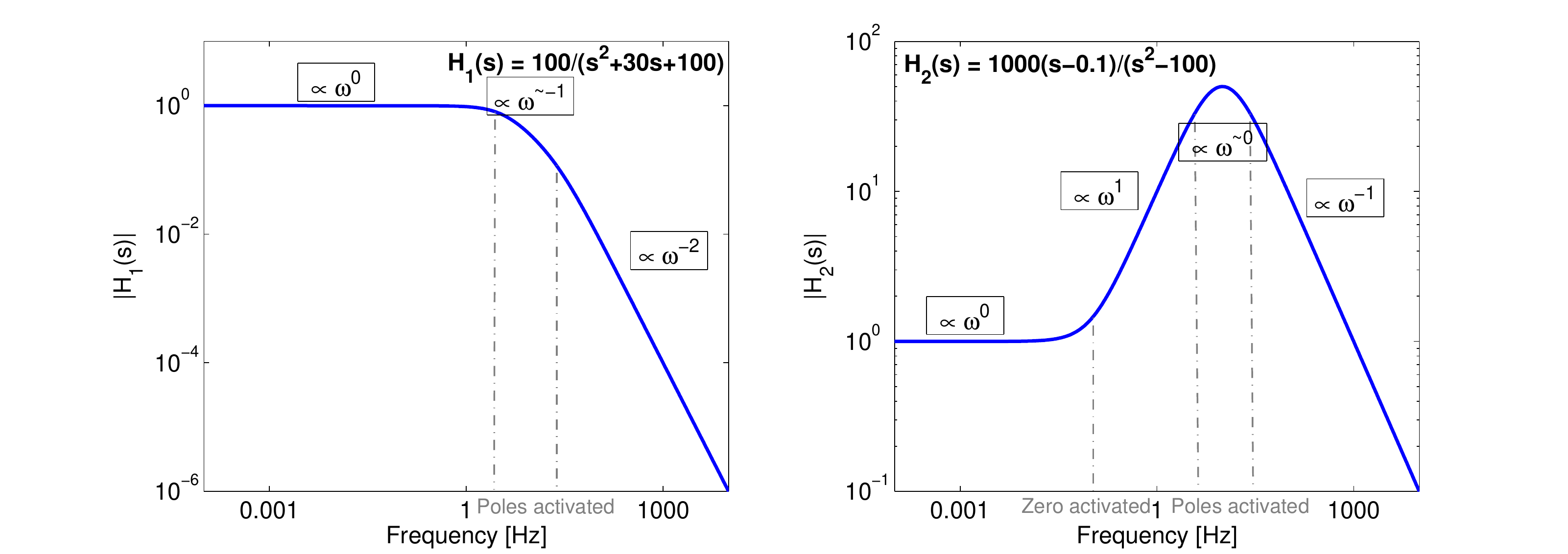}

  \caption[Typical behavior of rational functions.]{Typical behavior of rational functions. Left: Transfer function of a mass-spring-dashpot system with a natural frequency of 10 Hz and a damping ratio of 1.5. The transfer function is independent of the frequency at frequency lower than the system's natural frequency. At frequency larger than the natural frequency, the two poles of the transfer function are activated and the dependency of the transfer function on the frequency changes from $\propto \omega^0$ to $\propto \omega^{-2}$ (i.e., the exponent decreases by two). Right: Transfer function with a zero at 0.1 Hz and a pair of conjugated poles at $\pm$10 Hz. The effect of the zero is to increase the exponent of the transfer function dependency on the frequency by one, while the effect of the pair of poles is to decrease it by two.}
  \vspace{-0.0cm}
  \label{fig:rational_function}
\end{figure}

The transfer function of a system relates directly to the differential equations used to represent mathematically the system. Therefore, the poles and zeros of a transfer function characterize the behavior of a system. For example, the differential equation of a second order system is
\begin{equation}
\ddot{x}+2 \zeta \omega_n \dot{x}+ \omega_n^2 x=0,
\end{equation}
where $\zeta$ and $\omega_n$ are the damping ratio and the natural frequency of the system, respectively---e.g., for a mass-spring-dashpot system $\zeta = c/(1\sqrt{km})$ and $\omega_n = \sqrt{k/m}$ where $m$, $c$, and $k$ are the mass, the spring constant, and the damping coefficient, respectively. The transfer function of a second order system is 
\begin{equation}
H(s) = K \frac{\omega_n^2}{s^2+2 \zeta \omega_n s+ \omega_n^2} ,
\end{equation}
where $K$ is the system gain (K = 1 for the mass-spring-dashpot system) and $s = j\omega$ where $j$ is the imaginary unit (see Figure\,\ref{fig:rational_function} for more details). Therefore, at frequencies low compared to $\omega_n$, $H(j\omega) \approx K$, meaning that the response of a mass-spring-dashpot system to a low-frequency input, $i(t) = A\exp(j\omega t)$, is the input, i.e. $o(t) = i(t)$. At high frequencies, $H(j\omega) \approx K/(j\omega)^2 = -K/\omega^2$ meaning that the response is in opposition of phase with the input and its amplitude is proportional to $\omega^{-2}$. 

The position of a transfer function poles and zeros in the complex field ($\mathbb{C}$) define the domains of different dependency regimes for the transfer function (Figure\,\ref{fig:rational_function}). In the neighborhood of a zero, the exponent of the transfer function dependency to its variable increases by 1, while it decreases by 1 in the neighborhood of a pole. Hence, the overall shape of transfer function relates directly to its formulation. Therefore, from simulations or measurements, it is possible to derive the appropriate formulation to represent mathematically a system using a rational function.
\clearpage
\newpage

%% file: AppendixB/appb.tex
\chapter{Towards a Generalization of Transmission Spectrum Equations}

\label{sec:generalizedtransmissionspectra}

In this Appendix, we aim to extend the analytical derivations introduced in Section\,\ref{app:dependency}. In particular, we address here the case of a non-isothermal atmosphere. The case of a non-isocompositional atmosphere is derived analogously. Thus, we do not provide the derivation here.

\section{General Formulations for Atmospheric Quantities}

\subsection{Temperature}
The temperature profile of a planet's atmosphere can be formulated as
\begin{eqnarray}
 T(z) & = & T_0 \displaystyle\sum_{h=0}^{n} \beta_h z^h, 
\end{eqnarray}
where $T_0$ is the temperature at $R_{p,0}$ and the $\beta_h$ are the coefficient of the Maclaurin series expansion of the temperature profile. Note that $\beta_0 = 1$. For the rest of the demonstration, it is suitable to write $T(z)$ as follow
\begin{eqnarray}
 T(z) & = & T_0 f_T(z), 
\end{eqnarray}
because it allows to rewrite $T(z)$ as a function of the zeros of $f_T(z)$, $\omega$ : 
\begin{eqnarray}
 T(z) & = & T_0 \prod_{\{\forall \omega : f_T(\omega) = 0\}} \left(1-\frac{z}{\omega}\right). 
 \label{eq:general_T_profile}
\end{eqnarray}
Note that the smallest positive zero of $f_T(z)$ define the domain in which the series expansion is physically adequate---i.e., from 0 to $\omega_0$, $\left( \{ \omega_0 : \omega_0\leq \omega, \forall \omega > 0 \} \right)$. In addition, $f_T(z)$ cannot physically have zeros with a multiplicity greater than 1.

\subsection{Pressure}
The pressure profile of a planetary atmosphere can be derived from the equation of hydrostatic equilibrium,
\begin{eqnarray}
\text{d}\ln(p) & = & -\frac{1}{H} \text{d} z,\label{eq:hydro_eq_tz}\\
 	& = & -\frac{1}{H_0 f_T(z)} \text{d} z,
\end{eqnarray}
where $H_0$ is the atmospheric scale height at $R_{p,0}$. Because $f_T(z)$ has no zero with a multiplicity greater than 1, $f_T^{-1}(z)$ can be expressed as a sum of irreducible rational functions (see Appendix\,\ref{app:TF}):
\begin{eqnarray}
f_T^{-1}(z) = \sum_{\{\forall \omega : f_T(\omega) = 0\}} \frac{\text{R}_\omega}{(z - \omega)},
\end{eqnarray}
where $\text{R}_\omega$ are the residues of $f_T^-1(.)$ in the context of rational function expansion in a irreducible rational function. In particular, \begin{eqnarray}
\text{R}_\omega = \frac{1}{f_T'(\omega)} ,
\end{eqnarray}  
where $f_T'(.)$ is the derivative of $f_T^-1(.)$, because $f_T(.)$ has no zero with a multiplicity greater than 1.

The second term of Equation\,\ref{eq:hydro_eq_tz} can then be integrated by parts: 
\begin{eqnarray}
-\frac{1}{H_0} \int  \sum_{\{\forall \omega : f_T(\omega) = 0\}} \frac{1}{f_T'(\omega)(z - \omega)} \text{ d}z & = & \sum_{\{\forall \omega : f_T(\omega) = 0\}} \frac{-\log(z - \omega)}{H_0 f_T'(\omega)} + C_0,
\end{eqnarray}
where $C_0$ is the integration constant. As a result, one can express the pressure profile as
\begin{eqnarray}
p(z) & = & C_1 \prod_{\{\forall \omega : f_T(\omega) = 0\}} (z - \omega)^{-\frac{1}{H_0 f_T'(\omega)}},\\
& = & C_2 \prod_{\{\forall \omega : f_T(\omega) = 0\}} \left(1 - \frac{z}{\omega}\right)^{-\frac{1}{H_0 f_T'(\omega)}},
\end{eqnarray}
where $C_1$ and $C_2$ are integration constants. The integration constants are determined using the identity $p(z=0) = P_0$, where $P_0$ is the pressure at $R_{p,0}$. For example, $C_2 = P_0$. Therefore, the generalized formulation of the pressure profile is
\begin{eqnarray}
p(z) & = & p_0 \prod_{\{\forall \omega : f_T(\omega) = 0\}} \left(1 - \frac{z}{\omega}\right)^{-\frac{1}{H_0 f_T'(\omega)}}. 
\label{eq:general_p_profile}
\end{eqnarray}


\subsection{Number Densities}

The number-density profiles can be derived from the ideal gas law
\begin{eqnarray}
p V = n R T,
\end{eqnarray}
where $V$ is the volume of gas, $n$ is the gas number density, and $R$ is the ideal gas constant ($\sim 8.314 \text{J.K}^{-1}.\text{mol}^{-1}$). In particular, the generalized number-density profile can be derived from:
\begin{eqnarray}
 n(z) = n_0 \frac{T_0}{T} \frac{p}{p_0},
\end{eqnarray}
where $n_0$ is the atmospheric number density at $R_{p,0}$. Therefore, using Equations\,\ref{eq:general_T_profile} and\,\ref{eq:general_p_profile} one obtains
 \begin{eqnarray}
 n(z) = n_0 \prod_{\{\forall \omega : f_T(\omega) = 0\}} \left(1 - \frac{z}{\omega}\right)^{-1-\frac{1}{H_0 f_T'(\omega)}}. 
\label{eq:general_n_profile}
\end{eqnarray}

\subsection{Slant-Optical Depth}
\label{sec:slant_path_non_iso}
The slant-optical depth profile (Equation\,\ref{eq:optical_depth}) can now be written as 
 \begin{eqnarray}
 \tau(z,\lambda) & = & 2 n_0 \sigma_0(\lambda) T_0^\delta p_0^\epsilon \int_0^{x_\infty} \prod_{\{\forall \omega : f_T(\omega) = 0\}} \left(1 - \frac{x^2}{2 (R_{p,0}+z)\omega}\right)^{F(\omega)} \text{ d}x,
 \label{eq:tau_adiabat_general}
\end{eqnarray}
where $ x_\infty = \sqrt{2 (R_{p,0}+z)\omega_0}$ is the boundary of the series-expansion domain in the $x$-space and
 \begin{eqnarray}
 F(\omega) = \delta-1+\frac{\epsilon-1}{H_0 f_T'(\omega)},
\end{eqnarray}
\textit{assuming the cross section to be of the form}\footnote{For molecular absorption, the absorption coefficient (Equation\,\ref{line_alpha}) takes the form
\begin{eqnarray}
	\alpha(z,\lambda) = \Lambda_{\kappa} \sum_{m=0}^{n_T} a_{T,i,m} \prod_{\{\forall \omega : f_T(\omega) = 0\}} \left(1 - \frac{z}{\omega}\right)^{m-1-\frac{1}{H_0 f_T'(\omega)}} \frac{p(z)+a_{\kappa}}{p^2(z)+b_{\kappa}}.
\label{eq:alpha_for_non_isotherm_and_molecular}
\end{eqnarray}
The form of $\alpha(\lambda,T,p)$ provides an analytical form for $\tau(z,\lambda)$ for each of the 4 regimes introduced in Section\,\ref{app:molecules} via an integration by parts analogous to the derivation in Section\,\ref{sec:slant_path_non_iso}. However, the transmission spectrum equation (Equation\,\ref{eq:transmission_spectrum_h2}) cannot be expressed in terms of standard mathematical functions because these analyticals forms for $\tau(z,\lambda)$ have multiple zeros (see Section\,\ref{sec:tra_spec_equ}). Therefore, we do not address here the case of molecular absorption.} $\sigma(\lambda) \varpropto T^\delta p^\epsilon $. For example, for Rayleigh scattering and CIA, $\left\lbrace \delta=0,\epsilon=0 \right\rbrace $ and $\left\lbrace \delta\sim-1,\epsilon=1 \right\rbrace $, respectively.

We find no solution to the integral in Equation\,\ref{eq:tau_adiabat_general} using standard mathematical functions. However, we find analytical formulations for the three following simplified cases: 
\begin{enumerate}
\item the case where $f_T(z)$ has a single zero, $\omega_0 \text{ }(\omega_0 \in \mathbb{R}^+)$,
\item the case where $f_T(z)$ has two zeros, ${\omega_0,\omega_1} \text{ }(\omega_0 \in \mathbb{R}^+ \text{ and } \omega_1 \in \mathbb{R}^-)$,
\item the case where $f_T(z)$ has three zeros, ${\omega_0,\omega_1,\omega_2} \text{ }(\omega_0 \in \mathbb{R}^+, {\omega_1,\omega_2} \in \mathbb{R}^-)$.
\end{enumerate}

The general formulation for the slant-optical depth profile is 
 \begin{eqnarray}
 \tau(z,\lambda) & = & \tau(0,\lambda)\prod_{\{\forall \omega : f_T(\omega) = 0\}} \left(1 - \frac{z}{\omega}\right)^{F(\omega)}.
 \label{eq:tau_adiabat_general_short}
\end{eqnarray}

\subsubsection{\ref{sec:slant_path_non_iso}.1 $T(z)$ is a first-order polynomial} 
In this case, the temperature decreases linearly with the altitude, i.e.,
 \begin{eqnarray}
  f_T(z) & = & 1-\frac{z}{\omega_0}.
\end{eqnarray}
Hence, 
 \begin{eqnarray}
  f_T'(z) & = & -\frac{1}{\omega_0},
\end{eqnarray}
and,
 \begin{eqnarray}
  \tau(z,\lambda) & = & 2 n(z) \sigma_0(\lambda) T_0^\delta p_0^\epsilon \int_0^{x_\infty} \left(1 - \frac{x^2}{2 (R_{p,0}+z)\omega_0}\right)^{\delta-1-\frac{\omega_0}{H_0}(\epsilon-1)} \text{ d}x.
  \label{eq:tau_adiab_equ1_raw}
\end{eqnarray}

It can be shown that
 \begin{eqnarray}
  \int \left(1-\left(\frac{x}{A}\right)^2\right)^B \text{ d}x & = & x _2F_1\left(\frac{1}{2},-B;\frac{3}{2};\frac{x^2}{A^2}\right),
\end{eqnarray}
and,
\begin{eqnarray}	
	\left\{{
   \begin{array}{c c c}
     _2F_1(a,b;c;1) & = & \frac{\Gamma(c)\Gamma(c-a-b)}{\Gamma(c-a)\Gamma(c-b)}\\
     _2F_1(a,b;c;0) & = & 1
  \end{array}}
  	\right.
  	,
\end{eqnarray}
where $_2F_1(a,b;c;z)$ is the Gaussian hypergeometric function and $\Gamma(z)$ is the Euler integral of the second kind: the Gamma function. Therefore, Equation\,\ref{eq:tau_adiab_equ1_raw} can be written as
 \begin{eqnarray}
  \tau(z,\lambda) & = & \sqrt{2 \pi (R_{p,0}+z)\omega_0} \sigma_0(\lambda) T_0^\delta p_0^\epsilon n(z) \frac{\Gamma(\delta-\frac{\omega_0}{H_0}(\epsilon-1))}{\Gamma(\delta+\frac{1}{2}-\frac{\omega_0}{H_0}(\epsilon-1))},\\
  & = & \sqrt{2 \pi (R_{p,0}+z)\omega_0} \sigma_0(\lambda) T_0^\delta p_0^\epsilon n_0 \frac{\Gamma(\delta-\frac{\omega_0}{H_0}(\epsilon-1))}{\Gamma(\delta+\frac{1}{2}-\frac{\omega_0}{H_0}(\epsilon-1))} \left(1 - \frac{z}{\omega_0}\right)^{\delta-1-\frac{\omega_0}{H_0}(\epsilon-1)}.
  \label{eq:tau_adiab_equ1}
\end{eqnarray}

\subsubsection{\ref{sec:slant_path_non_iso}.2 $T(z)$ is a second-order polynomial} 
In this case, the temperature profile is a second-order polynomial with two zeros, one positive and the other negative---to be physically plausible.
 \begin{eqnarray}
  f_T(z) & = & \left(1-\frac{z}{\omega_0}\right)\left(1-\frac{z}{\omega_1}\right).
\end{eqnarray}
Hence, 
 \begin{eqnarray}
 \left\{{
   \begin{array}{c c l}
     f_T'(\omega_0) & = & \frac{\omega_1-\omega_0}{\omega_0 \omega_1}\\
     f_T'(\omega_1) & = & \frac{\omega_0-\omega_1}{\omega_0 \omega_1}
  \end{array}}
  	\right.
  	,
\end{eqnarray}
and,
 \begin{eqnarray}
  \tau(z,\lambda)  =  2 n(z) \sigma_0(\lambda) T_0^\delta p_0^\epsilon \int_0^{x_\infty}                  &\left(1 - \frac{x^2}{x_\infty}\right)^{F(\omega_0)} \left(1 - \frac{x^2}{2(R_{p,0}+z)\omega_1}\right)^{F(\omega_1)} \text{ d}x,
  \label{eq:tau_adiab_equ2_raw}
\end{eqnarray}
where
 \begin{eqnarray}
  n(z) & = & n_0 \left(1 - \frac{z}{\omega_0 }\right)^{F(\omega_0)} \left(1 - \frac{z}{\omega_1}\right)^{F(\omega_1)}.
\end{eqnarray}

It can be shown that
 \begin{eqnarray}
  \int \left(1-\left(\frac{x}{A}\right)^2\right)^B \left(1-\left(\frac{x}{C}\right)^2\right)^D \text{ d}x & = & x F_1\left(\frac{1}{2};-B,-D;\frac{3}{2};\frac{x^2}{A^2},\frac{x^2}{C^2}\right),\\
  & = & x F_1\left(\frac{1}{2};-D,-B;\frac{3}{2};\frac{x^2}{C^2},\frac{x^2}{A^2}\right),
\end{eqnarray}
and,
\begin{eqnarray}	
	\left\{{
   \begin{array}{c c l}
     F_1(a;b_1,b_2;c;z_1,0) & = & _2F_1(a,b_2;c;1) _2F_1(a,b_1;c-b_2;z_1)\\
     F_1(a;b_1,b_2;c;0,0) & = & 1
  \end{array}}
  	\right.
  	,
\end{eqnarray}
where $F_1(a,b;c;z)$ is the Appell hypergeometric function. Therefore, Equation\,\ref{eq:tau_adiab_equ2_raw} can be written as
 \begin{eqnarray}
 \begin{array}{l l}
  \tau(z,\lambda)  = & \sqrt{2 \pi (R_{p,0}+z)\omega_0} \sigma_0(\lambda) T_0^\delta p_0^\epsilon n(z)  \frac{\Gamma\left(\delta-\frac{\omega_0 \omega_1(\epsilon-1)}{H_0 (\omega_0-\omega_1)}\right)}{\Gamma\left(\delta+\frac{1}{2}-\frac{\omega_0 \omega_1(\epsilon-1)}{H_0 (\omega_0-\omega_1)}\right)} \\
  &\times\text{  }  _2F_1\left(\frac{1}{2},1-\delta+\frac{\omega_0 \omega_1(1-\epsilon)}{H_0 (\omega_0-\omega_1)};\delta+\frac{1}{2}-\frac{\omega_0 \omega_1(\epsilon-1)}{H_0 (\omega_0-\omega_1)};\frac{\omega_0^2}{\omega_1^2}\right).
  \end{array}
  \label{eq:tau_adiab_equ2}
\end{eqnarray}

\subsubsection{\ref{sec:slant_path_non_iso}.3 $T(z)$ is a third-order polynomial} 
In this case, the temperature profile is a third-order polynomial with three zeros, one positive and two negative---to be physically plausible. In addition, one of the zeros is particularly small $(|\omega_2| \ll z)$.
 \begin{eqnarray}
  f_T(z) & = & \left(1-\frac{z}{\omega_0}\right)\left(1-\frac{z}{\omega_1}\right)\left(1-\frac{z}{\omega_2}\right).
\end{eqnarray}
Hence, 
 \begin{eqnarray}
 \left\{{
   \begin{array}{c c l}
     f_T'(\omega_0) & = & \frac{-\omega_0^2+\omega_0\omega_1+\omega_0\omega_2-\omega_1\omega_2}{\omega_0 \omega_1 \omega_2}\\
     f_T'(\omega_1) & = & \frac{-\omega_1^2+\omega_1\omega_0+\omega_1\omega_2-\omega_0\omega_2}{\omega_0 \omega_1 \omega_2}\\
     f_T'(\omega_2) & = & \frac{-\omega_2^2+\omega_2\omega_1+\omega_2\omega_0-\omega_1\omega_0}{\omega_0 \omega_1 \omega_2}
  \end{array}}
  	\right.
  	,
\end{eqnarray}
and,
 \begin{eqnarray}
  \tau(z,\lambda) & = & 2 n(z) \sigma_0(\lambda) T_0^\delta p_0^\epsilon \int_0^{x_\infty}                  \prod_{i=0}^{2} \left(1 - \frac{x^2}{2(R_{p,0}+z)\omega_i}\right)^{F(\omega_i)} \text{ d}x,
  \label{eq:tau_adiab_equ3_raw}
\end{eqnarray}
where
 \begin{eqnarray}
  n(z) & = & n_0 \prod_{i=0}^{2} \left(1 - \frac{z}{\omega_i}\right)^{F(\omega_i)}.
\end{eqnarray}

It can be shown that
 \begin{eqnarray}
 \begin{array}{l l}
  \int \left(1-(\frac{x}{A})^2\right)^B \left(1-(\frac{x}{C})^2\right)^D \left(-(\frac{x}{E})^2\right)^F \text{ d}x = & \frac{x}{2F+1} (-\frac{x^2}{E^2})^F \\
  & \times F_1\left(F+\frac{1}{2};-D,-B;F+\frac{3}{2};\frac{x^2}{C^2},\frac{x^2}{A^2}\right).
  \end{array}
  \label{eq:three_binomials}
\end{eqnarray}

We can use Equation\,\ref{eq:three_binomials} for the present simplified case as we assumed that $|\omega_2| \ll z$.

Therefore, Equation\,\ref{eq:tau_adiab_equ3_raw} can be written as
 \begin{eqnarray}
 \begin{array}{l l}
  \tau(z,\lambda)  = &   \frac{2 n(z) \sigma_0(\lambda) T_0^\delta p_0^\epsilon}{2F(\omega_2)+1} x_\infty^{2F(\omega_2)+1}(-\frac{1}{\omega_2^2})^{F(\omega_2)} \frac{\Gamma\left(F(\omega_2)+\frac{1}{2}\right) \Gamma\left(F(\omega_0)+1\right)}{\Gamma\left(F(\omega_0)+F(\omega_2)+\frac{3}{2}\right)}\\  
  & \times  _2F_1\left(F(\omega_2)+\frac{1}{2},-F(\omega_1);F(\omega_0)+F(\omega_2)+\frac{3}{2};\frac{\omega_0^2}{\omega_1^2}\right).
  \end{array}
  \label{eq:tau_adiab_equ3}
\end{eqnarray}

\section{Transmission Spectrum Equations}
\label{sec:tra_spec_equ}
The transmission spectrum (Equation\,\ref{eq:transmission_spectrum_h2}) can now be rewritten as 
\begin{eqnarray}	
     \frac{c}{2} = \int_{0}^{\omega'_0}  (1+y) \left(1-e^{-\tau(0,\lambda)\prod_{\{\forall \omega : f_T(\omega) = 0\}} \left(1 - \frac{y}{\omega'}\right)^{F(\omega)}}\right) \text{ d}y,
  	 \label{eq:tau_full_demo}
\end{eqnarray}
where $y = z/R_{p,0}$ and $ \omega'_i = \omega_i/R_{p,0} $.

We find no solution to the general integral introduced in Equation\,\ref{eq:tau_full_demo} using standard mathematical functions. However, we find that the above integral can be expressed using standard mathematical functions if $f_T(.)$ has less than two zeros. If $f_T(.)$ has no zero, the atmosphere is isothermal. The isothermal case is discussed in detail in Section\,\ref{app:ray}. If $f_T(.)$ has a zero, the temperature is a linear function of the altitude. We introduce the derivation for such a case below. Note that the case of a linear temperature profile encompasses the case of a dry-adiabatic atmosphere in hydrostatic equilibrium addressed in Section\,\ref{sec:adiabitic_lapse}. 

If $f_T(.)$ has a zero, the integral introduced in Equation\,\ref{eq:tau_full_demo} takes the form 
\begin{eqnarray}	
     \frac{c}{2} = \int_{0}^{\omega'_0}  (1+y) \left(1-e^{-A_{\lambda} (1 - \frac{y}{\omega'_0})^F}\right) \text{ d}y,  
     & \mbox{where } & 
	\left\{{
   \begin{array}{c c l}
     A_{\lambda} & = & \tau(0,\lambda)\\
	F & = & \delta-1+\frac{\epsilon-1}{H_0 f_T'(\omega_0)}
  \end{array}}
  	\right.
  	 \label{eq:tau_adiabatic_demo}
\end{eqnarray}
The integral introduced in Equation\,\ref{eq:tau_adiabatic_demo} can be rewritten using the following identity:
\begin{eqnarray}
	\begin{array}{l l}
     \int  (1+y) (1-e^{-C (1 - \frac{y}{D})^{E}}) \text{ d}y = & \frac{y}{2} (y+2) \\ 
     &- \frac{D}{E} \frac{D+1}{\sqrt[E]{C}} \Gamma\left(\frac{1}{E},C (1 - \frac{y}{D})^{E}\right) \\ 
     &+ \frac{D}{E} \frac{D}{\sqrt[E/2]{C}} \Gamma\left(\frac{2}{E},C (1 - \frac{y}{D})^{E}\right) + constant,
     \end{array}
  	 \label{eq:tau_adiabatic_identity}
\end{eqnarray}
because $C$, $D$, $E$, and $y \in \mathbb{R}^+$ in the context of transmission spectroscopy. Therefore Equation\,\ref{eq:tau_adiabatic_demo} can be rewritten as 
\begin{eqnarray}
	\begin{array}{l l}	
     c = & \omega'^2_0+2\omega'_0 \\ 
     &- 2\frac{\omega'^2_0+\omega'_0}{F} A_{\lambda}^{\frac{-1}{F}} \left(\Gamma\left(\frac{1}{F}\right)-\Gamma\left(\frac{1}{F},A_{\lambda}\right)\right) \\ 
     &+ 2\frac{\omega'^2_0}{F} A_{\lambda}^{\frac{-2}{F}} \left(\Gamma\left(\frac{2}{F}\right)-\Gamma\left(\frac{2}{F},A_{\lambda}\right)\right).
     \end{array}
  	 \label{eq:tau_adiabatic_short}
\end{eqnarray}
We recall that
\begin{eqnarray}
 \begin{array}{l l}
 \Gamma(x) - \Gamma(x,a)&=  \int_0^a t^{x-1} e^{-t} \text{ d}t\\
  & = \gamma(x,a),
  \end{array}
\end{eqnarray}
the lower incomplete gamma function. In addition, 
\begin{eqnarray}
 \begin{array}{l l}
 a^{-x} \Gamma(x,a) &=  \int_1^\infty \frac{e^{-at}}{t^{1-x}} \text{ d}t\\
  &= E_{1-x}(a),
  \end{array}
\end{eqnarray}
the generalized exponential integral. For the present derivation, we introduce $e_{1-x}(a)$, the ``lower'' generalized exponential integral, which we define as
 \begin{eqnarray}
 	e_{1-x}(a) = \int_0^1 \frac{e^{-at}}{t^{1-x}} \text{ d}t.
 \label{eq:lower_gener_exponetial_integral}
 \end{eqnarray}
Using the lower generalized exponential integral, we can rewrite Equation\,\ref{eq:tau_adiabatic_short} as
\begin{eqnarray}
     c & = & \omega'^2_0+2\omega'_0 - 2\frac{\omega'^2_0+\omega'_0}{F} e_{1-\frac{1}{F}}(A_{\lambda}) + 2\frac{\omega'^2_0}{F} e_{1-\frac{2}{F}}(A_{\lambda}),\\
     & \simeq & 2 \frac{\omega'_0}{F} \left(F - e_{1-\frac{1}{F}}(A_{\lambda})\right),
  	 \label{eq:tau_adiabatic_short_e}
\end{eqnarray}
because $ \omega'_0 \ll 1$ and $ A_{\lambda} \gtrsim 5$---i.e., optically-active spectral bands, see the definition of $R_{p,0}$.
 Therefore, we obtain from Equation\,\ref{eq:def_yeff}
 \begin{eqnarray}
 	\begin{array}{r c l}
	y_{eff}(\lambda) & \simeq & -1 + \sqrt{1+2 \frac{\omega'_0}{F} \left(F - e_{1-\frac{1}{F}}(A_{\lambda})\right)},\\
	& \simeq & \frac{\omega'_0}{F} \left(F - e_{1-\frac{1}{F}}(A_{\lambda})\right),
	\end{array}
	\label{eq:y_eff_adiab_ana}
\end{eqnarray}
because $\frac{\omega'_0}{F} \left(F - e_{1-\frac{1}{F}}(A_{\lambda})\right) \ll 1$.

As a result, we provide an analytical form for the effective planetary radius (Equation\,\ref{eq:transmission_spectrum_h}):

\begin{eqnarray}
 R_p(\lambda) & \simeq & R_{p,0}+\frac{\omega_0}{F} \left(F- e_{1-\frac{1}{F}}(A_{\lambda})\right) \text{,} \label{eq:R_eff_adiabat}\\
 & \simeq & \left[r:\tau(r,\lambda) = A_{\lambda} \left(\frac{e_{1-\frac{1}{F}}(A_{\lambda})}{F}\right)^F  \triangleq \tau(y_{eff})\right].\label{eq:R_eff_adiabat_tau}
\end{eqnarray}

\begin{figure}
 \centering
  \begin{center}
    \includegraphics[trim = 00mm 00mm 00mm 00mm,clip,width=12cm,height=!]{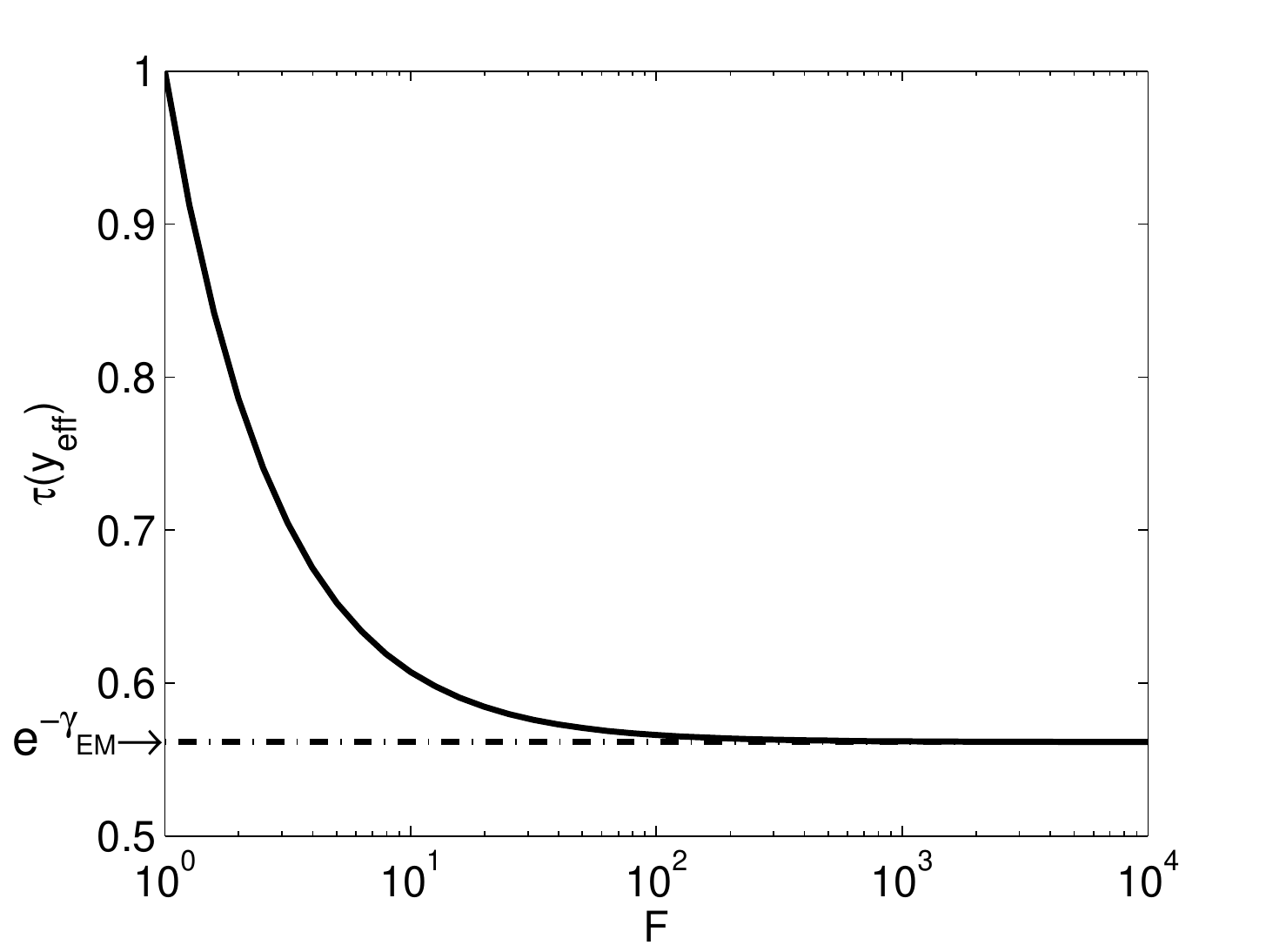}
  \end{center}
  \vspace{-0.7cm}
  \caption[Convergence of $\tau(y_{eff})$ to $e^{-\gamma_{EM}}$ for non-isothermal atmosphere.]{Convergence of $\tau(y_{eff})$ to $e^{-\gamma_{EM}}$ for non-isothermal atmosphere.}
  \vspace{-0.0cm}
  \label{fig:tau_eq_adiaba_to_ga}
\end{figure}

We note that as expected
 \begin{eqnarray}
	\lim_{F \rightarrow \infty} A_{\lambda} \left(\frac{e_{1-\frac{1}{F}}(A_{\lambda})}{F}\right)^F = e^{-\gamma_{EM}}.
\label{eq:tau_y_eff_adiab_ana_to_gam}
\end{eqnarray}
Indeed, as $F \rightarrow \infty$ the temperature profile tends to an isothermal. Hence, one retrieves the same formulation as the one found in Section\,\ref{app:ray} for $\tau_{eq}$ (see Equation\,\ref{eq:tau_y_eff_ana}). We emphasize this convergence in Figure\,\ref{fig:tau_eq_adiaba_to_ga}---note that for $ A_{\lambda} \gtrsim 5$, $\tau(y_{eff})$ is independent of $A_{\lambda}$.  

\section{The Case of Dry Adiabatic Lapse Rate}
\label{sec:adiabitic_lapse}

We focus in this Section on the formulation for atmospheric layers where the temperature profile is solely affected by an adiabatic lapse rate. An adiabatic lapse rate refers to the negative/positive change in temperature experienced by a parcel of air as it moves upwards/downwards by convection and expands/shrinks without exchanging heat with its surroundings. The balance between potential energy and kinetic energy can be written as
 \begin{eqnarray}
 c_p\text{ d}T = \frac{1}{\rho}\text{ d}p,
 \label{eq:adiabatic_balance}
\end{eqnarray}
where $c_p$ is the specific heat at constant pressure and $\rho$ is the gas density. In the context of hydrostatic equilibrium ($\text{d}p = -\rho g \text{ d}z$), Equation\,\ref{eq:adiabatic_balance} can be rewritten as
 \begin{eqnarray}
 \text{ d}_zT = -\frac{g}{c_p},
\end{eqnarray}
where $\frac{g}{c_p} \triangleq \Gamma_{LR}$ is the (dry) adiabatic lapse rate. 

Equations\,\ref{eq:general_T_profile},\,\ref{eq:general_p_profile}, and\,\ref{eq:general_n_profile} can thus be rewritten for the case of an adiabatic atmosphere in hydrostatic equilibrium. First, Equation\,\ref{eq:general_T_profile} is now
 \begin{eqnarray}
 f_T(z) = 1-\frac{\Gamma_{LR}}{T_0} z.
\end{eqnarray}
Therefore,
 \begin{eqnarray}
 \omega & = & \left\lbrace\frac{T_0}{\Gamma_{LR}}\right\rbrace \text{, and} \\
  f_T'(z) & = & -\frac{\Gamma_{LR}}{T_0}.
\end{eqnarray}
Hence, we obtain for the pressure and number-density profiles
 \begin{eqnarray}
 p(z) & = & p_0 \left(1 - \frac{\Gamma_{LR}}{T_0} z\right)^{-\frac{T_0}{H_0 \Gamma_{LR}}} \text{, and} \label{eq:barometric_adiabat}\\
 n(z) & = & n_0 \left(1 - \frac{\Gamma_{LR}}{T_0} z\right)^{-1-\frac{T_0}{H_0 \Gamma_{LR}}},
\end{eqnarray}
where Equation\,\ref{eq:barometric_adiabat} is the barometric formula for a non-zero lapse rate.

\begin{figure}[!h]
 \centering
  \begin{center}
    \vspace{-0.2cm}\includegraphics[trim = 00mm 00mm 00mm 05mm,clip,width=12cm,height=!]{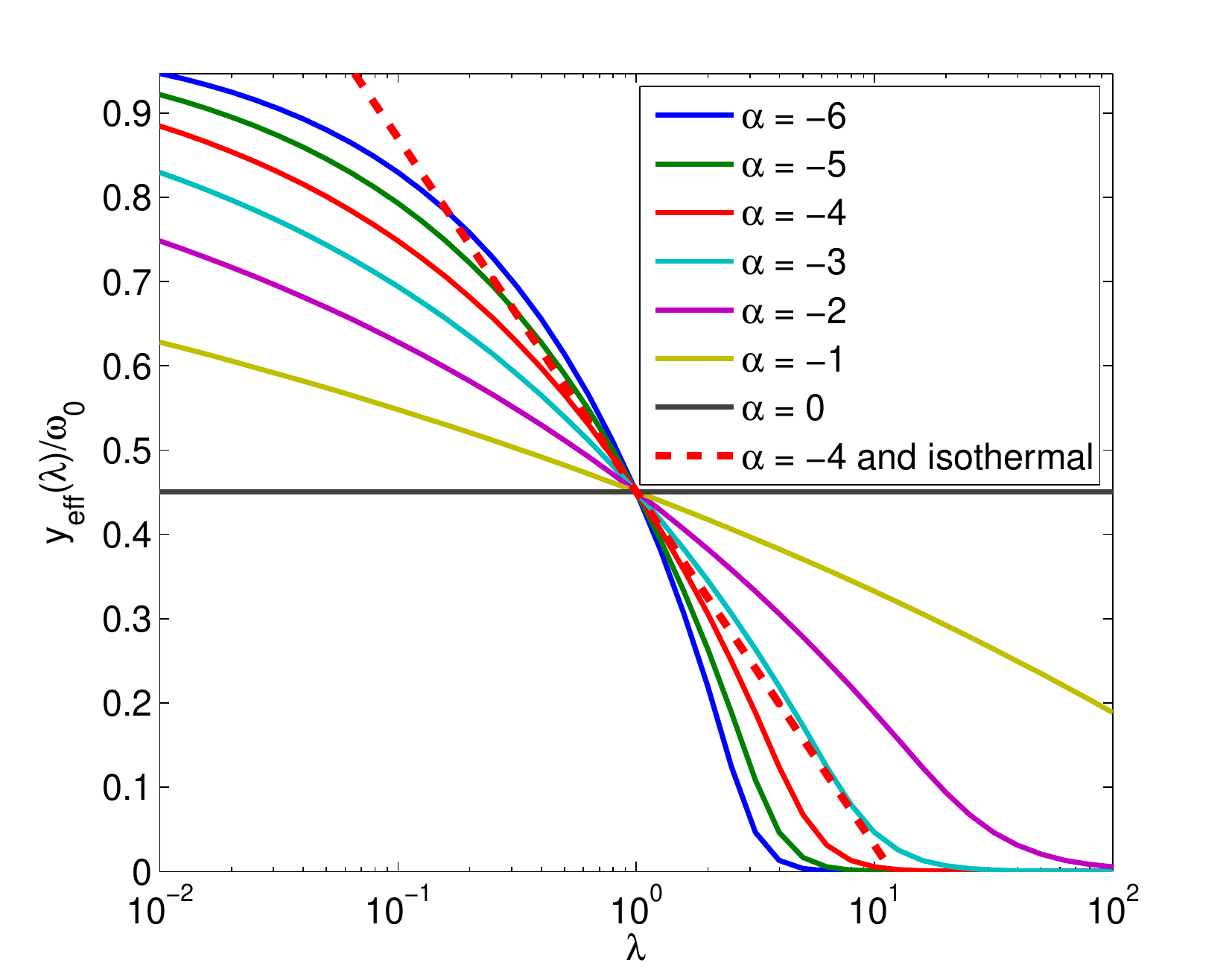}
  \end{center}
  \vspace{-0.7cm}
  \caption{Rayleigh ``slope'' for non-isothermal atmospheres.}
  \vspace{-2cm}
  \label{fig:rayleigh_slope_when_non_isothermal}
\end{figure}

Finally, the slant-path optical depth and the transmission spectrum equation can be written as
\begin{eqnarray}
\tau(z,\lambda) & = & \tau(0,\lambda) \left(1 - \frac{\Gamma_{LR}}{T_0} z\right)^{\frac{T_0-H_0\Gamma_{LR}}{H_0\Gamma_{LR}}}, \text{ and}\\
R_p(\lambda) & \simeq & R_{p,0}+\frac{T_0 H_0}{T_0-H_0\Gamma_{LR}} \left(\frac{T_0-H_0\Gamma_{LR}}{H_0\Gamma_{LR}} - e_{\frac{T_0}{T_0-H_0\Gamma_{LR}}}\left(\tau(0,\lambda)\right)\right), 
\end{eqnarray}
where
\begin{eqnarray}
\tau(0,\lambda) & \simeq & \sqrt{2 \pi (R_{p,0}+z)\frac{T_0}{\Gamma_{LR}}} \sigma_0(\lambda) n_0 \frac{\Gamma(\frac{T_0}{H_0\Gamma_{LR}})}{\Gamma(\frac{1}{2}+\frac{T_0}{H_0\Gamma_{LR}})},
\end{eqnarray}
assuming the extinction process to be Rayleigh scattering, i.e., $\left\lbrace \delta=0,\epsilon=0 \right\rbrace $ (Equations\,\ref{eq:tau_adiabat_general} and\,\ref{eq:tau_adiabatic_demo}). Using the scaling law function for the Rayleigh-scattering cross section, $\sigma_{sc}(\lambda) = \sigma_0(\lambda/\lambda_0)^\alpha$, we show in Figure\,\ref{fig:rayleigh_slope_when_non_isothermal} the effect of $\alpha$ and $\frac{T_0}{\Gamma_{LR}}$ on $R_p(\lambda)$. Figure\,\ref{fig:rayleigh_slope_when_non_isothermal} shows that $R_p(\lambda)$ does not present a Rayleigh-scattering linear slope---see Section\,\ref{sec:masspechd189733b} and Figure\,\ref{fig:weighing_WASP33b}---but rather a slope that gets steeper towards larger wavelength as these relate to larger atmospheric scale height---i.e., deeper and hotter atmospheric layers.

\clearpage
\newpage

%% file: AppendixC/appc.tex
\chapter{Simple Validation Tests for Transmission Spectrum Models}
\label{sec:testtransmissionspectrummodels}

We introduced here how the analytical derivations of Section\,\ref{app:dependency} and Appendix\,\ref{sec:generalizedtransmissionspectra} can be used to validate transmission spectrum models. In particular, the transmission spectrum of a synthetic atmosphere in hydrostatic equilibrium with a dominant scatterer has a simple analytical formulation. Hence, it is suitable case for transmission model validation. 

\textbf{(i)} If for such an atmosphere $T(z) = T_0$, the slant-path optical depth (Equation\,\ref{eq:optical_depth}) can be formulated as
\begin{eqnarray}
\tau(z,\lambda) & \simeq & \sigma(\lambda) n_{0} e^{-z/H} \sqrt{2\pi (R_{p,0}+z) H},
\end{eqnarray}
see Equation\,\ref{eq:tau_fortney}, and the effective planetary radius as 
\begin{eqnarray}
 R_p(\lambda) & \simeq & R_{p,0}+H\left(\gamma_{EM}+\ln{\left(n_{0} \sigma(\lambda) \sqrt{2\pi R_{p,0} H}\right)}\right),
\end{eqnarray}
see Equation\,\ref{eq:R_eff_tau_eq}, if $n_{0} \sigma_0(\lambda) \sqrt{2\pi R_{p,0} H}\gg1$---for validation tests, set $n_{0} \sigma(\lambda) \sqrt{2\pi R_{p,0} H}\geq10$, i.e., ensure the atmospheric layer at $R_{p,0}$ to be optically thick.
In other words, by setting up the profile of the atmospheric absorption coefficient to 
\begin{eqnarray}
 \alpha(r) & = & \alpha_{0} e^{-r/H},
\end{eqnarray}
where $\alpha_{0}$, $r$, and $H$ are respectively the absorption coefficient at the reference radius ($R_{p,0}$, e.g., the synthetic planet's surface), the radial distance, and the atmospheric scale height, one should be modeling a transmission spectrum of the form
\begin{eqnarray}
 \frac{\Delta F}{F} & \simeq & \left(\frac{R_{p,0}+H\left(\gamma_{EM}+\ln{ \left(\alpha_{0} \sqrt{2\pi R_{p,0} H}\right)}\right)}{R_\star}\right)^2.
\end{eqnarray}

\textbf{(ii)} If for such an atmosphere $T(z) \propto z$---i.e., $T(z) = T_0(1-\frac{z}{\omega_0})$, the slant-path optical depth (Equation\,\ref{eq:optical_depth}) can be formulated as
\begin{eqnarray}
\tau(z,\lambda) & \simeq & \sqrt{2 \pi (R_{p,0}+z)\omega_0} \sigma_0(\lambda) n_0 \frac{\Gamma(\frac{\omega_0}{H_0})}{\Gamma(\frac{1}{2}+\frac{\omega_0}{H_0})} \left(1 - \frac{z}{\omega_0}\right)^{-1+\frac{\omega_0}{H_0}},
\end{eqnarray}
see Equation\,\ref{eq:tau_adiab_equ1}, and the effective planetary radius as 
\begin{eqnarray}
R_p(\lambda) & \simeq & R_{p,0}+\frac{\omega_0 H_0}{\omega_0-H_0} \left(\frac{\omega_0-H_0}{H_0} - e_{\frac{\omega_0}{\omega_0-H_0}}\left(\tau(0,\lambda)\right)\right),\\
\tau(0,\lambda) & \simeq & \sqrt{2 \pi (R_{p,0}+z)\omega_0} \sigma_0(\lambda) n_0 \frac{\Gamma(\frac{\omega_0}{H_0})}{\Gamma(\frac{1}{2}+\frac{\omega_0}{H_0})},
\end{eqnarray}
see Equation\,\ref{eq:R_eff_adiabat}, if $\tau(0,\lambda)\gg1$---for validation tests, set $\tau(0,\lambda)\geq10$, i.e., ensure the atmospheric layer at $R_{p,0}$ to be optically thick.
In other words, by setting up the profile of the atmospheric absorption coefficient to 
\begin{eqnarray}
 \alpha(z) & = & \alpha_{0} \left(1 - \frac{z}{\omega_0}\right)^{-1+\frac{\omega_0}{H_0}},
\end{eqnarray}
where $\alpha_{0}$, $z$, and $H_0$ are respectively the absorption coefficient at the reference radius ($R_{p,0}$, e.g., the synthetic planet's surface), the altitude, and the atmospheric scale height at $R_{p,0}$, one should be modeling a transmission spectrum of the form
\begin{eqnarray}
 \frac{\Delta F}{F} & \simeq & \left(\frac{R_{p,0}+\frac{\omega_0 H_0}{\omega_0-H_0} \left(\frac{\omega_0-H_0}{H_0} - e_{\frac{\omega_0}{\omega_0-H_0}}\left(\sqrt{2 \pi R_{p,0}\omega_0} \alpha_{0} \frac{\Gamma(\frac{\omega_0}{H_0})}{\Gamma(\frac{1}{2}+\frac{\omega_0}{H_0})}\right)\right)}{R_\star}\right)^2.
\end{eqnarray}

\clearpage
\newpage

%% file: AppendixD/appd.tex
\chapter{Modeling Transmission Spectra by Solving $R_{p}(\lambda) = \left\lbrace r:\tau(r,\lambda) = e^{-\gamma_{EM}} \right\rbrace$}
\label{sec:quicktransmissionspectrummodels}

In Section\,\ref{app:dependency} and Appendix\,\ref{sec:generalizedtransmissionspectra}, we show the pivotal role of the Euler-Mascheroni constant \citep[$\gamma_{EM}$,][]{Euler1740} in transmission spectroscopy---see Equations\,\ref{eq:tau_eq_summ},\,\ref{eq:R_eff_adiabat_tau} and\,\ref{eq:tau_y_eff_adiab_ana_to_gam}. Here, we introduce a computationally efficient way to model transmission spectra. Solving  
 \begin{eqnarray}
 R_{p}(\lambda) & = & \left\lbrace r:\tau(r,\lambda) = e^{-\gamma_{EM}} \right\rbrace, \text{ or}\label{D0}\\
 h_{eff}(\lambda) & = & \left\lbrace z:\tau(z,\lambda) = e^{-\gamma_{EM}} \right\rbrace,\label{D1}
\end{eqnarray}
is computationally more efficient than a direct numerical integration of Equation\,\ref{eq:transmission_spectrum_h} as it solely requires solving Equation\,\ref{D0} or \,\ref{D1} knowing the dependency of each species' extinction cross-section on the pressure and the temperature parameters---e.g., $\Lambda_{\kappa}, a_{\kappa},$ and $b_{\kappa}$ (Equation\,\ref{line_alpha2}). These parameters are described by quantum physics or they can be measured in the lab \citep[e.g.,][]{Rothman2009}. This is why we advocate in this thesis that accurate extinction cross section databases covering various atmospheric conditions (i.e., temperature, pressure, composition) will be essential in the future to interpret consistently exoplanet transmission (and emission) spectra (Section\,\ref{app:cross_section_database}).

\subsubsection{Validation with Isothermal and Isocompositional Atmospheres}
\begin{figure}
 \centering
  \begin{center}
    \includegraphics[trim = 00mm 00mm 00mm 00mm,clip,width=12cm,height=!]{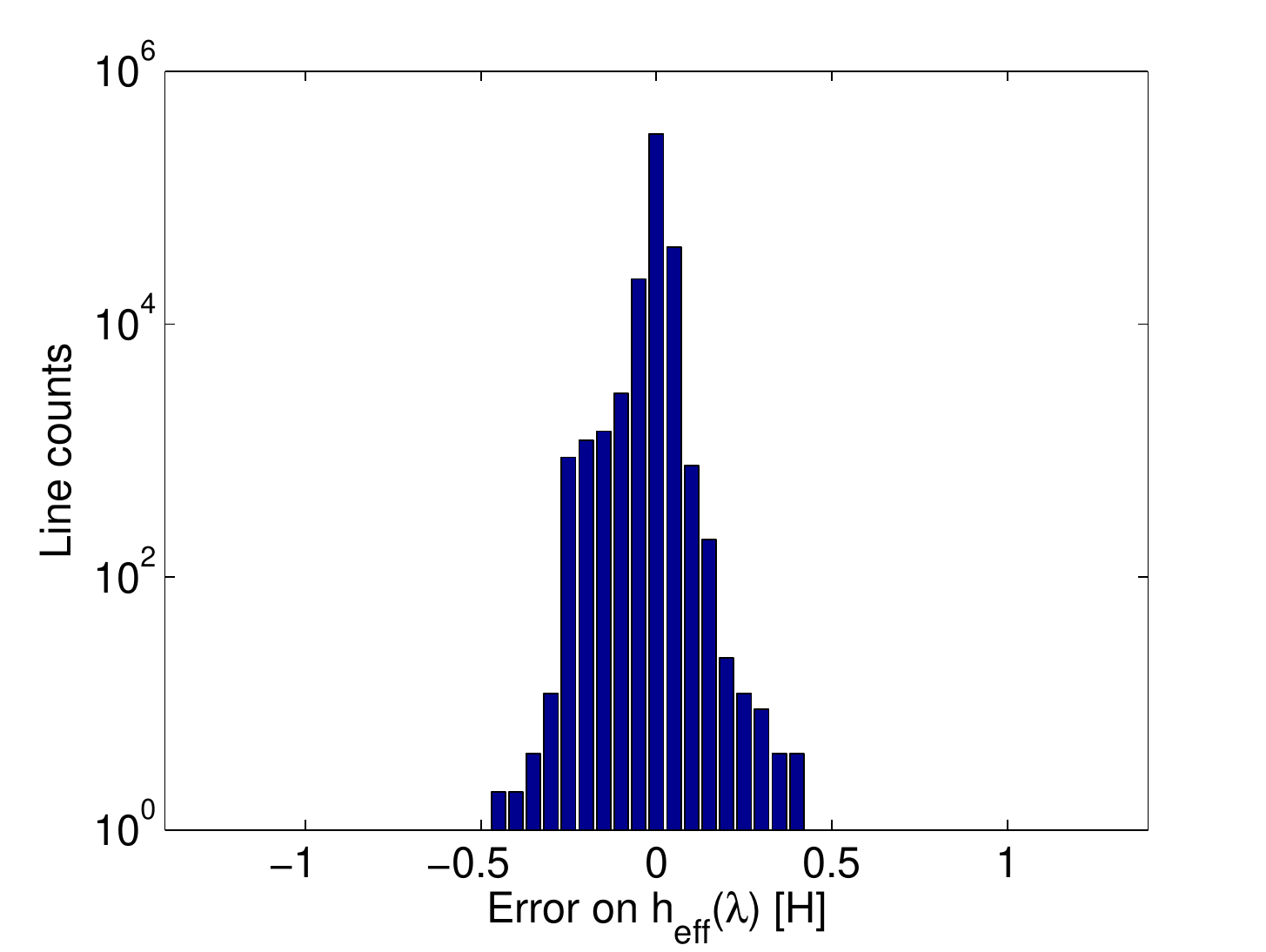}
  \end{center}
  \vspace{-0.7cm}
  \caption[Adequacy of $R_{p}(\lambda) =  \left\lbrace r:\tau(r,\lambda) = e^{-\gamma_{EM}} \right\rbrace$ to model high-resolution transmission spectra --- isothermal-isocomposition Earth.]{Distribution of the error on $h_{eff}(\lambda)$ (expressed in scale heights) when estimated from $h_{eff}(\lambda) = \left\lbrace z:\tau(z,\lambda) = e^{-\gamma_{EM}}\right\rbrace$ for an Earth-sized planet with a isothermal and isocompositional atmosphere with the same abundances as at Earth's surface. The error is below $3\%$ for 99.7$\%$ of the lines. An error below 3$\%$ corresponds to an error on the simulated effective height below 250 meters.}
  \vspace{-0.0cm}
  \label{fig:error_on_h_eff_iso_Earth}
\end{figure}

We show here that it is adequate to model a planet's transmission spectrum via $R_{p}(\lambda) = \left\lbrace r:\tau(r,\lambda) = e^{-\gamma_{EM}}\right\rbrace$ with an isothermal and isotermal atmosphere. In particular, we model here an Earth-sized planet with a isothermal and isocompositional atmosphere with the same abundances as at Earth's surface. Figure\,\ref{fig:error_on_h_eff_iso_Earth} shows the resulting error on $h_{eff}(\lambda)$ is below $3\%$ of the scale height for 99.7\% of the active spectral bins. An error on $h_{eff}(\lambda)$ below $3\%$ corresponds to an error of the simulated effective height below 250 meters---which is sufficient to advocate for the adequacy of modeling transmission spectra by solving $h_{eff}(\lambda) = \left\lbrace z:\tau(z,\lambda) = e^{-\gamma_{EM}}\right\rbrace$.

\subsubsection{Validation with Earth's Atmosphere}

\begin{figure}
 \centering
  \begin{center}
    \includegraphics[trim = 00mm 00mm 00mm 00mm,clip,width=12cm,height=!]{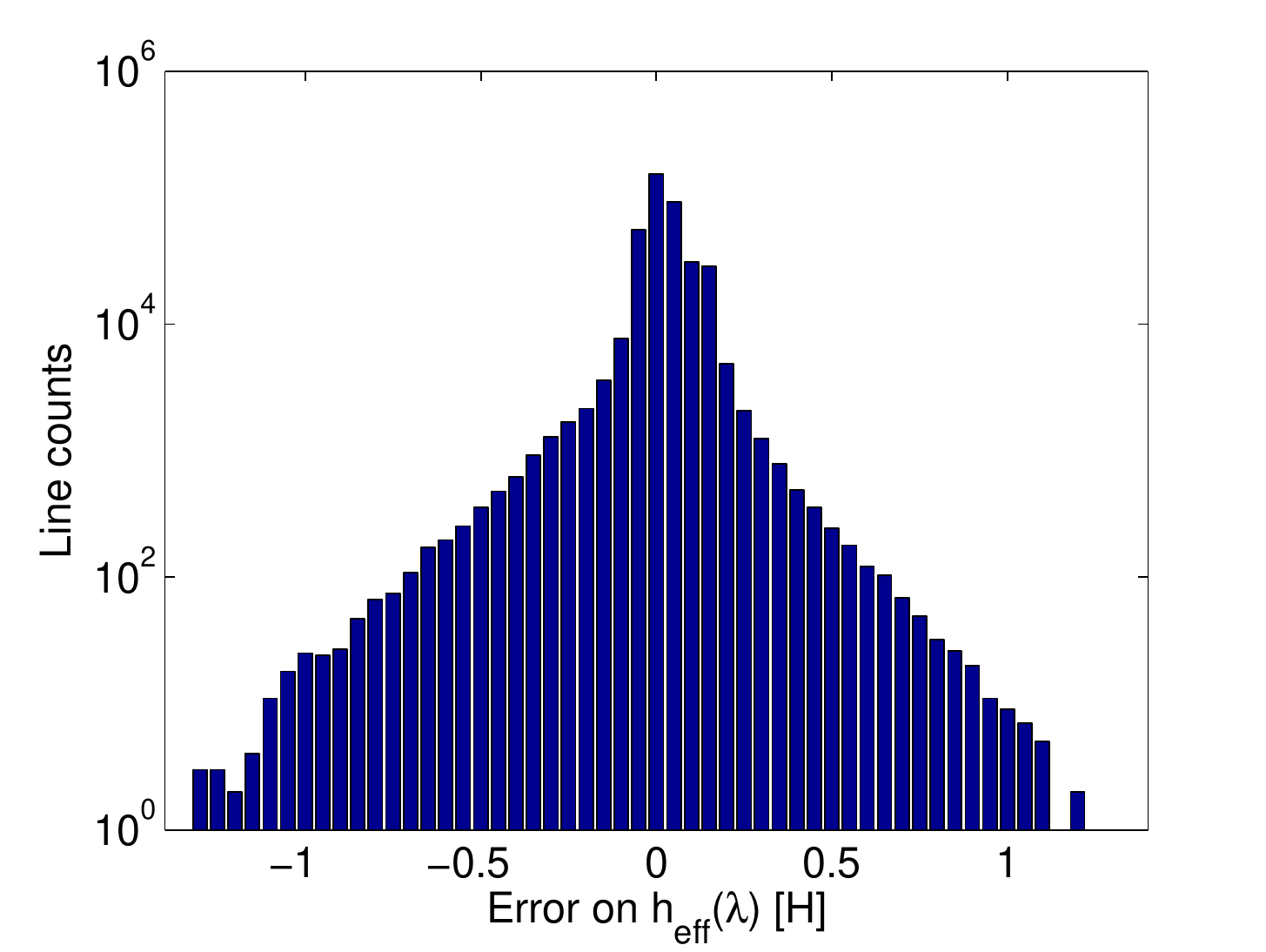}
  \end{center}
  \vspace{-0.7cm}
  \caption[Adequacy of $R_{p}(\lambda) = \left\lbrace r:\tau(r,\lambda) = e^{-\gamma_{EM}} \right\rbrace$ to model high-resolution transmission spectra---Earth.]{Distribution of the error on $h_{eff}(\lambda)$ (expressed in scale heights) when estimated from $h_{eff}(\lambda) = \left\lbrace z:\tau(z,\lambda) = e^{-\gamma_{EM}}\right\rbrace$ for Earth---temperature-pressure-mixing ratio profiles from \cite{Cox2000}. The error is below $18\%$ for 99.7$\%$ of the lines. An error below 18$\%$ corresponds to an error on the simulated effective height is below 1500 meters. The error is larger than for an isothermal-isocomposition Earth (Figure\,\ref{fig:error_on_h_eff_iso_Earth}) mainly because water's mixing ratio drops significantly in the troposphere (as highlight by the comparison of Figs\,\ref{fig:tau_eq_vs_a_lambda_Earth} and \ref{fig:tau_eq_vs_a_lambda_Earth_without_water}).}
  \vspace{-0.0cm}
  \label{fig:error_on_h_eff_Earth}
\end{figure}

We show in Figure\,\ref{fig:error_on_h_eff_Earth} the error on $h_{eff}(\lambda)$ when modeling Earth's transmission spectra using $h_{eff}(\lambda) = \left\lbrace z:\tau(z,\lambda) = e^{-\gamma_{EM}}\right\rbrace$. The error on $h_{eff}(\lambda)$ is below $18\%$ for 99.7\% of the active lines. An error on $h_{eff}(\lambda)$ below $18\%$ corresponds to an error of the simulated effective height below 1500 meters---which is sufficient to advocate that modeling transmission spectra based on solving $h_{eff}(\lambda) = \left\lbrace z:\tau(z,\lambda) = e^{-\gamma_{EM}}\right\rbrace$ is adequate. We note that while Earth's case deviates significantly from the assumptions underlying our derivations (atmospheric temperature and mixing ratios strongly dependent on the altitude), modeling its spectrum using $\tau_{eq} \simeq e^{-\gamma_{EM}}$ remains adequate. The main origin for the error on $h_{eff}(\lambda)$ comes from the strong water maxing ratio---as shown in Figures\,\ref{fig:tau_eq_vs_a_lambda_Earth} and \,\ref{fig:tau_eq_vs_a_lambda_Earth_without_water}. 

In summary, modeling transmission spectra via $R_{p}(\lambda) = \left\lbrace r:\tau(r,\lambda) = e^{-\gamma_{EM}}\right\rbrace $ is adequate if the vertical scale for temperature/composition changes are larger than the local scale height, as emphasized by  Equation\,\ref{eq:tau_y_eff_adiab_ana_to_gam} in Appendix\,\ref{sec:generalizedtransmissionspectra}.

\clearpage
\newpage